\documentclass[prd,aps,longbibliography,twocolumn,nofootinbib,superscriptaddress,showkeys,reprint]{revtex4-1}

\usepackage[linktocpage=true]{hyperref}
\hypersetup{colorlinks=true, linkcolor=blue, filecolor=magenta,  urlcolor=blue, citecolor=red,}

\usepackage{extramarks}
\usepackage{graphicx}
\usepackage{amsmath}
\usepackage[usenames,dvipsnames]{color}
\usepackage{listings}
\usepackage{courier}
\usepackage[normalem]{ulem}
\def\red{\textcolor{red}}
\def\bblue{\textcolor{Cerulean}}
\usepackage{float}
\usepackage{enumerate}
\usepackage{bigints}
\usepackage{comment}
\usepackage{macros}

\usepackage{amsfonts}
\usepackage{cancel}

\usepackage{diagbox}

\makeatletter
\renewcommand{\l@subsubsection}[2]{}
\makeatother

\begin{document}

\title{Modeling the gravitational wave spectrum from slowly-evolving sources in the early Universe: constant-in-time and coherent-decay models}

\author{Antonino S. Midiri}
\email{antonino.midiri@unige.ch}
\affiliation{D\'epartement de Physique Th\'eorique, Universit\'e de Gen\`eve,
CH-1211 Gen\`eve, Switzerland}

\author{Alberto Roper Pol}
\email{alberto.roperpol@unige.ch}
\affiliation{D\'epartement de Physique Th\'eorique, Universit\'e de Gen\`eve, CH-1211 Gen\`eve, Switzerland}

\author{Madeline Salom\'e}
\email{m.salome@ip2i.in2p3.fr}
\affiliation{D\'epartement de Physique Th\'eorique, Universit\'e de Gen\`eve,
CH-1211 Gen\`eve, Switzerland}
\affiliation{EPFL, CH-1015 Lausanne, Switzerland}
\affiliation{Universite Claude Bernard Lyon 1, CNRS/IN2P3,
IP2I Lyon, UMR 5822, Villeurbanne, F-69100, France}

\author{Chiara Caprini}
\email{chiara.caprini@cern.ch}
\affiliation{D\'epartement de Physique Th\'eorique, Universit\'e de Gen\`eve,
CH-1211 Gen\`eve, Switzerland}
\affiliation{Theoretical Physics Department, CERN, Gen\`eve, Switzerland}

\begin{abstract}
We study the production of gravitational wave (GW) backgrounds from
slowly-evolving
sources in the early Universe,
following the ``Constant-in-Time'' (CiT) model, presented
and validated against simulations
of decaying turbulence in \cite{RoperPol:2022iel}.
We extend the CiT model
to account for
the source decay using a coherent
description for the unequal-time correlator (UETC) of the source.
We present a generic framework that allows us
to study different types of sources
and provide analytical expressions
of the resulting GW background and its polarization.
We consider
different types of sources:
velocity and magnetic fields with a generic
fractional helicity,
velocity fields with different degrees of compressibility,
and scalar fields.
As examples, we discuss the applicability of this framework
to the production of GWs from decaying magnetohydrodynamic (MHD) turbulence,
sound waves, and bubble collisions in first-order phase transitions.
We show that the validity of the CiT and the coherent-decay
models presented in this work relies
on long decorrelation times of the source.
Therefore,
these models accurately describe the GW spectrum
of sound waves and MHD turbulence with subrelativistic
speed of sound and decorrelation velocities, respectively.
However, as the decorrelation timescale of the UETC becomes short,
a resonant term grows on top of the GW spectrum predicted
by the CiT model, as found, for example, in earlier analytical
studies of GW production from turbulence and sound waves.
This model is compatible with previous numerical results and explains
why sound waves in a radiation-dominated fluid are more efficient
than subrelativistic MHD turbulence in producing GWs,
which is no longer the case for relativistic turbulence.
The CiT model
predicts
a generic spectrum that scales as $k^3$ in the IR regime,
develops a linear $k$ at intermediate frequencies, and
decays as 
$k^{1-b_\Pi}$ at frequencies above the inverse correlation length scale of the source, where 
$(3-b_\Pi)$
is the small-scale slope of the anisotropic stress spectrum,
which is uniquely determined
by the type of source and the
UV slope of its spectrum.
For example, 
$b_\Pi = 11/3$
for Kolmogorov turbulence.
Since the $k^3$ and $k$ scalings always appear for
slowly-evolving sources and they
are commonly
found
in studies of GW production,
we motivate checking the
potential applicability of the CiT to other sources of GWs.
The resulting GW background can be accurately computed following the proposed
generic framework, which has been included 
in the open-source {\sc CosmoGW} code,
providing the GW community with a tool to study
different sources of GWs in the early Universe.
\end{abstract}

\maketitle

\tableofcontents

\section{Introduction}

Gravitational waves (GWs) produced by stochastic processes in the early Universe, such as first-order phase transitions \cite{Witten:1984rs,Hogan:1986qda}  
and primordial
magnetohydrodynamic (MHD) turbulence \cite{Kosowsky:1992rz,Kosowsky:2001xp}, 
are a unique probe of physics at high
energy scales, during the first microseconds of the evolution of the universe
(see refs.~\cite{Caprini:2018mtu,Caprini:2019egz} and references therein). 

Recent observations by Pulsar Timing Arrays (PTA) \cite{EPTA:2023fyk,EPTA:2023xxk,NANOGrav:2023gor,NANOGrav:2023hvm,Reardon:2023gzh,Xu:2023wog}
have reported strong evidence for a stochastic
GW background at nano-Hertz
frequencies that could be compatible with sourcing anisotropic stresses produced
around the Quantum Chromodynamics (QCD) scale \cite{Caprini:2024lxj}.
On the other hand,
the space-based detector LISA is planned to be launched
in the next decade \cite{LISA:2017pwj,LISA:2024hlh}, and it will be able to probe GW signals
produced at the electroweak scale
 \cite{Caprini:2015zlo,Caprini:2019egz,Caprini:2024hue}.
 Therefore, current
 and forthcoming GW observations make accurate theoretical predictions of the GW spectrum from early-Universe sources increasingly urgent.
 Otherwise, extracting
 clean information of the source and, hence, probing
early Universe physics from PTA observations and future data
 from LISA or other GW experiments, remains elusive.

In a first-order phase transition, GWs are directly produced
by three different contributions:
bubble collisions \cite{Kosowsky:1991ua,Kosowsky:1992rz,Kosowsky:1992vn,Caprini:2007xq,Huber:2008hg,Jinno:2017fby,Cutting:2018tjt,Cutting:2020nla}, sound waves  or acoustic turbulence
\cite{Hindmarsh:2013xza,Hindmarsh:2015qta,Hindmarsh:2016lnk,Hindmarsh:2017gnf,Hindmarsh:2019phv,Jinno:2020eqg,Jinno:2022mie,Caprini:2024gyk,RoperPol:2023dzg,RoperPol:2023bqa,Dahl:2021wyk,Dahl:2024eup,Correia:2025qif,Caprini:2026nnk}, and vortical (M)HD turbulence \cite{Kosowsky:1992rz,Kosowsky:2001xp,Gogoberidze:2007an,Caprini:2009fx,Caprini:2009yp,Niksa:2018ofa,RoperPol:2018sap,RoperPol:2019wvy,Kahniashvili:2020jgm,Brandenburg:2021tmp,Brandenburg:2021bvg,RoperPol:2021uol,RoperPol:2022iel,Auclair:2022jod}.
Due to the stochastic nature of these sources, the central object
characterizing
their emitted GW spectrum is the unequal-time correlator (UETC) of the anisotropic stresses of the source, $E_\Pi$.
The UETC
is in general unknown {\em a priori} for a particular source,
especially for highly non-linear systems, 
and its direct determination requires numerical simulations that are expensive and cover only a limited range of scales and durations \cite{Hindmarsh:2013xza,Hindmarsh:2015qta,Jinno:2022mie,Caprini:2024gyk,Correia:2025qif,Jinno:2017fby,Cutting:2018tjt,Cutting:2020nla,RoperPol:2019wvy,Kahniashvili:2020jgm,Brandenburg:2021tmp,Brandenburg:2021bvg,RoperPol:2021uol,RoperPol:2022iel,Auclair:2022jod}.
It is therefore essential to have analytical models for the UETC that capture the relevant physics and allow for fast, reliable,
and general predictions for the GW spectrum.

In particular, 
reference~\cite{RoperPol:2022iel}
introduced the ``Constant-in-Time'' (CiT) model for decaying
vortical turbulence, also validated in ref.~\cite{Auclair:2022jod},
and later
used in refs.~\cite{EPTA:2023xxk,RoperPol:2023bqa,Caprini:2024hue,Caprini:2024ofd}
for parameter estimation of turbulence production with PTA data and LISA sensitivity.
This model assumes that the dynamical evolution of the source is slower than the timescale
of GW production, which corresponds, at a wave number $k$, to the light crossing time,
$\delta t_{\rm GW} = 1/k$.
Therefore, when the timescales over which
the UETC evolves are larger than $\delta t_{\rm GW}$ for a mode $k$,
the UETC can be taken constant in time for an
effective
duration of the source. 
The resulting GW spectrum has a universal shape: a causal $k^3$ tail below the inverse source duration, a linear branch between the inverse duration and the peak, and a power law decay at large $k$
that depends on the nature of the source.
We argue that the typical spectral shape of the CiT model is observed for
a broad range of sources,
which motivates its validity for any
source of GWs
at characteristic wave numbers for which
the evolution of the UETC is slower
than the light crossing time,
$\delta t_{\rm GW} = 1/k$.

In this work, we provide a generic theoretical framework to describe
the GW spectrum and its circular degree
of polarization following the CiT model for different types of sources
related to fluid perturbations, magnetic fields, and scalar fields.
We also
address the following key question: for which sources,
and in which range of wave numbers, is the CiT model a valid description? 

For this purpose, we first consider sound waves and
stationary turbulence.
Since the UETC of sound waves is determined by the speed of sound, $\cs$,
assessing the validity of the CiT model is independent of the amplitude
of the fluid perturbations, which are assumed small for linear
sound waves.
We indeed find that, as we decrease the speed of sound, $\cs$,
the resulting GW spectrum reduces to the
one found in the CiT model.
However, for a radiation-dominated fluid with relativistic $\cs^2 = 1/3$,
even though the CiT model
can describe the GW spectrum
at small $k$ \cite{RoperPol:2023dzg,Sharma:2023mao},
the amplitude at the peak grows with the
source duration,
as found in previous work \cite{Hindmarsh:2019phv},
whenever the sourcing time, $\delta t_{\rm fin}$, is long $\cs \delta t_{\rm fin}
k_\ast \gg 1$ \cite{RoperPol:2023dzg}, where $k_\ast$ is the characteristic
wave number of the source.

On the other hand, the CiT model predicts an initial quadratic growth of the GW spectrum
with the source duration until each mode $k$ reaches saturation
by the time $\delta t_{\rm GW}$, as was found in
numerical simulations of vortical MHD turbulence
 \cite{RoperPol:2019wvy,RoperPol:2021uol,RoperPol:2022iel}.
This is in contradiction with earlier theoretical studies that
considered Kraichnan decorrelation \cite{1964SvA.....7..566I,1965PhFl....8.1385K}
to describe
stationary turbulence \cite{Kosowsky:2001xp,Gogoberidze:2007an}.
The authors
find that the GW spectrum should instead grow linearly
proportional to the source duration in flat Minkowski space-time, as also
argued in refs.~\cite{Hindmarsh:2013xza,Hindmarsh:2016lnk,Hindmarsh:2019phv}
for sound waves.
As shown in ref.~\cite{RoperPol:2023dzg}, the linear growth
with the source duration is a consequence of taking
the limit of long durations of the source (both for sound waves
and stationary turbulence), which is not valid at small wave numbers,
nor for small decorrelation velocities.
Therefore, we provide
an explanation for the discrepancy between the theoretical
and numerical results: for small
fluid or Alfv\'en velocities (hence, long decorrelation times),
the linear growth is not present in the GW spectrum, which can be
described by the CiT model, agreeing with the
simulations.
However, as the decorrelation velocity increases,
hence reducing the decorrelation timescale,
the component proportional to the source duration arises
for vortical MHD turbulence, as predicted in refs.~\cite{Kosowsky:2001xp,Gogoberidze:2007an}.

In summary, since the linear growth of the GW amplitude
is absent for subrelativistic
turbulence, this is precisely the reason why previous numerical
work found sound waves
or acoustic turbulence with relativistic equation of state, $\cs^2 = 1/3$,
to be more efficient than subrelativistic
vortical turbulence \cite{RoperPol:2019wvy,RoperPol:2021uol},
even if vortical sources are more efficient in producing anisotropic
stresses \cite{Brandenburg:2021bvg}.
However, we expect the linear growth
to also appear for larger decorrelation velocities,
$u_{\rm dcorr} \delta t_{\rm fin} k_\ast \gg 1$,
making turbulence as efficient
as sound waves in converting anisotropic stresses into GW energy density.

In order to account for the decay of the source,
we extend the CiT model by
introducing the coherent-decay model, which assumes that the UETC
is coherent in time.
We show that this assumption leads to modifications of the GW
spectral shape in the IR tail
(i.e., small $k$) of the spectrum when the decay time is large.
The deviations from the CiT model depend in general on the
time evolution
of the energy
density and
the characteristic length scale of the source.
When accounting for the decay of the source, we also show that if the decay timescale is longer than the GW light crossing time, we recover the CiT model.

The models presented in this work are completely
generic and can be applied to a broad range of GW sources in the early Universe.
After this manuscript is published,
these models and the computations presented in this work, will
be made available as part
of the {\sc CosmoGW}
Python package \cite{cosmogw_manual,CosmoGW_GH}.

\noindent
{\bf Structure of the paper.}
This paper is structured as follows.
In \Sec{sec:OmGW}, we review GW production in the early Universe,
explicitly showing the relation between the GW spectrum and the UETC of the source.
We comment on the computation of the GW spectrum from the results of
a numerical simulation using the Weinberg's formula in \Sec{weinberg}, generalizing
it to an expanding background.
Then, in \Sec{stress},
we provide a comprehensive description of statistically homogeneous and isotropic
velocity and magnetic fields, as well as scalar fields, and describe them
in terms of their spectral components (i.e., longitudinal, normal,
and helical components).
We introduce a smoothed broken power law
model
to describe a generic source of GWs in \Sec{sec:smoothed_dbpl},
and present analytical computation of its moments in \App{appendix_sbpl}.
The spectra of the anisotropic stresses produced by these sources are then described in \Sec{EPi_sec}, where broken power law fits for them are provided.
Their detailed computation is separately presented in \App{autocorr_EPi}.
The models for GW production from slowly decaying sources in the early Universe are presented starting from \Sec{cit_model_sec}, where we describe
in detail the CiT model for GW production.
We apply this model to various physical cases in the context of first-order phase transitions and primordial turbulence in \Sec{applications}.
In particular, we deal in \Sec{constant_model} with the applicability of the CiT model to decaying (M)HD turbulence, showing that, when the decay timescale is much longer than the GW saturation timescale, the effects of Kraichnan decorrelation become negligible as the decorrelation velocity becomes much smaller than the speed of light, recovering in this limit the CiT prediction. In \Sec{sound_wave_component} we show how, in a similar way, the GW spectrum from sound waves produced in the radiation-dominated era of the early Universe deviates from the CiT due to the sound speed being ultrarelativistic, recovering the CiT result in the limit when the sound speed is much smaller than the speed of light.
Moreover, in \Sec{other_sorc} we
discuss the potential applicability of the CiT
to bubble collisions in first-order phase transitions sourcing GWs via
the anisotropic stresses from the gradients of the scalar field.
In \Sec{beyond_cit} we then go beyond the CiT model, introducing
time decay of the source, in particular
time decay of
the amplitude and the peak position of the
source,
under the assumption that the UETC is coherent in time, which again is
applicable as long as the decorrelation time is short.
The introduction of this model
allows us to account for the decay of the source,
and it provides a physical description for the effective source duration
required in the CiT model.
Furthermore, it avoids the appearance of spurious effects in the GW spectrum
due to the artificial abrupt switching off  of the CiT model.
We finally present our conclusions in \Sec{conclusions}.

\noindent
{\bf Conventions.}
In this paper, we use the following Fourier convention for a generic field $u(\xx)$,
\begin{subequations}
\begin{align}
    u (\kk) &= \int u (\xx) \, e^{i \kk \cdot \xx} \dd^3 \xx\,,
    \\ u (\xx) &= \int \frac{u(\kk)}{(2 \pi)^3} \,
    e^{-i\kk \cdot \xx} \dd^3 \kk \,,
    \label{fourierconvention}
\end{align}
\end{subequations}
distinguishing fields in coordinate and Fourier space only by their arguments.
We use Greek letters to denote space-time components, e.g., $\mu = 0, 1, 2, 3$,
and Latin letters for space components, e.g., $i = 1, 2, 3$.

In general, we use the Friedmann-Lema\^itre-Robertson-Walker
as the background
metric tensor that describes the isotropic and homogeneous expanding
Universe, such that space-time coordinates $x^\mu = (\tau, x^i)$
refer to conformal time and comoving space coordinates,
and $\kk$ is the comoving wave vector.
The four-velocity of the
fluid in a  expanding
Universe is
\begin{equation}
    U^\mu = \gamma (1, u_i)/a\,, 
    \quad \text{with \ } \gamma^2 = \frac{1}{1 - u^2}\,,
\end{equation}
where $u_i \equiv \frac{\dd x^i}{\dd \tau}$
is the peculiar velocity and $\gamma$ is
the Lorentz factor.
Therefore, $u_i$ is defined as an Euclidean vector and, hence,
it should not be confused with the spatial components of the
four-velocity $U^i$.
We also introduce the Euclidean vector $X_i \equiv \sqrt{w} \,
\gamma \, u_i$, relevant to describe the GW production when the fluid
fluctuations are large \cite{Xvariable}.
In the limit of small velocities and enthalpy
perturbations, $X_i \simeq \sqrt{\bar w}
\, u_i$, where $\bar w$ is the mean fluid
enthalpy, such that the statistical description
of $u_i$ is directly inherited by $X_i$.

\section{Gravitational wave production in the early Universe}
\label{sec:OmGW}

In this section, we review the production of GWs
from stochastic sources in the early Universe that occurs during
the epoch of radiation domination, and show that the
GW spectrum at present time can be computed from the unequal-time
correlator (UETC) of the source of GWs, $E_\Pi (\tau_1, \tau_2, k)$,
evaluated at conformal times $\tau_1$ and $\tau_2$.
We show in \Sec{weinberg}
how the ``Weinberg'' formula \cite{Weinberg:1972kfs},
applied to GW computations from phase transitions
in Minkowski space-time
\cite{Kosowsky:1991ua,Kosowsky:1992vn,Kamionkowski:1993fg,Huber:2008hg,Konstandin:2017sat,
Jinno:2020eqg,Jinno:2022mie,Caprini:2024gyk}
(in particular, see App.~A of ref.~\cite{Caprini:2024gyk} for details),
can be applied to compute the
GW spectrum from the realizations of the source, modeled as a stochastic variable,
in an expanding Universe with different possibilities for the
background evolution.

\subsection{Tensor-mode perturbations}

GWs are defined as the tensor perturbations $\ell_{ij}$
over the Friedmann-Lema\^itre-Robertson-Walker background metric that characterizes the
isotropic and homogeneous expanding Universe with
the scale factor $a$,
\begin{equation}
    \label{GWconf}
\dd s^2 = a^2(\tau) \, \left[ - \dd \tau^2 + (\delta_{ij} + \ell_{ij}) \, \dd x^i \dd x^j \right] \,,
\end{equation}
where $\tau$ corresponds to conformal time.
The evolution of the scale factor $a$ with conformal time is
determined by the background equation of state $\omega$,
which relates the background pressure and energy density
$\bar p = \omega \bar \rho$.
When $\omega$ is constant in time,
\begin{equation}
    a(\tau) = a_\ast \biggl(\frac{\tau}{\tau_\ast}\biggr)^{1 + \nu}\,,
    \quad \nu = \frac{1 - 3\omega}{1 + 3 \omega}\,.
\end{equation}

The GW equation for the scaled strains
$h_{ij}=a \, \ell_{ij}$ in Fourier space\footnote{Note that we distinguish functions in real and Fourier space only by their arguments, e.g., $h_{ij}(\tau, \xx)$ and $h_{ij}(\tau, \kk)$.} is \cite{Grishchuk:1974ny,Caprini:2018mtu}
\begin{equation}
    \biggl(\partial_{\tau}^2 + \kk^2 - \frac{a''}{a}\biggr)\, h_{ij} (\tau, \kk) = 
    6\, \Pi_{ij}(\tau, \kk) \, a \, \HH^2\,, \label{GW_eqn_generic}
\end{equation}
where we use a prime to denote derivative with respect to conformal time,
$\kk$ is the comoving wave vector, $\HH = a'/a$ the conformal Hubble rate,
and $\Pi_{ij}$
is the normalized anisotropic stress tensor sourcing the GWs.
It can be expressed as
\begin{equation}
\label{def_pi_ij}
    \Pi_{ij} (\tau, \kk) = a^{2} \Lambda_{ijlm} 
    (\hat \kk) \frac{T^{lm} (\tau, 
    \hat \kk)}{\bar \rho (\tau)}\,,
\end{equation}
where $\bar \rho = 3 H^2 \Mpl^2$ is the
background energy density of the Universe,
$H = \HH / a$ the physical Hubble rate,
$\Mpl$ is the reduced
Planck mass,
and $\Lambda_{ijlm}$
is the traceless and transverse (TT) projection operator \cite{Misner:1973prb},
\begin{equation}
    \Lambda_{ijlm} (\hat \kk) = \half ( P_{il} P_{jm} + P_{im} P_{jl} - P_{ij} P_{lm})\,,
    \label{projection}
\end{equation}
with $P_{ij} = \delta_{ij} - \hat k_i \hat k_j$ and $\hat \kk = \kk / |\kk|$.
Since $h^{0\mu} = 0$ in the TT gauge, we have restricted the TT projector to Euclidean geometry
and, hence, we do not distinguish between lower and upper
indices in $\Pi_{ij}$.

During the radiation-dominated era, $\omega = 1/3$ and the scale factor evolves linearly
with conformal time, $a = a_\ast (\tau/\tau_\ast) = a_\ast \tau \, \HH_\ast$,
such that $a'' = 0$ and $\HH = 1/\tau$.
Setting $a_\ast = 1$,
the GW equation during radiation domination becomes
\begin{equation}
    \bigl(\partial_\tau^2 + \kk^2\bigr) h_{ij} (\tau, \kk) = 6 \, \Pi_{ij} (\tau, \kk)
    \frac{\HH_\ast}{\tau}\,. \label{GW_eqn_rad}
\end{equation}

Assuming that the source is active for a 
finite duration $\delta \tfin = \tfin - \tini$,
the solution to \Eq{GW_eqn_generic} with initial conditions $h_{ij}(\tini, \kk) = h'_{ij}(\tini, \kk) = 0$
can be expressed
as
\begin{align}
     h_{ij} (\tau <  \tfin, \kk) = 6  \int_{\tini}^\tau &\, \dd \tau_1 \, a (\tau_1)\, \HH^2 (\tau_1)
     \nonumber \\ &\, \!\!\! \times
     \Pi_{ij}(\tau_1, \kk)
    \, G (\tau, \tau_1, k)\,,
	\label{sol1_gen}
\end{align}
where $G (\tau, \tau_1, k)$ is the Green's function associated to the wave operator
in \Eq{GW_eqn_generic}.
Matching this solution to
the freely propagating regime at $\tau > \tfin$,
the solution to the GW equation at all times becomes
\begin{align}
    h_{ij} (\tau,  \kk) =
    6
    \int_{\tini}^{\bar \tau} &\,
    \dd \tau_1 \, a (\tau_1) \, \HH^2 (\tau_1) \nonumber \\  &\, \times
    \Pi_{ij} (\tau_1, \kk) \, G (\tau, \tau_1, k) \,, \label{GW_sol_rd}
\end{align}
where $\bar \tau = \min(\tfin, \tau)$. Its derivative with respect to conformal time is 
\begin{align}
    h'_{ij} (\tau,  \kk) =
    6  \int_{\tini}^{\bar \tau} &\, 
   \dd \tau_1  \, a (\tau_1) \, \HH^2 (\tau_1) \nonumber \\  &\, \times
   \Pi_{ij} (\tau_1, \kk) \, \partial_\tau G
   (\tau, \tau_1, k) \,, \label{GW_sol_der}
\end{align}
since $G (\tau, \tau, k) = 0$ 
(see, e.g., eq.~(2.11) of ref.~\cite{Giombi:2024kju}).

In \Secss{cit_model_sec}{beyond_cit}, we restrict our analysis to sources
produced during the radiation-dominated era.
However, we note that our description concerns the dynamical modeling of the source
and, hence, it can be directly applied to different cosmologies using \Eq{GW_sol_der}.
For radiation domination,
the Green's
function is $G = \frac{1}{k}
\sin [k (\tau - \tau_1)]$
and \Eq{GW_sol_der}
becomes
\begin{align}
    h'_{ij} (\tau, &\,  \kk) = \nonumber \\ &\, 
    6 \, \HH_\ast \int_{\tini}^{\bar \tau} 
    \frac{\dd \tau_1}{\tau_1} 
    \Pi_{ij} (\tau_1, \kk) \, \cos [k (\tau- \tau_1)] \,. \label{GW_sol_rd_der}
\end{align}

\subsection{Gravitational wave energy and polarization spectra}

The GW energy density, normalized to the critical energy density
at present cosmic time $t_0$,
$\bar \rho_0 = 3 H_0^2 \Mpl^2$,
is
\begin{align}
\OmGW (t_0)  = &
\frac{\Mpl^2}{4 \, \bar \rho_0}
\big\langle \partial_t \ell_{ij} ({\bf x},t_0) \,
\partial_t \ell_{ij} ({\bf x},t_0) \big\rangle\,,
\label{eq:OmGWdef}
\end{align}
where the brackets indicate an ensemble average over stochastic
realizations.
The GW energy density
can be expressed as \cite{Dufaux:2007pt,Caprini:2018mtu}
\begin{align}
    \OmGW  (t_0) &\, =
    \frac{1}{12 H_0^2}   \biggl(\frac{a_\ast}{a_0}\biggr)^{4} \, \big\langle [h'_{ij} - \HH \, h_{ij} ]^2 \big\rangle
    \nonumber \\ &\, \!\!\!\!  \simeq 
    \frac{1}{12 H_0^2}  \biggl(\frac{a_\ast}{a_0}\biggr)^{4} \bra{h'_{ij} (\xx, t_0)\, h'_{ij} (\xx, t_0)}
    \,, \label{OmGW_t0}
\end{align}
where $h'_{ij}$ corresponds to the solution to the GW
\Eq{GW_sol_rd_der} when one sets $a_\ast = 1$ 
\cite{RoperPol:2018sap},
and we have omitted the terms proportional to $\HH/k$ in the second
line since, at present time, all relevant wave numbers of causal
signals produced  in the early Universe are deep inside
the horizon, $k \tau_0 \gg 1$.

We express the spectral properties of $h'_{ij}$
through its two-point correlation function in Fourier space,
which
for a statistically homogeneous and isotropic field,
becomes
\cite{MY75,Caprini:2003vc,Kahniashvili:2005qi,RoperPol:2018sap,
Brandenburg:2018ptt,RoperPol:2021uol}
\begin{subequations}
\label{twopoint_0}
\begin{align}
    \!\!\!\! \bra{h'_{ij} (\kk) \, {h'}^\ast_{\!\!lm} (\kk')} =
     (2 \pi)^6  \, \delta^3 (\kk - \kk') \frac{{\cal F}_{ij, lm}^{h'} (\kk)}
     {4 \pi k^3}\,, \label{twopoint}
\end{align}
where ${\cal F}_{ij, lm}^{h'} (k)$ is
\begin{align}
    {\cal F}_{ij, lm}^{h'} &\, (\kk) = \frac{1}{2} \nonumber \\ 
     &\, \!\!\!\!\!\!\!\! \times
    \bigl[\Lambda_{ijlm} (\hat \kk) \, S_{h'} (k)
    + i \, {\cal A}_{ijlm} (\hat \kk) \, A_{h'} (k) \bigr]\,. \label{twopoint_b}
\end{align}
\end{subequations}
The spectral functions
$S_{h'} (k)$ and $A_{h'} (k)$ are,
respectively, the symmetric and antisymmetric (or chiral) spectra
per logarithmic unit of $k$ of the strain derivatives $h'_{ij}$.
The chiral spectrum $A(k)$ characterizes the asymmetry
of the stochastic field $h_{ij}'$
under mirror reflections, and vanishes
when its statistical properties satisfy parity symmetry.
The projection tensor $\bf{\Lambda}$ is defined in \Eq{projection},
and $\pmb{\cal A}$ is \cite{Caprini:2003vc}
\begin{align}
    {\cal A}_{ijlm} (\hat \kk) = &\,
    \fourth 
    \hat \kk_q \bigl(  P_{jm} \varepsilon_{ilq} +  P_{il}
    \varepsilon_{jmq} \nonumber \\ &\, \hspace{15mm}  + P_{im} \varepsilon_{jlq}  + P_{jl} \varepsilon_{imq} \bigr)\,. \label{calA_ijlm}
\end{align}
When applied to a generic tensor, both $\pmb{\Lambda}$ and $\pmb{\cal A}$
project onto the corresponding symmetric and antisymmetric TT components
of the tensor.

Multiplying \Eqq{twopoint_0} by $\delta_{il} \delta_{jm}$
to obtain the trace $\bra{h_{ij}' h_{ij}'}$,
which describes the GW energy density [see \Eq{OmGW_t0}],
we find
\begin{align}
    \OmGW (\tau, k) \equiv &\,
    \frac{\dd \OmGW (\tau)}
    {\dd \ln k}
    \nonumber \\ \simeq &\,
    \frac{1}{12 H_0^2} \biggl(\frac{a_\ast}{a_0}\biggr)^{4} 
    \, S_{h'} (\tau, k)\,. \label{OmGW_hp}
\end{align}

The degree of circular polarization of the GW
background is determined
by the ratio of the chiral to the symmetric GW spectra,
\begin{equation}
\label{degree_circ_pol}
    {\cal P}_{\rm GW} (\tau, k) = \frac{\XiGW (\tau, k)}{\OmGW (\tau, k)}\,,
\end{equation}
where
$\XiGW (\tau, k)$ is found contracting \Eq{twopoint_0} with ${\cal A}_{ijlm} (\hat \kk)$
\cite{Kahniashvili:2005qi,Kisslinger:2015hua,Kahniashvili:2020jgm,RoperPol:2021uol},
\begin{equation}
    \XiGW (\tau, k) \simeq \frac{1}{12 H_0^2} \biggl(\frac{a_\ast}{a_0}\biggr)^{4} 
    \, A_{h'} (\tau, k)\,.
    \label{AGW_hp}
\end{equation}
We have
again assumed $k \tau \gg 1$, such that contributions
proportional to $1/(k \tau)$ are omitted as they vanish
at present time \cite{RoperPol:2018sap}.

The UETC of the stochastic
anisotropic stresses $\Pi_{ij}$ sourcing GWs
can also be decomposed into a symmetric, $E_\Pi (k)$,
and an antisymmetric, $H_\Pi (k)$, spectral function
with a decomposition equivalent to that of
\Eqq{twopoint_0},
\begin{subequations}
\label{twopoint_EPis}
\begin{align}
    \langle \Pi_{ij} (\tau_1, \kk) & \, \Pi_{lm}^{\ast} (\tau_2, \kk') \rangle = \nonumber \\ &\,
    (2 \pi)^6 \delta^3 (\kk - \kk') \frac{{\cal F}_{ij, lm}^{\Pi}
    (\tau_1, \tau_2, \kk)}{4 \pi k^3}
    \,, \label{twopoint_EPi}
\end{align}
with
\begin{align}
    {\cal F}_{ij, lm}^{\Pi} (\tau_1, \tau_2, \kk)  &\,  = \frac{1}{2}\bigl[
    \Lambda_{ijlm} (\hat \kk) \, E_\Pi (\tau_1, \tau_2, k)
    \nonumber \\ &\, \ + i \, {\cal A}_{ijlm} (\hat \kk) \, H_\Pi (\tau_1, \tau_2, k) \bigr]\,. \label{twopoint_EPi_b}
\end{align}
\end{subequations}
In this case, $\Pi_{ij}$ is evaluated at different times $\tau_1$
and $\tau_2$, and different $\kk$ and $\kk'$.

Inserting the solution for GWs produced during radiation domination,
given in \Eq{GW_sol_rd_der}
and the UETC $E_\Pi (\tau_1, \tau_2, k)$ of \Eqq{twopoint_EPis}
into \Eqq{twopoint_0},
we find $S_{h'} (\tau, k)$,
which yields the following expression for the GW spectrum
\cite{Caprini:2018mtu}
\begin{align}
    &\, \OmGW  (\tau, k) = 3 \, F_{\rm GW}^0  \int_{\tini}^{\bar \tau}
    \frac{\dd \tau_1}{\tau_1} \int_{\tini}^{\bar \tau}
     \frac{\dd \tau_2}{\tau_2} \nonumber \\
    &\, \ \, \times  E_\Pi (\tau_1, \tau_2, k)
    \cos [k (\tau - \tau_1)] \cos[k(\tau - \tau_2)]\,. \label{eq:OmGWt}
\end{align}
$\XiGW (\tau, k)$ can be computed using
\Eq{eq:OmGWt} if we substitute $E_\Pi \to H_\Pi$.
The transfer function $F_{\rm GW}^0$ represents the
redshift of the GW spectrum,
\begin{align}
    F_{\rm GW}^0  = \biggl({a_\ast \over a_0}\biggr)^4 &\, \biggl({H_\ast \over H_0}\biggr)^2      \nonumber \\  
    \simeq &\,  \frac{1.6 \times 10^{-5}}{h^2} \biggl(\frac{100}{g_\ast}\biggr)^{\tfrac{1}{3}}  \,,
\end{align}
where $g_\ast$ is the number of relativistic degrees of freedom at the
time of generation $\tau_\ast$ and $h$ characterizes the Hubble
rate at present time, $H_0 = 100 \, h \, \km / (\s \Mpc)$.

\subsection{Present-day gravitational wave spectrum and Weinberg's formula}
\label{weinberg}

In this section, we provide an expression
analogous to the Weinberg's formula for the power emitted
by isolated binaries,
following the description of
refs.~\cite{Jinno:2020eqg,Jinno:2022mie,Caprini:2024gyk},
where this expression was used in flat Minkowski space-time.
We extend their results to include the effects of the expansion of the
Universe during the process of GW production.
These expressions explicitly show the impact of computing the
GW spectrum from numerical simulations and how
these can be interpreted to make predictions of the
observable GW spectrum.

In first place, we are interested in evaluating the
GW spectrum at present time.
For subhorizon
modes $k \tau_0 \gg 1$, we can average the product of Green's functions in \Eq{eq:OmGWt} over fast frequency
oscillations to find
\begin{align}
    \OmGW^0 (k) =\frac{3}{2}F_{\rm GW}^0  &\,  \int_{\tini}^{\tfin} \frac{\dd \tau_1}{\tau_1} \int_{\tini}^{\tfin} \frac{\dd \tau_2}{\tau_2} \nonumber \\ &\, \times
    E_\Pi (\tau_1, \tau_2, k) \cos (k \tau_-) \,, \label{OmGW_aver}
\end{align}
where $\tau_- = \tau_2 - \tau_1$.
Since $\Pi_{ij}$ is a
statistically homogeneous and isotropic 
stochastic field, we can replace
its ensemble average, using the ergodic assumption,
by an average
over the solid angle at fixed $k$ in Fourier space, thus obtaining
(see App.~A of ref.~\cite{Caprini:2024gyk} for details)
\begin{align}
E_\Pi (\tau_1,  \tau_2, k) &\, 
 \nonumber \\ =  \frac{k^3}{2 \pi^2 V}  &\, \int_{\Omega_k} \frac{\dd \Omega_k}{4\pi} \, \Pi_{ij} (\tau_1, \kk) \, \Pi_{ij}^* (\tau_2, \kk) \,.
\label{EPi_Pi}
\end{align}
Then, substituting \Eq{EPi_Pi} into \Eq{OmGW_aver},
we find 
\begin{align}
    \OmGW^0 (k) = \frac{3 k^3}{4 \pi^2 V}  &\,   F_{\rm GW}^0   \, \nonumber \\ \times \int_{\Omega_k}
    \frac{\dd \Omega_k}{4\pi}  &\,
    \int_{\tini}^{\tfin}
    \frac{\dd \tau_1}{\tau_1} \, \Pi_{ij} (\tau_1, \kk)\, e^{ik\tau_1}   \nonumber \\  \times
    &\, \int_{\tini}^{\tfin}
    \frac{\dd \tau_2}{\tau_2} \, \Pi_{ij}^\ast (\tau_2, \kk)\, e^{-ik\tau_2} \,.
\end{align}
Defining the following integral over the stress-energy
tensor $T_{ij} (\tau, \kk)$,
\begin{equation}
T_{ij} (q, \kk) =  \int_{\tini}^{\tfin} \frac{T_{ij} (\tau, \kk)}{\tau}
\, e^{iq\tau} \dd \tau \,, \label{Tij_qk}
\end{equation}
we find
\begin{align}
    \OmGW^0 (k) =  \frac{3 k^3}{4 \pi^2 V \bar \rho^2} &\,  \, F_{\rm GW}^0
    \int_{\Omega_k}  \frac{\dd \Omega_k}{4\pi} \, \Lambda_{ijlm} (\hat \kk)
    \nonumber \\   \times &\,  \bigl[T_{ij} (q, \kk)\, T_{lm}^\ast (q, \kk)\bigr]_{q=k} \,.
\label{OmGW}
\end{align}
The expansion of the Universe is directly
incorporated in our expression
by including $1/\tau$ in
the proposed definition of $T_{ij} (q, \kk)$ in \Eq{Tij_qk},
compared to the one used in previous
work \cite{Kosowsky:1991ua,Kosowsky:1992vn,Kamionkowski:1993fg,Huber:2008hg,Konstandin:2017sat,
Jinno:2020eqg,Jinno:2022mie,Caprini:2024gyk},
which only applied to
flat Minkowski space-time.

We note that this expression can be extended
to an expanding background described by a generic
constant
equation of state $\omega$, using the solution to \Eq{GW_eqn_generic}
to define $T_{ij} (q, \kk)$ as
\begin{align}
    T_{ij} ( q, \kk) = \HH_\ast
    \int_{\tini}^{\tfin}
    \biggl(\frac{\tau}{\tau_\ast}\biggr)^{\nu - 1} \,
    \!\!\! T_{ij} (\tau, \kk)
    \, e^{iq\tau} \dd \tau \,.
    \label{Tij_qk_gen}
\end{align}
This expression
applies only in the late-time limit
$k \tau \gg 1$, when $a''/a$ can be neglected
and the Green's function of 
\Eq{GW_eqn_generic} reduces
to $G  = \frac{1}{k}\sin [k (\tau - \tau_1)]$
(see App.~A of ref.~\cite{Giombi:2024kju}).

\section{Statistical description of vector and scalar fields}
\label{stress}

The stress tensor producing GWs can include, among others,
contributions from a perfect fluid,
from electromagnetic fields, and from scalar fields \cite{Figueroa:2026esg}
\begin{align}
    T_{ij} \supset &\, w  \,
    U_i U_j - E_i E_j - B_i B_j + \partial_i \phi \, \partial_j \phi\,,
    \label{Tij}
\end{align}
where $U_\mu = \gamma \, a \, (-1, u_i)$,
$u_i \equiv \dd x^i/\dd \tau$
is the peculiar velocity, $\gamma^2 = 1/(1 - u^2)$ the squared Lorentz factor,
$w$ is the fluid enthalpy,
and $E_i$, $B_i$, and $\phi$ respectively indicate the electric, magnetic,
and scalar fields.

Furthermore, we have excluded diagonal terms in the stress-energy tensor, as they
vanish after applying the TT projection and, hence, they
do not produce GWs.
The contributions from the velocity field can be expressed as $X_i X_j$,
where $X_i \equiv \sqrt{w} \, \gamma \, u_i$  \cite{Caprini:2026nnk,Xvariable}.
When the enthalpy and velocity
perturbations are small, one can express the UETC of
$X_i \simeq \sqrt{\bar w} \, u_i$, where $\bar w$ is the mean fluid
enthalpy, as being proportional to the UETC of the velocity field.

We note that the fluid enthalpy, $w$, in the case of a radiation-like fluid, and the electromagnetic energy
densities, $\rho_E = \half E_i E^i$ and $\rho_B = \half B_i B^i$,
are all proportional to $a^{-4}$
if the fields are not dynamically coupled
and just evolve following the Universe expansion.
Therefore, since we have defined $\Pi_{ij} = a^2
\Lambda_{ijlm} T^{lm} / \bar \rho =
\Lambda_{ijlm} T^{l}_{\ m} / \bar \rho$ in \Eq{def_pi_ij},
and we focus our analysis on the radiation-dominated era, such that the average fluid energy density follows
$\bar \rho \sim a^{-4}$, the resulting components of $\Pi_{ij}$ are
comoving with the Universe, i.e., they remain
constant in time as the Universe expands, unless they dynamically evolve
due to, e.g., decaying MHD turbulence \cite{Brandenburg:1996fc,Christensson:2000sp,
Brandenburg:2017neh,RoperPol:2025lgc}.
On the other hand, the gradient energy density of the scalar field,
$\rho_G^\phi = \half \partial^i \phi \, 
\partial_i \phi$, does not,
in general, evolve proportional to $a^{-4}$.
Therefore, the dynamical evolution of
$\rho_G^\phi / \bar \rho$
depends on the equation of state of the scalar field.

For each source, the stresses $T_{ij}$ are expressed
as a bilinear form in the vector field
$\varphi_i = \{a X_i, i E_i, i B_i, 
\partial_i \phi\}$, which we describe in next \Sec{EPi_sec}.
In the following, we first provide the statistical description
of stochastic velocity $\uu$ (or $\XX$) and
magnetic $\BB$ fields in \Sec{sec:vel}, including
both longitudinal and normal spectral components, and a generic fractional
helicity, allowing us to describe
a broad range of hydromagnetic sources.
We omit the contribution from electric fields, as we expect the conductivity
to be very large during the radiation-dominated era
\cite{Ahonen:1996nq,Brandenburg:1996fc,Arnold:2000dr,Caprini:2006jb,Durrer:2013pga,RoperPol:2025lgc},
such that $\EE \simeq - \uu \times \BB$ is suppressed
for subrelativistic fluid
velocities.
However, their inclusion in \Eq{Tij} and
the extension of the results of this work to describe electrical fields
is straightforward.
Then, we provide the statistical description of the gradients of
scalar fields sourcing GWs in \Sec{scalar_Ephi}.
We introduce in \Sec{sec:smoothed_dbpl} a
generic smoothed broken power law model for the
spectral components.
Analytical calculations of integrals involving
this generic model are provided in \App{appendix_sbpl}.

\subsection{Velocity and magnetic fields}
\label{sec:vel}

The two-point correlation function of a statistically homogeneous and
isotropic velocity field is  expressed in Fourier space as
\begin{subequations}
\label{two_point_Ev}
\begin{align}
    \bra{u_i (\kk) \, u_j^\ast (\kk')} = (2 \pi)^6   \,\delta^3
   (\kk   &\,  - \kk') \frac{{\cal F}_{ij}^{u} (\kk)}{4 \pi k^3}\,, 
\label{twoppcorr_velocity}
\end{align}
where ${\cal F}_{ij}^u (\kk)$
presents the following unique decomposition \cite{MY75}
\begin{align}
    \label{spectral_funcs}
    {\cal F}_{ij}^u (\kk) =  \half P_{ij} (\hat \kk) E^u_\N (k)&\,
    + \hat k_i \hat k_j E^u_\L (k) \nonumber \\ &\,
    + \tfrac{i}{2} \varepsilon_{ijl}
    \hat k_l H_u(k)\,.
\end{align}
\end{subequations}

The spectral functions $E^u_{\rm N} (k)$ and 
$E^u_{\rm L} (k)$ correspond respectively
to the normal and longitudinal components of
the velocity field $\uu$.
In the case of magnetic fields, since they are
solenoidal ($\nab \cdot \BB = 0$),
the longitudinal component vanishes, $E_\L^B(k) = 0$,
\begin{equation}
    {\cal F}_{ij}^{B} (\kk) = \half
    P_{ij} (\hat \kk) \, E_B (k) + \tfrac{i}{2} 
    \varepsilon_{ijl} \hat k_l H_B (k)\,. \label{two_point_B}
\end{equation}

As in \Eq{OmGW_hp}, the spectral
function ${\cal F}_{ij} (\kk)$ is defined
such that its trace is the velocity or the magnetic
spectrum:
\begin{equation}
\label{norm_plus_long}
    E_u (k) \equiv
    {\cal F}_{ii}^{u} (\kk) = E^u_\N (k) + E^u_\L (k)\,,
\end{equation}
and $E_B (k) = {\cal F}_{ii}^B (\kk)$.
Integrated over $\ln k$, these spectra
correspond to the
ensemble average of the squared field, i.e.,
$u_{\rm rms}^2$ and $B_{\rm rms}^2$.
Under the ergodic assumption, this is
approximated as a volume average \cite{MY75}
\begin{subequations}
\label{squared_values}
\begin{align}
    u_{\rm rms}^2 =
    \bra{u^2 (\xx)}_V = &\, \int_{-\infty}^\infty E_{u} (k) \dd \ln k\,, \label{velocity_squared} \\ 
    B_{\rm rms}^2
    = \bra{B^2 (\xx)}_V
    = &\, 
    \int_{-\infty}^\infty E_B (k) \dd \ln k\,.
    \label{magnetic_squared}
\end{align}
\end{subequations}

We also characterize the degree of compressibility in the velocity
field as the ratio of the
longitudinal component to the total spectrum,
\begin{equation}
    q (k) \equiv
    \frac{E^u_{\rm L} (k)}{E^u_\L(k) + E^u_\N (k)} \in [0, 1] \,.
    \label{fraction_q}
\end{equation}
Then, the longitudinal and normal
components can be expressed as
fractions of the velocity spectrum,
\begin{subequations}
\label{components_vel_E}
\begin{align}
    E^u_{\rm L} (k) = &\, q (k) \, E_u (k)\,, \\
    E^u_{\rm N} (k) =  &\, [1 - q(k)] \, E_u (k)\,.
\end{align}
\end{subequations}
Note that for magnetic fields, we always have $q (k) = 0$.

Causality implies that the spectrum of a generic
vector field becomes
proportional to $k^{a_E}$ with odd $a_E \geq 3$ in the $k \to 0$
limit \cite{MY75},
\begin{equation}
    E(k) = E_\N (k) + E_\L (k) \sim k^{a_E}
    \quad \text{with} \ a_E \geq 3\,.
\end{equation}
However,
analyticity of \Eq{spectral_funcs} implies that the term
proportional to $\hat k_i \hat k_j = k_i k_j / k^2$
should be proportional to
$k^{a_E}$ with odd $a_E \geq 5$ when $k \to 0$ \cite{Durrer:2003ja},
\begin{equation} \label{k5}
    |E_\L (k) - \half E_\N (k)|
    \sim k^{a_E} \quad \text{with} \ a_E \geq 5\,.
\end{equation}
Therefore, the spectrum of the source, ${\cal F}_{ii}(k)$ becomes proportional to $k^5$ or steeper
when the field satisfies one of the following two conditions: {\em i)} the field is
irrotational ($\oom = \nab \times \uu = \pmb{0}$),
such that $E_\N (k) = 0$,
or {\em ii)} the field
is solenoidal ($\nab \cdot \uu = 0$ or $\nab \cdot \BB = 0$), such that $E_\L (k) = 0$ \cite{Durrer:2003ja,MY75}
(see also App.~C of ref.~\cite{Caprini:2026nnk} for details).

Helical spectra $H(k)$
are associated to mirror symmetry breaking
in the statistical distribution
of the vector field and, hence,
$H (k) = 0$ when the statistics of a vector
field satisfy parity symmetry.
They characterize the kinetic
helicity $\HH_u \equiv \bra{\oom \cdot \uu}$ or the magnetic
helicity $\HH_A \equiv \bra{\AAA \cdot \BB}$ 
\cite{1978mfge.book.....M},
\begin{subequations}
    \begin{align}
        {\cal H}_u \equiv &\, \bra{\oom (\xx) \cdot \uu (\xx)}_V = \int_{-\infty}^\infty
        k\, H_u (k)\dd \ln k
        \,, \label{kin_hel} \\
        {\cal H}_A \equiv &\, \bra{\AAA (\xx) \cdot 
        \BB (\xx)}_V = \int_{-\infty}^\infty \frac{H_B (k)}{k}
        \dd \ln k\,,
        \label{mag_hel}
    \end{align}
\end{subequations}
where $\AAA$
is the vector potential, such that $\BB = \nab \times \AAA$.
Note that the extra $k$ and $k^{-1}$ in the integrals compared to \Eqq{squared_values}
appear to match, respectively,
the dimensions of $\oom (\kk) = - i \, \kk \times
\uu (\kk)$ and $\BB (\kk) = - i \, \kk \times \AAA (\kk)$.
Again, analyticity in \Eq{spectral_funcs} requires that the
helical spectrum of a causal field becomes proportional
to $k^{a_H}$ with even $a_H \geq 4$
when $k \to 0$ \cite{Durrer:2003ja,Brandenburg:2018ptt},
\begin{equation}
    H(k) \sim k^{a_H} \quad \text{with} \ a_H \geq 4\,.
\end{equation}

In ideal MHD, the current helicity is defined as
\begin{equation}
    \HH_B \equiv \bra{\BB (\xx) \cdot 
    \JJ (\xx)}_V = \int_{-\infty}^\infty
    k \, H_B(k) \dd \ln k\,, \label{curr_hel}
\end{equation}
where $\JJ = \nab \times \BB$ is the current density under the
assumption of infinite conductivity.
We note that we are using the generic definition
of the helicity of a vector field $\varphi_i (\xx)$
as
\begin{align} \label{helicity}
    \HH_\varphi \equiv \bigl\langle \pmb{\varphi}(\xx) \cdot
    \bigl[\nab \times &\,  \pmb{\varphi} (\xx)\bigr] \bigr\rangle
    \nonumber \\ = &\,
    \int_{-\infty}^\infty k \, H_\varphi (k) \dd \ln k\,.
\end{align}

Based on the kinetic and the magnetic helicities, $\HH_u$
and $\HH_A$, we can define helical length scales
$\xi_u^\H$ and $\xi_A^\H$ as
\begin{subequations}
\label{helicities_length_scales}
\begin{align}
    \xi_u^\H \equiv &\, \frac{\int E_u(k) \dd \ln k}{\int k \, H_u(k) \dd \ln k}
    = \frac{u_{\rm rms}^2}{\HH_u}
    \,,
    \label{length_scale_u} \\ 
    \xi_A^\H \equiv &\, \frac{\int k^{-1} H_B (k) \dd \ln k}{\int E_B (k) \dd \ln k}
    = \frac{\HH_A}{B_{\rm rms}^2}
    \,.
    \label{helical_A_H}
\end{align}
\end{subequations}
The helicity of a vector field can always be expressed as
$\HH_\varphi = \varphi_{\rm rms}^2 / \xi_\varphi^\H$
with $\xi_\varphi^\H$ defined as in \Eq{length_scale_u}.
Note that \Eq{helical_A_H} should not be confused
with the magnetic
integral length scale used in MHD
literature to describe the decay of primordial magnetic fields \cite{MY75,Campanelli:2004wm,Kahniashvili:2008pe,
Brandenburg:2014mwa,Brandenburg:2017neh,RoperPol:2018sap},
\begin{equation}
    \xi_B = \frac{\int k^{-1} E_B (k) \dd \ln k}{\int E_B (k) \dd \ln k}\,,
    \label{integral_scale_M}
\end{equation}
which is not directly related to magnetic helicity.

The realizability condition bounds the value of the helical spectrum
\cite{1978mfge.book.....M,Caprini:2003vc,Kahniashvili:2005qi,
Kisslinger:2015hua,Kahniashvili:2020jgm,RoperPol:2021uol},
\begin{equation}
    |H (k)|  \leq
    E_\N (k)\,,
    \label{helical_fract}
\end{equation}
and the fractional helicity determines the ratio to exhaust the inequality,
\begin{equation}
    \label{helical_eps}
    \varepsilon (k) = \frac{H(k)}{E_\N (k)} \in [-1, 1]\,,
\end{equation}
defined when $E_\N (k) \neq 0$.
Note that $E_\N (k) = 0$ iff $\oom = \nab \times \uu = {\pmb 0}$
\cite{MY75}, then
$\HH = 0$ and, hence, $H(k) = 0$.
The realizability condition also applies for
the helical and integral length scales defined
in \Eqs{helical_A_H}{integral_scale_M},
$\xi_B \geq \xi_A^\H$ or, equivalently, $\HH_A \leq \xi_B B_{\rm rms}^2$.
Using the fractional helicity, we can rewrite the
helical spectra of velocity and magnetic fields as
\begin{subequations}
\begin{align}
    H_u (k) = &\, \varepsilon_u (k) [1 - q(k)] \, E_u (k) \,, \\
    H_B (k) = &\, \varepsilon_B (k) \, E_B (k)\,.
\end{align}
\end{subequations}

The realizability condition in \Eq{helical_fract}
also implies that $a_H \geq a_E$ in the $k \to 0$
limit, which, together with causality (or analyticity
of the correlation function), requires that 
$H(k \to 0) \sim k^{a_H}$ with even $a_H \geq 6$
when the field is solenoidal
with $E_\N (k \to 0) \sim k^{a_E}$ and odd $a_E \geq 5$
\cite{Durrer:2003ja}.
Since this is always the case for magnetic fields,
we have
\begin{equation}
    H_B(k \to 0) \sim k^{a_H} \quad \text{with \ }
    a_H \geq 6\,.
\end{equation}
In general, we expect $\varepsilon_B
(k) \sim k$ (or steeper)
when $k \to 0$.

As mentioned above,
when the enthalpy and velocity
perturbations are small,
the velocity stresses can be approximated as $X_i X_j = w \, \gamma^2\, u_i \, u_j
\simeq \bar w  \, u_i u_j$.
Hence, the spectral function ${\cal F}_{ij}^{X} (\kk)$,
defined as
\begin{align}
    \bra{X_i (\kk) \, X_j^\ast (\kk')} = (2 \pi)^6   \,\delta^3
   (\kk   - \kk') \frac{{\cal F}_{ij}^{X} (\kk)}{4 \pi k^3}\,, 
\end{align}
relevant to determine the source of GWs, can be approximated
as
\begin{equation}
    {\cal F}_{ij}^{X} (\kk) \simeq
    \bar w \, {\cal F}_{ij}^{u} (\kk)\,.
\end{equation}
Similarly, the mean fluid kinetic energy density is
$\bar \rho_{\rm kin} = \bra{X^2} = X_{\rm rms}^2 \simeq \bar w \bra{u^2} = \bar w \, u_{\rm rms}^2$.
In many parts of this work, we 
take this approximation to describe
the source of GWs in terms of the velocity field as this has been
commonly done in the literature.
However,
we note that our description can in general be applied
to large enthalpy and velocity
fluctuations using the spectral
components of $X_i$ instead of $u_i$ \cite{Xvariable}.

\subsection{Scalar fields}
\label{scalar_Ephi}

Scalar fields source GWs through their kinetic term,
proportional to their gradients,
$T_{ij} \supset \partial_i \phi \, \partial_j \phi$.
The two-point correlator of this source is then purely longitudinal,
\begin{align}
    \langle \partial_i \phi (\kk) \, \partial_j \phi^\ast
    (\kk') \rangle 
    =   k & \,  k' \hat{k}_i \hat{k}'_j \bra{\phi(\kk)\,\phi^\ast
    (\kk')}
    \nonumber \\ =  (2 \pi)^6 &\, \delta^3  (\kk - \kk') \, \hat k_i \hat k_j
    \frac{E_\phi (k)}{4 \pi k}\,, \label{E_Phi}
\end{align}
such that $E_\L^\phi (k) = k^2 E_\phi (k)$.
Therefore, this source can be described with a single longitudinal
spectral function, $E_\phi^\L (k)$.
In this case, causality implies that $E_\L^\phi (k)
= k^2 E_\phi (k) \sim k^{2 + a_E}$ with odd $a_E \geq 3$.

The
mean squared value of the scalar field and the mean
gradient energy density are
defined as
\begin{subequations}
\begin{align}
    \phi_{\rm rms}^2 \equiv  &\,
    \bra{\phi^2 (\xx)}_V = 
    \int_{-\infty}^\infty E_\phi (k) \dd \ln k\,, \label{scalar_squared} \\ 
    \bar \rho_G^\phi \equiv &\,  \frac{1}{2a^2}
    \bra{|\nab \phi|^2 (\xx)}_V \nonumber \\& 
    \hspace{15mm} =
    \frac{1}{2a^2}
    \int_{-\infty}^\infty E^\phi_\L (k) \dd \ln k\,.
    \label{gradient_energy_density}
\end{align}
\end{subequations}
We can define
a squared length scale $\xi_\phi^2$ that characterizes their
ratio as
\begin{equation}
    \label{squared_xi_phi}
    \xi_\phi^{2}
    = \frac
    {\int k^{-2} E_{\L}^\phi (k) \dd \ln k}
    {\int E_{\L}^\phi (k) \dd \ln k} = \frac{\phi_{\rm rms}^2}
    {2 a^2 \rho_G^\phi} \,.
\end{equation}

\subsection{Smoothed broken power law model}
\label{sec:smoothed_dbpl}

A common description for the
spectral shape of a causal field sourcing GWs (e.g.,
velocity, electromagnetic, or scalar field)
is a broken power law in $k$, presenting different scaling
at small and large $k$.

As discussed in previous sections, due to causality,
it is expected to find a Saffman
spectrum at large scales \cite{Saffman1967}, i.e.,
$E(k \to 0) \sim k^{a_E}$ with odd $a_E \geq 3$, where $a_E = 3$ corresponds to white noise.
This condition applies to the spectra of all of the sourcing
fields, as well as to the one of the anisotropic
stresses, $E_\Pi (k)$, and of the GWs, $\OmGW (k)$ \cite{MY75,Caprini:2009fx}.
Since values of $a_E < 3$
would violate causality at large
scales, they cannot be found in fields generated by causal processes during the
radiation-dominated era.
However, $a < 3$ can be found in
fields generated during inflation as,
for example, inflationary magnetic fields
\cite{Turner:1987bw,Brandenburg:2018ptt,Caprini:2003vc,Brandenburg:2020vwp}.
On the other hand, for the helical components of the spectra,
$H(k)$, as well as the chiral $H_\Pi (k)$ and $\Xi_{\rm GW} (k)$, the
general causality condition requires them to be proportional to
$k^{a_H}$,
with even $a_H \geq 4$, in the $k \to 0$ limit \cite{Durrer:2003ja}.

In the particular case of vector fields spectra, this causality condition can be
more restrictive.
Indeed, velocity and magnetic fields spectra
are required to follow
a Batchelor or steeper
spectrum  ($a_E \geq 5$)
at large scales
when the field is irrotational or solenoidal [see \Eq{k5}].
Due to the realizability condition,
$|H (k)| \leq E_\N (k)$
[see \Eq{helical_fract}], when
the field is solenoidal, causality implies that $H (k \to 0) \sim k^{a_H}$
with even $a_H \geq 6$ \cite{Durrer:2003ja}.

At larger wave numbers, i.e., smaller scales, the
field becomes correlated.
This is commonly characterized by the integral length scale $\xi$,
defined in \Eq{integral_scale_M}.
For $k \gg k_\ast$, where $k_\ast \sim \xi^{-1}$ is the spectral peak,
the spectrum decays with a slope
that depends on the properties of the source.
For example, velocity fields in
stationary Kolmogorov turbulence\footnote{Note that the Kolmogorov spectrum is only
well established for purely vortical hydrodynamic
turbulence \cite{1941DoSSR..30..301K}.
In MHD turbulence different predictions have been proposed
(see review
\cite{Schekochihin:2020aqu} and references therein), starting
with the Iroshnikov-Kraichnan model,
which predicts a small-scale
spectrum $E_B (k)
\sim k^{-1/2}$ \cite{1964SvA.....7..566I,1965PhFl....8.1385K}, also found 
in the Boldyrev alignment model \cite{Boldyrev:2005ms}.
However, we might still
find that the magnetic
field presents Kolmogorov-like spectra, e.g.,
in the Goldreich-Sridhar model \cite{Goldreich:1994zz},
also found in numerical simulations of decaying
turbulence
\cite{Christensson:2000sp,2000PhRvL..84..475M,Brandenburg:2014mwa,Brandenburg:2016odr,Brandenburg:2017neh,Brandenburg:2017rnt}.} are expected to
follow $E_u (k) \sim k^{-2/3}$ at $k \gg k_\ast$, while the
spectrum of acoustic perturbations (i.e., purely
compressional velocity field with $\oom = 0$)
is expected to be $E_u (k) \sim k^{-1}$ at $k \gg k_\ast$ \cite{Hindmarsh:2013xza,Hindmarsh:2015qta,Hindmarsh:2016lnk,Hindmarsh:2017gnf,Niksa:2018ofa,Hindmarsh:2019phv,Jinno:2022mie,Caprini:2024gyk,RoperPol:2023dzg,Dahl:2021wyk,Dahl:2024eup,Caprini:2026nnk}.

Eventually, at larger wave numbers,
the spectra are expected to follow an exponential
damping due to diffusive processes, for example,
viscosity and
magnetic diffusivity dissipating respectively
velocity and magnetic fields at small
scales.
This is a generic feature that is also expected for
acoustic turbulence at wave numbers above the inverse
scale that characterizes the thickness of the shocks \cite{Dahl:2021wyk,Dahl:2024eup,Jinno:2022mie,Caprini:2024gyk,Correia:2025qif} and for
sound waves in first-order phase transitions for scales
above the inverse bubble wall thickness \cite{Caprini:2026nnk}.

In the case of a scalar field in a first-order phase transition,
its spectrum is less well understood
as it has been less explored numerically.
The simulations of refs.~\cite{Cutting:2018tjt,Cutting:2020nla}
show that $E_\phi (k)$
becomes proportional to $k^3$
at small $k \ll k_\ast$, as expected from
causality, and it becomes proportional to
$k^{-b_\phi}$ with $2.5 \lesssim b_\phi \lesssim 3$
in the thin-wall limit, i.e., when the two vacua
of the first-order phase transition are nearly 
degenerate (see, for example,
Figs.~6--7 in ref.~\cite{Cutting:2018tjt}).

For generality,
in the present work we provide the theoretical
framework to study smoothed broken power law spectra
of the fields sourcing GWs,
$\varphi = \{ X, B, \phi \}$,
which transition from $k^a$ at $k \ll k_\ast$ to $k^{-b}$
at $k \gg k_\ast$,
for generic slopes $a, b > 0$,
\begin{align}
    E (K) = &\, E_\ast
    \frac{(b + a)^{1\over \alpha}\, K^a}
    {\bigl[ b + a \,  K^{\alpha(a + b)}
    \bigr]^{1 \over \alpha}} = E_\ast \, \zeta (K) \,, \label{powerlaw_model}
\end{align}
where $K \equiv k/k_\ast$ is the normalized wave number
and $\alpha$ is the smoothness between slopes
around $k_\ast$.
The normalized spectral shape $\zeta (K)$ is defined such that $\zeta(1)=1$ and, hence, $E(1) = E_\ast$.
The prefactor $(b + a)^{1 \over \alpha}$
and the values of $a$ and $b$ in the
denominator have been included to ensure that the
spectral maximum is reached at $k_\ast$ with
the normalized value $\zeta (1) = 1$.

The model function of \Eq{powerlaw_model} has already been considered in previous studies
of GW production from turbulence
\cite{RoperPol:2022iel,Auclair:2022jod,RoperPol:2023bqa,Caprini:2024ofd,EPTA:2023xxk,Caprini:2024hue}
and sound waves \cite{RoperPol:2023dzg}.
In previous work, it was sometimes alternatively
expressed as in \Eq{power_law_model_alp2},
which is equivalent to \Eq{powerlaw_model}
taking $\alpha_2 = \alpha (a + b)$.
The detailed analytical calculations in this section
are presented in 
\App{appendix_sbpl}.
For numerical values
presented in this section, unless noted
otherwise, we
consider as an example the von K\'arm\'an spectrum with $a = 5$ and $b = 2/3$, 
describing
vortical Kolmogorov turbulence.

The average value of the squared field
is
\begin{align}
    \varphi_{\rm rms}^2 = & \, \int_{-\infty}^{\infty} E (k) \dd \ln k =
    E_{\ast} \, {\cal A}_{ab} (\alpha)\,,
    \label{integ_OmM}
\end{align}
with $\varphi_{\rm rms}^2 / a^2 \equiv \tilde \varphi_{\rm rms}^2
\equiv \{ \bar \rho_{\rm kin}, 2 \bar \rho_B,
2 \bar \rho_G^\phi \}$,
where $\bar \rho_{\rm kin} = X_{\rm rms}^2$, $\bar \rho_B \equiv \half \, a^{-2} B_{\rm rms}^2$
and $\bar \rho_G^\phi$ are the
mean kinetic,
magnetic, and scalar gradient
energy densities.
In general, we assume that both $a$ and $b$ are positive,
which ensures that $\varphi_{\rm rms}^2$ is finite.
The geometric factor
${\cal A}_{ab}$ is [see \Eq{Om_I_Ei}]
\begin{align}
    {\cal A}_{ab}   (\alpha) \equiv &\,
    \int_{-\infty}^\infty \zeta(K) \dd \ln K 
    = I_{ab; -1} (\alpha)
    \nonumber \\ = &\,
    \biggl(\frac{a + b}{b} \biggr)^{1\over\alpha}
    \biggl(\frac{a}{b}\biggr)^{-{a\over \alpha (a+b)}}
    \nonumber \\ &\, \hspace{18mm} \times
    \frac{\Gamma\bigl[{{a} \over {\alpha (a+b)}}
    \bigr] \Gamma\bigl[{{b} \over {\alpha (a+b)}}\bigr]}{
    \alpha (a+b)
    \, \Gamma\bigl[{1 \over \alpha}
    \bigr]}\,, \label{AA_alpha}
\end{align}
shown in \Fig{AA_ab_alp},
where $I_{ab; n}$ corresponds to the moments of $\zeta(K)$
[see \Eq{Iabn_def}],
\begin{equation} \label{Iabn_general}
    I_{ab; n} (\alpha) = \int_0^\infty \zeta(K)
    \, K^n \dd K\,.
\end{equation}

\begin{figure}[t]
    \centering
    \includegraphics[width=\columnwidth]{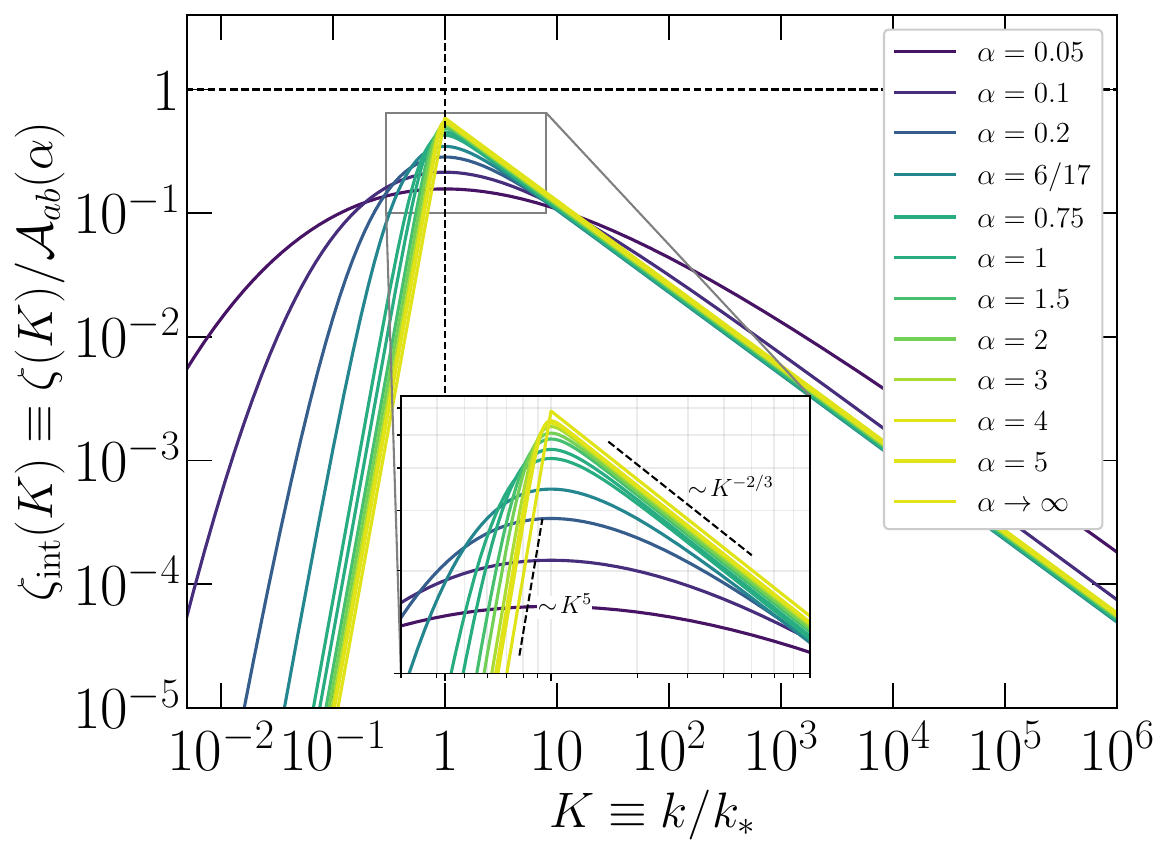}
    \caption{Normalized smoothed broken power law $\zeta_{\rm int} (K)
    \equiv \zeta (K) / {\cal A}_{ab} (\alpha)$ describing a
    von K\'arm\'an spectrum (i.e., $a = 5$
    and $b = 2/3$),
    for different values of the smoothing
    parameter $\alpha$.
    For reference, the models of refs.~\cite{RoperPol:2022iel}
    and \cite{Niksa:2018ofa,Auclair:2022jod,RoperPol:2023bqa,Caprini:2024hue} use $\alpha = 2$ and
    $\alpha = 6/17$, respectively.
    }
    \label{EM_alphas}
\end{figure}

The geometric factor ${\cal A}_{ab}$ allows us to define the spectral shape in terms
of the integrated spectrum, i.e., the mean squared $\varphi_{\rm rms}^2$,
by substituting \Eq{integ_OmM}
into \Eq{powerlaw_model},
\begin{align}
    E (K) =  \varphi_{\rm rms}^2
    \, \zeta_{\rm int}  & (K)\,, \nonumber \\ &\,
    \text{with } \ 
    \zeta_{\rm int} (K) \equiv \frac{\zeta(K)}{{\cal A}_{ab} (\alpha)} \,.
    \label{EM_OmM}
\end{align}
Unlike $\zeta (K)$, 
the spectrum $\zeta_{\rm int} (K)$
is normalized by its integral, $\int \zeta_{\rm int} (K) \dd \ln K = 1$.
It is
shown in \Fig{EM_alphas}
for $a = 5$, $b = 2/3$, and different values of $\alpha$.
Larger values of $\alpha$ correspond to
sharper transitions and, hence,
the corresponding values at the peak increase with $\alpha$ to keep
the same integrated values.

For velocity fields with longitudinal (L), normal (N), and helical (H) components,
each component can be expressed in terms of $\zeta_{\rm int}^u (K)$ as
\begin{subequations}
\label{norm_generic}
\begin{align} \label{norm_EL}
    E_\L^u(K) = &\, u_{\rm rms}^2 \, q(K) \, \zeta_{\rm int}^u (K) \,, \\ 
    E_\N^u(K) = &\, u_{\rm rms}^2 \, [1 - q (K)]
    \, \zeta_{\rm int}^u (K)
    \,, \\ 
    H_u (K) = &\, u_{\rm rms}^2\,
    \varepsilon_u (K) [1 - q (K)] \, 
    \zeta_{\rm int}^u (K) \,,
    \label{hel_spec}
\end{align}
\end{subequations}
where $q(K)$ and $\varepsilon_u (K)$ are
the degree of compressibility and fractional helicity
introduced in \Sec{sec:vel}.
Note that $\zeta_{\rm int}^u (K)$ corresponds to the
normalized velocity spectrum,
such that
\begin{align}
\label{tot_spec_vel}
    E_u(K) \equiv E_\N^u (K) + E_\L^u (K)
    = u_{\rm rms}^2 \, \zeta_{\rm int}^u (K) \,.
\end{align}

The spectral components of the magnetic field are
normalized as in \Eqq{norm_generic} taking $q = 0$,
while the spectrum of the scalar field gradients is normalized
taking $q = 1$, such that $E_\N(k) = 0 = H(k)$.
Note that, since we are interested in sources of GWs, we denote as
$\zeta_\phi (K)$ the normalized spectrum
of the scalar field gradients, instead of the normalized
spectrum of the scalar field, $E_\phi (K)$.

If the helical 
spectrum in \Eq{hel_spec}, proportional to $\varepsilon (K) [1 - q (K)] \zeta (K)$,
can also be expressed
as a smoothed broken power law with slopes $a_H$, $b_H$, and smoothness
$\alpha_H$, and with peak amplitude $H_\ast = E_\ast (1 - q_\ast) \, \varepsilon_\ast$,
where $\varepsilon_*$ and $q_*$ are respectively
the fractional helicity and degree of compressibility at the peak,
the total helicity of the field is
\begin{align}
    {\cal H} = &\, \int_{-\infty}^\infty k\,
    H(k) \dd \ln k =  H_\ast
    k_\ast I_{a_Hb_H; 0} (\alpha_H) \nonumber \\ 
    = &\, \frac{H_*}{E_*}
    \varphi_{\rm rms}^2 \, k_\ast \frac{I_{a_Hb_H; 0} (\alpha_H)}{{\cal A}_{ab} (\alpha)} \,.
\end{align}
Note that this expression directly applies for
helicities defined as in \Eq{helicity} (i.e., kinetic and
current helicities),
while the magnetic helicity includes a $k^{-1}$ term in the integral [see \Eq{mag_hel}],
\begin{align}
    \HH_A = \frac{\HH_B}{k_\ast^2} 
    \frac{I_{a_Hb_H; -2} (\alpha_H)}{I_{a_Hb_H; 0} (\alpha_H)} &\,  \nonumber \\ 
    = \frac{H_*}{E_*}
    \frac{B_{\rm rms}^2}{k_\ast} &\, \frac{I_{a_Hb_H; -2} (\alpha_H)}{{\cal A}_{ab} (\alpha)}
    \,.
\end{align}

The integral length scale of the field, given in \Eq{integral_scale_M}, is
\begin{align}
    k_\ast \xi = &\, \frac{I_{ab; -2} (\alpha)}{{\cal A}_{ab} (\alpha)}
    =
    \BBB_{ab} (\alpha)\,,
    \label{xiM}
\end{align}
where the geometric factor ${\cal B}_{ab} (\alpha)$ can be expressed as [see \Eq{BB_ab}]
\begin{align}
    {\cal B}_{ab} (\alpha) = \biggl({a\over b}
    \biggr)^{
    \frac{1}{\alpha (a+b)}}
    \frac{\Gamma\bigl[{{a - 1} \over {\alpha (a+b)}}
    \bigr]}
    {\Gamma\bigl[{{a} \over {\alpha (a+b)}}
    \bigr]} \frac{\Gamma\bigl[{{b + 1} \over {\alpha (a+b)}}
    \bigr]}{\Gamma\bigl[{{b} \over {\alpha (a+b)}}
    \bigr]}\,.
    \label{B_alp}
\end{align}
The factor ${\cal B}_{ab}(\alpha)$ is shown in \Fig{BB_ab_alp} of \App{appendix_sbpl}.

For scalar fields,
the length scale $\xi_\phi$ defined in \Eq{squared_xi_phi}
relates
the mean gradient energy density to the
mean squared value of $\phi$.
It is given as
\begin{equation}
    k_\ast \xi_\phi = \Biggl[\frac{I_{ab; -3} (\alpha)}
    {{\cal A}_{ab} (\alpha)} \Biggr]^{1\over 2}
    = {\cal B}_{ab; 2} (\alpha)\,. \label{length_scalar}
\end{equation}
The coefficient ${\cal B}_{ab; n}$ is defined for a generalized length scale in
\App{appendix_sbpl} [see, in particular, \Eq{generalized_Bab}].
Setting $n = 2$, we find
\begin{equation} \label{Babm2}
    \! {\cal B}_{ab; 2} =  \biggl({a\over b}\biggr)^{1 \over 
    \alpha (a+b)}
    \Biggl(\frac
    {\Gamma\bigl[ {a - 2 \over \alpha (a+b)
    } \bigr]}
    {\Gamma\bigl[ {a \over \alpha (a+b)} \bigr]}
    \frac
    {\Gamma\bigl[ {b + 2 \over \alpha (a+b)} \bigr]}
    {\Gamma\bigl[ {b \over \alpha (a+b)} \bigr]}
    \Biggr)^{{\frac{1}{2}}}\,.
\end{equation}
Note that the slopes $a,b$, and smoothness $\alpha$ refer to the
spectrum of the scalar field gradients $E_\L^\phi
(K)$ in \Eq{norm_EL}, taking $q (K) = 1$.

\section{Spectra of the anisotropic stresses}
\label{EPi_sec}

The GW spectrum and its circular polarization are
determined by the integrals over $\tau_1$
and $\tau_2$ of the symmetric and antisymmetric spectral
functions $E_\Pi(\tau_1, \tau_2, k)$ and $H_\Pi(\tau_1, \tau_2, k)$,
respectively,
as shown in \Eqs{eq:OmGWt}{OmGW_aver}.
These functions represent the two-point unequal time 
correlator (UETC) of the anisotropic stresses, $\Pi_{ij}$,
as defined in \Eqq{twopoint_EPis}.
In this section, we present the relation between
the UETC of the different sources (velocity, electromagnetic, and
scalar fields) with that of the anisotropic stresses that
source GWs.
We omit the $\tau_1$ and $\tau_2$ dependence in the following for brevity,
as this dependence does not affect the results presented in
this section.

We start by describing the two-point correlation function
of the stress-energy tensor,
\begin{widetext}
\begin{align}
    \bra{T_{ij} (\kk) \, T_{lm}^\ast (\kk')} =
    \int \frac{\dd^3 \pp}{(2 \pi)^3} \int \frac{\dd^3 \qq}{(2 \pi)^3}  \sum_{a, b} \bra{\varphi_i^a(\tilde \pp) \, \varphi_j^a(\pp) \, \varphi_l^{b\ast} (\tilde \qq)
    \, \varphi_m^{b\ast} (\qq)}\,,
    \label{Tij_Fourier}
\end{align}
\end{widetext}
where $\varphi^{a, b}_i (\kk) = \{a X_i, i B_i, i E_i,
- i k_i \phi\}$
corresponds
to each of the sourcing
fields, in Fourier space, that contribute to the
stress-energy tensor,
\begin{equation} \label{biproduct_Tij}
    T_{ij} (\xx) = \displaystyle\sum_a \varphi_i^a (\xx) \, \varphi_j^a (\xx)\,.
\end{equation}
We have defined $\tilde \pp = \kk - \pp$ and $\tilde \qq = \kk' - \qq$
to describe the convolution in \Eq{Tij_Fourier}.
For simplicity, we assume that
the cross-correlations between different sources are
negligible
and focus on GW production by each of the sources separately.
In reality, however, when two or more sources are present with similar amplitudes,
the cross-correlations between the sourcing fields should be taken
into account as they are non-vanishing when the sourcing
fields are dynamically coupled.

Under the assumption that the cross-correlations of different
fields are vanishing,
and omitting the contributions from electric fields,
which is justified given the very large conductivity
of the primordial plasma \cite{Arnold:2000dr}
(see the discussion at the beginning of \Sec{stress}),
we can express the symmetric and antisymmetric spectra
of the anisotropic stresses as
\begin{subequations}
\begin{align}
    E_\Pi (k) =  &\, E_\Pi^{X} (k) + E_\Pi^{B} (k) + E_\Pi^{\phi}
    (k)\,, \\ 
    H_\Pi (k) = &\, H_\Pi^X (k) + H_\Pi^B (k)\,.
\end{align}
\end{subequations}
We denote $E_\Pi^\varphi (k)$ and $H_\Pi^\varphi (k)$
with $\varphi = \{X, B, \phi\}$ as the 
component corresponding to the sourcing field $\varphi_i = \{a X_i, 
i B_i, - i k_i \phi \}$, respectively.

When the statistical distribution of the
sourcing field $\varphi_i$
is Gaussian, we can apply Isserlis'
(or Wick's)
theorem to the
four-point correlation of \Eq{Tij_Fourier}, and express
it as
a superposition of products of two-point correlation functions.
Introducing the spectral decomposition of the velocity field
of \Eqq{two_point_Ev},
and applying the TT operators $\Lambda_{ijlm}$
and ${\cal A}_{ijlm}$, we 
find the spectral functions describing
the statistics of the anisotropic stresses $\Pi_{ij}
= a^{-2} \Lambda_{ijlm} T_{lm} / \bar \rho$,
defined in \Eqq{twopoint_EPis} (see details in \App{autocorr_EPi}).
In general, we note that even if the initial
perturbations of a field are statistically
Gaussian, non-Gaussianities can develop due
to the non-linear dynamics of the source.
When the field is non-Gaussian, then its four-point correlation function cannot be decomposed
applying Isserlis' theorem and numerical simulations
describing the non-linear dynamics are, in general,
required.
We then restrict our theoretical
analysis to Gaussian fields in the present work.

\subsection{Symmetric spectrum of the anisotropic stresses $E_\Pi$}\label{subsec:symm_EPi}

Let us first focus on the different components
of the symmetric spectrum of the anisotropic
stresses, $E_\Pi^\varphi (k)$,
computed
in \App{autocorr_EPi} [see \Eq{EPi_xy}], which can be
decomposed in up to
four terms,
\begin{equation}
    E_\Pi^\varphi (k) = \sum_{i} E_\Pi^{\varphi, i} (k)\,,
\end{equation}
where $i = \{$vort, comp, mix, hel\}
accounts
for the vortical (NN), compressional (LL), mixed
vortical-compressional (NL),
and helical (HH) components.

When the field is purely vortical
(e.g., $\nab \cdot \uu = 0$ or $\nab \cdot \BB = 0$), such that $E_\L (k) = 0$,
and non-helical with $H (k) = 0$,
then $E_\Pi (k)$
only contains the ``vort'' component
\cite{Kosowsky:2001xp,Caprini:2001nb},
given in \Eq{EPi_vort},
\begin{align}
    a^4 E_\Pi^{\rm vort}  (k) &\,  = \frac{k^2}{8 \bar \rho^2}
    \int_0^\infty \frac{\dd p}{p^2} E_\N (p) \nonumber \\ \times
    &\, \int_{|k - p|}^{k + p}  \frac{\dd \tilde p}{\tilde p^2} E_\N (\tilde p)
    (1 + z^2) (1 + \tilde z^2)
    \,,
    \label{EPivort_txt}
\end{align}
where $\tilde p = |\tilde \pp|
= \sqrt{p^2 + k^2 - 2 pk z}$,
and $z$ and $\tilde z$ are, respectively, the dot products
$\hat \kk \cdot \hat{\pp}$ and $\hat \kk \cdot \hat{\tilde \pp}$.
If the purely vortical field has a non-zero helicity,
then an additional ``hel'' term, proportional to the helical spectrum, contributes to $E_\Pi (k)$ \cite{Caprini:2003vc}, given in \Eq{EPi_hel},
\begin{align}
    a^4 E_\Pi^{\rm hel} (k)    = \frac{k^2}{2 \bar \rho^2}
    \int_0^\infty &\, \frac{\dd p}{p^2} H (p) \nonumber \\ \times
    &\,
    \int_{|k - p|}^{k + p} \frac{\dd \tilde p}{\tilde p^2} H (\tilde p)
    \, z \tilde z \,. \label{EPihel_txt}
\end{align}

On the other hand, if the field is purely compressional
(e.g., $\nab \times \uu = 0$ or $\nab \times \nab \phi = 0$), such that
$E_\N (k) = 0$,
then $E_\Pi (k)$ presents only
the ``comp'' component \cite{Caprini:2007xq}, given in \Eq{EPi_comp},
\begin{align}
    a^4 E_\Pi^{\rm comp}  (k)  &\,   = \frac{k^2}{2 \bar \rho^2}
    \int_0^\infty \frac{\dd p}{p^2} E_\L (p) \nonumber \\ \times &\,
    \int_{|k - p|}^{k + p} \frac{\dd \tilde p}{\tilde p^2} E_\L (\tilde p)
    (1 - z^2) (1 - \tilde z^2)\,.
    \label{EPicomp_txt}
\end{align}

More generally, for velocity fields with both non-vanishing curl and divergence, the
spectrum of the anisotropic stresses becomes the sum
of the ``vort'', the ``comp'', and the ``hel''
components, together with an additional
mixed ``mix''  term.
This mixed term is given in \Eq{EPi_mix},
\begin{align}
    a^4 E_\Pi^{\rm mix} (k) =  \frac{k^2}{2\bar \rho^2}&\, 
    \int_0^\infty \frac{\dd p}{p^2} E_\L (p) \nonumber \\ \times &\,
    \int_{|k - p|}^{k + p} \frac{\dd \tilde p}{\tilde p^2} E_\N (\tilde p)
    \bigl(1 - z^2 \tilde z^2 \bigr) \,.
    \label{EPimix_txt}
\end{align}

With the exception of $E_\Pi^{\rm hel} (k)$,
all components $E_\Pi^i$ are positive-definite, and this
term satisfies
$|E_\Pi^{\rm hel} (k)| \leq E_\Pi^{\rm vort} (k)$,
ensuring that the resulting $E_\Pi (k)$
is always positive.
The integrands ${\cal K} (p, \tilde p)$
that multiply the
spectra
in \Eqss{EPivort_txt}{EPimix_txt} 
are all symmetric functions of $p$ and $\tilde p$, with the exception
of the mixed term that is computed from the product $E_\L (p) E_\N (\tilde p)$.
This product is, in general, not symmetric in
$(p, \tilde p)$ unless the degree of compressibility $q$
is a constant.
However, the integrals over $E_\L(\tilde p) E_\N (p)$ and $E_\N(\tilde p) E_\L (p)$ can be shown to yield the
same result and, hence, they have been combined
in \Eq{EPimix_txt}.

In general, velocity fields can contain
contributions from all four components
$E_\Pi^i$,
while the anisotropic stresses of magnetic fields can only contain a vortical
and a helical component.
On the other hand,
the gradients of scalar fields only present the
compressional component.

Following the normalization of refs.~\cite{RoperPol:2022iel,RoperPol:2023dzg,RoperPol:2023bqa},
we re-express each component $i = \{$vort, comp, mix, hel\}
of the anisotropic stresses spectrum in terms of normalized functions.
In first place, we introduce the normalization of \Eqq{norm_generic}
to each spectral component
of the source, $E_\L^\varphi (K)$, $E_\N^\varphi (K)$, and $H_\varphi(K)$.
Then, each of the components of $E_\Pi^\varphi (K)$
is expressed as [see \Eq{EPi_app}]
\begin{align}
    E_\Pi^{\varphi, i} (K) = &\, \frac{\tilde \varphi_{\rm rms}^4}{\bar \rho^2} \frac{{\cal C}_i^\varphi}
    {{\cal A}_{ab}^2 (\alpha)} \, K^3
    \, \zeta_\Pi^{\varphi, i} (K) \,,
    \label{EPi_norms}
\end{align}
where $\tilde \varphi_{\rm rms}^2 \equiv \varphi_{\rm rms}^2 /a^2 \equiv
\{ \bar \rho_{\rm kin},
2 \bar \rho_B, 2 
\bar \rho_G^\phi \} $.
For small fluid perturbations, one can approximate $\bar \rho_{\rm kin} = X_{\rm rms}^2
\simeq \bar w \, u_{\rm rms}^2$, such that $\tilde \varphi_{\rm rms}^2 / \bar \rho \simeq \Gamma \, u_{\rm rms}^2$,
where $\Gamma \equiv \bar w/\bar \rho$ is the adiabatic
ratio, which takes the
value $\Gamma = 4/3$ for a pure radiation fluid.

\begin{figure*}
    \centering
    \includegraphics[width=.49\linewidth]{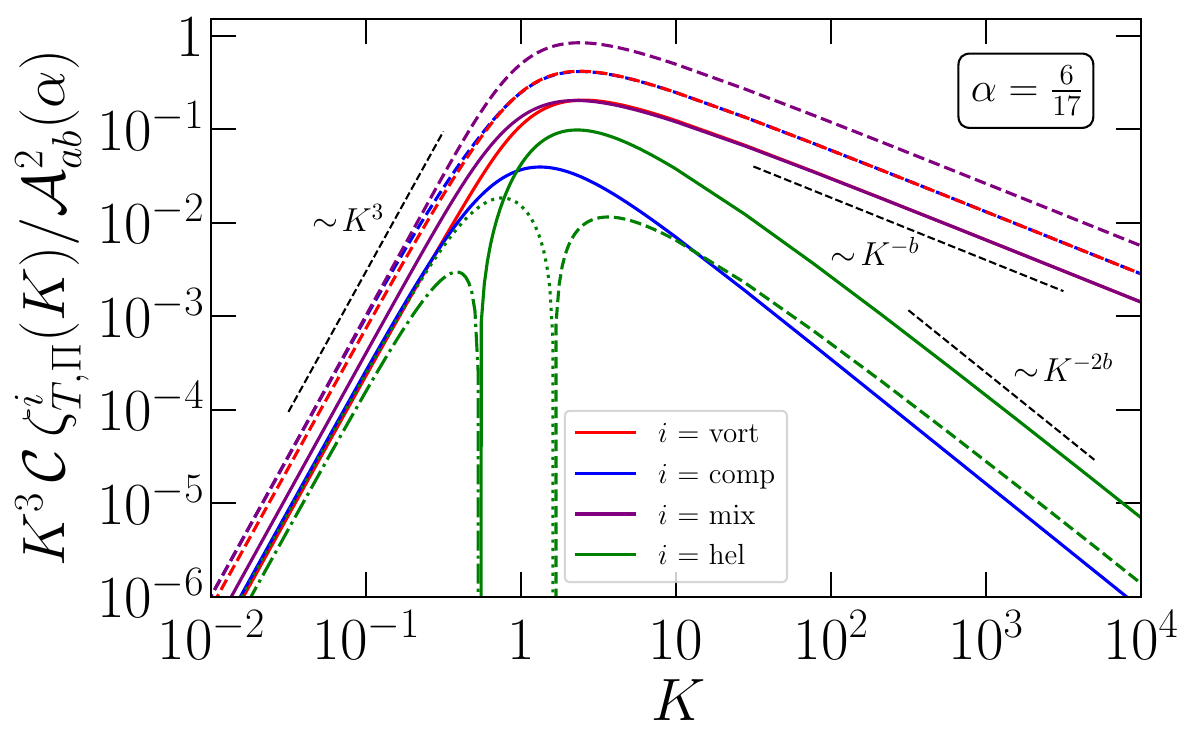}
    \includegraphics[width=.49\linewidth]{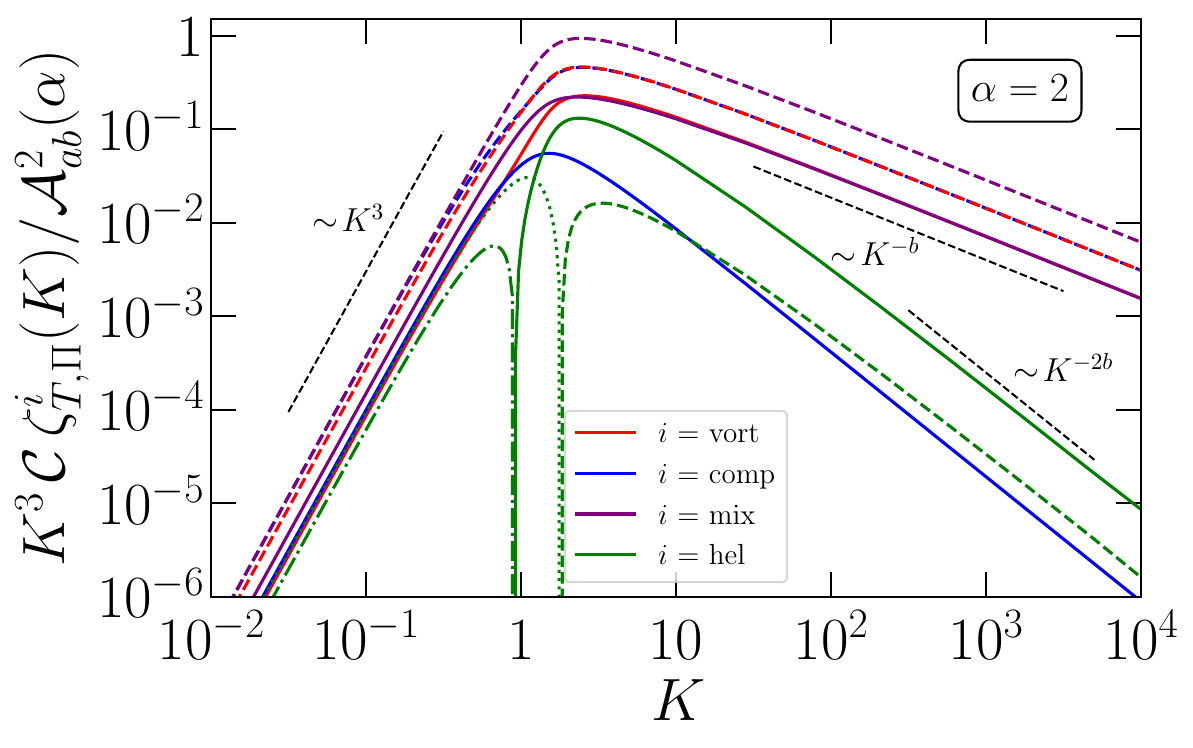}
    \caption{
    Normalized components of the spectrum
    of the anisotropic and unprojected stresses,
    $K^3 \, {\cal C}_i \, \zeta_\Pi^i (K)$ and
    $K^3 \, {\cal C}_T^i \, \zeta_T^i (K)$,
    respectively in solid and dashed lines, with $i = $ vort (red), comp (blue),
    mix (purple), and hel (green),
    divided by ${\cal A}_{ab}^2(\alpha)$,
    for the von K\'arm\'an spectrum with $a = 5$,
    $b = 2/3$, and $\alpha = 6/17$ (left panel) and $2$ (right panel).
    Since the helical component can become negative,
    we show separately negative values of the
    anisotropic stresses with a dash-dotted line
    and of the unprojected stresses with a dotted
    line.
    For the plot, we are omitting the prefactors multiplying each of the
    components of the spectrum of the source as in \Eqq{norm_generic}, so the lines correspond
    to the maximum possible amplitudes of each of the components, with the exception
    of the mixed term, which is bounded by $\tfrac{1}{4}$ times the values in the plots,
    arising from maximizing $q (1 - q) \leq \tfrac{1}{4}$.
    We show in dashed black lines the different
    scalings with $K$ for each component:
    $\sim K^3$ at small
    $K \ll 1$ for all components; $K^{-b}$ at $K \gg 1$
    for the vortical and mixed components and the unprojected compressional
    component;  $K^{-2b}$ at $K \gg 1$
    for the projected compressional and for the helical components.
    The latter scaling $K^{-2b}$
    dominates at large $K$ for $b < 1$
    (helical components) and $b < 2$
    (compressional anisotropic component), but it needs
    to be changed for larger values of $b$
    as we show in \Sec{bpl_anisotropic}.
    The locations of the scaling lines are the same in both plots
    for a direct comparison of the relative amplitude of $\zeta_\Pi (K)$ for the two values of $\alpha$.
    \label{p_pi_alpha}
    }
\end{figure*}

The different
components of the
normalized anisotropic stresses spectra, $\zeta_\Pi^{\varphi, i} (K)$,
are given in \Eqq{normalized_zetaPis}.
The geometrical factors ${\cal C}_i^\varphi > 0$
are defined such that $|\zeta_\Pi^{\varphi, i} (K)| \to 1$
in the $K \to 0$ limit [see \Eqq{C_pis}], 
\begin{subequations}
\label{c_components}
\begin{align}
    {\cal C}_{\rm vort}  = &\, \frac{7}{15} \int_0^\infty [1 - q (P)]^2 \,
    \frac{\zeta^2 (P)}{P^4} \dd P\,,
    \\ 
    {\cal C}_{\rm comp} = &\, 
    \frac{8}{15} \int_0^\infty q^2 (P) \,
    \frac{\zeta^2 (P)}{P^4} \dd P\,,
    \\ 
    {\cal C}_{\rm mix} = &\, 
    \frac{4}{5} \int_0^\infty q (P)
    \, 
    [1 - q(P)] \,
    \frac{\zeta^2 (P)}{P^4} \dd P\,,
    \\
    {\cal C}_{\rm hel} = &\,
    \frac{1}{3} \int_0^\infty
    \varepsilon^2 (P) \, [1 - q (P)]^2 \,
    \frac{\zeta^2 (P)}{P^4} \dd P\,.
\end{align}
\end{subequations}
In the particular case when the degree of compressibility and the fractional
helicity are constant, all these
coefficients can be expressed as
\begin{align} \label{Cuu_2}
    {\cal C}_i = &\, c_i
    \int_0^\infty \frac{\zeta^2 (P)}
    {P^4} \dd P = {\cal C}_{ab}^i (\alpha) \equiv c_i \,
    I_{2a, 2b; -4}
    (\alpha/2) \nonumber \\ = &\, 
    c_i
    \biggl( \frac{a + b}{b}  \biggr)^{2\over\alpha}
    \biggl(\frac{a}{b}\biggr)^{-{2a - 3 \over 
    \alpha(a+b)
    }} \nonumber \\ &\, \hspace{25mm} \times
    \frac{\Gamma\bigl[{{2 a  - 3} \over {\alpha(a+b)
    }}
    \bigr] \Gamma\bigl[{{2 b + 3} \over {\alpha(a+b)
    }}\bigr]}{\alpha(a+b)
    \, \Gamma\bigl[{2 \over \alpha}
    \bigr]}\,,
\end{align}
where we have assumed that $\zeta(K)$
is the smoothed broken power law of \Eq{powerlaw_model}, and
$I_{ab;n} (\alpha)$ is defined
in \Eq{Iabn_general}.
The prefactor $c_i$ is
\begin{align} \label{c_i_norm}
    c_i = \biggl\{
    \frac{7}{15} (1 - q)^2 ,  \
    &\,\frac{8}{15}  q^2 , \nonumber \\ &\, \frac{2}{5}
    \,2 q (1 - q), \ \frac{1}{3} \varepsilon^2 (1 - q)^2
    \biggr\}
\end{align}
for $i = \{$vort, comp, mix, hel$\}$.
The dependence of ${\cal C}^i_{ab} (\alpha)/c_i$ with $\alpha$ is shown
in \Fig{CC_ab_alp}.

We show in \Fig{p_pi_alpha} the normalized anisotropic stresses $K^3 \, {\cal C}_i^\varphi 
\, \zeta_\Pi^{\varphi, i} (K) / {\cal A}_{ab}^2 (\alpha)$ that characterize $E_\Pi^{\varphi, i} (K)$
in \Eq{EPi_norms}
for each component using the smoothed broken power law of \Eq{powerlaw_model}, taking
the von K\'arm\'an spectrum with
$a = 5$ and $b = 2/3$ for all terms.
In reality, we note that for the helical spectrum $a = 5$
is not compatible with causality and the realizability
condition,
which require $a$ to be even with $a \geq 4$,
or $a \geq 6$
when the source is fully vortical (e.g., magnetic fields)
\cite{Durrer:2003ja}.
For simplicity, we assume that the fractional helicity 
$\varepsilon (K)$ remains constant
until sufficiently small $K$, when it eventually becomes linear,
satisfying the causality condition.
Therefore, we omit its contribution at small $K$
(see also footnote \ref{constant_eps}) and take $a = 5$
also for the helical spectrum, such that we can directly
compare all terms.
Note that for generality, we omit the prefactors
that depend on the degree of compressibility and
fractional helicity for each spectral
component [see \Eqq{norm_generic}].
Therefore, to compute $E_\Pi^{\varphi} (K)$, the
prefactors need to be reincorporated for 
particular choices of $q$ and $\varepsilon$.
We also note that the mixed term is always proportional to $q ( 1 - q)$,
which takes a maximum value of $\tfrac{1}{4}$ for $q = \tfrac{1}{2}$.
Therefore, this term is always suppressed with respect to the one
in \Fig{p_pi_alpha} by a factor of 4 or larger, depending on the value
of $q$.
 
For comparison, we also show in \Fig{p_pi_alpha} the spectral components of the unprojected stresses,
\begin{equation}
    E_T^{\varphi, i} (K) = \frac{\tilde \varphi_{\rm rms}^4}{\bar \rho^2} \, \frac{{\cal C}_T^{\varphi, i}}{{\cal A}_{ab}^2}
    K^3\, \zeta_T^{\varphi, i} (K)\,,
\end{equation}
which are computed in \App{autocorr_EPi} [see \Eqss{ET_components}{CT_components}].

\begin{figure}
    \centering
     \includegraphics[width=\linewidth]{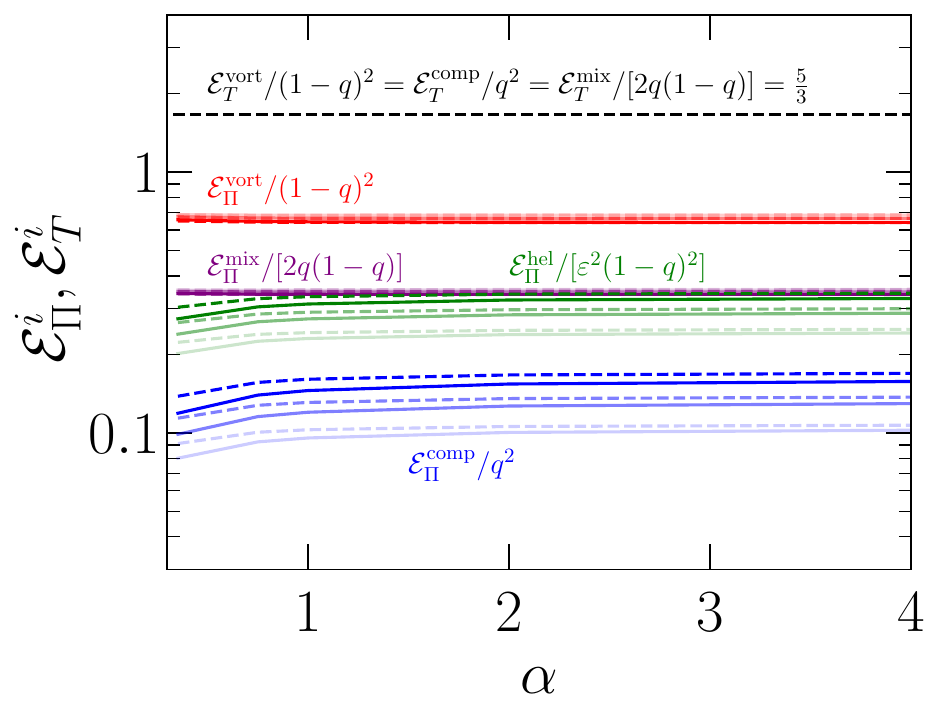}
    \caption{Efficiency factor ${\cal E}_\Pi^i$, defined in \Eq{rho_Pi},
    for each component $i = \{$vort, comp,
    mix, hel$\}$ contributing
    to $E_\Pi^i (K)$, divided respectively
    by $(1 - q)^2$, $q^2$, $2q (1 - q)$,
    and $\varepsilon^2 (1 - q)^2$ assuming constants
    $q$ and $\varepsilon$.
    For comparison, we show
    the analogous ratio for
    the unprojected stresses,
    ${\cal E}_T^\varphi \equiv
    \bar \rho_T^\varphi
    / \tilde
    \varphi_{\rm rms}^4 = \tfrac{5}{3}$
    [see \Eq{efficiencies_ET}]. 
    Dashed (solid) lines correspond to $a = 5$ (3),
    and three values of $b = \{2/3, 1, 1.5\}$ are overplotted with larger opacity
    for larger $b$.}
    \label{efficiency_alpha}
\end{figure}

Taking the integral
of $E_\Pi^{\varphi, i} (k)$ over $\ln k$, we can find the mean squared
anisotropic stresses, quantifying
the amount of energy
density
that is transferred from the source into anisotropic
stresses and, hence, is available to produce GWs.
Let us integrate \Eq{EPi_norms} to find,
\begin{equation}
\label{rho_Pi}
    \int_{-\infty}^\infty 
    E_\Pi^{\varphi, i} (K) \dd \ln K
    = \frac{\tilde \varphi_{\rm rms}^4}
    {\bar \rho^2} {\cal E}_\Pi^{\varphi, i} \,,
\end{equation}
where we have defined the efficiency factor ${\cal E}_\Pi^{\varphi, i}$ as
\begin{equation}
\label{eff_factors}
    {\cal E}_\Pi^{\varphi, i} = \frac{{\cal C}_i^\varphi}
    {{\cal A}^2_{ab} (\alpha)}
    \int_0^\infty K^3 \,
    \zeta_\Pi^{\varphi, i}
    (K) \dd \ln K \,.
\end{equation}
This expression allows us to define
a spectrum of the anisotropic
stresses normalized by its integral,
\begin{equation} \label{norm_int_zetaPi}
    \zeta_{\Pi, \rm int}^{\varphi, i} (K) \equiv
    \frac{
    {\cal C}_i^\varphi \, 
    }{{\cal A}_{ab}^2 (\alpha) \,
    {\cal E}_\Pi^{\varphi, i}}
     K^3  \,\zeta_\Pi^{\varphi, i} (K)\,,
\end{equation}
such that
\begin{equation}
    E_\Pi^{\varphi, i} (K)
    = \frac{\tilde \varphi_{\rm rms}^4}
    {\bar \rho^2} \, {\cal E}_\Pi^{\varphi, i} \, \zeta_{\Pi,
    \rm int}^{\varphi, i} (K)\,.
\end{equation}

The efficiency factors,
${\cal E}_\Pi^{\varphi, i}$, are shown in 
\Fig{efficiency_alpha} for a
range of
$\alpha = \{ 6/17, 0.75, 1, 2, 4\}$,
$a = \{3, 5\}$, and $b = \{2/3, 1, 1.5\}$.
Their mean values over the considered
range of parameters 
are
\begin{subequations}
\label{efficiencies_EPi}
\begin{align}
    {\cal E}_\Pi^{\rm vort} \simeq &\, (1 - q)^2
    \,
    (0.67 \pm 0.02) \,,
    \\ 
    {\cal E}_\Pi^{\rm comp} \simeq &\, q^2 
    \, (0.12 \pm 0.045)\,,
    \\ 
    {\cal E}_\Pi^{\rm mix} \simeq &\, 2q\, (1 - q)\, 
    (0.35 \pm 0.008)\,, \label{EPi_mix_eff}
    \\ 
    {\cal E}_\Pi^{\rm hel} \simeq &\, 
    \varepsilon^2 (1 - q)^2 \, (0.27 \pm 0.07)\,,
\end{align}
\end{subequations}
where we have assumed constant $q$ and $\varepsilon$.
The
errors indicate the variation of the efficiency
factors over the parameters $a$, $b$,
and $\alpha$.

For comparison, the mean squared value of the unprojected
stresses
is [see details in \App{autocorr_EPi},
in particular \Eq{mean_inhom_rhoT}]
\begin{equation}
    \bar \rho_T^\varphi = a^{-4} \int E_T^\varphi (K) \dd \ln K = \frac{5}{3}
    \tilde \varphi_{\rm rms}^4\,,
\end{equation}
such that the equivalent split of $\bar \rho_T^{\varphi}$ into components
as in \Eq{rho_Pi} is
\begin{equation} \label{efficiencies_ET}
    \frac{{\cal E}_T^{\rm vort}}{(1-q)^2}= 
    \frac{{\cal E}_T^{\rm comp}}{q^2} 
    =
    \frac{{\cal E}_T^{\rm mix}}{2q \,( 1 - q)}
    = \frac{\bar \rho_T^\varphi}{\tilde \varphi_{\rm rms}^4}
    =\frac{5}{3}\,,
\end{equation}
and
${\cal E}_T^{\rm hel} = 0$.
Unlike the efficiencies for the unprojected stresses, the anisotropic efficiencies ${\cal E}_\Pi^{\varphi, i}$ depend on the spectral
shape due to the appearance of the dot products
$z = \hat \pp \cdot \hat \kk$ and $\tilde z = \hat{\tilde \pp} \cdot \hat{\kk}$, introduced
by the $\pmb{\Lambda}$ projection tensor.
Furthermore, the projection tensor
breaks the symmetry that makes
the contribution from helicity to the unprojected
stresses to vanish, such that
helicity contributes to the mean squared value
of the anisotropic stresses. 
Similarly, we see in \Fig{p_pi_alpha} that the helical contribution
to the anisotropic stresses spectrum is also larger than
its contribution to the unprojected stresses spectrum
at wave numbers $K \gtrsim 1$.
For the vortical and mixed\footnote{Note that the mixed term is always suppressed by the
factor $2 q \, (1 - q) \leq \tfrac{1}{2}$ [see \Eq{EPi_mix_eff}], so it can be, at most,
half of the values shown in \Fig{efficiency_alpha}.} components, we find
that the efficiency is larger and
much less sensitive to the spectral
shape, with variations up to $\sim \! 30$ \%, while the 
compressional and helical components are much more
sensitive, with variations up to $\sim \! 40$ \%.
We find that steeper (larger $a$ or larger $b$) and sharper (larger $\alpha$) spectra lead to a
larger value of ${\cal E}_\Pi$.
On the other hand, even though the helical
component is negative at small $K$, hence,
reducing the amplitude of the total spectral branch proportional
to $K^3$, it becomes positive around the peak, and its net contribution to the
total mean squared stresses is always positive, hence yielding overall larger
efficiencies for helical fields.
This result is consistent with a larger
amplitude in the GW spectrum found in numerical simulations for
magnetic fields with a larger
fractional helicity \cite{RoperPol:2019wvy,Kahniashvili:2020jgm,Brandenburg:2021bvg,RoperPol:2021uol}.

Interestingly, as already shown
in ref.~\cite{Brandenburg:2021bvg},
vortical modes are significantly more efficient at
producing anisotropic stresses than compressional ones by a factor of ${\cal E}_\Pi^{\rm vort} (q = 0) / {\cal E}_\Pi^{\rm comp} (q = 1) \sim 6$.
This seems in contradiction with the 
results of numerical simulations of GW production,
which show that compressional motion is
more efficient
in producing GWs \cite{RoperPol:2018sap,RoperPol:2019wvy,Kahniashvili:2020jgm,Brandenburg:2021bvg,RoperPol:2021uol}.
The difference in the GW amplitude can then only be
due to the
dynamics of the UETC of the source and it needs to
compensate
for the different efficiency at producing anisotropic
stresses (see \Fig{efficiency_alpha}).
We address this issue in
\Sec{applications}, where the 
CiT model \cite{RoperPol:2022iel}, described in
\Sec{cit_model_sec}, is compared
to the GW production from sound waves and stationary
turbulence. There,
we show that the actual
dynamics of the UETC of the
source impacts the resulting
GW production when either the speed of propagation or of decorrelation in time is large,
leading to an enhancement of the GW signal with
respect to the predictions of the CiT model.
Therefore, when we consider sound waves in a pure radiation fluid, i.e., with an ultrarelativistic speed of sound $\cs^2 = 1/3$, they transfer
kinetic energy into GWs more efficiently
than decaying vortical turbulence with subrelativistic
decorrelation velocity $u_{\rm dcorr} \ll 1$.
These results clarify the origin of the different
efficiencies found when
converting the energy of the source
into GWs.
Furthermore, the presented model shows how
relativistic fluid velocities
can lead to a significant enhancement, similar
to the one found in sound waves, of the
GW signal from vortical turbulence, as we show in \Sec{applications}.

The
purely vortical or compressional components
have been extensively considered in the literature 
\cite{Kosowsky:2001xp,Caprini:2003vc,Kahniashvili:2005xe,Kahniashvili:2005qi,Gogoberidze:2007an,Caprini:2006jb,Gogoberidze:2007an,Caprini:2009yp,Caprini:2007xq,Caprini:2009fx,Hindmarsh:2013xza,Hindmarsh:2016lnk,Hindmarsh:2017gnf,Niksa:2018ofa,Hindmarsh:2019phv,RoperPol:2021uol,Dahl:2021wyk,Dahl:2024eup,Brandenburg:2021bvg,Kahniashvili:2020jgm,RoperPol:2022iel,Auclair:2022jod,RoperPol:2023dzg,RoperPol:2023bqa,Correia:2025qif}.
Let us now consider the additional contributions
to the anisotropic stresses spectrum when the velocity
field\footnote{Note that all the results presented for velocity fields apply
to describe magnetic fields if we omit the longitudinal
components (i.e., $q = 0$)
and to describe the
gradients of the scalar fields if we omit the normal
component (i.e., $q = 1$).}
has both normal and longitudinal components.
In this case, assuming small fluid perturbations,
\begin{equation}
    E_\Pi^u (K) =
    \Gamma^2 u_{\rm rms}^4
    \frac{{\cal C}_u}{{\cal A}_{ab}^2 (\alpha)}\, K^3 \, \zeta_\Pi^u (K)\,,
    \label{EPi_u}
\end{equation}
where
\begin{align}
    {\cal C}_u\, \zeta_\Pi^u  (K)  = 
    \sum_{i} {\cal C}_i^u \, \zeta_\Pi^{u, i} (K)\,,
    \label{zetapi_nohel}
\end{align}
with $i$ = \{vort, comp, mix, hel\}.
As before, the parameter ${\cal C}_u$ determines the value of the velocity
anisotropic stresses spectrum in the $K \to 0$ limit,
such that $|\zeta_\Pi^u(K \to 0)| = 1$.

For simplicity, let us now assume that the degree of compressibility
$q$ in \Eq{fraction_q} is a constant and that the velocity field
is non-helical (i.e., $\varepsilon_u = 0$).
Then, ${\cal C}_u$ becomes
[see \Eqs{Cuu_2}{c_i_norm}]
\begin{subequations}
\label{Cuuu3}
\begin{align}
    {\cal C}_u^{\rm nohel} =   &\,  c_u^{\rm nohel}
     \int_0^\infty   \frac{\zeta_u^2 (P)}{P^4} \dd P\,,
    \label{Cuuu} 
    \\
     &\, {\rm where \ \ }
    c_u^{\rm nohel} =
    \frac{4}{9} + \frac{1}{5} \biggl(q-\frac{1}{3}\biggr)^2\,.
\end{align}
\end{subequations}
The value $c_u^{\rm nohel}$ recovers
$c_{\rm vort} = 7/15$ and $c_{\rm comp} = 8 / 15$ for $q = 0$ and 1,
respectively,
and has a minimum value $4 / 9$ at
$q = 1 / 3$.

Similarly,
if the helicity fraction $\varepsilon_u$, defined in \Eq{helical_eps},
is a non-zero constant,
the resulting $c_u$, defined as in \Eqq{Cuuu3}, becomes\footnote{As discussed above, a constant $\varepsilon$
is not possible at all $K$, since the
helicity spectrum becomes proportional to $K^4$ or
$K^6$ in the $K \to 0$ limit, respectively for a mixed or fully vortical vector field, while the normal spectrum respectively goes as $K^3$ or $K^5$,
such that $\varepsilon(K) \sim K$ when $K \to 0$.
The deviation between the two spectra at small $K$
would lead to a small correction in front
of the $\frac{1}{3} \varepsilon_u^2 (1 - q)^2$
that we neglect in the following for simplicity. \label{constant_eps}}
\begin{equation}
    c_u = c_u^{\rm nohel} - \frac{1}{3} \varepsilon_u^2 (1 - q)^2\,,
\label{cu_hel_eq}
\end{equation}
where the negative sign comes from $\zeta_\Pi^{\rm hel}(K)$ being negative at small $K$ (see \Fig{efficiency_alpha}).
As the helicity fraction increases, $c_u$ decreases until
it reaches its minimum value for $\varepsilon_u^2 = 1$,
\begin{equation}
    c_u (\varepsilon_u^2 = 1) = \frac{2}{3}
    - \frac{2}{15} (2 - q)^2 \,,
\end{equation}
which varies from $2/15$ for $q = 0$ to $8/15$ for $q = 1$. 
Note that the helical spectrum for the velocity field is proportional to $\varepsilon_u(1-q)$ [see \Eq{hel_spec}], hence for $q=1$ there is no helicity, and the values of $c_u$ and $c_u^{\rm nohel}$ coincide and are both given by $c_{\rm comp}$.

\begin{figure}
    \centering
    \includegraphics[width=.9\columnwidth]{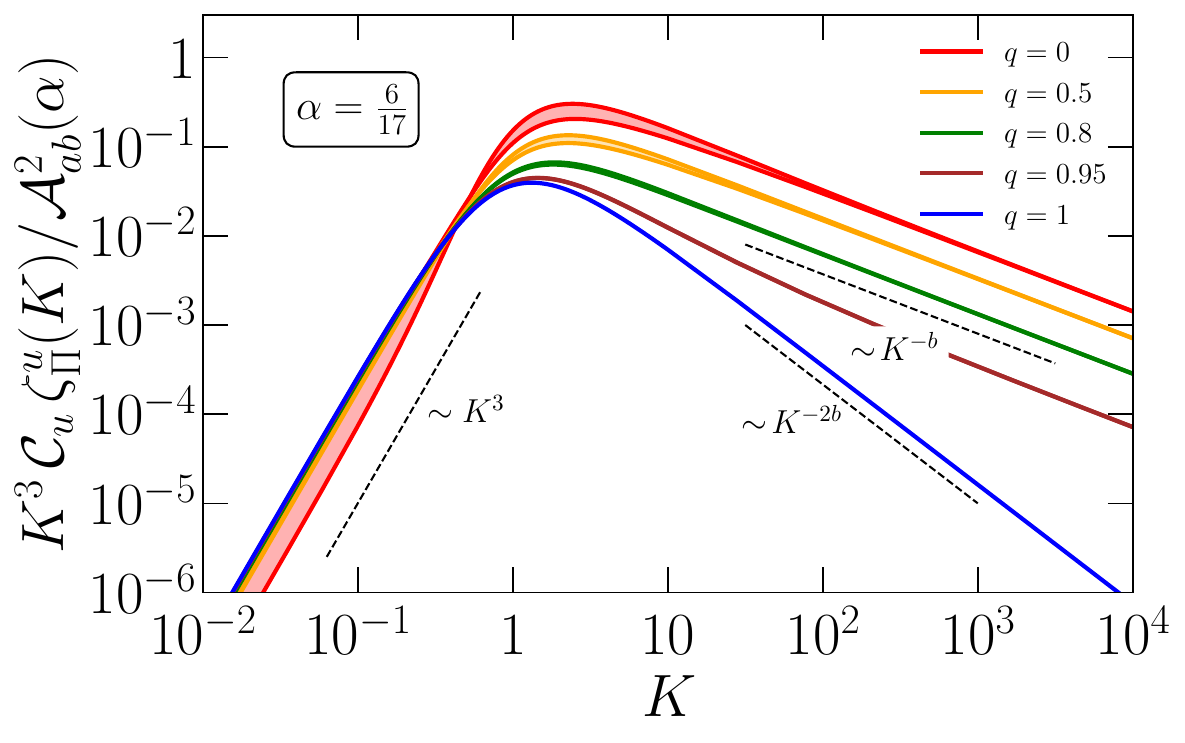}
    \includegraphics[width=.9\columnwidth]{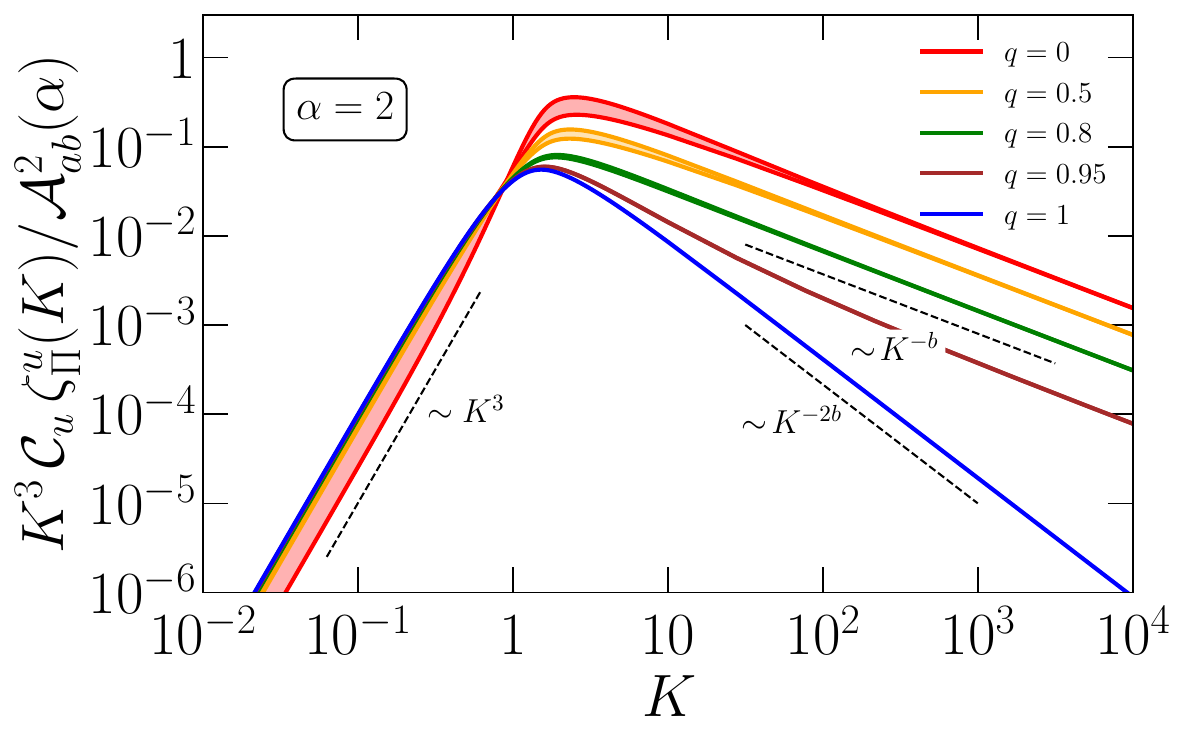}
    \caption{
    Normalized spectrum of the anisotropic stresses for
    a velocity von K\'arm\'an spectrum $\zeta_u (K)$ with $a = 5$, $b = 2/3$, and $\alpha = 6/17$ (upper panel) and $\alpha = 2$
    (lower panel)
    for different constant values of
    the degree of compressibility $q$
    (different colors)
    and fractional helicity $|\varepsilon_u| \in [0, 1]$
    (shaded regions).
    {\em Lower panel:} Efficiency factor ${\cal E}_\Pi^u$ characterizing the transfer
    of energy into anisotropic stresses,
    as a function of $q$ for
    different values of $|\varepsilon_u| \in [0, 1]$
    (shaded red region),
    indicating the largest efficiency for $q = 0$
    (vortical) and lowest one for $q = 1$ (compressional)
    as well as a generic trend of larger efficiency for
    larger $|\varepsilon_u|$.
    The value of $5/3$ found for the unprojected stresses
    is shown for comparison.
    All panels apply to magnetic fields with different fractional helicities $\varepsilon_B \in [0, 1]$
    when $q = 0$.
    }
    \label{zetaPivel}
\end{figure}

\begin{figure}
    \centering
    \includegraphics[width=\columnwidth]{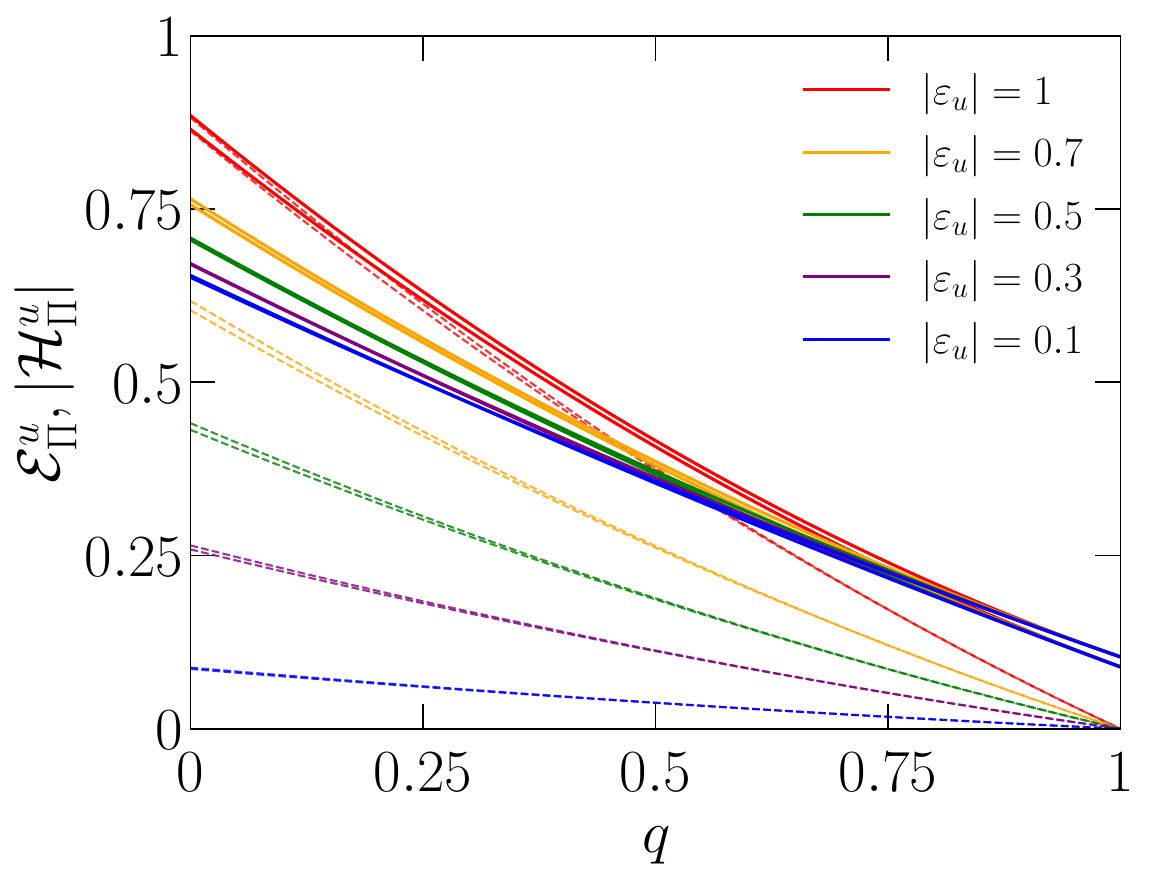}
    \caption{
    Efficiency factors ${\cal E}_\Pi^u$ (solid lines)
    and ${\cal H}_\Pi^u$ (dashed lines)
    characterizing the transfer
    of energy from unprojected into anisotropic stresses
    for the von K\'arm\'an spectrum with $a = 5$,
    $b = 2/3$ and two values of $\alpha = 6/17$ and 2,
    as a function of $q$ for
    different values of $|\varepsilon_u| \in (0, 1]$,
    indicating the largest efficiency for $q = 0$
    (vortical) and lowest one for $q = 1$ (compressional)
    as well as a generic trend of larger efficiency for
    larger $|\varepsilon_u|$.
    Note that the realizability condition requires
    $|{\cal H}_\Pi^u| < {\cal E}_\Pi^u$.
    Stresses from magnetic fields correspond
    to $q = 0$.
    }
    \label{efficiency_Pi_HPi}
\end{figure}

The normalized anisotropic stress spectra $K^3 \, {\cal C}_u \, \zeta_\Pi^u (K)$, divided by ${\cal A}_{ab}^2(\alpha)$, for
velocity fields with different degrees of compressibility $q = \{0, 0.5, 0.8, 0.95, 1\}$
and different fractional kinetic helicities $|\varepsilon_u| \in [0, 1]$
are shown in \Fig{zetaPivel} for the von K\'arm\'an spectrum with $a = 5$,
$b = 2/3$, and 
$\alpha = 6/17$ and $\alpha = 2$.
We note that even when the vorticity is subdominant
in the system, i.e., the degree of compressibility is
very close to one but not exactly one (e.g., 0.95),
then the spectrum of the stresses is
modified at large $K$, since $E_\Pi^{\rm comp} (K) \sim K^{-2b}$ decays faster than $E_\Pi^{\rm vort} (K) \sim K^{-b}$
at large $K$.
The shaded regions represent variation of $|\varepsilon_u|$ from 0 to 1.
For low $K$, the helical contribution is negative
and, hence, smaller amplitudes correspond to larger
fractional helicities.
At wave numbers at around and above the peak, $K \gtrsim 1$, the helicity contribution becomes positive.
In this range, the maximum amplitude corresponds
to $\varepsilon_u = \pm 1$.
In \Fig{efficiency_Pi_HPi}, we show the integrated
efficiency ${\cal H}_\Pi^u$ [see \Eq{eff_factors}].

\subsection{Antisymmetric spectrum of the anisotropic stresses $H_\Pi$}
\label{HPi_sec}

In the previous section, the symmetric UETC of
the anisotropic stresses, $E_\Pi (k)$, has been computed.
In this section, we compute the antisymmetric UETC of
the anisotropic stresses, $H_\Pi (\tau_1, \tau_2, k)$,
which sources the chiral spectrum of GWs
[see description below \Eq{eq:OmGWt}].
For this purpose, we now project \Eq{Tij_Fourier} with the
tensor ${\cal A}_{ijlm}$ (see \App{autocorr_EPi} for
details).

We find that the antisymmetric spectrum of a vector source $\varphi = \{X, B\}$,
$H_\Pi^\varphi (k)$, is only non-zero when the fractional helicity of the
source is non-zero.
For velocity fields,
it can be decomposed into a compressional ($i = $ comp)
and a vortical ($i =$ vort) component,
while it has a unique vortical component for magnetic fields,
\begin{subequations}
\begin{align}
    H_\Pi^u (k) = &\, H_\Pi^{u, {\rm vort}} (k) +  H_\Pi^{u, {\rm comp}} (k)\,,
    \\ H_\Pi^B (k) = &\, H_\Pi^{B, {\rm vort}} (k)\,.
\end{align}
\end{subequations}
Even though $H_\Pi^u (k)$ contains a compressional component,
it vanishes for a purely compressional field, since then
$\varepsilon_u \to 0$.
Therefore, the presence
of a compressional component always requires that
the vortical component is non-zero.

The two components of the antisymmetric anisotropic
stresses spectrum are given in \Eqs{HPi_vort}{HPi_comp}
(see also ref.~\cite{Caprini:2003vc}
for $H_\Pi^{\rm vort} (k))$,
\begin{subequations}
\label{HPi_hel}
\begin{align}
    a^4  H_\Pi^{\rm vort}  (k) =
    \frac{k^2}{2 \bar \rho^2}
    &\, \int_0^\infty   \frac{\dd p}{p^2} E_\N (p) \nonumber \\  \times &\,
    \int_{|k - p|}^{k + p} \frac{\dd \tilde p}{\tilde p^2} H (\tilde p) \, z \, (1 + \tilde z^2) 
    \,,
    \label{HPi_vort_txt} \\
    a^4 H_\Pi^{\rm comp}  (k) =  \frac{k^2}
    {\bar \rho^2}
     &\, \int_0^\infty\frac{\dd p}{p^2} E_\L (p) \nonumber \\ \times&\, 
    \int_{|k - p|}^{k + p} \frac{\dd \tilde p}{\tilde p^2} H (\tilde p)
    \, z  \,(1 - \tilde z^2)\,.
    \label{HPi_comp_txt} 
\end{align}
\end{subequations}

The anisotropic stresses satisfying the TT properties
can be decomposed into
two polarization modes as
\begin{equation}
    \Pi_{ij} (\kk) = \sum_{\lambda = \pm} e_{ij}^\lambda \Pi_\lambda (\kk)\,, 
\end{equation}
in the basis \cite{Caprini:2003vc,Hu:1997hp}
\begin{equation}
    e_{ij}^\pm = (\ee_1 \pm  i \, \ee_2)_i
    \times (\ee_1 \pm i\, \ee_2)_j\,.
\end{equation}
Then, ref.~\cite{Caprini:2003vc} shows that the ratio of the
antisymmetric to symmetric
spectral functions of the anisotropic stresses is
\begin{align} \label{polar_Pi}
    {\cal P}_\Pi (k) \equiv &\, \frac{H_\Pi(k)}{E_\Pi (k)}
    \nonumber \\ = &\, \frac{|\Pi_+ (\kk)|^2 - |\Pi_- (\kk)|^2}
    {|\Pi_+ (\kk)|^2 + |\Pi_- (\kk)|^2} \in [-1, 1]\,.
\end{align}
Similarly to the realizability condition for the sourcing
fields, $|H(k)| \leq E(k)$, this implies that $|H_\Pi (k)|
\leq E_\Pi (k)$.
The polarization ${\cal P}_\Pi (k)$ characterizes the
degree of circular polarization of the GWs,
${\cal P}_{\rm GW} (k)$.
In particular, both functions become identical
in the CiT model, as we show in \Sec{circ_pol}.
 
Using the normalization of \Eqq{norm_generic} for the
spectral components of the source,
as we have
done in \Eq{EPi_norms} for the components of the symmetric spectrum of the anisotropic
stresses,
we find [see \Eq{normalization_H_comps}]
\begin{align}
    H_\Pi^{\varphi, i} (K) = \frac{\tilde \varphi_{\rm rms}^4}
    {\bar \rho^2}
    \frac{{\cal C}_{i}^{\varphi, \H}}{{\cal A}_{ab}^2} K^4 \,
    \zeta_\Pi^{\varphi, \H, i} (K)\,,
    \label{normalized_HPi_i}
\end{align}
with $i =$ \{vort, comp\}.
We note that, unlike the symmetric stress tensor,
the antisymmetric stress tensor components are proportional to $K^4$ 
in the limit $K \to 0$,
as expected from causality  \cite{Durrer:2003ja}.

The normalized $\zeta_\Pi^{\H, i} (K)$ are given in \Eqs{zetaPi_Hvort}{zetaPi_Hcomp}, and the
prefactors ${\cal C}^\H_i$ are again chosen such
that $|\zeta_\Pi^{\H, i} (K)| \to 1$ when $K \to 0$,
given in \Eqq{zero_K_ampl_H}
\begin{subequations}
\begin{align}
    {\cal C}^\H_{\rm vort} = &\, 
    \frac{2}{15} \biggl| \int_0^\infty [1 - q(P)]^2 \,
    \varepsilon (P) \, \frac{\zeta^2 (P)}{P^5} \dd P
    \biggr| \,, \label{CH_vort} \\ 
    {\cal C}^\H_{\rm comp} = &\, \frac{2}{5}
    \biggl| \int_0^\infty q (P) \, [1 - q(P)]  \nonumber \\& 
    \hspace{25mm}
    \times \, \varepsilon (P)\,
    \frac{\zeta^2 (P)}{P^5} \dd P \biggr|\,.
\end{align}
\end{subequations}
When $q$ and $\varepsilon$ are both constants,
and assuming that $\zeta(P)$ is a smoothed broken
power law as given in \Eq{powerlaw_model}, these coefficients can be expressed
as
\begin{align} \label{CH_0}
    {\cal C}^\H_i = &\, c^\H_i \int_0^\infty \frac{\zeta^2 (P)}{P^5}
    \dd P = c_\H^i I_{2a, 2b; -5} (\alpha/2)
    \nonumber \\ = &\, c^\H_i 
    \biggl(\frac{a +b}{b} \biggr)^{2 \over \alpha}
    \biggl(\frac{a}{b}\biggr)^{-\frac{2a - 4}{\alpha(a + b)}}
    \nonumber \\ &\, \hspace{20mm} \times 
    \frac{\Gamma\bigl[\frac{2a - 4}{\alpha (a + b)}
    \bigr] \Gamma\bigl[\frac{2b + 4}{\alpha (a + b)}
    \bigr]}{\alpha (a + b) \, \Gamma\bigl[\frac{2}{\alpha}\bigr]}\,,
\end{align}
with
\begin{equation} \label{ch_u}
    c^\H_i = \frac{2}{15} |\varepsilon| \, (1 - q) \,
    \Bigl\{ 1 - q,
    \ 3 q \Bigr\}
\end{equation}
for $i  = \{$vort, comp$\}$.
The dependence of ${\cal C}_{ab}^{i, \H}(\alpha)/c_i^\H$ is
shown in \Fig{CC_H_ab_alp}.

In analogy to \Eq{eff_factors}, we define an efficiency
factor characterizing the conversion
of the mean stress of the source into
an antisymmetric component of the anisotropic stresses,
\begin{equation}
    {\cal H}_\Pi^{\varphi, i} = \frac{{\cal C}_i^{\varphi, \H}}
    {{\cal A}_{ab}^2 (\alpha)} \int_0^\infty 
    K^3 \, \zeta_\Pi^{\varphi, \H, i} (K) \dd \ln K\,,
\end{equation}
which also allows us to define a function $\zeta_{\Pi, \rm int}^{\varphi, \H, i} (K)$ normalized by its integral,
as in \Eq{norm_int_zetaPi}, such that $\int \zeta_{\Pi, \rm int}^{\varphi, \H, i} (K) \dd \ln K = 1$.
Due to the property in \Eq{polar_Pi}, 
$|\HH_\Pi^\varphi| < {\cal E}_\Pi^\varphi$ after summing over the corresponding
components.
We note that the equality can never be exhausted because
causality requires $\varepsilon \sim K$ when $K \to 0$.

The helical spectrum of the anisotropic stresses 
for a generic velocity field
is
\begin{equation}
    H_\Pi^u (K) = 
    \Gamma^2 u_{\rm rms}^4 \frac{{\cal C}_u^\H}{{\cal A}_{ab}^2 (\alpha)}
    K^4 \, \zeta_{\Pi}^{u, \H} (K) \,,
    \label{HPi_u}
\end{equation}
where
\begin{align}
    {\cal C}_u^\H \, \zeta_{\Pi}^{u, \H} (K) = 
    \sum_i {\cal C}_i^\H \, \zeta_\Pi^{u, \H, i} (K)\,,
    \label{zetaHPi_u}
\end{align}
with $i = $ \{vort, comp\}.
Note that the helical spectrum of the anisotropic
stresses is proportional to the helicity fraction
and the vortical component of the source.
Therefore, if the source has
a constant
positive or negative helicity at all modes $k$, it will induce
positive or negative stresses $H_\Pi^u (K)$,
and GW polarization
\cite{Kahniashvili:2005qi,Kahniashvili:2020jgm,RoperPol:2021uol}.
For constant $q$ and $\varepsilon_u$, then ${\cal C}_u^\H$ is
given by \Eq{CH_0} with $c^\H_u = \frac{2}{15} |\varepsilon_u|
(1 - q)(1 + 2 q)$, obtained adding up the two components
in \Eq{ch_u}.

\begin{figure}
    \centering
    \includegraphics[width=\linewidth]{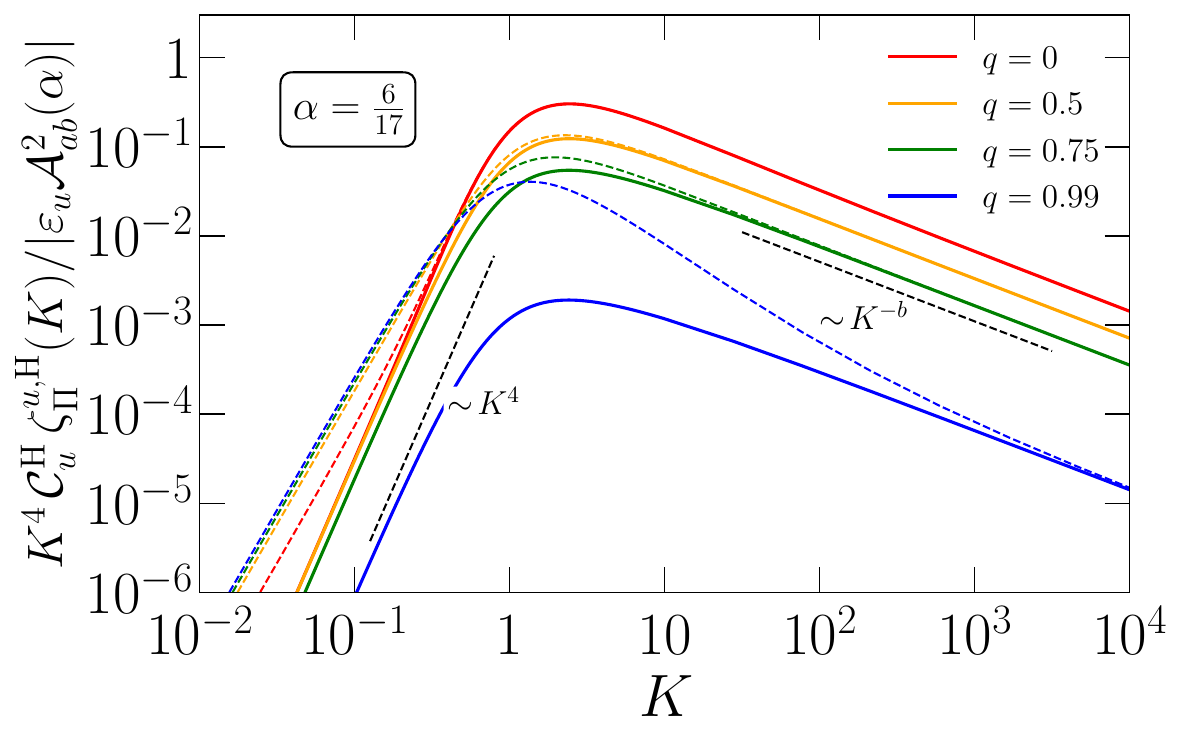}
    \includegraphics[width=\linewidth]{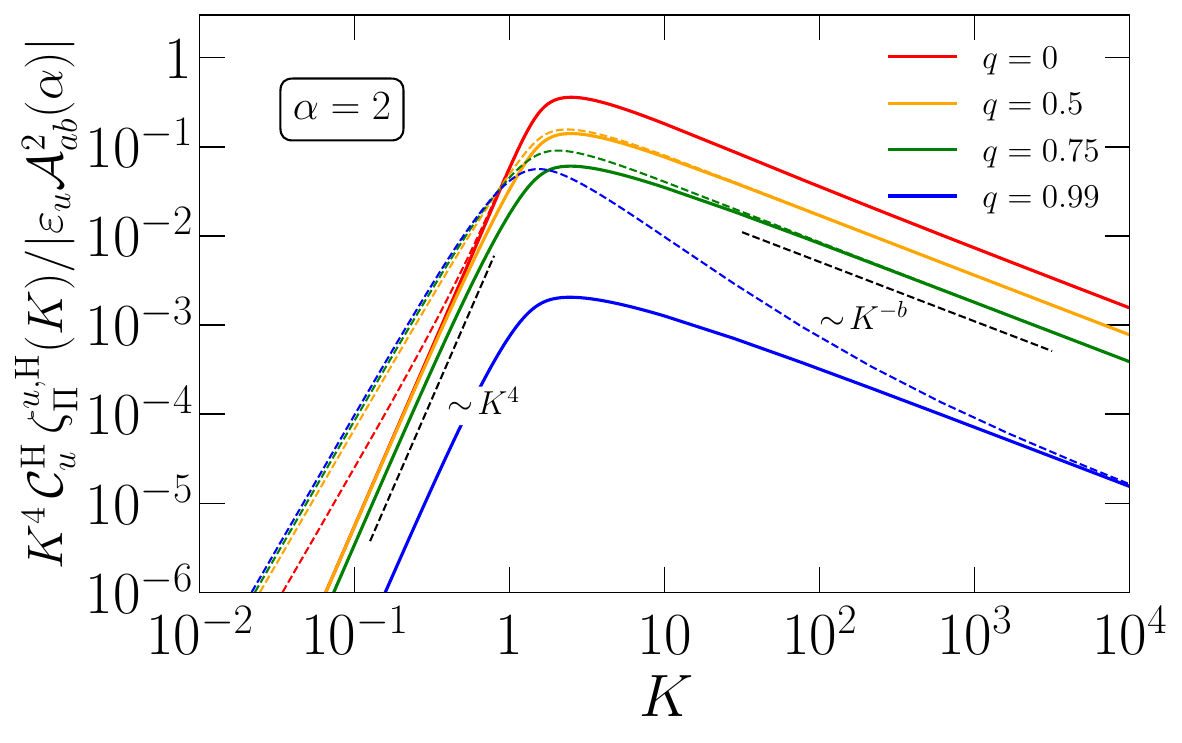}
    \caption{
    Normalized helical spectra of the anisotropic stresses,
    divided by a constant fractional helicity,
    $K^{4} \, \zeta_\Pi^{u, \H} (K) / |\varepsilon_u|$
    and normalized with ${\cal A}_{ab}^2 (\alpha)$,
    corresponding to a velocity field
    with a constant degree of compressibility
    $q \in [0, 1)$.
    Notice that for $q=1$, since there is no helicity [see \Eq{hel_spec}], we have $\zeta_\Pi^{u, \H}
    (K) = 0$.
    The velocity spectrum is
    a smoothed broken power law with $a = 5$, $b = 2/3$, and $\alpha = 6/17$ (upper panel)
    and $\alpha = 2$ (lower panel).
    The helicity spectrum for a magnetic field source, $\zeta_\Pi^{B, \H} (K)$,
    corresponds to the
    purely vortical case, i.e., $q = 0$.
    Dashed lines correspond to the symmetric
    spectra $K^3 \, {\cal C}_u^\H 
    \, \zeta_\Pi^u (K) / {\cal A}_{ab}^2 (\alpha)$
    shown in \Fig{zetaPivel},
    with $|\varepsilon_u| = 1$.
    Scaling lines $\sim K^4$ and $\sim K^{-b}$
    are shown at the same positions in both panels
    to allow for a direct comparison.
    }
    \label{zetaPihel_vel}
\end{figure}

The resulting normalized
spectra $K^4 \, {\cal C}_u^\H \,
\zeta_\Pi^{u, \H} (K)$, divided by  ${\cal A}^2$
and $|\varepsilon|$,
for velocity fields
with different degrees of compressibility are shown
in \Fig{zetaPihel_vel}.
The efficiency factor $|{\cal H}^u_\Pi|$ for different
values of $q$ and $\varepsilon_u$ for the von K\'arm\'an
spectrum with $a = 5$, $b = 2/3$, and two values
of $\alpha = 6/17$ and 2, is shown in \Fig{efficiency_Pi_HPi}.
The spectra and efficiencies for
magnetic fields correspond to the fully vortical case
($q = 0$).

\subsection{Smoothed broken power law fits}
\label{bpl_anisotropic}

As we have discussed, the spectrum of the anisotropic stresses
sourcing GWs (energy density and polarization) can in general be decomposed
as a superposition of terms 
reflecting the contributions from different components
of the spectrum of the source (compressional, vortical, helical).
Each of these terms
can be described by normalized spectral functions
$\zeta_\Pi (K)$ and
$\zeta_\Pi^\H (K)$ following the normalization in \Eqs{EPi_norms}{normalized_HPi_i},
which take values of one when $K \to 0$.

\begin{figure*}[t]
    \centering
    \includegraphics[width=.329\textwidth]{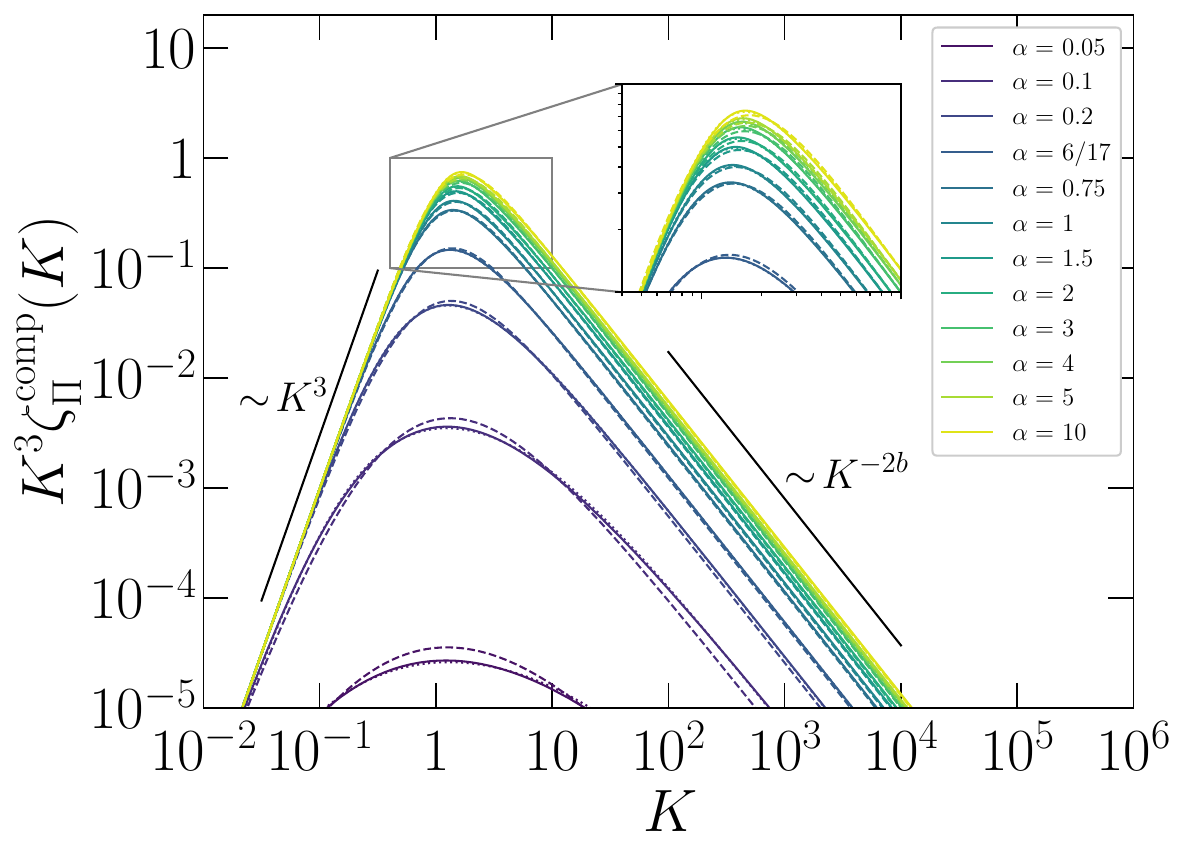}
    \includegraphics[width=.329\textwidth]{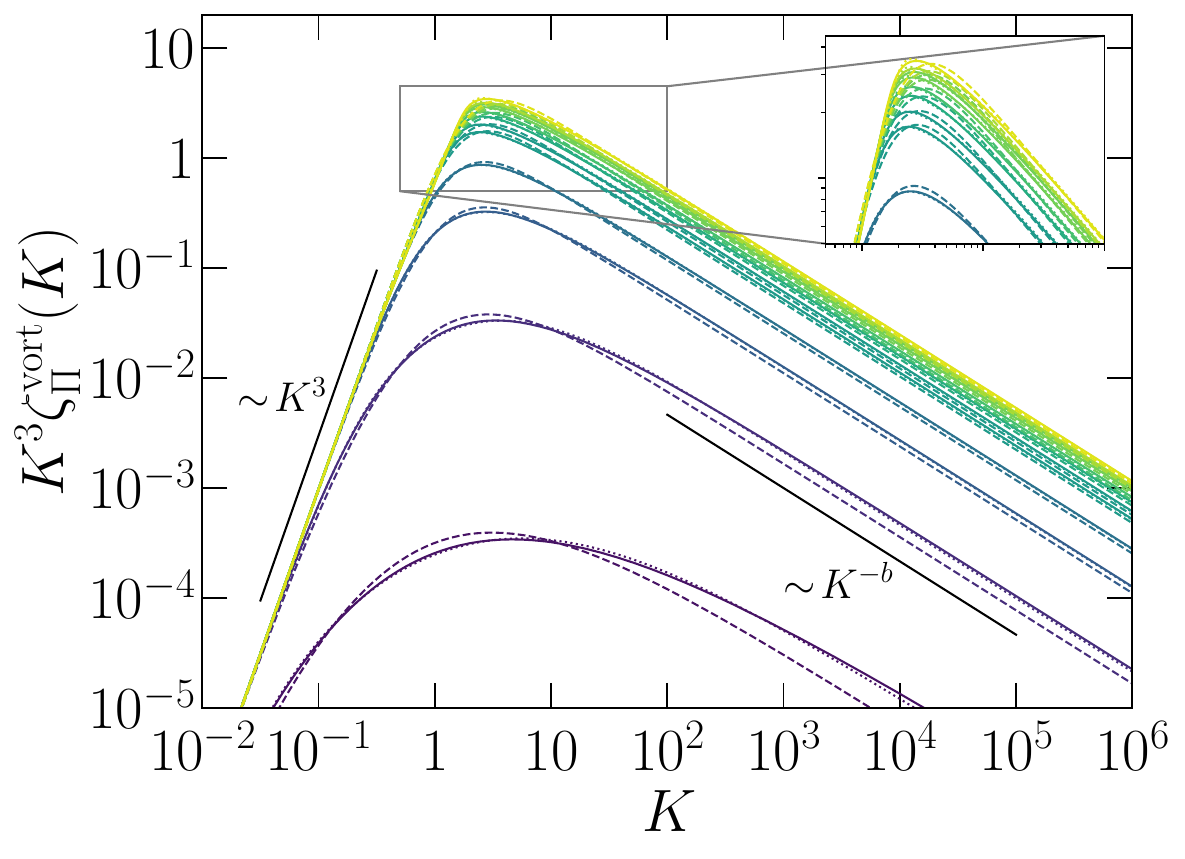}
    \includegraphics[width=.329\textwidth]{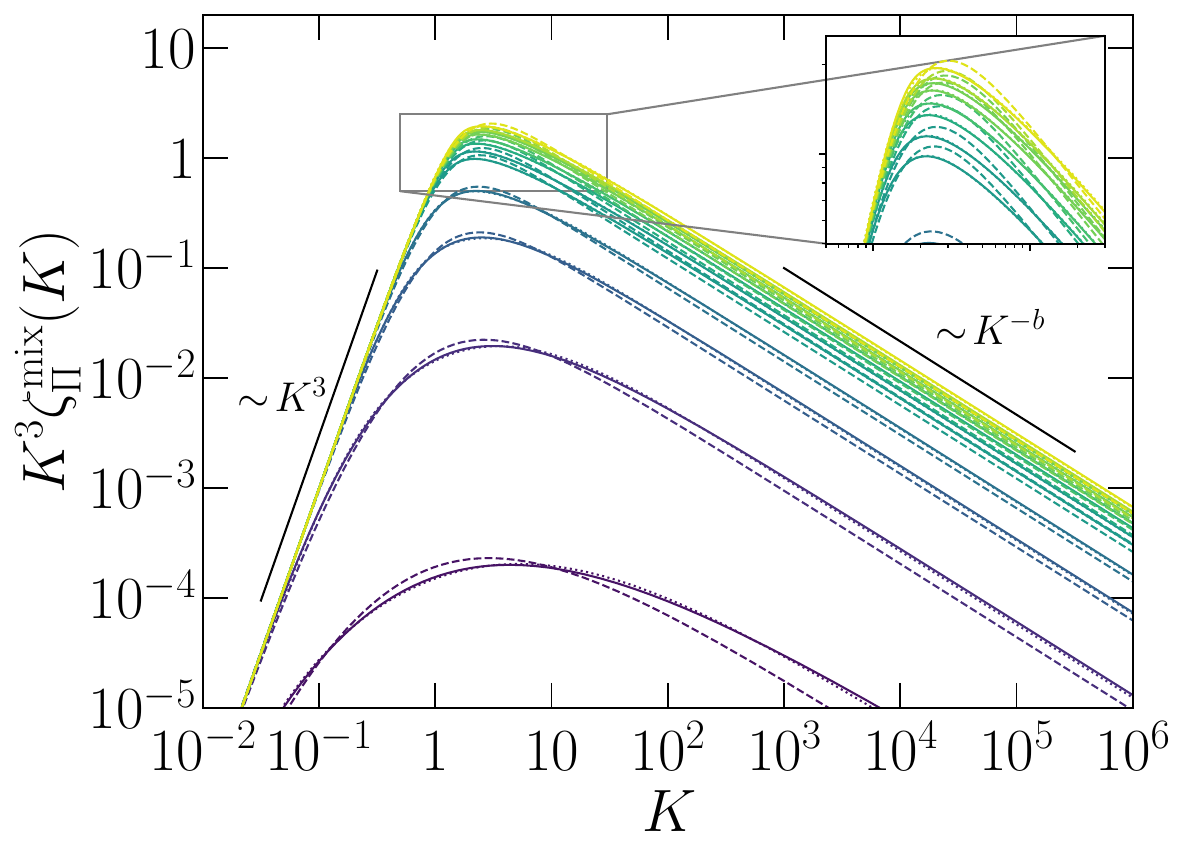}
    \includegraphics[width=.329\textwidth]{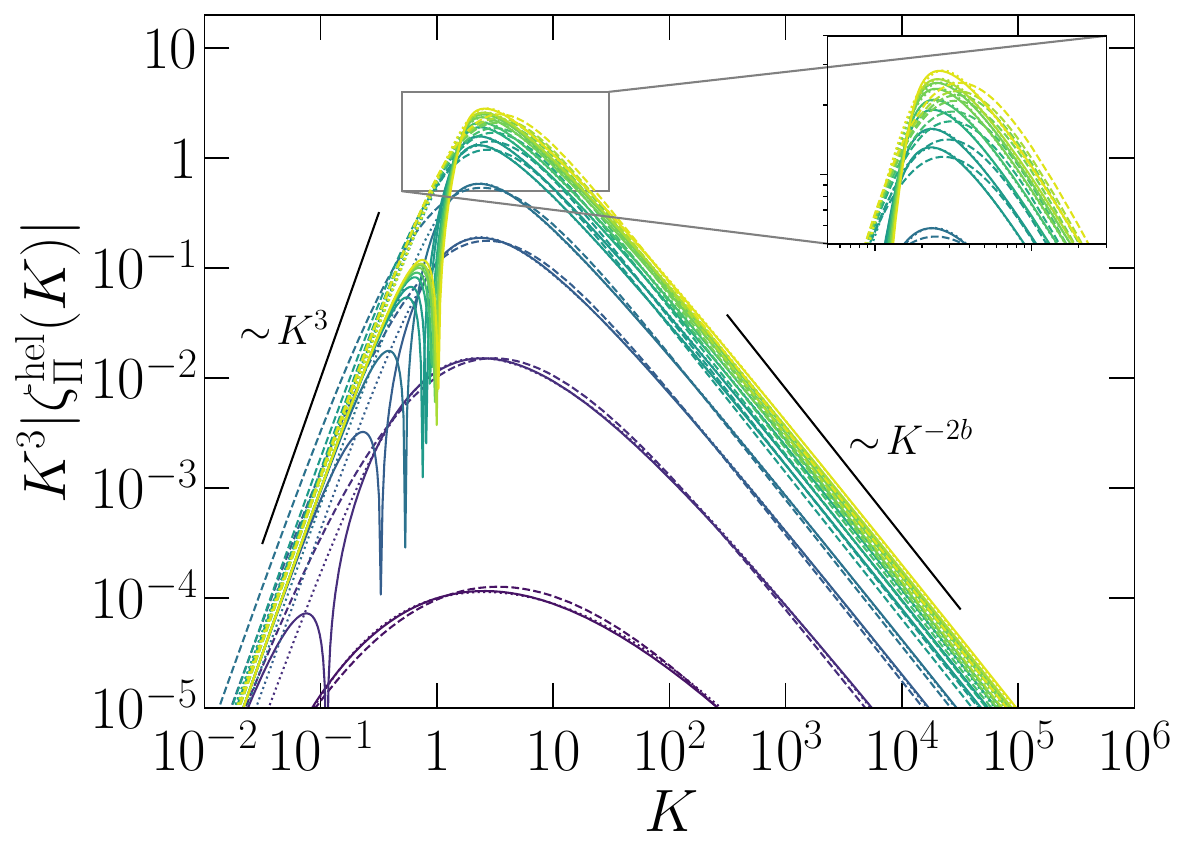}
    \includegraphics[width=.329\textwidth]{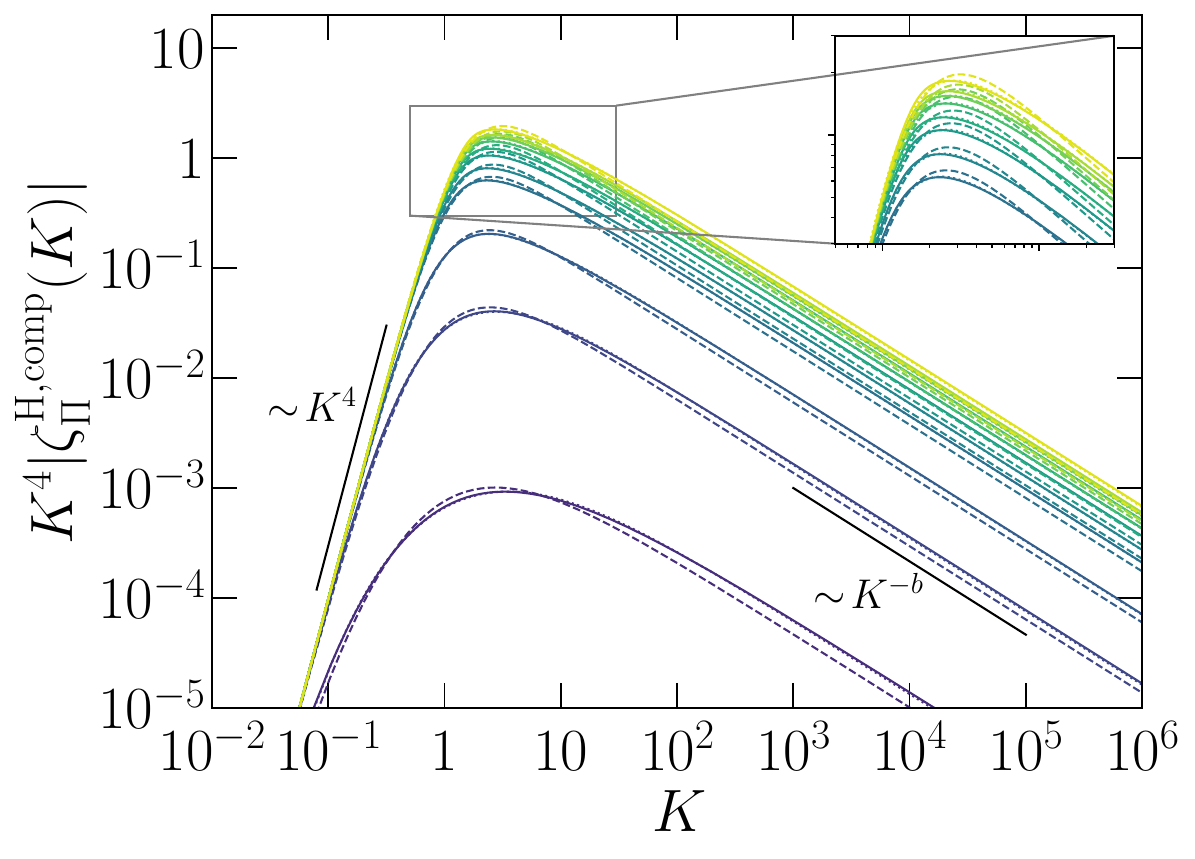}
    \includegraphics[width=.329\textwidth]{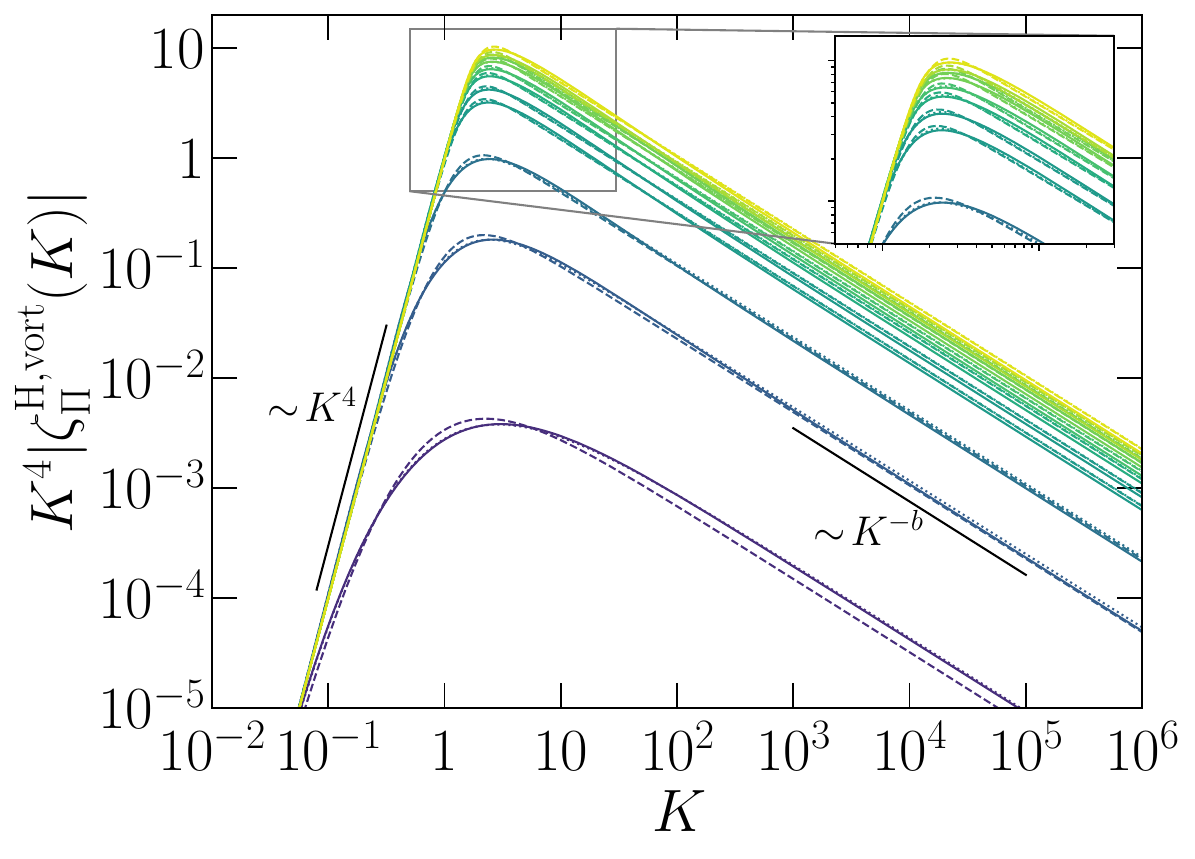}
    \caption{
    Compressional (top left panel), vortical
    (top middle panel), mixed (top right panel), helical (bottom left panel)
    components of the normalized spectra
    of the anisotropic stresses $K^3 |\zeta_\Pi^i (K)|$, and vortical (bottom middle panel) and compressional (bottom right panel) components of the
    normalized helical spectra of the
    anisotropic stresses $K^4 |\zeta_\Pi^{i, \H} (K)|$.
    We take a smoothed broken power law for 
    the spectrum of the source with $a = 5$ and $b = 2/3$, and different values of $\alpha$.
    Solid lines indicate the numerical values [computed from \Eqq{normalized_zetaPis}] while dashed lines are the fits provided using
    \Eqq{zetaPi_fits} and the values in \Tab{parameters_fits}.
    For increased accuracy, a double power
    law fit is included in dotted lines.}
\label{zetaPi_fits_fig}
\end{figure*}

As it can be seen respectively in 
\Figs{p_pi_alpha}{zetaPihel_vel},
all of the contributions to the symmetric 
and antisymmetric 
spectra, $\zeta_\Pi (K)$ and $\zeta_\Pi^\H(K)$, become
constant at $K \ll 1$ and then decay proportional to a power law
at $K \gg 1$ whose exponent only depends on the power law
decay exponent of the spectrum of the source, $b$.
This motivates us to consider the following broken power law fits, analogous
to the model given in \Eq{powerlaw_model}, but adapted for $a = 0$,
to describe each component separately,
\begin{subequations}
\label{zetaPi_fits}
\begin{align}
    |\zeta_\Pi^i (K)| &= \bigl[1 + (K/K_\Pi^i)^{\alpha_\Pi^i}\bigr]^{-b_\Pi^i/\alpha_\Pi^i}\,, \\
    |\zeta_\Pi^{\H, i}(K)| & = \bigl[1+(K/K_\Pi^{\H,i})^{\alpha^{\H, i}_\Pi} \bigr]^{-b_\Pi^{\H,i}/\alpha_\Pi^{\H,i}}\,,
\end{align}
\end{subequations}
where the large-$K$ slope of the stresses spectrum is
\begin{subequations}
\label{b_pi_definitions}
\begin{align}
     &\, b_\Pi^{\rm vort, mix} = b + 3\,, \\ 
     &\, b_\Pi^{\rm comp} = \min(2b + 3, b + 5)
     \quad b \neq 2 \,,
    \\ 
    &\, b_\Pi^{\rm hel} = \min(2b + 3, b + 4)
    \ \ \, \quad b \neq 1 \,,
    \\ 
    &\, b_\Pi^{\H, \rm comp, vort} =  b + 4\,.
\end{align}
\end{subequations}

The resulting fits found for the various 
components of the stresses are shown in \Fig{zetaPi_fits_fig} for the von K\'arm\'an
spectrum with $a = 5$, $b = 2/3$, and different
values of $\alpha$.
Note that $b_\Pi^i$ and $b_\Pi^{\H,i}$ are completely determined by the small-scale broken power law slope $b$, as in \Eqq{b_pi_definitions}, hence the only fitting parameters are $K_\Pi^i, \alpha_\Pi^i$ and $K_\Pi^{\H,i}, \alpha_\Pi^{\H,i}$.
The values
of the fitting parameters $\alpha_\Pi^i$ and $K_\Pi^i$
for each component are provided
in \Tab{parameters_fits} for $a = 5$, $b = 2/3$
and $1$, describing von K\'arm\'an turbulence
and sound waves, respectively, and a range of values
of $\alpha$.
For all the fits, we are assuming that $q$ and $\varepsilon$ are constants,
and each spectral component is divided by the prefactors
in \Eqq{norm_generic}, as we have done in \Fig{p_pi_alpha}.
In particular, for $\alpha = 6/17$ the values of 
$\alpha_\Pi^{\rm vort}$ and $K_\Pi^{\rm vort}$
to fit the vortical component
are approximately 
2 and 1.3, respectively.
This fit\footnote{Note that in previous work, the
spectral shape is defined in unit of $k$, instead
of $\ln k$. Hence, the value reported
in previous work is $\tilde K_\Pi \equiv K_\Pi/K_{-1}^\ast \simeq 2.2$ [see details in \App{appendix_sbpl}, in particular \Eq{Kstar_m1}].} 
has been used in refs.~\cite{RoperPol:2023bqa,Caprini:2024hue,EPTA:2023xxk},
where the value of $\tilde K_\Pi \equiv K_\Pi / K_{-1}^\ast
\simeq 2.2$ was used (see Eq.~17 in ref.~\cite{RoperPol:2023bqa}).
The helical component $\zeta^{\rm hel}_\Pi(K)$
takes negative
values at $K \lesssim 1$ and then changes sign around $K \sim 1$, hence we provide a fit for the envelope of its absolute value.

\begin{figure}
    \centering
    \includegraphics[width=\linewidth]{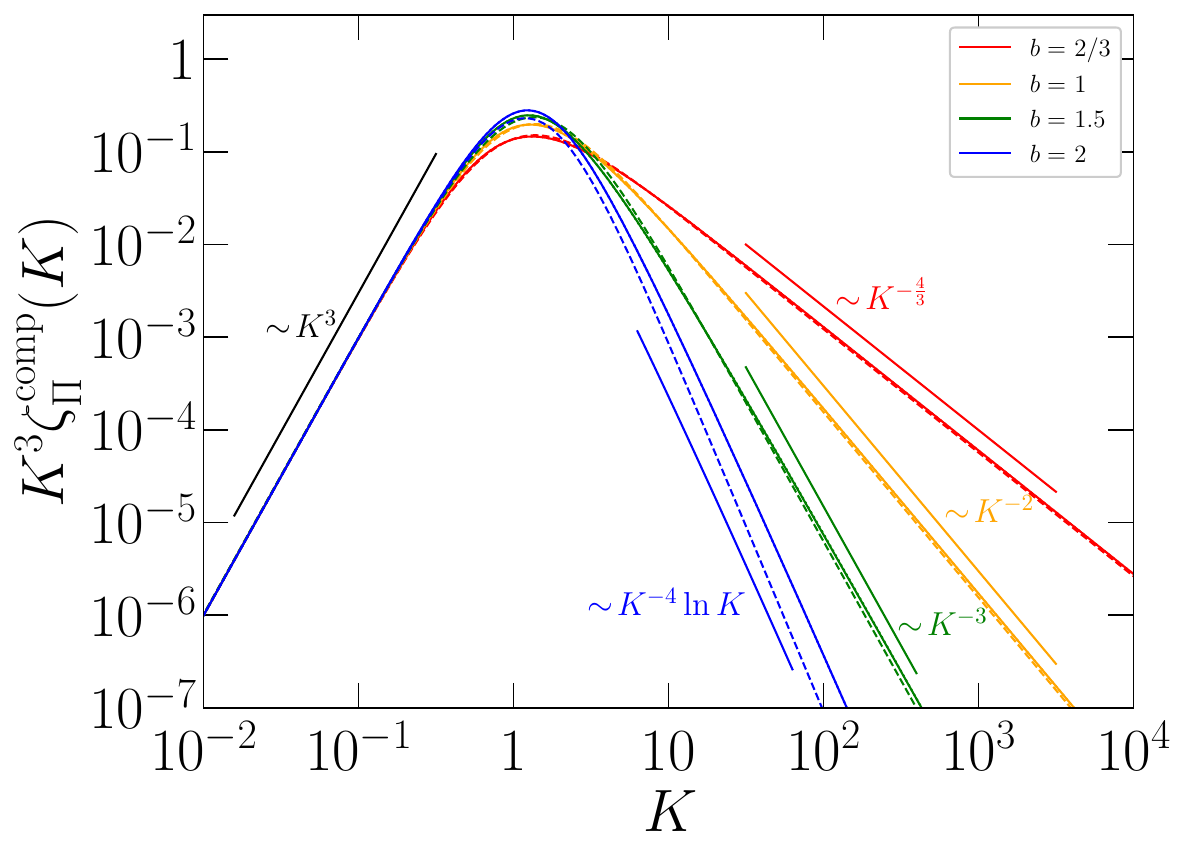}
    \includegraphics[width=\linewidth]{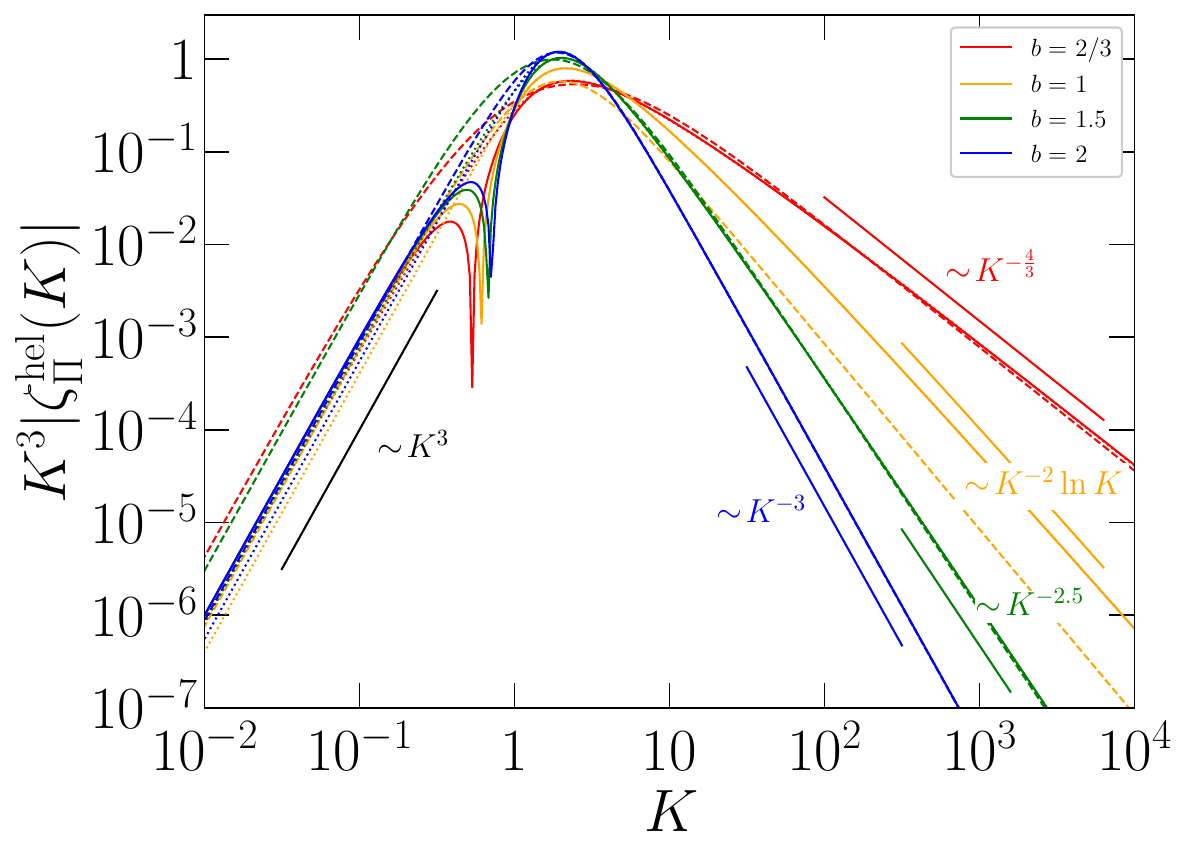}
    \caption{Normalized anisotropic stresses
    compressional (upper panel)
    and helical (lower panel) components
    for different values of $b$.
    At large $K$, the compressional component
    shows a scaling $K^3 \zeta_\Pi^{\rm comp} (K) \sim\!K^{-2b}$ for $b < 2$, $\sim\!K^{-4} \ln K$ for $b = 2$,
    and $\sim\!K^{-b-2}$ for $b > 2$.
    The helical component scales as $K^3
    \zeta_\Pi^{\rm hel} (K) \sim\!K^{-2b}$
    for $b < 1$, $\sim\!K^{-2} \ln K$ for $b = 1$,
    and $\sim\!K^{-b - 1}$ for $b > 1$.}
    \label{asymptotic_log_corr}
\end{figure}

As mentioned in the caption of \Fig{p_pi_alpha},
the asymptotic scaling of $\zeta_\Pi^i (K)$
for $i=\{$comp, hel$\}$
at large
$K$ is not a power law when 
$b_{\rm comp} = 2$ and $b_{\rm hel} = 1$
due to the appearance of logarithmic
corrections $\propto \ln K$.
The scaling at large $K$ is then modified
from an asymptotic power law with $b_\Pi^i = 2b + 3$
for smaller $b$ to $b_\Pi^{\rm comp} = b + 5$
and $b_\Pi^{\rm hel} = b + 4$ at larger $b$
as we show numerically in \Fig{asymptotic_log_corr}.
We show the compressional and helical components
for different values of $b = \{2/3, 1, 1.5, 2\}$.

The fits with a single power law in \Eqq{b_pi_definitions}, although qualitatively
accurate, can present large relative
errors of up to order one at some values of $K$
when we scan the fits over several values of $a = 3$
and $5$, $b = \{2/3, 1, 1.5, 2\}$ and $\alpha \in [0.05, \infty)$.
Extending this fit to a double broken power law,
multiplied by the function $1 + \ln (1 + K / K_L)$,
allows us to add a $\ln K$ correction at large
$K \gg K_L$ for the special cases
$b_{\rm comp} = 2$ and $b_{\rm hel} = 1$,
and reduces the relative
error of the fit below 10\% for all
scanned parameters.
The numerical computations for all components,
the scan for all values of the fits, and values
of the fitting parameters are included in the
public Python package \href{https://github.com/AlbertoRoper/cosmoGW}{\sc CosmoGW} \cite{cosmogw_manual,CosmoGW_GH}, with a tutorial
to reproduce these results that will be publicly
available
after the publication of this manuscript.

\begin{table*}
    \centering
    \begin{tabular}{|c|c|c||c|c||c|c||c|c||c|c||c|c||c|c|}
    \hline
    $a$ & $b$ &
    $\alpha$ & $K_\Pi^{\rm vort}$ & $\alpha_\Pi^{\rm vort}$ & $K_\Pi^{\rm comp}$ & $\alpha_\Pi^{\rm comp}$  & $K_\Pi^{\rm mix}$ & $\alpha_\Pi^{\rm mix}$  & $K_\Pi^{\rm hel}$ & $\alpha_\Pi^{\rm hel}$ & $K_{\Pi}^{\rm H, vort}$ & $\alpha_{\Pi}^{\rm H, vort}$ & $K_{\Pi}^{\rm H,  comp}$ & $\alpha_{\Pi}^{\rm H, comp}$  \\ 
    \hline \hline
\ \ 5\ \  & \ 2/3 \ & \ $0.05$ \ & \ 0.194 \ & \ 0.543 \ &  \ 0.207 \ & \ 0.451 \ & \ 0.167 \ & \ 0.528 \ &
\ 0.393 \ & \ 0.379 \ & \ 0.084 \ & 0.534 & 0.067 & 0.478 \\
& &$0.1$ & 0.593 & 0.951 & 0.472 & 0.788 & 0.506 & 0.913 & 0.850 & 0.631 & 0.388 & 1.017 & 0.303 & 0.838 \\
& &$0.2$ & 1.022 & 1.552 & 0.723 & 1.294 & 0.869 & 1.476 & 1.123 & 0.920 & 0.845 & 1.895 & 0.649 & 1.338 \\
& &$6/17$ & 1.290 & 2.060 & 0.876 & 1.779 & 1.093 & 1.967 & 1.144 & 1.069 & 1.180 & 2.978 & 0.896 & 1.777 \\
& &$0.75$ & 1.541 & 2.428 & 1.024 & 2.293 & 1.299 & 2.378 & 1.562 & 1.441 & 1.503 & 4.582 & 1.131 & 2.176 \\
& &$1$ & 1.616 & 2.457 & 1.069 & 2.407 & 1.359 & 2.436 & 1.733 & 1.562 & 1.594 & 5.047 & 1.198 & 2.248 \\
& &$1.5$ & 1.706 & 2.444 & 1.124 & 2.498 & 1.431 & 2.458 & 1.857 & 1.621 & 1.699 & 5.466 & 1.275 & 2.294 \\
& &$2$ & 1.761 & 2.419 & 1.157 & 2.528 & 1.475 & 2.452 & 1.922 & 1.639 & 1.760 & 5.620 & 1.320 & 2.304 \\
& &$3$ & 1.825 & 2.385 & 1.197 & 2.544 & 1.527 & 2.436 & 1.990 & 1.646 & 1.829 & 5.714 & 1.372 & 2.303 \\
& &$4$ & 1.863 & 2.366 & 1.220 & 2.546 & 1.557 & 2.425 & 2.060 & 1.666 & 1.868 & 5.736 & 1.401 & 2.300 \\
& &$5$ & 1.887 & 2.355 & 1.236 & 2.546 & 1.577 & 2.418 & 2.060 & 1.651 & 1.893 & 5.741 & 1.420 & 2.297 \\
& &$10$ & 1.942 & 2.336 & 1.271 & 2.544 & 1.622 & 2.405 & 2.133 & 1.658 & 1.949 & 5.741 & 1.462 & 2.290 \\
\hline
5 & 1 & $0.05$ & 0.487 & 0.598 & 0.640 & 0.477 & 0.423 & 0.580 & 1.127 & 0.476 & 0.233 & 0.599 & 0.190 & 0.530 \\
& &$0.1$ & 0.931 & 1.059 & 0.881 & 0.814 & 0.802 & 1.009 & 1.490 & 0.861 & 0.650 & 1.144 & 0.515 & 0.918 \\
& &$0.2$ & 1.241 & 1.768 & 1.018 & 1.319 & 1.067 & 1.660 & 1.595 & 1.457 & 1.067 & 2.196 & 0.831 & 1.482 \\
& &$6/17$ & 1.399 & 2.398 & 1.090 & 1.786 & 1.198 & 2.249 & 1.623 & 1.953 & 1.317 & 3.589 & 1.012 & 1.987 \\
& &$0.75$ & 1.539 & 2.878 & 1.165 & 2.260 & 1.310 & 2.762 & 1.674 & 2.444 & 1.535 & 5.895 & 1.168 & 2.454 \\
& &$1$ & 1.581 & 2.917 & 1.192 & 2.362 & 1.343 & 2.836 & 1.733 & 2.615 & 1.595 & 6.637 & 1.211 & 2.539 \\
& &$1.5$ & 1.634 & 2.898 & 1.226 & 2.442 & 1.385 & 2.863 & 1.857 & 2.885 & 1.664 & 7.351 & 1.261 & 2.593 \\
& &$2$ & 1.667 & 2.865 & 1.249 & 2.468 & 1.412 & 2.856 & 1.922 & 2.997 & 1.704 & 7.630 & 1.290 & 2.604 \\
& &$3$ & 1.707 & 2.820 & 1.276 & 2.482 & 1.444 & 2.834 & 1.990 & 3.081 & 1.750 & 7.812 & 1.324 & 2.602 \\
& &$4$ & 1.731 & 2.795 & 1.293 & 2.484 & 1.463 & 2.820 & 1.990 & 3.032 & 1.776 & 7.858 & 1.344 & 2.598 \\
& &$5$ & 1.746 & 2.780 & 1.305 & 2.484 & 1.476 & 2.811 & 1.990 & 2.997 & 1.793 & 7.872 & 1.357 & 2.594 \\
& &$10$ & 1.783 & 2.756 & 1.331 & 2.482 & 1.506 & 2.795 & 2.060 & 3.066 & 1.832 & 7.875 & 1.386 & 2.587 \\
\hline
    \end{tabular}
    \caption{
    Numerical parameters of the fits
    in \Eqq{zetaPi_fits} found to describe the various components of the anisotropic stresses
    spectrum as a broken power law
    for the von K\'arm\'an and sound waves
    spectra with $a = 5$ and $b = 2/3$ 
    and 1 respectively.
    \FFig{zetaPi_fits_fig} shows the numerical and the fitted $\zeta_\Pi^i (K)$ and $\zeta_\Pi^{\H, i} (K)$ functions.
    }
    \label{parameters_fits}
\end{table*}

\section{Constant-in-Time model for gravitational wave production}
\label{cit_model_sec}

In this work, we propose that the
production of gravitational waves (GWs) by slowly evolving sources
is accurately described by the ``Constant-in-Time'' (CiT) model,
proposed in ref.~\cite{RoperPol:2022iel} and validated with numerical simulations
of decaying MHD and hydrodynamical turbulence in refs.~\cite{RoperPol:2022iel,Auclair:2022jod}.
This model of GW production from decaying turbulence
has been used for parameter
estimation with LISA and PTA in refs.~\cite{RoperPol:2023bqa,EPTA:2023xxk,Caprini:2024hue,Caprini:2024ofd}.
In the present work, we give a detailed derivation of the
CiT model of ref.~\cite{RoperPol:2022iel} and argue
that this model can be universally applicable
to any source of GWs with evolution timescales (of the source)
that are slower
than the characteristic GW production time $\delta \tau_{\rm GW} = 1/k$
\cite{RoperPol:2022iel,Caprini:2024ofd}.

When the source decays slowly, such that one can assume
that it remains constant in time for an effective
duration period $\tau \in [\tini, \tfin]$, the GW production at a certain wave number $k$ 
occurs on a characteristic time interval
$\delta \tau_{\rm GW} = \tau_{\rm GW} - \tini \simeq 1/k$,
as can be inferred from the Green's function of the wave 
equation [see \Eq{eq:OmGWt}].
Therefore, if the source starts to decay at a time $\tau_{\rm dc} = \tau_\ast + \delta \tau_{\rm dc}$,
all modes with $k > 1/\delta \tfin \sim 1/\delta \tau_{\rm dc}$
have already reached saturation and just oscillate around their saturated
amplitude,
being almost unaffected by the decay of the source (see, e.g., the results of numerical
simulations for MHD turbulence in refs.~\cite{RoperPol:2019wvy, Kahniashvili:2020jgm,
RoperPol:2021uol, Brandenburg:2021bvg, Brandenburg:2021tmp,RoperPol:2022iel}).
Hence, slowly-evolving sources with an effective finite duration $\delta \tfin \sim {\cal O} (\delta \tau_{\rm dc})$
can be studied using the CiT model to accurately describe the GW spectrum at wave numbers
$k > 1/\delta \tfin$.
Note that in this model,
$\delta \tfin$ is introduced as an effective parameter
to separate the sourcing phase
from the phase of GW free propagation,
found to be $\delta \tfin \simeq 2 \, \delta \tau_{\rm dc}$
for the decaying MHD turbulence simulations
of ref.~\cite{RoperPol:2022iel}.
Indeed, $\tfin$ does not really represent the time at which the source disappears, as the source is physically expected
to continue to evolve
at $\tau > \tau_{\rm dc}$.
To take the source decay phase into account,
we go beyond the CiT model
in \Sec{beyond_cit} by introducing a coherent
unequal time correlator (UETC) of the source,
and we study the impact of the decay, finding modifications in the IR tail of the GW spectrum, while wave numbers around
and above the peak remain well described by the CiT
model for moderate decay rates and large enough decay timescales (see \Figs{shapeGW_vort_coh_deltatau_dc}{shapeGW_comp_coh_deltatau_dc}).

At later times, once the GWs enter the free-propagation regime
at $\tau > \tfin$, the rescaled tensor-mode perturbations
$h_{ij}$
generated from $\tini$ to $\tfin$ just propagate
proportionally to $\cos (k \tau)$ and $\sin (k \tau)$ as long as the 
Hubble friction can be neglected, i.e., for modes
with $k^2 \gg a''/a$ [see \Eq{GW_eqn_generic}].
Even when the Universe
is no longer dominated by radiation,
we can still safely ignore the friction as long as the relevant GW modes were produced deeply
inside the radiation-dominated era.

Even though the validity of the CiT model has been previously considered
for decaying MHD turbulence \cite{RoperPol:2022iel,Auclair:2022jod,RoperPol:2023bqa,Caprini:2024hue}, it is a generic model that applies to
all types of slowly-varying sources of GWs \cite{Caprini:2024ofd}.
We focus on examples of such sources in \Sec{applications}.

In particular, in \Sec{constant_model} we review
its applicability to decaying MHD
turbulence when the turbulent decay timescale,
$\tau_{\rm dc}$,
is large, and the characteristic velocity of decorrelation in time, $u_{\rm dcorr}$, is small.
We consider a Gaussian decay of the UETC in the time
difference $\tau_- = \tau_2 - \tau_1$, proportional
to $\exp(-\half u_{\rm dcorr}^2 k^2 \tau_-^2)$,
to describe
stationary turbulence following Kraichnan
decorrelation \cite{1965PhFl....8.1385K}.
In reality, $u_{\rm dcorr}$ is a function of
times $\tau_1, \tau_2$ and $k$
\cite{1965PhFl....8.1385K,1993PhFlA...5.2835K,Kosowsky:2001xp,Gogoberidze:2007an,Caprini:2009yp,Niksa:2018ofa,Auclair:2022jod}
but here we estimate it to be constant
for a direct comparison with the CiT model.
For small decorrelation velocities $u_{\rm dcorr}$
(i.e., long
decorrelation times), the CiT model is found to accurately
describe the GW spectrum,
contrary to previous analytical estimates
describing vortical turbulence
(see, for
example, refs.~\cite{Kosowsky:2001xp,Gogoberidze:2007an}).
However,
as $u_{\rm dcorr}$ becomes large, we recover the GW
spectrum found in previous theoretical
work \cite{Kosowsky:2001xp,Gogoberidze:2007an},
together with
an additional low-$k$ tail that is described by the
CiT model.
We then estimate how large $u_{\rm dcorr}$
is required to be to modify the GW spectrum
with respect to the one
found following the CiT model.
The low-$k$ tail is analogous to the term introduced in
ref.~\cite{RoperPol:2023dzg} for sound waves,
where the applicability to Kraichnan decorrelation
is also discussed (see their Sec.~V B).
This finding explains
why the CiT model accurately describes
the GW spectrum found in simulations
of decaying MHD turbulence \cite{RoperPol:2019wvy,RoperPol:2021uol,RoperPol:2022iel,Auclair:2022jod}, since the
decorrelation velocity in these simulations
is subrelativistic.

In \Sec{sound_wave_component},
we study the applicability of the 
CiT model
to sound waves.
In this case, the UETC becomes proportional to
$\cos (\cs k \tau_-)$, where $\cs$ is the
speed of sound and plays the role of $u_{\rm dcorr}$
in determining a decorrelation velocity.
Therefore,
when $\cs$ is relativistic
(e.g., $\cs^2 = 1/3$),
an additional resonant component of the GW spectrum
grows at $k \sim \cs k_\ast$ on top of the one  predicted by the CiT
model, and dominates the GW spectrum
when the source duration is long $\cs \delta \tfin k_\ast \gg 1$
\cite{Hindmarsh:2016lnk,Hindmarsh:2019phv,RoperPol:2023dzg,Sharma:2023mao}.
We show that the component that dominates
the GW spectrum at small $k$,
found in refs.~\cite{RoperPol:2023dzg,Sharma:2023mao},
is accurately described by the CiT model.
This is the same low-$k$ tail that we find for
Kraichnan decorrelation, such that the CiT model
can be used to predict this contribution
for both types of sources.
We show that both the amplitude
and spectral peak of the resonant
component decrease as $\cs$ decreases
(see also ref.~\cite{Giombi:2024kju})
and eventually the GW spectrum from sound waves
reduces to the one found in the CiT at all $k$
in the $\cs \to 0$ limit.

These results manifest that both turbulence and sound
waves are well described by the CiT model when the
corresponding decorrelation
velocity of the UETC is subrelativistic,
$u_{\rm dcorr} k_\ast \delta \tfin \ll 1$
or $\cs k_\ast \delta \tfin \ll 1$,
while they both present a similar
resonant growing mode around the peak
as these velocities become larger, such that the
UETC time variation becomes of a similar order
to the time variation of the Green's function
\cite{RoperPol:2022iel,RoperPol:2023dzg}.
Since the speed of sound is relativistic in the
primordial 
radiation plasma, even if the fluid velocities
are subrelativistic, we expect sound waves to be
more efficient than subrelativistic
vortical turbulence in producing GWs \cite{RoperPol:2019wvy,RoperPol:2023bqa},
even though the compressional
anisotropic stresses are more suppressed than the
vortical ones (see \Sec{EPi_sec} and ref.~\cite{Brandenburg:2021bvg}).

In \Sec{other_sorc}, we describe  
the potential applicability of the CiT model
to sources of GWs arising from scalar fields:
bubble collisions in the bulk flow model \cite{Konstandin:2017sat,Jinno:2017fby,Jinno:2022mie} or feebly interacting
particles \cite{Jinno:2022fom}.
In general, we point out that a very particular feature
of the CiT model for GW production is the cubic,
$\OmGW (k) \sim k^3$,
and linear, $\OmGW (k) \sim k$, scaling
of the GW spectrum
at low and intermediate frequencies,
which are very commonly found in the literature.
It remains to be separately
explored which of these sources
is accurately described by the CiT, and under which
assumptions, by studying their UETCs.

\subsection{Model}
\label{cit_model}

The GW spectrum at any time $\tau$ is computed
in \Eq{eq:OmGWt} as the integral over $\tau_1$ and
$\tau_2$ of the UETC of the anisotropic stresses and the product of Green's functions.
Under the assumption that the UETC of the source $E_\Pi (\tau_1, \tau_2, k)$
can be approximated as constant
(hence, the Constant-in-Time or CiT model)
for
an effective period
of GW sourcing, $\tau \in [\tini, \tfin]$, i.e., $E_\Pi (\tau_1, \tau_2, k)
\approx E_\Pi (\tini, \tini, k) 
\equiv E_\Pi^\ast (k)$,
the GW spectrum becomes \cite{RoperPol:2022iel}
\begin{align}
    \OmGW  (\tau, k) = 3\,
    E_\Pi^* (k) \,
    F_{\rm GW}^0
    \, \Delta^2 (\tau, k)\,,
\label{OmGW_constant_tfin}
\end{align}
where $\Delta^2$ is the time kernel, corresponding to the integral over
times $\tau_1$ and $\tau_2$ of the product
of Green's functions in \Eq{eq:OmGWt}.
Since the integral over $\tau_1$ and $\tau_2$ becomes separable for 
the CiT model, the result is expressed as the square of 
the kernel $\Delta$,
\begin{align}
    \Delta (\tau,k) & \,  =  \int_{\tini}^{\bar \tau}
    \frac{\dd \tau'}
    {\tau'} \cos [k  (\tau - \tau')]
    \nonumber \\
    = &\, 
    \cos (k \tau) \Delta \Ci (k \bar \tau) + \sin (k\tau) \Delta \Si (k \bar \tau)\,,
    \label{G_constant}
\end{align}
where $\bar \tau = \min(\tau, \tfin)$, and
\begin{subequations}
\label{Delta_Ci_Si}
\begin{align}
    \Delta \Ci (k \tau') \equiv &\, \Ci(k \tau') - \Ci (k \tau_\ast)\,, \\ 
    \Delta \Si (k \tau') \equiv &\, \Si (k \tau')
    - \Si (k\tau_\ast)\,,
\end{align}
\end{subequations}
being $\Ci(x)$ and $\Si(x)$ the cosine and sine integral
functions.
This kernel corresponds to a source produced during the radiation-domination
era of the Universe.
It can be extended to different cosmological backgrounds using
the GW solution given in \Eq{GW_sol_der}, i.e., substituting $1/\tau$ by
$a(\tau) \, \HH^2(\tau) / \HH_\ast$, and by using the corresponding Green's function,
\begin{align}
    \Delta (\tau, &  \, k) = \nonumber \\ &\, \HH_\ast
    \int_{\tini}^{\bar \tau}
    \dd \tau'
    \frac{\HH^2 (\tau')}{\HH_\ast^2} \, a(\tau')
    \, \partial_\tau G(\tau, \tau', k)\,.
\end{align}
Note that, in general, we can normalize all times
with the initial time $\tau_\ast$, and wave numbers and the Hubble
rate with $1/\tau_\ast$.
We recall that for radiation domination $\tau_\ast \HH_\ast = 1$.

When the source duration is short compared to the Hubble time,
the expansion of the Universe can be ignored, and
the kernel in \Eq{G_constant} becomes
\begin{align}
    \Delta_{\rm flat} (\tau, k) =  &\, \frac{1}{\tini} \int_{\tini}^{\bar \tau}
    \dd \tau' \cos [k (\tau - \tau')]\,,
\end{align}
which has the two following branches:
\begin{subequations}
\begin{align}
    \label{Delta_flat_1}
    \Delta_{\rm flat} (\tau < \tfin, k) =\frac{1}{k\tini}  &\,  \sin (k \delta \tau)\,,
    \\ \Delta_{\rm flat} (\tau > \tfin, k) = \frac{1}{k\tini}  &\, \bigl(
    \sin (k \delta \tau) \nonumber \\ &\, - \sin
    [k (\tau - \tfin)] \bigr) \,,
\end{align}
\end{subequations}
where $\delta \tau = \tau - \tini$.

\begin{figure*}
    \centering
    \includegraphics[width=\columnwidth]{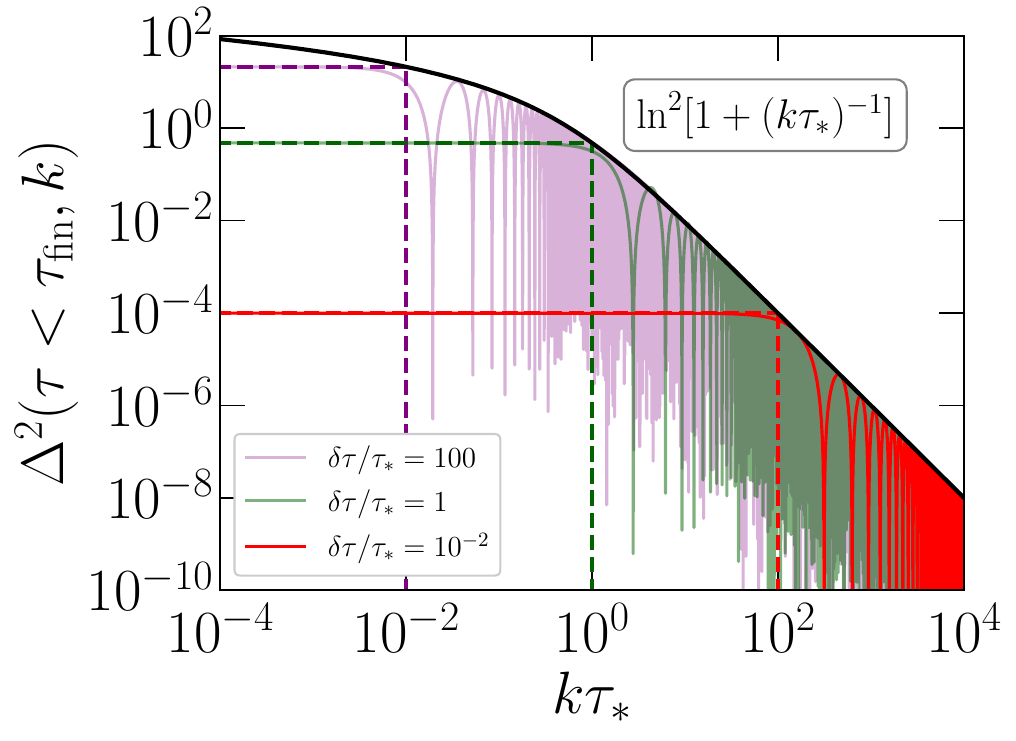}
    \includegraphics[width=\columnwidth]{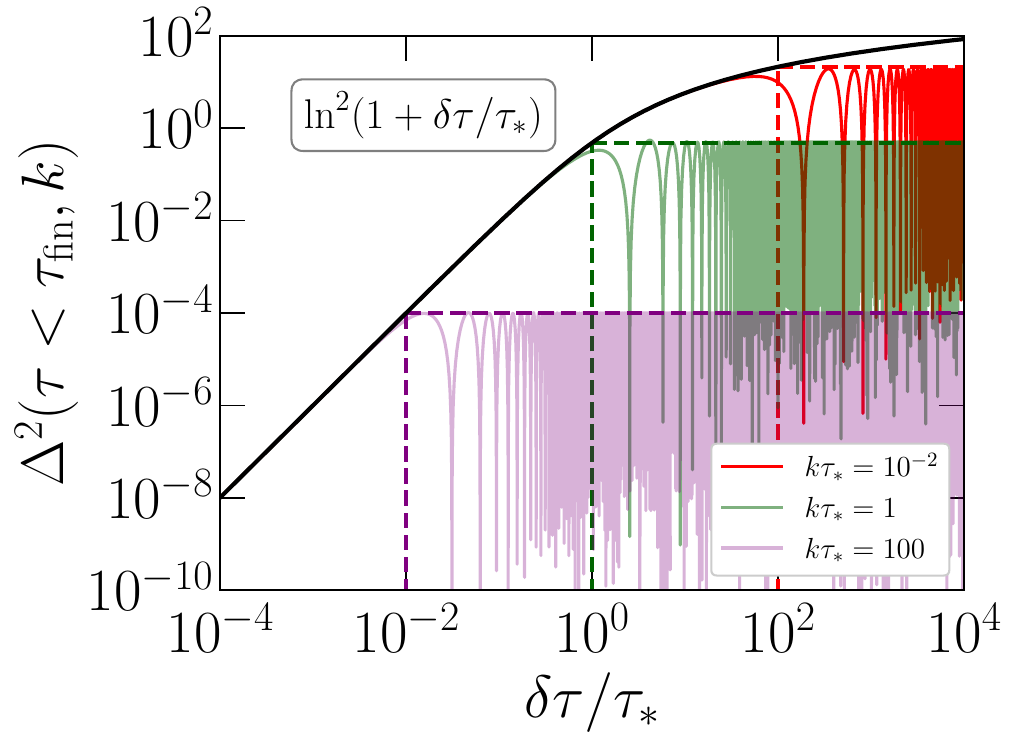}
    \caption{Dynamical evolution of the kernel $\Delta^2
    (\tau, k)$ for times $\tau \leq \tfin$ as a function
    of $k$ for different values of $\delta \tau$ (left panel),
    and as a function of
    $\delta \tau$ for different values of $k$ (right panel).
    Vertical lines correspond to the wave numbers and times when
    saturation is reached, $k \delta \tau_{\rm GW} = 1$.
    The thick black line in the right panel corresponds to the envelope of the kernel as time
    evolves, $\ln^2(1 + \delta \tau/\tini)$ [see \Eq{growing_delta2_cit}], until saturation is reached
    at $\delta \tau_{\rm GW} = 1/k$. The thick black line in the left panel shows the envelope of $\Delta^2(\tau, k)$ at saturation [see \Eq{saturated_delta2_cit}], as a function of $k$.}
    \label{G2_kt}
\end{figure*}

The dynamical evolution of the squared kernel, $\Delta^2(\tau \leq \tfin, k)$,
for an expanding radiation-dominated Universe is shown in \Fig{G2_kt}.
This spectrum oscillates in time and wave numbers,
and the frequency of oscillations in time is proportional 
to $k$, while the frequency of oscillations in wave numbers is proportional to $\delta \tau$.
The amplitude of the oscillations is studied in \App{dynamical_G}
in its asymptotic limits and it
allows us to characterize the resulting GW spectrum.

We observe two regimes in the dynamical evolution
of $\Delta^2 (\tau \leq \tfin, k)$:
\begin{enumerate}[\em i)]
    \item  $k \delta \tau \equiv k(\tau - \tini) \leq 1$: the
    GW modes grow with time,
    \begin{equation}
    \label{growing_delta2_cit}
        \Delta^2 (\tau, k) = \ln^2
        \bigl(1 + \delta \tau/\tini \bigr)\,,
    \end{equation}
    which does not depend on $k$ and grows quadratically
    at early times $\delta \tau \ll \tau_\ast$.
    \item $k \delta \tau > 1$: the GW modes have reached
    saturated amplitudes and oscillate in time.
    The envelope of  $\Delta^2 (\tau, k)$ becomes
    \begin{equation}
    \label{saturated_delta2_cit}
        \Delta^2_{\rm env} (\tau, k) = 
        \ln^2 \bigl[1 + (k \tini)^{-1} \bigr]\,,
    \end{equation}
    which is constant in
    time and decreases with $k$, quadratically
    in the limit $k \gg 1/\tau_\ast$.
\end{enumerate}

Hence, at each mode $k$, we observe a growing phase of the
GW spectrum given by $\ln^2(1 + \delta \tau/\tini)$
for a duration $\delta \tau_{\rm GW} = 1/k$, i.e.,
until the spectral mode reaches its
saturation value: $\ln^2 \bigl[1 + (k \tini)^{-1} \bigr]$
(regime $i$).
At later times, but still $\tau < \tfin$,
the kernel function $\Delta^2$ oscillates
around its saturated amplitude (regime $ii$).
Omitting the expansion of the Universe, these functions become
$(\delta \tau/\tini)^2$ and $(k \tini)^{-2}$.

The decay of the kernel proportional to $k^{-2}$
in the $k \to \infty$ limit
is a consequence of the discontinuity introduced in the sourcing,
assuming that it starts instantaneously at $\tini$ \cite{Caprini:2009fx}.
In reality, we expect that the GW spectrum is modified by the
dynamics during the period of sourcing, $\Delta \tau_{\rm s}$,
at wave numbers $k \gtrsim 1/\Delta \tau_{\rm s}$ 
and, eventually, decays exponentially if the
sourcing process is smooth.
Therefore, the model we present is valid for
$k \ll 1/\Delta \tau_{\rm s}$, and it needs to include the dynamics
of the sourcing process at $k \gtrsim 1/\Delta \tau_{\rm s}$.

To show this explicitly, let us rewrite \Eq{G_constant}
using the function $g(\tau) = \Theta(\tau-\tau_*)-\Theta(\tau-\tau_{\rm fin})$ that takes into
account the switching on and off of the source at $\tau_\ast$
and $\tau_{\rm fin}$,
\begin{align}
    \Delta (\tau, k) = & \,
    \int_{-\infty}^{\tau}
    \frac{\dd \tau'}
    {\tau'} \cos [k  (\tau - \tau')]  \,
    g (\tau') \,.
    \label{delta_modified}
\end{align}
Integrating by parts, we find
\begin{align}
    \Delta (\tau, k) =& 
    \sum_{\tau_s = \tini, \bar \tau} \Delta_s 
    \biggl(\frac{g(\tau')}{k \tau'} \sin [k (\tau - \tau')] \biggr)
    \nonumber \\
     & \!
     + \int_{\tini}^{ {\bar \tau}
     } \frac{[g(\tau')/\tau']'\dd \tau'}{k} \sin [k(\tau - \tau')]\,,
     \label{int_by_parts}
\end{align}
where $\Delta_s$ corresponds to the jump of the
function $g(\tau')/(k \tau') \sin [k(\tau - \tau')]$
at $\tau_s = \tini, \bar \tau$.
Since $g(\tini^+)=g(\bar \tau^-)=1$,
$g(\tini^-) = 0$, and $g(\bar \tau^+) = 1 - \Theta(\bar \tau - \tfin)$,
the leading-order term of $\Delta$ in the $k \to \infty$ limit becomes
\begin{align}
    \lim_{k \to \infty} \Delta ( \tau, k) &= 
     \frac{1}{k \tini} \sin(k\delta \tau) \nonumber \\ &\,
    -  \frac{1}{k \bar \tau}
    \sin [k(\tau - \bar \tau)] + {\cal O}(k^{-2})\,.
    \label{Delta_cit}
\end{align}
Notice that the second term
only starts contributing to the envelope of $\Delta$
at $\tau \geq \tau_{\rm fin}$.
\EEq{Delta_cit} shows that $\Delta \sim k^{-1}$
in the $k \to \infty$ limit
and allows us to evaluate
its amplitude in this limit.

\begin{figure}
    \centering
    \includegraphics[width=\linewidth]{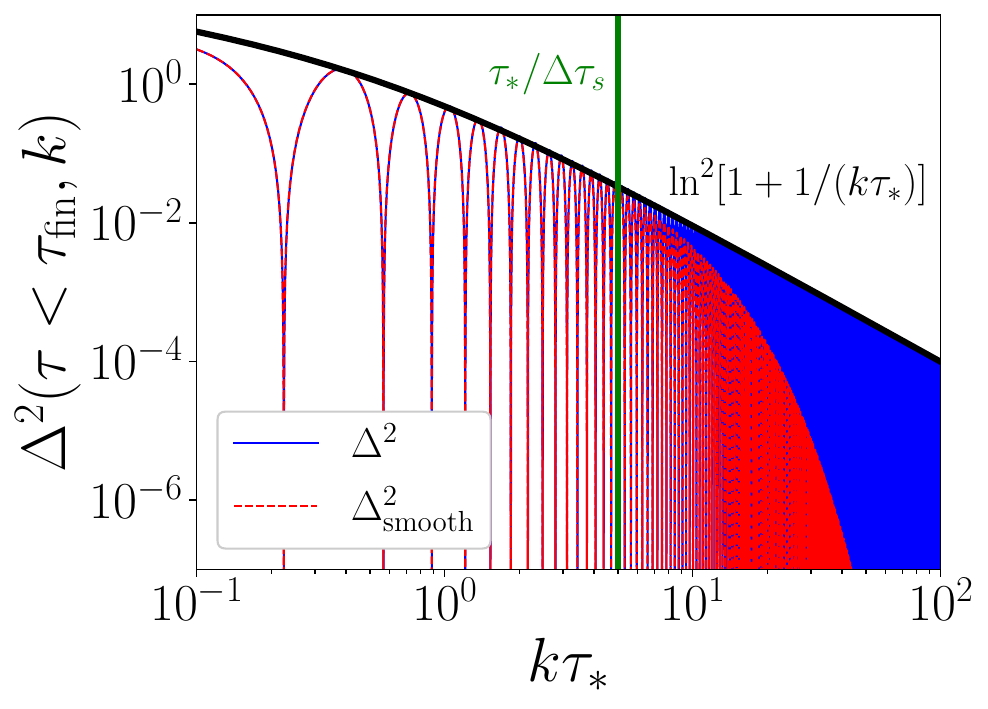}
    \caption{
    Kernel $\Delta^2 (\tau, k)$ found in the CiT
    model when the source is assumed to start instantaneously
    at $\tau_\ast$ (blue) compared to the one found when the
    growth is determined by a ${\cal C}^\infty$ 
    function with a width $2\Delta \tau_s$ around $\tau_\ast$ (red). The kernels are evaluated at $\delta \tau = 10 \tau_\ast$ and $\Delta \tau_s = 0.2 \tau_\ast$.}
    \label{kernel_instant_smooth}
\end{figure}

If, on the other hand, we consider a ${\cal C}^\infty$ function $g(\tau)$ describing a smooth sourcing
process instead of an instantaneous one, and we assume for simplicity that
the UETC of the source
is a separable function proportional to $g(\tau_1)\,
g(\tau_2)$, the kernel $\Delta$ can still be expressed
as in \Eq{delta_modified}.
For example, let us consider
the following $g$ function,
\begin{equation}
    g(\delta \tau) = \frac{1}{2} \biggl[\tanh\biggl(\frac{2 \, \delta \tau}{\Delta \tau_s}\biggr) + 1 \biggr]\,,
\end{equation}
where $\Delta \tau_s$ determines the interval of time over which the
sourcing function $g$ changes from $0$ to $1$, which is approximately $2 \Delta \tau_s$.
Iteratively
applying the same procedure as the one
that led to \Eq{int_by_parts},
and taking into account that $g$ is
${\cal C}^\infty$,
one would find that $\Delta(\tau, k)$ goes to zero
faster than any power law when $k \to \infty$.
We show the resulting kernel $\Delta^2$ in \Fig{kernel_instant_smooth},
compared to the one found using \Eq{G_constant}, which assumes instantaneous
growth of the source.
Indeed, both follow the same dependence
with $k$ at all $k < 1/\Delta \tau_s$.

\begin{figure}
    \centering
    \includegraphics[width=\columnwidth]{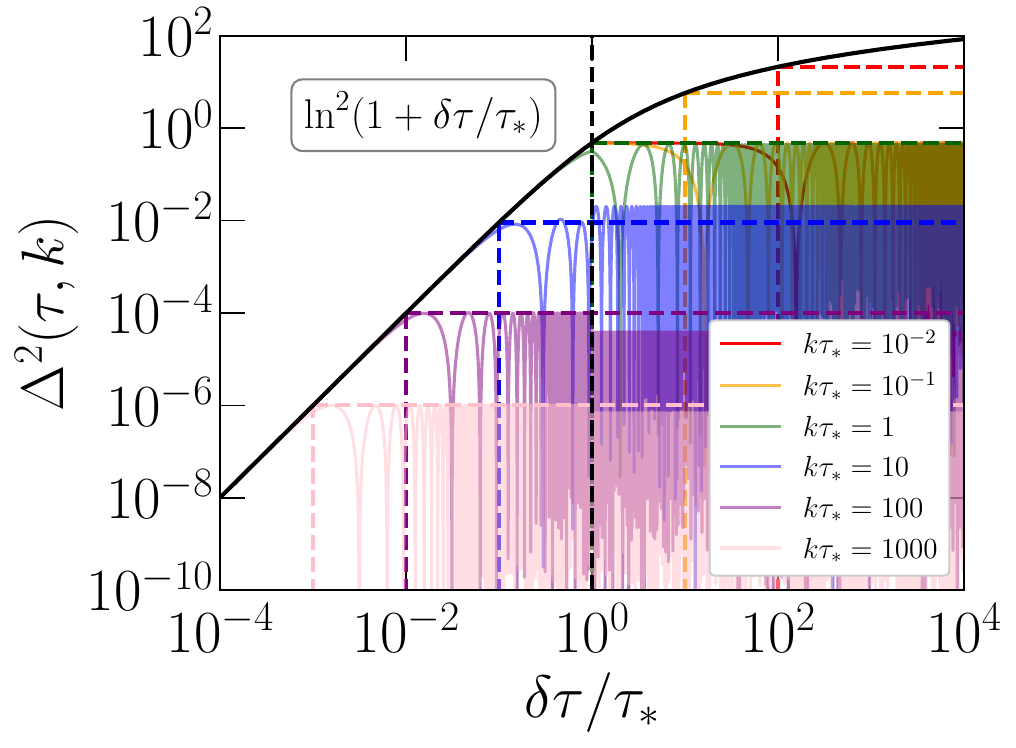}
    \includegraphics[width=\columnwidth]{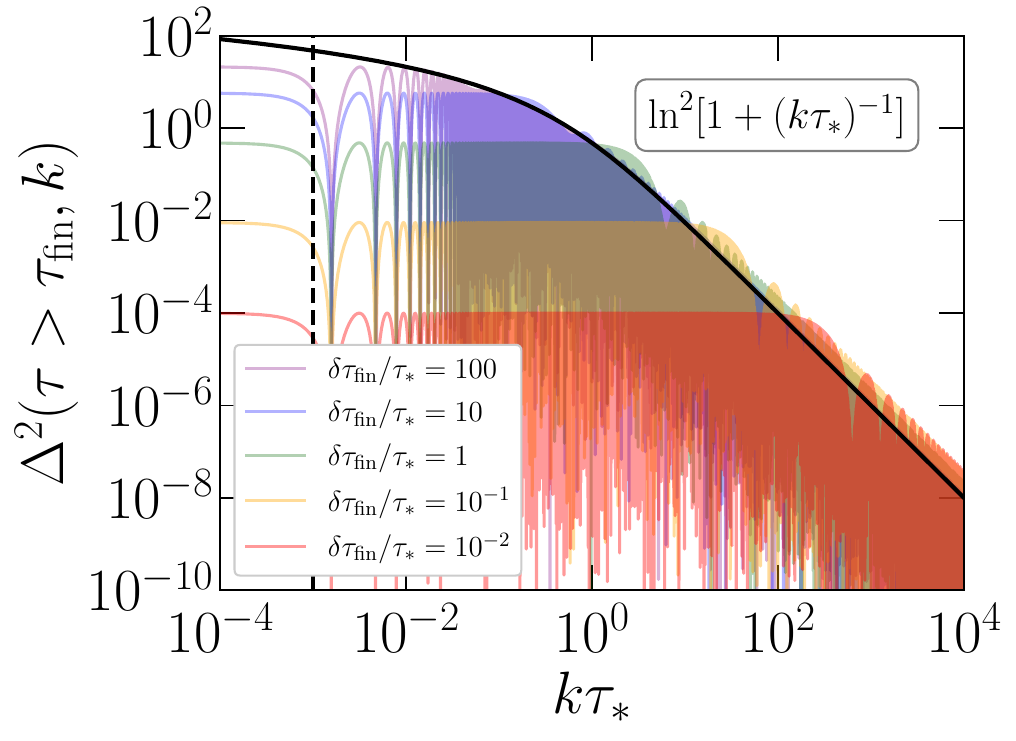}
    \caption{{\em Upper panel:} Kernel $\Delta^2 (\tau, k)$
    evaluated for different values of $k$ as a function of
    $\tau$ for $\delta \tfin = \tau_*$.
    The vertical and horizontal lines are the same as in the right panel of \Fig{G2_kt}
    for comparison.
    {\em Lower panel:} Kernel $\Delta^2 (\tau, k)$ evaluated
    using
    different source durations $\delta \tfin/\tini =
    \{10^{-2}, 10^{-1}, 1, 10, 10^2\}$ at a time $\delta \tau/\tini = 10^3 \gg \tfin/\tini$.
    The vertical line indicates $k = 1/\delta \tau$.
    }
    \label{D2_kt_fint}
\end{figure}

On the other hand, at times after
$\tfin$, $\tau > \tfin$, the dynamical behavior of $\Delta^2(\tau, k)$
presents a modification at large $k$, as it can already be seen
in \Eq{Delta_cit}, as a consequence of the discontinuity of the source at $\tfin$.
The upper panel of \Fig{D2_kt_fint} shows the kernel
$\Delta^2$ for $\delta \tfin = \tini$ as a function
of time $\delta \tau = \tau - \tini$ at different wave numbers, while
the lower panel shows the kernel
evaluated at
a late time, $\delta \tau = 10^3 \, \tau_\ast$, obtained if the sourcing is active for finite
durations $\delta \tfin = \{10^{-2}, 10^{-1}, 1, 10, 10^2\} \, \tini  < \delta \tau$.
After $\tau$ reaches $\tfin$, we find two additional regimes, depending on whether
the modes have already saturated by this time, i.e., $k > 1/\delta \tfin$, or not:
\begin{enumerate}
    \item[\em iii)] $k \delta \tfin \equiv k (\tfin - \tini) < 1$:
    the GW modes had not reached saturation by
    the time $\tfin$.
    They stop growing at $\tfin$ with an envelope
    \begin{equation}
        \Delta^2_{\rm env} (\tau, k) = \ln^2(1 + \delta \tfin/\tini)\,.
    \end{equation}
    Their amplitude remains constant.
    After they have been processed by the Green's function at
    $\delta \tau \gtrsim \delta \tau_{\rm GW} = 1/k$, these modes oscillate, keeping the same envelope amplitude.
    \item[\em iv)]  $k \delta \tfin > 1$: the GW modes had already saturated
    before $\tfin$ is reached and were in the oscillatory regime {\em ii)} with
    an amplitude $\ln^2\bigl[1 + (k \tini)^{-1}]$.
    At $\tfin$, the amplitude of the kernel $\Delta^2$ presents a discontinuity
    that depends on the wave number and the values of $\tini$ and $\tfin$
    (see \Eq{Delta_cit} and \Fig{D2_kt_fint}),
    and then the function continues to oscillate in time with a saturated amplitude.
\end{enumerate}

In the regime $k \geq 1/\delta \tfin$, the spectrum shows
the characteristic $\ln^2 \bigl[1 + 1/(k\tini)\bigr]$ that corresponds to
all the saturated values reached at earlier times $\tau \leq \tfin$
\cite{RoperPol:2022iel}.
However, the abrupt switch off of the
source enhances the envelope amplitude of $\Delta^2$ at
$\tau \geq \tfin$ and $k > 1/\delta \tfin$,
as we have explicitly shown in \Eq{Delta_cit}.
The exact modification of the spectrum depends on the
specific values of
$k$, $\tau$, and $\tfin$, with an upper bound over $k$ (envelope)
determined by the factor
${\cal F} (\tfin)$, 
computed in \App{dynamical_G} [see, in particular, \Eq{F_factor}],
\begin{equation}
    {\cal F} (\tfin) = 1 + \frac{\tini}{\tfin} \in (1, 2]\,.
    \label{factor_F}
\end{equation}
This factor is equivalent to the envelope of $|\Delta|$ found in the $k\to \infty$
in \Eq{Delta_cit}, and
asymptotically approaches one as $\tfin$
increases, so it is mostly relevant
for short source durations $\tfin$.
Indeed, we see in \Fig{D2_kt_fint} (upper panel) a jump on $\Delta^2$ that
can either be positive or negative depending on $k$ at $\delta \tfin$,
while we observe a resulting enhancement of the envelope amplitude
in the lower panel at $k > 1/\delta \tfin$, whose amplitude
decreases as $\delta \tfin$ increases.
This factor becomes a constant in the limit of
small source durations $\delta \tfin \ll \tini$.
In this limit,
${\cal F}$ takes its maximum value, ${\cal F} (\tini) = 2$.
Therefore, this modification becomes maximal and independent of $\tfin$ in
flat space-time (i.e., when one omits the expansion of the Universe).

Taking into account the description discussed above for the
CiT model,
the envelope of the GW spectrum at times $\tau \geq \tfin$ can be used
to model the GW spectrum
at present time
\cite{RoperPol:2022iel,EPTA:2023xxk,RoperPol:2023bqa,Caprini:2024hue}
\begin{align}
    \OmGW^{\rm env}  (\delta \tfin, k) = 3 \, E_\Pi^* (k) \, F_{\rm GW}^0 \,
    {\cal T}_{\rm GW} (\delta \tfin, k) \,,
    \label{OmGW0_kt}
\end{align}
where the function ${\cal T}_{\rm GW}$, introduced in ref.~\cite{RoperPol:2022iel}, is
\begin{align}
     {\cal T}_{\rm GW} &\, (\delta \tfin, k) \nonumber \\  &\, \!\!\!\!\!\!\!\! =
     \left\{  \begin{array}{ll}
        \ln^2 (1 + \delta \tfin / \tini)\,,
        & \text{ if } k\,\delta \tfin^* < 1,\\
        {\cal F}^2
        \ln^2 \bigl[1 + (k\tau_\ast)^{-1} \bigr]\,,
        & \text{ if } k\, \delta \tfin^*
        \geq 1,
        \end{array} \right. \label{TGW_0}
\end{align}
with ${\cal F}$ corresponding to the enhancement found in \Eq{factor_F}.
The separation between branches corresponds to the source duration $\delta \tfin$, modified
to ensure continuity of the envelope ${\cal T}_{\rm GW}$ \cite{RoperPol:2022iel},
\begin{equation}
    \frac{\delta \tfin^*}{\tini} = 
    \biggl(1 + \frac{\delta \tfin}{\tini}\biggr)^{1/{\cal F}} - 1 \,.
\end{equation}
Its position reduces to $\delta \tfin$
when ${\cal F} \rightarrow 1$.
Hence, if the duration of the source is short and it is abruptly
switched off, the envelope of the GW spectrum presents
an enhancement by a factor ${\cal F}^2$ at large $k \geq 1/\delta \tfin^*$.
In the limit of short durations $\delta \tau_{\rm fin} \ll \tau_\ast$, we have ${\cal F} \rightarrow 2$ and $\delta \tfin^\ast \to \half \delta \tfin$.

\begin{figure}
    \centering
    \includegraphics[width=\linewidth]{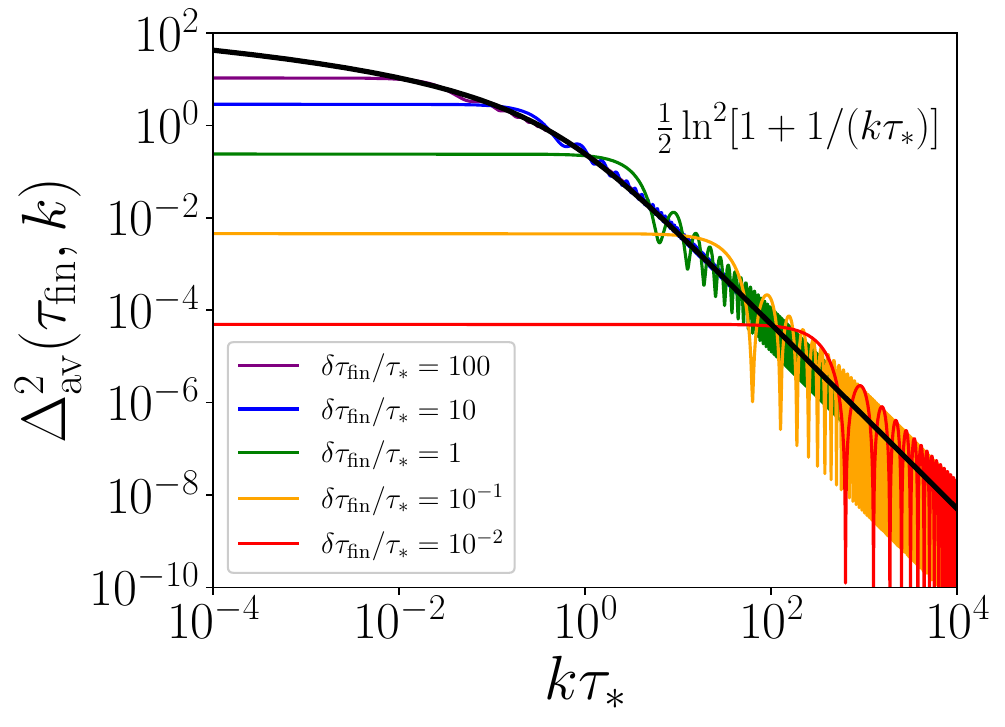}
    \caption{Kernel 
    $\Delta^2_{\rm av}$ 
    averaged over fast frequency oscillations at late times $\tau \gg \tfin$, given in \Eq{Delta_aver} for an expanding Universe, for different
    source durations $\delta \tfin$.}
    \label{kernel2_0}
\end{figure}

At present time, the oscillations over $k$ with frequency proportional to $\tau_0$ 
can be averaged out at all
$k \tau_0 \gg 1$, as done to obtain the GW spectrum in \Eq{OmGW_aver}.
Following the CiT model, the time kernel found when 
these fast oscillations 
are averaged out becomes
\begin{equation}
    \Delta^2_{\rm av} (\tfin, k) = \half [\Delta \Ci^2 (k \tfin) +
    \Delta \Si^2 (k \tfin)]\,,
    \label{Delta_aver}
\end{equation}
which is shown in \Fig{kernel2_0}.
We note that the resulting kernel still presents
some relic oscillations in $k$ with frequency proportional to $\tau_{\rm fin}$.
The envelope of \Eq{Delta_aver} is $\half {\cal T}_{\rm GW} (\delta \tfin, k)$,
justifying the $\half$ factor introduced in
refs.~\cite{RoperPol:2023bqa,Caprini:2024hue,EPTA:2023xxk}.
In flat space-time, the kernel averaged over fast oscillations becomes
\begin{equation} \label{Delta0_flat}
    \Delta^2_{
    {\rm av}, \rm flat} (\tfin, k) = \frac{1}{(k \tini)^2}
    \bigl[1 - \cos (k \delta \tau_{\rm fin})\bigr]\,,
\end{equation}
whose envelope is also $\half {\cal T}_{\rm GW} (\delta \tfin, k)$
taking the short-duration limit of the $\ln^2$
functions, i.e.,
\begin{align}
    {\cal T}_{\rm GW, flat} & (\delta \tfin, k) = \nonumber \\ & \left \{ 
    \begin{array}{ll}
        \bigl({\delta \tfin}/{\tini} \bigr)^2 \,, & \quad \text{ if } k\,\delta \tfin < 2\,,\\
       4 \, (k \tini)^{-2} \,, & \quad \text{ if } k\,\delta \tfin \geq 2\,.
    \end{array}
    \right.
\end{align}

\EEqs{Delta_aver}{Delta0_flat} explicitly indicate that if one applies the procedure
described in \Sec{weinberg} to compute the GW
spectrum, where the GWs are already averaged over fast frequency oscillations at late
times, it is crucial to make sure that the simulations are covering
all times up to $\tau_{\rm fin}$, i.e., to ensure that each mode $k$
has reached saturation, to recover the correct results
(see also discussion in App.~A of ref.~\cite{Caprini:2024gyk}).
Otherwise, the final GW spectrum is biased by the particular choice
of the final time of the simulation.

In ref.~\cite{RoperPol:2022iel}, this model was used and compared
with numerical simulations of decaying MHD turbulence.
The numerical results were accurate at $k \delta \tau_{\rm dc} \gtrsim 1$
with the model described 
above, with the exception of the factor ${\cal F}$, which does not
appear in the numerical results.
Since this is a consequence of the abrupt switching off of the
source, which is not realistic for decaying turbulence (or any
other time-decaying source), we introduce a coherent UETC in \Sec{coherent_model} that allows us to
describe the time decay of the source at late times.
We find that the GW spectrum at wave numbers that have reached saturation
before the source starts to decay, $k \delta \tau_{\rm dc} \gtrsim 1$,
remains unmodified and described
by the CiT model of
\Eq{OmGW0_kt} with ${\cal F} \to 1$, while smaller wave numbers are
affected by the decay.

\subsection{Gravitational wave template}
\label{gw_template_cit}

Introducing the normalization of $E_\Pi^\ast (k)$ described in \Sec{subsec:symm_EPi}
[see \Eq{EPi_norms}] into
\Eq{OmGW_constant_tfin}, the GW spectrum at present time, averaged
over fast frequency oscillations, becomes \cite{RoperPol:2022iel,RoperPol:2023bqa,EPTA:2023xxk,Caprini:2024hue}
\begin{align}
     \OmGW^0  (K) = \frac{3}{2} \frac{{\tilde \varphi}_{\rm rms}^4}{\bar \rho^2}
     \frac{{\cal C}}{{\cal A}^2}  F_{\rm GW}^0 \, K^3 \zeta_\Pi^\ast (K) \,
     {\cal T}_{\rm GW} \,,
        \label{OmGW_envelope}
\end{align}
where $K = k/k_\ast = k {\cal R}_\ast$, ${\tilde \varphi_{\rm rms}^2
=\{ \bar \rho_{\rm kin}, 2 \bar \rho_B, 2 \bar \rho_G^\phi\}}$ is the mean squared
amplitude of the different type
of sources considered
$\tilde \varphi_i = \{ X_i, i B_i / a, - i k_i \phi / a \}$.
Note that $X_{\rm rms}^2/\bar{\rho} \simeq \Gamma u_{\rm rms}^2$ in the limit
of small perturbations with $\Gamma = \bar w / \bar \rho$.
The envelope of the time kernel is described by the function ${\cal T}_{\rm GW}$ in \Eq{TGW_0}
and the average over fast oscillations is taken care of by the extra $1/2$ factor
with respect to \Eq{OmGW_constant_tfin} [see \Eq{Delta_aver}].
A main assumption of this model is that $\tilde \varphi_{\rm rms}$ remains
approximately constant over the sourcing period.
If the dynamics of the source are slow, this assumption is always justified
in flat space-time.
Taking into account the expansion of the Universe,
this assumption can be justified over cosmological times
for radiation-like sources that
dynamically evolve slowly over time, such that
the ratio $\tilde \varphi_{\rm rms}^2 / \bar \rho
\approx$ {\em const.}
For fluid or scalar fields with generic equation of state $\omega$,
this ratio will be proportional to $\rho_G^\phi / \bar \rho \sim a^{1 - 3 \omega}$
for a radiation-dominated Universe and, hence, our model is generally applicable
if the sourcing period is much shorter than one Hubble time.
On the other hand, if the source of GWs corresponds to the dominant contribution to the energy density of the
Universe, then both numerator and denominator scale with the same
power law of $a$, and the ratio can be taken to be approximately constant.

As expected, whether choosing the normalization
for the spectrum of the source,
$\zeta(K)$, following the convention in this work,
or following instead the convention of previous related work, $\zeta_{-1} (\tilde K)$
 [see discussion around \Eq{rescaling_Zeta}], does not affect the
resulting GW spectrum, which becomes
\begin{align}
    \OmGW^0  & (\tilde K) = \frac{3}{2} \frac{\tilde \varphi_{\rm rms}^4}{\bar \rho^2}
    \frac{\tilde {\cal C}}{\tilde {\cal A}^2}
    \, F_{\rm GW}^0 \tilde K^3 {\zeta}_\Pi^\ast (\tilde K) \, {\cal T}_{\rm GW}\,.
    \label{OmGW_envelope_2}
\end{align}
Indeed, using \Eqs{tilde_A}{normal_Cuu}, and $\tilde K = K / K_{-1}^\ast$, one finds
the equivalence between the normalization factors,
\begin{equation}
    \frac{\tilde {\cal C}}{\tilde {\cal A}^2} \tilde K^3 = 
    \frac{{\cal C}}{{\cal A}^2} K^3\,.
\label{norm_identity}
\end{equation}

The $k$ dependence of the GW background in \Eq{OmGW_envelope} has been
previously
described in ref.~\cite{RoperPol:2022iel}.
In summary, it presents:
\begin{enumerate}[\em i)]
    \item a causal $k^3$ branch at $k < 1/\delta \tfin^\ast$;
    \item a logarithmic branch proportional to
    \begin{equation}
        \OmGW^0 (K) \sim K^3 \ln^2 \bigl[1 + ({\cal R}_\ast \HH_\ast)/K \bigr]\,,
    \end{equation}
    in the range ${\cal R}_\ast/\delta \tfin^\ast < K < K_{\rm GW}$,
    where we have used $1/\tini = \HH_\ast$, and defined the
    comoving length scale ${\cal R}_\ast \equiv 1/k_\ast$.
    $K_{\rm GW}$ corresponds to the position of the GW spectral peak as
    we discuss below (see also discussion in ref.~\cite{RoperPol:2022iel}).
    For subhorizon modes with $k\tini = K / ({\cal R}_\ast \HH_\ast) \gg 1$, the logarithmic branch becomes the one in
    flat space-time, proportional to
    \begin{equation}
        \lim_{k \tini \gg 1}
        \OmGW^0 (K) \sim K ({\cal R}_\ast \HH_\ast)^{2}\,; \label{scaling_Regime2}
    \end{equation}
    \item a small-scale range proportional to
    \begin{equation}
    \label{largescale_omgw_cit}
        \OmGW^0 (K) \sim K^{3 - b_\Pi} \ln^2 \bigl[1 + ({\cal R}_\ast \HH_\ast) /K \bigr]
        \,,
    \end{equation}
    when $K \gtrsim K_{\rm GW}$,
    due to the decay of the normalized anisotropic stresses $\zeta_\Pi^\ast (K)$
    [see 
    \Eqq{zetaPi_fits}]. In \Eq{largescale_omgw_cit}, $-b_\Pi$ is the small-scale slope of $\zeta_\Pi(K)$, e.g., $b_\Pi = b + 3$ for a purely vortical
    vector field (see \Eqq{b_pi_definitions} for the slopes associated to the various components).
    As we generally consider causal sources with ${\cal R}_\ast \HH_\ast \ll 1$,
    and $K_{\rm GW} \gtrsim 1$,
    then this branch satisfies $K_{\rm GW} / ({\cal R}_\ast \HH_\ast)  \gg 1$, and
    the GW spectrum becomes proportional to
    \begin{equation}
        \lim_{k\tini \gg 1} \OmGW^0
        (K) \sim K^{1 - b_\Pi} ({\cal R}_\ast \HH_\ast)^{2}
        \,.
    \end{equation}
\end{enumerate}
We are omitting the presence of an exponential decrease of
the spectrum of the source below a certain dissipative scale, i.e., at $K > K_{\rm D}$, for example
due to viscosity, diffusivity, or shocks (see, e.g., refs.~\cite{Kosowsky:2001xp,Kahniashvili:2008pe,Caprini:2009yp,Cutting:2020nla,Caprini:2024ofd}) and
the dynamics of the production of the source over a period of time
$2 \, \Delta \tau_s$, which modifies the kernel at $k > 1/ \Delta \tau_s$
(see \Fig{kernel_instant_smooth}).
In both cases, the GW spectrum presents an exponential decay
at $K > K_{\rm exp} = \min\bigl(\frac{1}{k_\ast \Delta \tau_s}, K_{\rm D}\bigr)$.
As long as $K_{\rm exp} \gg K_{\rm GW}$ is satisfied, this correction
can be safely omitted to describe the relevant modes of the GW spectrum.

Since
GW spectra being proportional to $K^3$ and $K$ at small and intermediate wave numbers and decaying
at large wave numbers are commonly found in GW
studies conducted in flat space-time
and expanding backgrounds
(see, e.g., refs.~\cite{Jinno:2017fby,Konstandin:2017sat, RoperPol:2019wvy,Jinno:2020eqg,Jinno:2022mie, Jinno:2022fom, RoperPol:2021uol, RoperPol:2022iel, RoperPol:2023dzg,RoperPol:2023bqa, Sharma:2023mao,Blas:2026yqb}),
the CiT model seems to be in general applicable
to slowly-decaying sources of GWs \cite{Caprini:2024ofd}.

Following a common normalization for studies of GW backgrounds
(see, e.g., refs.~\cite{Caprini:2019egz,RoperPol:2019wvy,RoperPol:2021uol,Hindmarsh:2019phv,RoperPol:2022iel,RoperPol:2023bqa,
RoperPol:2023dzg,Caprini:2024hue,Caprini:2024gyk,Caprini:2024ofd,Stomberg:2025kxf}),
we can write the GW spectrum of \Eq{OmGW_envelope} as
\begin{equation}
    \OmGW^0 = 3 \, F_{\rm GW}^0 \, \tilde \Omega_{\rm GW} \,
    \frac{\tilde \varphi_{\rm rms}^4}
    {\bar \rho^2} \, ({\cal R}_\ast \HH_\ast)^2 \,
    {\cal S} (K)\,. \label{OmGW_template}
\end{equation}
The spectral shape is determined by the function ${\cal S} (K)$,
\begin{equation}
    {\cal S} (K) = {\cal N} \, 
    K^3 \zeta_\Pi^\ast (K) \, {\cal T}_{\rm GW} (K)\,, \label{OmGW_shape}
\end{equation}
where ${\cal N}$ is a normalization constant that ensures that $\int {\cal S} (K) \dd \ln K = 1$, i.e.,
\begin{equation}
\label{norm_factor_cit}
    {\cal N}^{-1} = \int_{-\infty}^\infty K^3 \zeta_\Pi^\ast (K) \,
    {\cal T}_{\rm GW} (K) \dd \ln K \,.
\end{equation}
The efficiency factor $\tilde \Omega_{\rm GW}$
determines the integrated amplitude of the GW spectrum \cite{Caprini:2019egz,RoperPol:2023bqa,Caprini:2024gyk},
\begin{align}
    \tilde \Omega_{\rm GW}  =  &\, \frac{{\cal C}}{2{\cal A}^2}
    \frac{{\cal N}^{-1}}{({\cal R}_\ast \HH_\ast)^2} \,.
    \label{OmGW_efficiency}
\end{align}
We note that the factor ${\cal C}/(2{\cal A}^2)$ has been used in
refs.~\cite{RoperPol:2023dzg,Caprini:2024hue} to describe $\tilde \Omega_{\rm GW}$,
based on the results of ref.~\cite{RoperPol:2022iel}.

We have introduced a quadratic scaling with the source length scale in \Eq{OmGW_template}
to take into account the dependence of the GW amplitude at its peak $k_{\rm GW}$ ~\cite{RoperPol:2022iel}
[see also \Eq{scaling_Regime2}].
To find the exact position of the peak of the GW spectrum, we thus need
to find the value of $K$ that maximizes $K \zeta_\Pi^\ast (K)$ since the peak is in the
regime where the kernel's envelope, ${\cal T}_{\rm GW} (K)$,
becomes proportional to $K^{-2}$.
Using the fits of \Eqq{zetaPi_fits}, this implies that
\begin{equation}
    K_{\rm GW} = \max_{K} [K \zeta_\Pi^\ast (K)]
    = \frac{K_\Pi}{( b_\Pi - 1)^{{1 \over \alpha_\Pi}}}
    \,. \label{KGW_fit}
\end{equation}
This value was used, e.g., in ref.~\cite{RoperPol:2023bqa} [see their Eq.~(19a)]
to characterize $s_{\rm turb} = f_{\rm turb} {\cal R}_\ast$ as the normalized
peak frequency of the GW spectrum from MHD turbulence.

\begin{figure}
    \centering
    \includegraphics[width=0.5\textwidth]{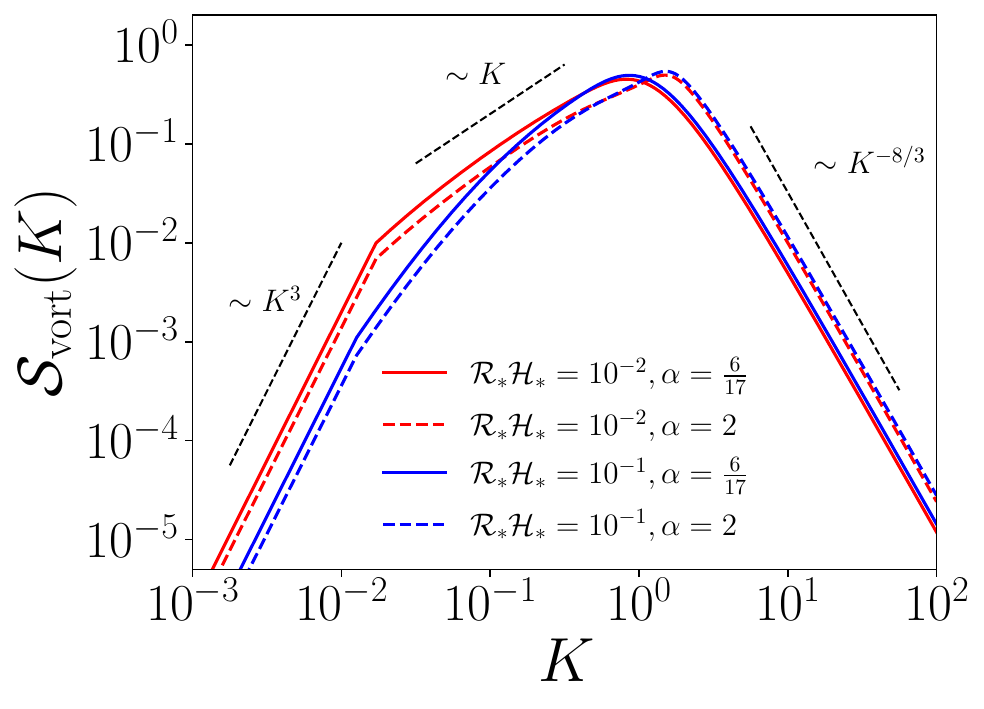}\\
    \includegraphics[width=0.5\textwidth]{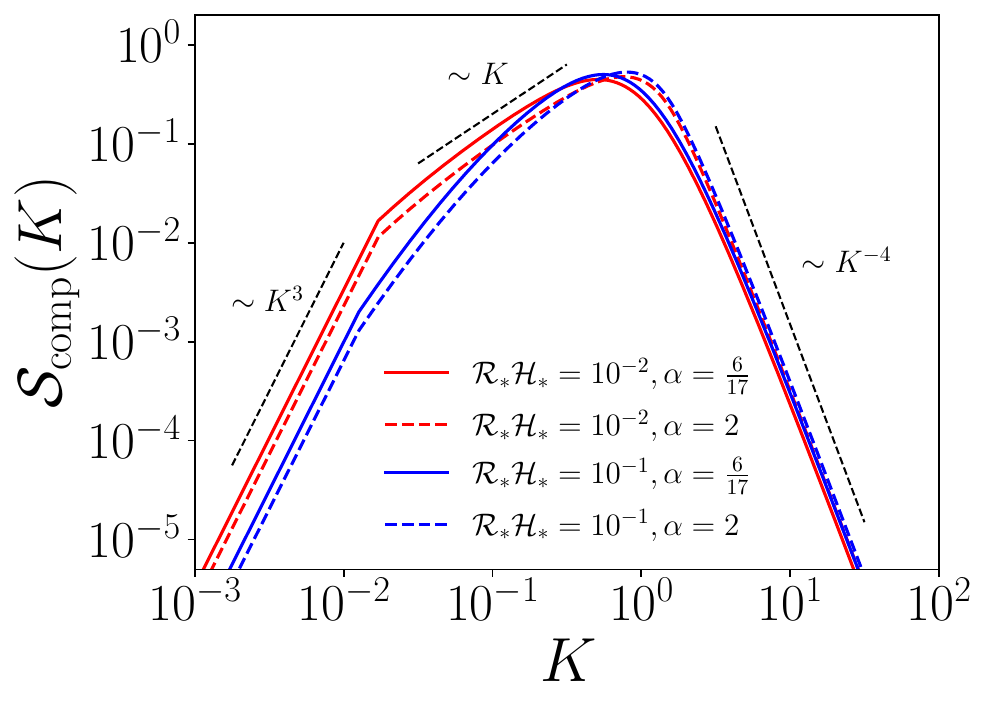}
    \caption{Normalized spectral shape of the GW spectrum for a purely
    vortical source with $a = 5$, $b = 2/3$
    (upper panel) and for a purely compressional one with
    $a=5, b=1$ (lower panel), for both $\alpha = 6 / 17$ (solid lines)
    and $2$ (dashed lines).
    We consider two values of ${\cal R}_\ast \HH_\ast = 10^{-2}$ (red lines)
    and $10^{-1}$ (blue lines) with sourcing duration $\delta \tau_{\rm fin}/{\cal R}_\ast = 100$.}
    \label{shapeGW_vortcomp}
\end{figure}

The normalized shape ${\cal S}(K)$ of the GW spectrum is shown in \Fig{shapeGW_vortcomp}
for a purely vortical source
with $a = 5, b=2/3$ (upper panel) and for a compressional
source with $a=5, b = 1$ (lower panel),
for two values of $\alpha$:  $\alpha = {6 \over 17}$ and $\alpha = 2$.
Their spectral peaks $K_{\rm GW}$ and GW
efficiency parameters $\tilde \Omega_{\rm GW}$
are given in \Tab{KGW_values}
for a range of values of $\alpha$.
As anticipated in \Eqq{efficiencies_EPi}, we find that $\tilde \Omega_{\rm GW}$
is in general
smaller for compressional sources than for vortical sources
when we compare their values for the same slopes $a$ and $b$,
and smoothness $\alpha$.
$K_{\rm GW}$ are
computed using \Eq{KGW_fit} and the
fitting values listed in \Tab{parameters_fits}. 
For generic velocity, magnetic, and scalar fields sourcing GWs, all
the components (vortical, compressional, mixed, and helical) need to 
be taken into account and included in \Eqs{OmGW_shape}{OmGW_efficiency}
to find the resulting
GW spectrum following the CiT model.

\begin{table}[]
    \centering
    \begin{tabular}{|c|c|c|c|c|c|c|}
    \hline 
    \rule{0pt}{3ex}
        $\alpha$ & $K_{\rm GW}^{\rm vort}$ & $K_{\rm GW}^{\rm comp}$ & $\tilde{\Omega}_{\rm GW}^{{\rm vort}, 2/3}$ & $\tilde{\Omega}_{\rm GW}^{{\rm comp}, 2/3}$ & $\tilde{\Omega}_{\rm GW}^{{\rm vort}, 1}$ & $\tilde{\Omega}_{\rm GW}^{{\rm comp}, 1}$ \\
        \hline
         0.05 & 0.032 & 0.035 & 2.591 & 0.993 & 1.345 & 0.571 \\
         0.1 & 0.211 & 0.160 & 0.491 & 0.237 & 0.390 & 0.197 \\
         0.2 & 0.543 & 0.356 & 0.196 & 0.108 & 0.208 & 0.119\\
         6/17 & 0.801 & 0.502 & 0.129 & 0.077 & 0.158 & 0.098 \\
         0.75 & 1.029 & 0.631 & 0.091 & 0.060 & 0.127 & 0.087 \\
         1 & 1.084 & 0.663 & 0.082 & 0.055 & 0.119 & 0.084 \\
         1.5 & 1.142 & 0.695 & 0.073 & 0.051 & 0.110 & 0.080 \\
         2 & 1.174 & 0.712 & 0.069 & 0.048 & 0.105 & 0.078 \\
         3 & 1.210 & 0.730 & 0.064 & 0.045 & 0.100 & 0.075 \\
         4 & 1.231 & 0.740 & 0.061 & 0.044 & 0.097 & 0.073 \\
         5 & 1.244 & 0.747 & 0.059 & 0.043 & 0.095  & 0.072 \\ 
         10 & 1.276 & 0.761 & 0.056 & 0.041 & 0.091 & 0.069 \\
         \hline
    \end{tabular}
    \caption{
    Values of the GW spectral peak $K_{\rm GW}$ [see \Eq{KGW_fit}] and of the GW efficiency in the CiT model $\tilde{\Omega}_{\rm GW}$ for a purely vortical non-helical
    and a purely compressional source with slopes $a = 5$ and both $b = 2/3$
    and $1$,
    and a range of values of the smoothness parameter $\alpha$. The values of $\tilde{\Omega}_{\rm GW}$ correspond to the case ${\cal R}_\ast {\cal H}_\ast = 10^{-2}$ and are computed at $\delta \tau_{\rm fin}/\tau_\ast = 100$. The values $K_{\rm GW}^{\rm vort}$
    and $K_{\rm GW}^{\rm comp}$ respectively correspond to $b=2/3$ and $b=1$.}
    \label{KGW_values}
\end{table}

\subsection{Circular polarization of GWs}
\label{circ_pol}

The degree of circular polarization $\mathcal{P}_{\rm GW} (k)$, as already defined above \Eq{degree_circ_pol},
is given by the ratio of the chiral $\Xi_{\rm GW}(\tau, k)$ to the symmetric $\Omega_{\rm GW}(\tau, k)$ GW energy density spectrum.
Applying the CiT model as in previous sections,
$\Xi_{\rm GW}(\tau, k)$ becomes [see \Eq{OmGW_constant_tfin}]
\begin{align}
    \Xi_{\rm GW}(\tau, k) = 3 \, H_\Pi^*(k) \, F_{\rm GW}^0
    \, \Delta^2 (\tau, k).
\end{align}
Then, the degree of circular polarization becomes
independent of time $\tau$ and it corresponds to the ratio
of the antisymmetric to the symmetric spectra
of the anisotropic stresses, ${\cal P}_\Pi (k)$ [see \Eq{polar_Pi}],
\begin{align}
    \mathcal{P}_{\rm GW} (k) \equiv \frac{\Xi_{\rm GW}(\tau, k)}
    {\Omega_{\rm GW} (\tau, k)} = {\cal P}_\Pi(k)  = \frac{H_\Pi^*(k)}{E_\Pi^*(k)}.
\end{align}

\begin{figure}
    \centering
    \includegraphics[width=\columnwidth]{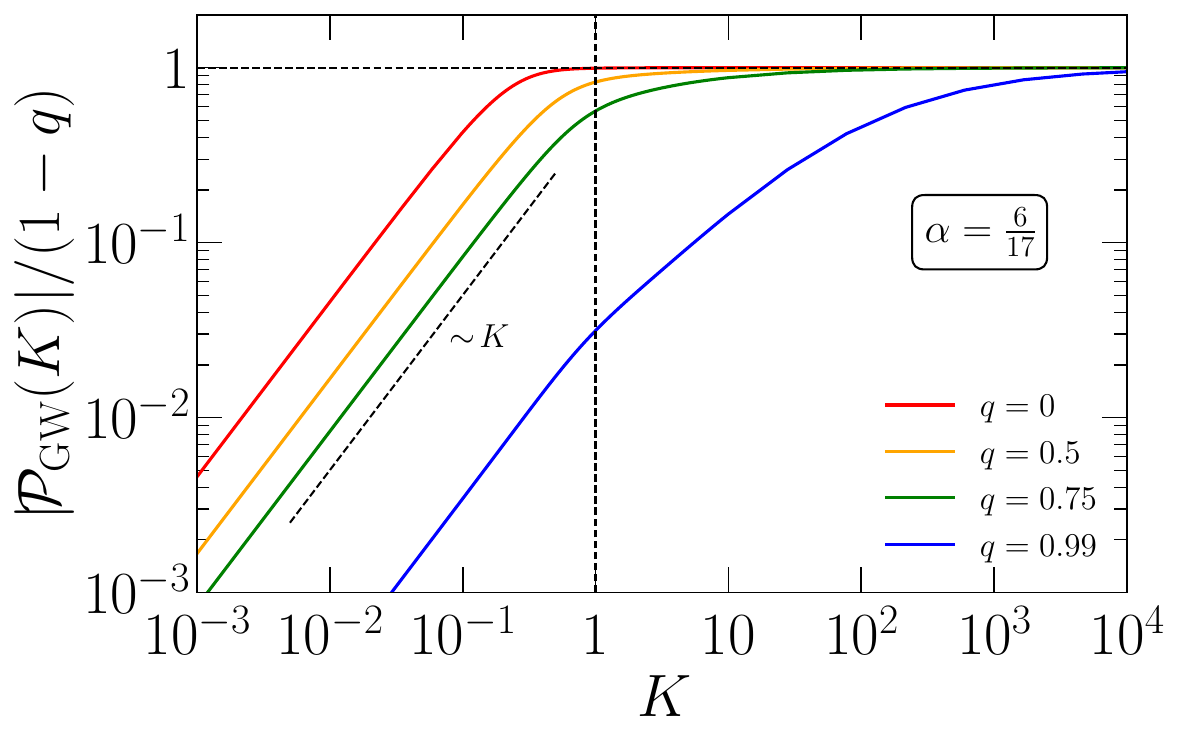}
    \includegraphics[width=\columnwidth]{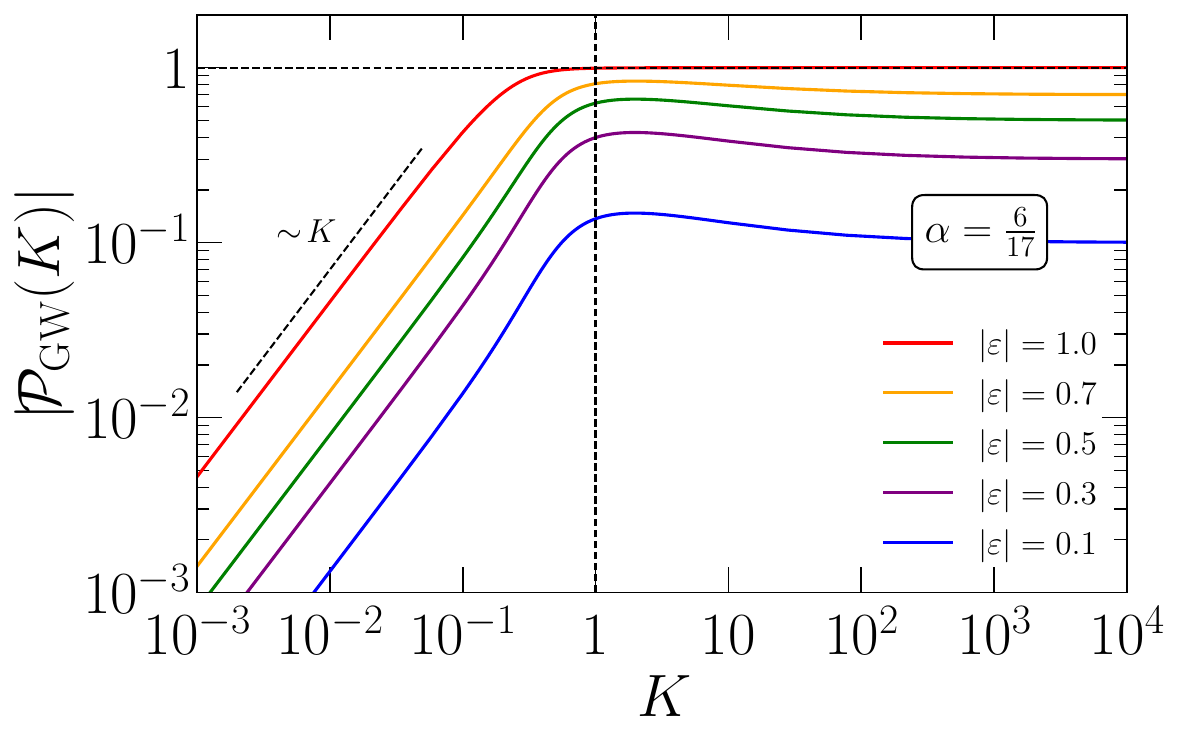}
    \caption{
    {\em Upper panel:} Degree of circular polarization $|{\cal P}_{\rm GW} (K)|$
    for different values of the degree of compressibility $q \in [0, 1]$
    of a fully helical velocity field with $\varepsilon_u = \pm 1$.
    {\em Lower panel:} Same as upper panel but for a fully vortical field ($q=0$) with different values of fractional helicity.
    In both panels, we take $a=5$ and $b=2/3$.}
    \label{PGW_q}
\end{figure}

For generic velocity field, using \Eqs{EPi_u}{HPi_u}
we find
\begin{align}
    \mathcal{P}_{\rm GW}
    ^u (K) = &\, \frac{H_\Pi^u(K)}{E_\Pi^u(K)} =
    K
    \frac{{\cal C}_u^\H \zeta_{\Pi}^{u, \H} (K)}{{\cal C}_u \,\zeta_\Pi^u (K)}\,,
\end{align}
where each spectral component
contributing
to the normalized anisotropic stresses can be determined using
\Eqq{normalized_zetaPis}.
Since ${\cal P}_{\rm GW} (k)$ is constructed from each component, accounting compressional,
vortical, and helical components of the velocity spectrum,
the degree of circular polarization depends on
the degree of compressibility $q(k)$ and on
the fractional helicity $\varepsilon_u (k)$,
as shown in \Fig{PGW_q},
taking constant values for both $\varepsilon_u$ (upper panel)
and $q$ (lower panel).

Similarly, the degree of circular polarization for a magnetic field can be expressed using \Eqss{normalized_HPi_i}{ch_u} for $H_\Pi$ and \Eqss{EPi_u}{cu_hel_eq} for $E_\Pi$, assuming constant fractional helicity, as 
\begin{align} \label{pol_GW_B}
    \mathcal{P}_{\rm GW}
    ^B (K)=  &\, 
    \frac{H_\Pi^B(K)}{E_\Pi^B(K)} = K 
    \frac{{\cal C}_B^\H \, \zeta_{\Pi}^{B, H}(K)}{{\cal C}_B \,\zeta_\Pi^B (K)}
    \nonumber \\ = &\, K \frac{2 \,\varepsilon_B}{7 - 5 \, \varepsilon_B^2}
    \frac{{\cal J}}{{\cal I}} \frac{\zeta_\Pi^{B, \H} (K)}
    {\zeta_\Pi^{B} (K)}
    \,,
\end{align}
where ${\cal I} \equiv \int
P^{-4} \zeta_B^2(P) \dd P$
and ${\cal J} \equiv \int P^{-5} \zeta_B^2(P) \dd P$ are integrals
that determine ${\cal C}_B$ and ${\cal C}_B^\H$, respectively, up to constants
that appear multiplying in \Eq{pol_GW_B}
if we assume constant fractional helicity [see \Eqs{Cuu_2}{CH_vort}].
 
In \Fig{PGW_q},
we see that
the degree of compressibility $q$ impacts the
scales at which the spectrum becomes maximally polarized, delaying it towards smaller scales
(larger $K$) as the compressibility is increased,
reaching a plateau at ${\cal P}_{\rm GW} (K) = (1 - q) \,\varepsilon$
(see upper panel of \FFig{PGW_q}).
Note that for magnetic fields, $q = 0$ and all wave numbers above $K \gtrsim 1$
in the GW spectrum are fully polarized for $\varepsilon_B = \pm 1$.

\begin{figure}
    \centering
    \includegraphics[width=\columnwidth]{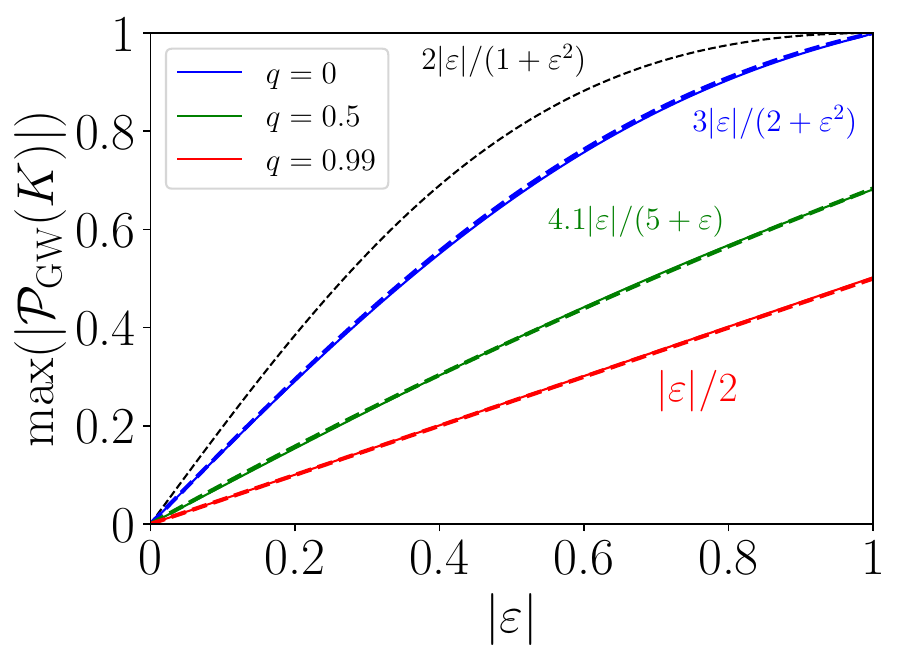}
    \caption{
    Maximum value of the circular polarization as a function of the helicity 
    of the field (solid lines) for different values of $q$, and $a=5, b=2/3, \alpha=6/17$. In dashed lines we show simple fits for each case with a relative error of less than $1\%$.}
    \label{PGW_eps}
\end{figure}

The lower panel of \Fig{PGW_q} shows the spectrum of circular polarization for
a vortical field with $q = 0$ and for different
values of the helicity fraction.
For a fully helical field,
a plateau appears at large wave numbers $K \gtrsim 1$,
consistent with numerical results of helical decaying MHD turbulence
(see Fig.~5 of ref.~\cite{RoperPol:2021uol})
and previous theoretical estimates \cite{Kahniashvili:2005qi,Kisslinger:2015hua}.
This plateau reaches a maximum value of one for ${\cal P}_{\rm GW} (K \gg 1)$ when the source
is fully helical.

The maximum value of $|{\cal P}_{\rm GW} (K)|$ as a function
of $|\varepsilon_B|$ is 
shown in \Fig{PGW_eps} for different values of $q$, together with some simple fits.
The fitting function for $q=0$ has a similar functional dependence to the one found for a deterministic magnetic
Beltrami field sourcing GWs (see App.~A of ref.~\cite{RoperPol:2021uol}),
${\cal P}_{\rm GW}^{\rm Bel} = 2 \, \varepsilon_B/(1 + \varepsilon_B^2)$, shown with a dashed black line in \Fig{PGW_eps}.
Furthermore, \Eq{pol_GW_B} explains the approximately linear dependence of ${\cal P}_{\rm GW} (K)$
with the fractional helicity of the source
(see Fig.~6 of ref.~\cite{RoperPol:2021uol}),
since ${\cal C}_B^\H \propto \varepsilon_B$, while ${\cal C}_B$ varies more slowly with $\varepsilon_B$.
However, the dependence of ${\cal C}_B \zeta_\Pi^B$ with $\varepsilon_B$
(see also \Fig{zetaPivel})
can lead to relevant corrections as $\varepsilon_B^2$ approaches one, up to a factor of $3.5$.
Finally, we note that positive or negative helicity of the source
only impacts the overall sign of the degree of circular polarization, as expected,
but not its absolute value.

\section{Applications of the Constant-in-Time model}
\label{applications}

In this section, we briefly discuss the applicability of the CiT
model to a few example sources.
As discussed above, this model had been developed \cite{RoperPol:2022iel} and validated with numerical
simulations of decaying MHD turbulence
\cite{RoperPol:2022iel,Auclair:2022jod}.
However, the most notable predictions of this model are the $k^3$ and $k$ scalings
of the GW spectrum at small and intermediate wave numbers,
followed by a decay $k^{1-b_\Pi}$  at large
wave numbers
that depends on the particular source.
This type of spectrum is broadly found in the literature on GW production in the early Universe 
\cite{Jinno:2017fby,Konstandin:2017sat, RoperPol:2019wvy, Jinno:2022fom, RoperPol:2022iel, RoperPol:2023dzg, Sharma:2023mao}.

It goes beyond the scope of this work to make an exhaustive list
of these cases and check the applicability of the CiT model to each of them,
so we just provide in the following a few examples relevant for first-order
phase transitions: stationary turbulence, sound waves, bubble collisions, 
and feebly interacting particles.
Our aim is to motivate
the community to test their GW predictions against the CiT model
when studying GW sources that vary slowly in time and explore in the future
its applicability to other sources.
Since this model allows us to make simple and robust predictions for the GW background,
identifying the sources that can be accurately described by this model
would allow preparing
pipelines for GW searches in current and future GW experiments like Pulsar
Timing Arrays or LISA.

\subsection{MHD turbulence}
\label{constant_model}

\subsubsection*{Turbulent decay}

The characteristic time of the decaying MHD turbulent
cascade is the eddy turnover time
$\delta \te = \te - \tini = (u_* \kf)^{-1}$,
where $u_*$ is a characteristic speed of the turbulent flow.
We assume this speed to be the maximum, at the initial time, between the enthalpy-weighted rms velocity
of the fluid, $U_{\rm rms} = \sqrt{\bra{w u^2}/\bar w} \simeq u_{\rm rms}$
and the
Alfv\'en speed, which, during the radiation-domination era,
is given by $\vA^2 = 2 \bar \rho_B/\bar w = \tfrac{3}{2} \bar \rho_B/\bar \rho$ \cite{Brandenburg:1996fc,RoperPol:2023bqa,RoperPol:2025lgc}.
The eddy turnover time corresponds to the time it takes for
the eddies of size $\lambda_* = 2\pi/k_*$ to be processed by turbulence.

Comparing this timescale $\delta \tau_e$ with the one found in the GW dynamics,
$\delta \tau_{\rm GW} = 1/k$, and taking into account that both $\vA$ and $u_{\rm rms}$
are smaller than one due to causality,
one finds that
the GW dynamics is faster than the dynamical timescale of
magnetic and velocity fields evolution
for all wave numbers
$k > u_* \kf$, as proposed in ref.~\cite{RoperPol:2022iel}.
For subrelativistic speeds, the range of wave numbers
where this condition holds is larger and it only breaks down at the largest scales.
There are, however, some discrepancies between the CiT model and the numerical results
in the wave numbers below the spectral peak (i.e., for $k < k_*$),
which increase when the
eddy turnover time becomes smaller.
This is precisely because at these wave numbers, the
timescale of GW production is larger than the eddy turnover time and
the decay of the source has an impact on the
GW production.
In addition, the CiT model requires
to introduce an effective
finite duration after which the source is abruptly
switched off in order to incorporate the
effect of the turbulent decay.
To go beyond these limitations, we introduce
in \Sec{beyond_cit} a coherent-decay model for the UETC
that accounts for the decay of the source.
We find indeed that when the decay occurs at a characteristic
timescale $\delta \tau_{\rm dc}$ that is longer than that of the relevant
GW dynamics, the impact on the GW background is restricted to $k < \delta \tau_{\rm dc}^{-1}$,
allowing us to apply the CiT model at $k> \delta \tau_{\rm dc}^{-1}$, and
incorporating the effects of the decay at $k < \delta \tau_{\rm dc}^{-1}$.
Here, $\delta \tau_{\rm dc}^{-1} = u_* k_*$
when the decay timescale is prescribed by
the eddy turnover time $\delta \tau_e$.

\subsubsection*{Kraichnan decorrelation in stationary turbulence}

Furthermore, turbulent motion 
is expected to decorrelate with time
due to its non-linear nature \cite{1965PhFl....8.1385K}.
This has been extensively studied and accounted for in turbulence theory and
used for studies of GW production from stationary turbulence \cite{Kosowsky:2001xp,Kahniashvili:2008pe,Caprini:2009yp,Niksa:2018ofa,Auclair:2022jod}.
In the following, we compute the impact of including a decorrelation
velocity $u_{\rm dcorr}$ in the UETC of the velocity or magnetic fields and how the
resulting GW spectrum compares to the one found in the CiT model.
This regime was also explicitly studied in ref.~\cite{RoperPol:2023dzg},
where the usual assumption of linear growth with the source duration considered
for stationary sources \cite{Gogoberidze:2007an} is only found for $k$ above
a threshold value that depends on the decorrelation velocity (see Fig.~8 in ref.~\cite{RoperPol:2023dzg}).
This indeed shows that our model allows us to compute the GW background when the velocities
are relatively small, such that both the decay and the decorrelation timescales are larger
than that of GW production.
Since our focus is to test the universality of the CiT model and its extension
to include decay of the source,
we restrict our analysis in this section to stationary turbulence, and defer the
study of decaying turbulence, including both the decay and the decorrelation
of the UETC
to a separate publication \cite{kraichnan}.

An important caveat of our proposed model is that it omits the GW production
during the process of turbulence development.
As we pointed out in \Sec{cit_model} (see \Fig{kernel_instant_smooth}), our model is applicable at all wave numbers
below the inverse duration of the sourcing phase.
If this sourcing phase occurs on a timescale similar to the dynamical
timescale of the source, it could have an impact on the GW background
at smaller wave numbers, even around the peak \cite{Caprini:2009fx,Caprini:2009yp}.
Therefore,
the sourcing phase of turbulence needs to be studied in detail for accurate and fully reliable predictions for the GW background from realistic
setups of MHD turbulence in the early Universe.

Omitting the decay of the source, the inertial range of the UETC of the velocity field in Kolmogorov stationary turbulence is expected to follow a Kraichnan decorrelation in time,  
\cite{1965PhFl....8.1385K,1993PhFlA...5.2835K,Niksa:2018ofa,Auclair:2022jod}
\begin{align}
\label{UETC_kraichnan_formula}
& E_u(\tau_1, \tau_2, k) = E_u^\ast (k) \exp \Bigl[ -\frac{1}{2}k^2 u_{\rm dcorr}^2(k) \, \tau_-^2 \Bigr]\,, \nonumber \\
& \text{where} \  E_u^* \equiv E_u(\tau_\ast, k) \ \text{and} \ \tau_- \equiv \tau_2-\tau_1\,.
\end{align}
Here we have introduced the decorrelation velocity $u_{\rm dcorr}(k)$, which is close
to a constant deep inside the inertial range
\begin{align}
    u_{\rm dcorr}^2(k) &= \frac{1}{3} C_u^2 u_{\rm rms}^2\,, \ \text{with} \ C_u \sim {\cal O}(1) \nonumber \\
    &\text{for} \ k_\ast \ll k \ll k_{\rm D}\,,
\end{align}
where $k_\ast$ and $k_{\rm D}$ are respectively the spectral
peak and the dissipative scales.
The decorrelation velocity can be generalized to a scale-dependent decorrelation velocity, as proposed in
ref.~\cite{1993PhFlA...5.2835K},
already considered in the context of GW production from turbulence in refs.~\cite{Niksa:2018ofa,Auclair:2022jod},
\begin{equation}
    u_{\rm dcorr}^2 (k) = \int_0^\infty h(p/k)  \, E_u (p) \dd \ln p\,, \label{vdc}
\end{equation}
where \cite{1993PhFlA...5.2835K}
\begin{align}
    h(x) = &\, \frac{13 - 8x^2 + 3x^4}{48} \nonumber \\ &\, +
    \frac{(1 - x^2)^3}{32x} \ln \frac{1 + x}{|1 - x|}\,.
\end{align}

For the von K\'arm\'an spectrum considered in ref.~\cite{Auclair:2022jod}, the dependence of
the decorrelation velocity with $k$ can be
approximated by the following fit
\begin{equation}
    u_{\rm dcorr}^2 (K) \simeq \frac{u_{\rm rms}^2}{3} \biggl(\frac{1 + 0.2 K}{\sqrt{5/2} + 0.2 K}
    \biggr)^2\,.
\label{kraichnan_dcorr_fit}
\end{equation}
This function takes values from $u_{\rm dcorr}^2 =  \frac{2}{15} \, u_{\rm rms}^2$
when $K \to 0$,
to $\onethird u_{\rm rms}^2$ for large $K$, deep in the inertial range.

In the following,
we want to show in which limit the Kraichnan decorrelation can be neglected for the computation of the GW spectrum from stationary turbulence and, hence, understand the regime of validity of the CiT model for this type of sourcing process.
We will assume
that the decorrelation velocity is a constant, such that the kernel $\Delta^2$
does not
depend on the spectrum of the source and we can provide general results
for this type of sources.

The GW spectrum at present time, found
after averaging over fast frequency oscillations, then becomes,
for a purely vortical velocity field [cf.~\Eq{zetaPi_vort}],
\begin{align}
    \Omega_{\rm GW}^0 (k) = &\,\frac{3}{8} \, \Gamma^2 \, u_{\rm rms}^4 \, K^2 \,  F^0_{\rm GW} \, \frac{{\cal C}}{{\cal A}^2}
    \nonumber \\
    \times & \,
    \int_0^{\infty} \frac{\dd P}{P^2} \int_0^{\infty} \frac{\dd \tilde{P}}{\tilde{P}^2} (1+z^2)(1+\tilde{z}^2) \nonumber \\
    \times &  \, \zeta_u(P) \, \zeta_u(\tilde{P})
    \, \Delta_{\rm turb, av}^2 (K, P, \tilde P)\,,
\label{om_gw_kraichnan_one}
\end{align}
where $\Delta_{\rm turb, av}^2$ is the kernel that includes the
time decorrelation of the UETC and the averaged Green's function,
\begin{align}
    \Delta_{\rm turb, av}^2 =&\, \frac{1}{2}
    \int_{\tau_\ast}^{\tau_{\rm fin}}  \frac{\dd \tau_1}{\tau_1} 
    \int_{\tau_\ast}^{\tau_{\rm fin}} \frac{\dd \tau_2}{\tau_2} 
    \cos (\tau_- k_\ast K) \nonumber \\ \times &\, 
    \exp \Bigl[ - \frac{1}{2} \tau_-^2 u_{\rm dcorr}^2 k_\ast^2  (P^2 +\tilde{P}^2)
    \Bigr]\,.
\end{align}
We note that this kernel can be integrated analytically over $\tau_-$ after
a change of variable, reducing it to one integral over time $\tau_1$ \cite{kraichnan}.

Similarly to what we did in \Sec{gw_template_cit} for the CiT model
[see \Eq{OmGW_template}], we can define a normalized spectral shape
${\cal S}_{\rm turb}(K)$ and a GW
efficiency $\tilde{\Omega}_{\rm GW}^{\rm turb}$ such that we can write the GW spectrum at present time as
\begin{subequations}
\begin{align}
    \Omega_{\rm GW}^0
    (k) = 3  & \, \Gamma^2 u_{\rm rms}^4 \, F^0_{\rm GW} \,
    \tilde{\Omega}_{\rm GW}^{\rm turb} \, \nonumber \\ \times &\,
    ({\cal R}_\ast {\cal H}_\ast)^2 \, {\cal S}_{\rm turb}(K)\,,
\end{align}
with
\begin{align}
    \tilde{\Omega}_{\rm GW}^{\rm turb} = &\, \frac{\cal C}{{\cal A}^2} \frac{{\cal N}_{\rm turb}^{-1}}{({\cal R}_\ast {\cal H}_\ast)^2}\,, 
    \label{tilde_om_gw_kraichnan}\\
    {\cal S}_{\rm turb}(K) =   & \, {\cal N}_{\rm turb} \frac{K^2}{8}  \nonumber \\
    \times &\,
    \int_0^{\infty} \frac{\dd P}{P^2} \int_0^{\infty} \frac{\dd \tilde{P}}{\tilde{P}^2} 
    (1+z^2)(1+ \tilde{z}^2)
    \nonumber \\ \times &\,
    \zeta_u(P) \, \zeta_u(\tilde{P}) \,
    \Delta_{\rm turb, av}^2 (K, P, \tilde P)\,,
    \label{cal_s_gw_kraichnan}
\end{align}
and where we have again
defined ${\cal N}_{\rm turb}$
such that $\int {\cal S}_{\rm turb} (K)
\dd \ln K = 1$.
\end{subequations}

\begin{figure}
    \centering
    \includegraphics[width=\columnwidth]{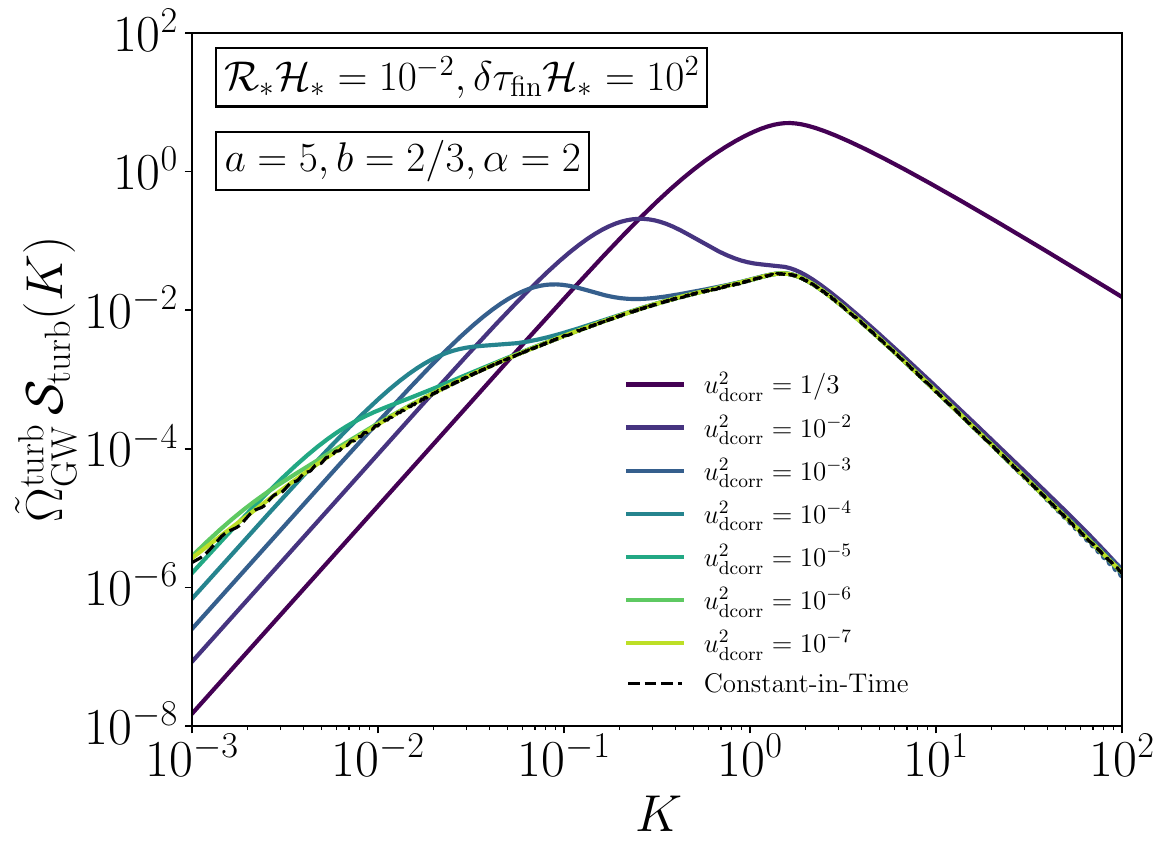}    \includegraphics[width=\columnwidth]{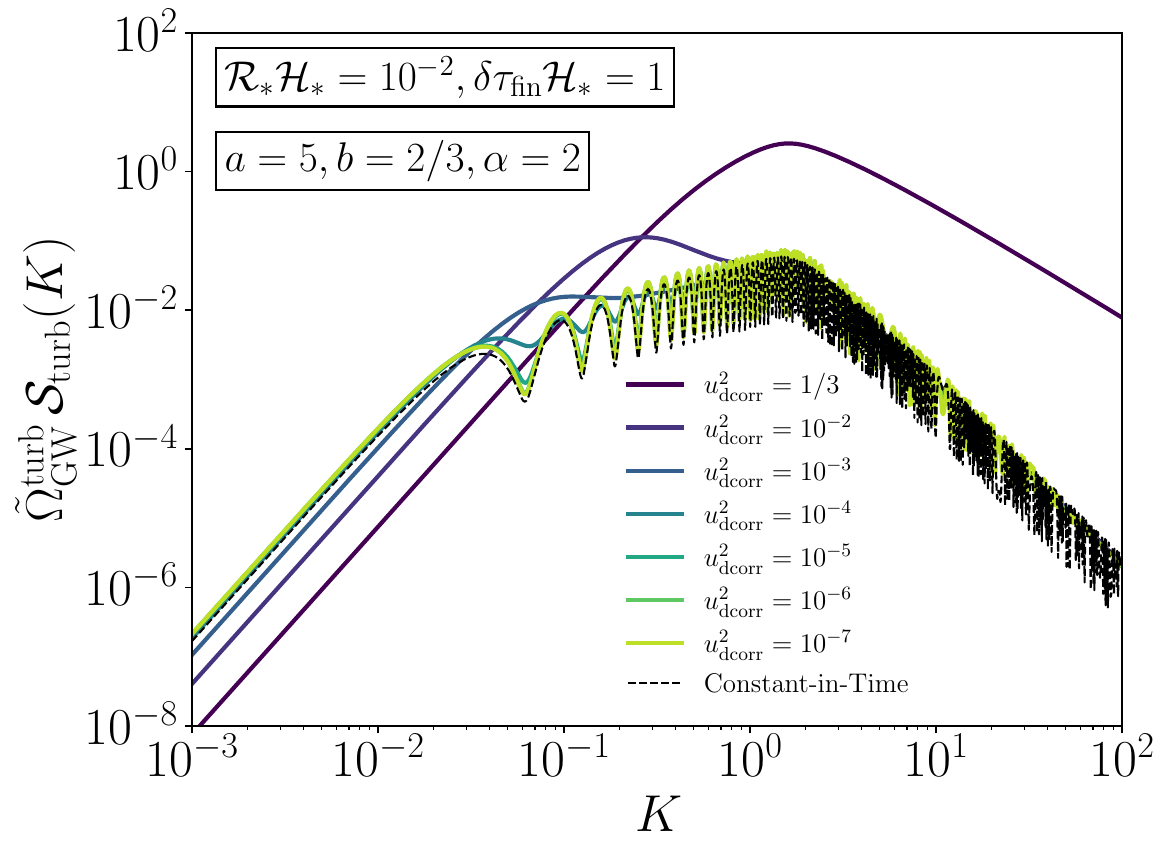}
    \caption{
    Normalized
    GW spectrum $\tilde{\Omega}_{\rm GW}^{\rm turb}\, {\cal S}_{\rm turb}(K)$
    in the case of purely vortical turbulence including Kraichnan decorrelation, for different constant decorrelation
    velocities $u_{\rm dcorr}$.
    In both panels
    we consider a von K\'arm\'an spectrum with $a = 5$, $b = 2/3$, and $\alpha =2$.
    The characteristic lengthscale of the source
    is ${\cal R}_\ast {\cal H}_\ast = 10^{-2}$ and we consider
    two source durations:
    $\delta \tau_{\rm fin} {\cal H}_\ast = 10^2$ (upper panel) and $\delta \tau_{\rm fin} {\cal H}_\ast = 1$ (lower panel).
    For comparison we plot in dashed black lines the spectrum in the CiT case,
    which is approached as $u_{\rm dcorr}$ decreases.}
    \label{om_gw_kraichnan}
\end{figure}

In \Fig{om_gw_kraichnan},
we show the normalized
GW spectrum $\tilde \Omega_{\rm GW}^{\rm turb} {\cal S}_{\rm turb}(K)$
for different values of $u_{\rm dcorr}$ and two
different source durations, $\delta \tfin = \tau_\ast$ (lower panel)
and $\delta \tfin = 100 \tau_\ast$ (upper panel).
We see that the GW spectrum of a field whose UETC includes
Kraichnan decorrelation accurately
recovers the CiT model, for ${\cal R}_\ast {\cal H}_\ast = 10^{-2}$,
when $u_{\rm dcorr}^2 \lesssim 10^{-4}$
corresponding to subrelativistic speeds $u_{\rm rms} \lesssim 10^{-2}$.
In general, we expect to recover the CiT model in the limit
$u_{\rm dcorr} k_\ast \delta \tfin \ll 1$.

\begin{table}[htbp]
    \centering
    \renewcommand{\arraystretch}{1.3}
    \begin{tabular}{|c|c|c|c|c|c|}
        \hline
        \diagbox{$\ \ \ \ \ u_{\rm dcorr}^2$}{$\vspace{-2mm} \delta\tau_{\rm fin}/\tau_*$}
        & $1$ & $5$ & $10$ & $50$ & $100$ \\
        \hline
        $\hspace{-2mm} 1/3$ & $4.589$ & $7.630$ & $8.321$ & $8.972$ & $9.061$ \\
        \hline
        $10^{-2}$ & $0.234$ & $0.318$ & $0.340$ & $0.360$ & $0.363$ \\
        \hline
        $10^{-3}$ & $0.095$ & $0.096$ & $0.097$ & $0.099$ & $0.099$ \\
        \hline
        $10^{-4}$ & $0.071$ & $0.072$ & $0.073$ & $0.073$ & $0.073$ \\
        \hline
        $10^{-5}$ & $0.069$ & $0.069$ & $0.070$ & $0.070$ & $0.070$ \\
        \hline
        $10^{-6}$ & $0.068$ & $0.069$ & $0.069$ & $0.070$ & $0.070$ \\
        \hline
        $10^{-7}$ & $0.068$ & $0.069$ & $0.069$ & $0.069$ & $0.069$ \\
        \hline
    \end{tabular}
    \caption{
    Values of $\tilde{\Omega}_{\rm GW}^{\rm turb}$ for different
    $u_{\rm dcorr}^2$ and $\delta\tfin/\tau_*$ computed from \Eq{tilde_om_gw_kraichnan} for a von K\'arm\'an
    vortical spectrum with $a=5$, $b=2/3$, and $\alpha = 2$. We take
    ${\cal R}_\ast {\cal H}_\ast = 10^{-2}$ as in \Fig{SSM_CinT}.
    Note that for $u_{\rm dcorr}^2 \lesssim 10^{-4}$ we recover
    the efficiency ${\tilde \Omega}_{\rm GW} \simeq 0.069$
    found in the CiT model (see \Tab{KGW_values}) for all values of $\delta \tfin$ considered.}
    \label{tab:tilde_omega_gw_kraichnan_tab}
\end{table}

In \Tab{tab:tilde_omega_gw_kraichnan_tab} we show the values of $\tilde{\Omega}_{\rm GW}^{\rm turb}$ for different source durations $\delta \tau_{\rm fin}/\tau_\ast$ and different values of the decorrelation velocity $u_{\rm dcorr}$, showing how the GW efficiency approaches the value
found in the CiT model (i.e., $\tilde \Omega_{\rm GW} \simeq 0.069$; see \Tab{KGW_values})
as $u_{\rm dcorr}$ decreases, i.e., as the motion becomes subrelativistic. As the decorrelation velocity increases we see the growth of the peak of the
resonant term at a wave number $k$ proportional to $k_\ast u_{\rm dcorr}$.
The resonant
peak appears when 
$u_{\rm dcorr}\, k_\ast \tau_\ast \Upsilon \gtrsim 1$,
and grows
proportional to $\Upsilon = \delta \tfin \HH_\ast / (1 + \delta \tfin \HH_\ast)$
for radiation domination \cite{RoperPol:2023dzg}, and linear in
$\delta \tfin$ in flat space-time.
For smaller 
$u_{\rm dcorr} k_\ast \tau_\ast \Upsilon$, 
the resonant peak does not appear,
the GW spectrum reduces to the one predicted by the CiT model, and the spectral
peak
becomes $K_{\rm GW}$ in \Eq{KGW_fit}.
The amplitude of the resonant peak is proportional to $u_{\rm dcorr}^2$.
At large $u_{\rm dcorr}$ the peak dominates and leads to a much larger GW efficiency than in the CiT model, as we can see in \Tab{tab:tilde_omega_gw_kraichnan_tab}.
However,
in the relativistic regime of velocities,
we need to 
also take
into account the decay of the source due to
the non-linear dynamics of the fluid.
Furthermore, we need to generalize the source of GWs to $X_{\rm rms}^2 = \bar \rho_{\rm kin}$
to also take into account large fluid enthalpy fluctuations.
We leave the
realistic prediction of the GW spectrum for turbulence with both Kraichnan
decorrelation and decay of the source
for future work \cite{kraichnan}.

\subsection{Sound waves}
\label{sound_wave_component}

In studies of GW production from linearized sound waves, the UETC of the
purely compressional velocity
field is commonly
approximated as $E_u (\tau_1, \tau_2, k) \simeq E_u^\ast (k) \cos (k \cs \tau_-)$, with
$\tau_- = \tau_2 - \tau_1$, $E_u^\ast (k) \equiv E_u (\tini, \tini, k)$,
and $\cs$ being
the speed of sound
\cite{Hindmarsh:2016lnk,Hindmarsh:2019phv,RoperPol:2023dzg,Sharma:2023mao,Giombi:2024kju,Dahl:2021wyk,Dahl:2024eup,Caprini:2024gyk}.
Assuming that the compressional velocity field follows a Gaussian distribution and
introducing the UETC of the anisotropic stresses [see \Eqs{EPi_u}{zetaPi_comp}]
into the GW spectrum
at present time [see \Eq{OmGW_aver}],
we find \cite{RoperPol:2023dzg}
\begin{align}
\label{omgw_sw_formula}
    \OmGW^{0} (K) = &\, 3 \, \Gamma^2 \, u_{\rm rms}^4
    \, K^2 F_{\rm GW}^0 \, \frac{\cal C}{{\cal A}^2} \nonumber \\ &\, \times 
    \int \frac{\dd P}{P^2} \int \frac{\dd \tilde P}{\tilde P^2} (1 - z^2) (1 - \tilde z^2)
    \nonumber \\
    &\, \times
    \zeta_u (P) \, \zeta_u (\tilde P) \,
    \Delta_{\rm sw, av}^2 (K, P, \tilde P) \,,
\end{align}
where $\zeta_u (K)$ is the normalized spectrum of the velocity field [cf.~\Eq{def_comp_zeta}],
as in \Sec{sec:smoothed_dbpl} (see also ref.~\cite{RoperPol:2023dzg}).
The time kernel for stationary
sound waves, averaged over fast frequency oscillations at late times,
$\Delta_{\rm sw, av}^2$,
can be split into four components \cite{RoperPol:2023dzg,Sharma:2023mao},
\begin{equation}
\label{delta_sw_decomp}
    \Delta_{\rm sw, av}^2 = \frac{1}{4} \sum_{m, n = \pm 1} \Delta^2_{{\rm sw},mn}\,,
\end{equation}
with 
\begin{align}
    \Delta_{{\rm sw},mn}^2 (\hat p_{mn}) = \frac{1}{2} \Bigl[ &\Delta \Ci^2 (\hat p_{mn} \tfin) \nonumber \\ 
    &\, + \Delta \Si^2 (\hat p_{mn} \tfin) \Bigr]\,,
\end{align}
where $\hat p_{mn} = (p + m \tilde p)\, \cs + nk$,
and the $\Delta \Ci$ and $\Delta \Si$ functions are defined as in \Eqq{Delta_Ci_Si}.

The dominant contribution initially considered in ref.~\cite{Hindmarsh:2016lnk}
(see also refs.~\cite{RoperPol:2023dzg,Sharma:2023mao}),
corresponds to the resonant $m = 1$ and $n = - 1$, and the resulting
GW spectrum behaves
very differently than the one found in the CiT model,
as described in refs.~\cite{RoperPol:2023dzg,Caprini:2024gyk,Caprini:2024ofd}.
On the other hand, the terms corresponding to $m = -1$ and $n = \pm 1$ become
identical after integration (see Fig.~7 in ref.~\cite{Sharma:2023mao}).
In this case, we can observe that if the spectrum of the source is sharply peaked
around $k_\ast$, such that $\zeta_u (K) = \delta(K - 1)$, then the only contribution
is coming from $P = \tilde P = 1$.
Therefore, $\hat p_{-1,\pm 1} = \pm k$, such that
\begin{equation}
    \Delta^2_{{\rm sw},-1, \pm 1} = \frac{1}{2} \Bigl[\Delta \Ci^2(k \tfin)
    +\Delta \Si^2 (k \tfin)\Bigr]\,,
\end{equation}
which is equivalent to the result found in the CiT model [see \Eq{Delta_aver}].
Hence, for sharply peaked velocity
spectra, these contributions to the
sound-wave production of GWs
are expected to be well described by the CiT model.
Indeed, this corresponds to the part of the GW spectrum found in ref.~\cite{RoperPol:2023dzg}
below the spectral peak, which presents $K^3$ scaling at $K < 1/\delta \tfin$
and becomes linear at $K > 1/\delta \tfin$.
This scaling can be explained following the CiT model.
Eventually, the dominant contribution ($m = 1$, $n= - 1$) takes over, leading
to a sharp bump around the spectral peak that grows with the source duration.
This resonant term is analogous to the one found for stationary
turbulence in previous section (see also refs.~\cite{Gogoberidze:2007an,Kahniashvili:2008pe}).

\begin{figure}
    \centering
    \includegraphics[width=\columnwidth]{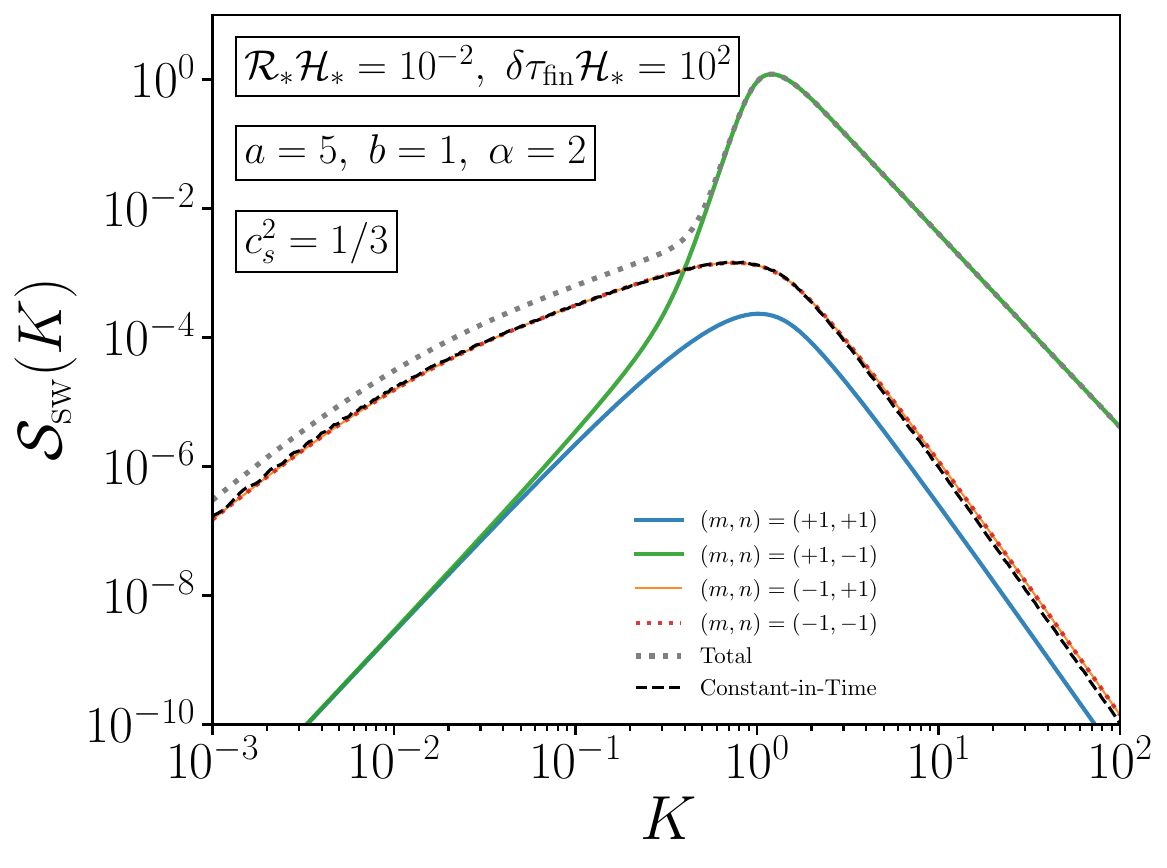}\\
    \includegraphics[width=\columnwidth]{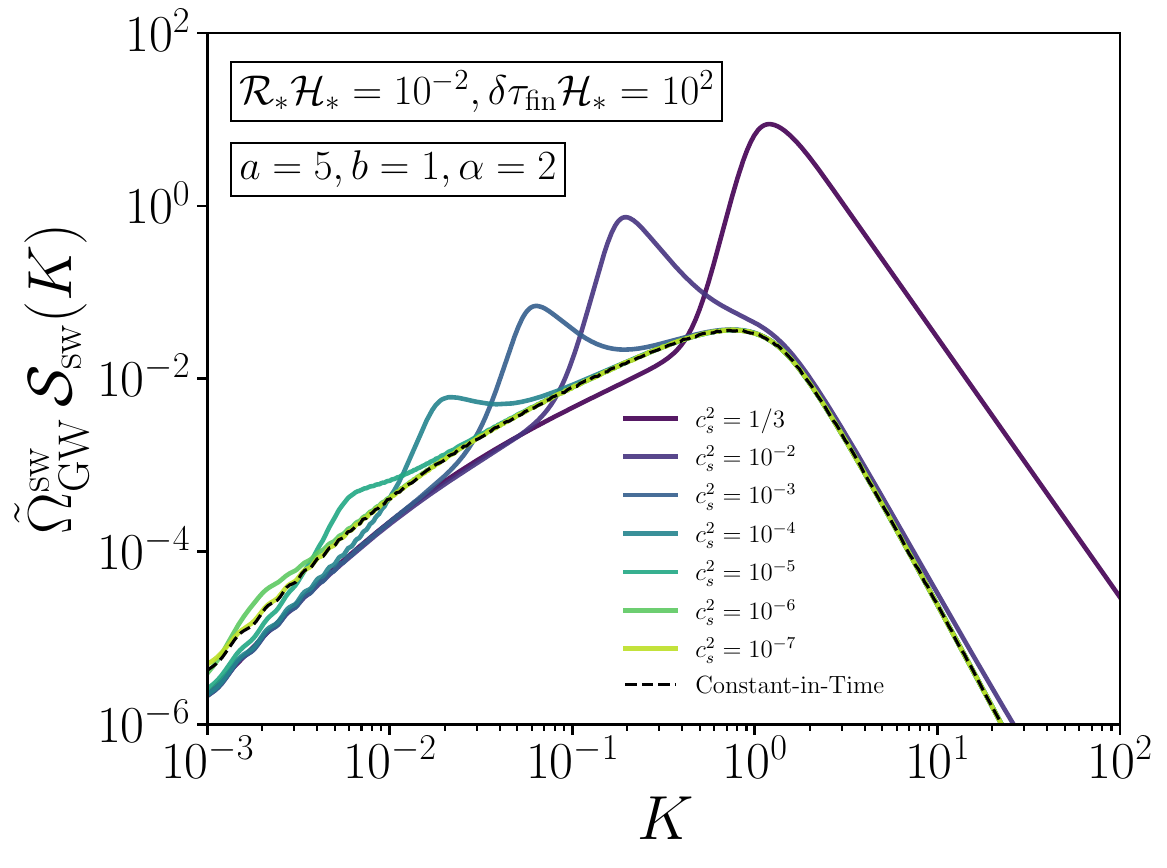}
    \caption{
    {\em Upper panel:} Four components of the normalized GW spectrum from sound waves
    with $\cs^2 = 1/3$
    \cite{RoperPol:2023dzg,Sharma:2023mao},
    following the sound-shell model \cite{Hindmarsh:2016lnk,Hindmarsh:2019phv},
    with
    $m=1, n=-1$ [see \Eq{delta_sw_decomp}] being the resonant contribution,
    and the terms $m=-1, n=\pm1$ being
    described by the CiT model.
    {\em Lower panel:} Normalized GW spectrum
    for different values of $\cs$,
    showing
    how it converges to the GW spectrum
    predicted by the CiT model in the $\cs \to 0$ limit.
    In both panels, the broken power law parameters for $\zeta_u(K)$ 
    are $a=5$, $b=1$, and $\alpha=2$.
    We use ${\cal R}_\ast {\cal H}_\ast = 10^{-2}$ and $\delta \tau_{\rm fin} {\cal H}_\ast = 10^2$ in both panels.}
    \label{SSM_CinT}
\end{figure}

Following \Sec{gw_template_cit}, we again
define the normalized spectral function ${\cal S}_{\rm sw}(K)$, such that
$\int{\cal S}_{\rm sw}(K) \dd \ln K = 1$, as
\begin{align}
\label{norm_omgw_sw_1}
    {\cal S}_{\rm sw}(K) = &\, {\cal N}_{\rm sw} K^2 \int \frac{\dd P}{P^2} \int \frac{\dd \tilde{P}}{\tilde{P}^2} (1-z^2)(1-\tilde{z}^2)  \nonumber \\
    & \times
    \zeta_u(P) \,\zeta_u(\tilde{P})\, \Delta_{\rm sw, av}^2(K, P, \tilde{P})\,.
\end{align}
By defining 
\begin{align}
\label{norm_omgw_sw_3}
    \tilde{\Omega}_{\rm GW}^{\rm sw} =  \frac{\cal C}{2{\cal A}^2} \frac{{\cal N}_{\rm sw}^{-1}}{({\cal R}_\ast \HH_\ast)^2}\,,
\end{align}
we can then rewrite \Eq{omgw_sw_formula} as
\begin{align}
\label{norm_omgw_sw_4}
    \OmGW^0(K) = &\, 3 \, \Gamma^2 u_{\rm rms}^4 \, F^0_{\rm GW} \,
    \tilde{\Omega}_{\rm GW}^{\rm sw} \nonumber \\ &\, \times ({\cal R}_\ast \HH_\ast)^2
    \, {\cal S}_{\rm sw}(K)\,.
\end{align}
We show in \Fig{SSM_CinT}, a comparison of the normalized GW spectrum,
$\tilde \Omega_{\rm GW}^{\rm sw} \, {\cal S}_{\rm sw}(K)$,
split into its different four contributions from \Eq{delta_sw_decomp},
with the normalized spectral shape ${\cal S}(K)$ in the CiT model [see \Eqss{OmGW_template}{OmGW_efficiency}],
for different values of the speed of sound, $\cs$.
In \Tab{tab:tilde_omega_gw_sw_tab} we show the values of the efficiency factor $\tilde{\Omega}_{\rm GW}$ for different final times and different values of $\cs^2$.
We see that for all values of $\cs^2$ the GW spectrum reaches saturation
due to the expansion of the Universe with an amplitude proportional
to $\Upsilon = \delta \tfin \HH_\ast / (1 + \delta \tfin \HH_\ast)$
when $\cs k_\ast \delta \tfin \gg 1$
\cite{RoperPol:2023dzg}.
Moreover, as the sound speed decreases,
also the efficiency quickly decreases, and for smaller values of $\cs^2$
saturation is reached earlier,
as expected in the CiT model, for which saturation is predicted to occur
by the time $\delta \tau_{\rm GW}^\ast \simeq 1/ k_\ast$.
We note that as the sound speed increases,
the resonant peak starts to dominate the total GW spectrum when  
$\cs \, k_\ast \tau_\ast \Upsilon
\gtrsim 1$, 
and causes a large increase of the GW efficiency
for the ultra-relativistic equation of state $\cs^2 = 1/3$, compared to smaller
values of $\cs$.
This can explain why previous simulations of acoustic turbulence with $\cs^2 = 1/3$
always find a larger efficiency producing GWs compared to vortical turbulence
with subrelativistic fluid and/or Alfv\'en velocities \cite{RoperPol:2019wvy,RoperPol:2021uol,RoperPol:2023bqa}.

\begin{table}[htbp]
    \centering
    \renewcommand{\arraystretch}{1.3}
    \begin{tabular}{|c|c|c|c|c|c|}
        \hline
        \diagbox{$\ \ \ \ \ c_s^2$}{$\vspace{-2mm} \delta\tau_{\rm fin}/\tau_*$}
        & $1$ & $5$ & $10$ & $50$ & $100$ \\
        \hline
        $\hspace{-2mm} 1/3$ & $3.665$ & $6.120$ & $6.677$ & $7.201$ & $7.272$ \\
        \hline
        $10^{-2}$ & $0.352$ & $0.510$ & $0.548$ & $0.585$ & $0.590$ \\
        \hline
        $10^{-3}$ & $0.125$ & $0.125$ & $0.127$ & $0.130$ & $0.130$ \\
        \hline
        $10^{-4}$ & $0.083$ & $0.083$ & $0.083$ & $0.083$ & $0.083$ \\
        \hline
        $10^{-5}$ & $0.078$ & $0.078$ & $0.078$ & $0.079$ & $0.079$ \\
        \hline
        $10^{-6}$ & $0.077$ & $0.077$ & $0.078$ & $0.078$ & $0.078$ \\
        \hline
        $10^{-7}$ & $0.076$ & $0.076$ & $0.077$ & $0.078$ & $0.078$ \\
        \hline
    \end{tabular}
    \caption{
    Values of $\tilde{\Omega}_{\rm GW}^{\rm sw}$ for different
    $c_s^2$ and $\delta\tfin/\tau_*$ computed from \Eq{norm_omgw_sw_3} for a purely compressional source
    with a spectrum characterized by
    $a=5$, $b=1$, and $\alpha = 2$. We take ${\cal R}_\ast {\cal H}_\ast = 10^{-2}$ as in \Fig{SSM_CinT}.
    Note that for
    $\cs^2 \lesssim 10^{-4}$
    we recover
    the efficiency ${\tilde \Omega}_{\rm GW} \simeq 0.078$
    found in the CiT model (see \Tab{KGW_values}) for all values of $\delta \tfin$ considered.}
    \label{tab:tilde_omega_gw_sw_tab}
\end{table}

We note that when the fluid is the dominant component of the Universe,
it is not consistent to vary $\cs$ while keeping the evolution of the scale
factor determined for a radiation-dominated Universe.
In general, as $\cs$ varies,
the dynamics of the sound waves also vary \cite{Giombi:2024kju,RoperPol:2025lgc}
and
the Green's function of the GW equation
is modified \cite{Giombi:2024kju}.
However, this simpler treatment allows us to
understand what is the impact of $\cs$ in the UETC, showing how $\cs$ acts
as a decorrelation velocity, such that the CiT model is recovered for long
decorrelation times $\delta \tau_{\rm dcorr} = (\cs k_\ast)^{-1} \gg \tau_\ast$, and fails to explain the full
GW spectrum
when $\cs k_\ast \tau_\ast \gg 1$,
e.g., when $\cs^2 = 1/3$.

We have shown that for a stationary UETC, approximated using a constant
correlation velocity ($u_{\rm dcorr}$ for turbulence with Kraichnan
decorrelation and $\cs$ for sound waves), there are two distinguished regimes.
For small $u_{\rm dcorr} k_\ast \delta \tfin \ll 1$ and $\cs k_\ast \delta \tfin \ll 1$,
the GW spectrum sourced by
both sound waves and stationary turbulence can be accurately described by the CiT
model, which is consistent with the findings of numerical simulations for
decaying MHD turbulence \cite{RoperPol:2019wvy,RoperPol:2021uol,RoperPol:2022iel,Auclair:2022jod}.
However, as the correlation velocity increases,
$u_{\rm dcorr} k_\ast \tau_\ast \Upsilon \gtrsim 1$ and $\cs k_\ast \tau_\ast \Upsilon \gtrsim 1$, with $\Upsilon = \delta \tau_{\rm fin}{\cal H}_\ast/(1+\delta \tau_{\rm fin} {\cal H}_\ast)$,
the CiT model describes one component of the GW spectrum, which dominates
at small $k$, but an additional resonant term grows proportional to the source
duration, $\OmGW (K) \sim \Upsilon = \delta \tfin \HH_\ast / (1 + \delta \tfin \HH_\ast)$
\cite{RoperPol:2023dzg},
which dominates around the peak and at larger $K$.
In this case, the location
of the peak moves proportional to $u_{\rm dcorr}$ \cite{Gogoberidze:2007an} or $\cs$ \cite{Hindmarsh:2019phv,Giombi:2024kju},
with an amplitude proportional to $u_{\rm dcorr}^2$ or $\cs^2$.
Therefore, we in general find that the presence of decorrelation in the UETC enhances
the efficiency in producing GWs \cite{Caprini:2009fx,RoperPol:2022iel},
such that the CiT model provides a conservative estimate of the GW spectrum
of these sources \cite{RoperPol:2022iel,Caprini:2024hue,EPTA:2023xxk}.

\subsection{Bubble collisions}
\label{other_sorc}

The GW production from scalar bubble collisions in first-order phase transitions was
first studied following the envelope approximation \cite{Kosowsky:1992rz,Kamionkowski:1993fg,Huber:2008hg,Weir:2016tov,Jinno:2016vai},
where the scalar bubbles
are assumed to vanish instantaneously after collisions.
The GW production occurs when the thin shells of scalar field
gradients collide.
This model predicts a GW background proportional to $k^3$ at $k \ll k_\ast$
and to $k^{-1}$ at $k \gg k_\ast$, respectively. 

Going beyond the envelope approximation, the bulk flow model was proposed
for the GW production \cite{Jinno:2017fby,Konstandin:2017sat},
where the shells of gradient
energy density are assumed
to continue producing GWs even after collisions,
instead of instantaneously vanishing, and their decay is taken
into account.
Simulations and analytical estimates following the
bulk flow model predict a linear scaling $\sim \! k$ at low wave numbers,
and a power-law decay $\sim \! k^{-3}$ at large wave numbers \cite{Jinno:2017fby,Konstandin:2017sat}.
A similar GW spectrum (with $k$ and $k^{-2}$ scalings) has been found
from feebly interacting particles during a phase
transition in the dark sector \cite{Jinno:2022fom}.

The linear scaling with $k$ strongly resembles
the predictions of the CiT model.
In particular,
in the bulk flow model,
the scalar field gradients sourcing GWs are assumed to decay
over a timescale $\delta \tau_{\rm dc}$.
This timescale can be roughly taken to be the source duration,
$\delta \tau_{\rm dc} \simeq \delta \tfin$.
Then, the linear-in-$k$ scaling is expected for $k \geq \delta \tfin$,
recovering the $k^3$ scaling at $k < 1/ \delta \tfin$,
as predicted by the CiT model.
Applying the CiT model, the GW spectrum decays 
at $k \gtrsim k_\ast$ proportionally to $k^{1 - b_\Pi^{\rm comp}}$,
where $b_\Pi^{\rm comp} = 2 b + 3$ (if $b < 2$) [cf.~\Eqq{b_pi_definitions}]
depends on the UV-slope $-b$ of the spectrum of the scalar field gradients
$E_\L^\phi (k) = k^2 E_\phi (k)$,
and $k_\ast$
is its characteristic wave number.

\begin{figure}
    \centering
    \includegraphics[width=0.5\textwidth]{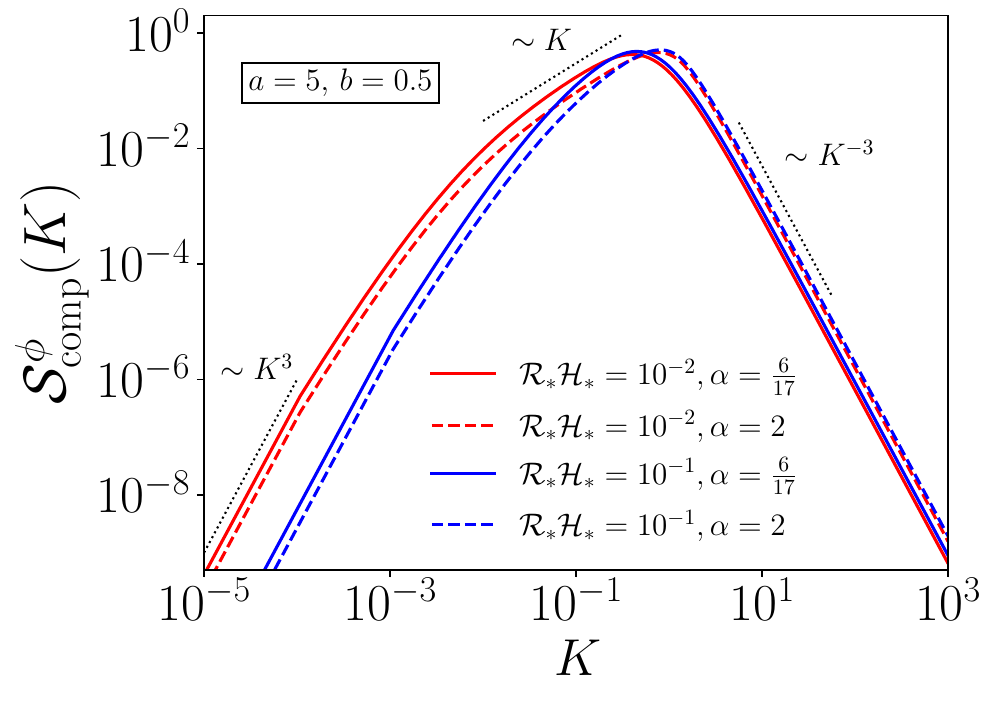}
    \caption{Normalized spectral shape of the GW spectrum
    produced by a scalar field, whose gradients are described by a spectrum
    with
    slopes $a = 5$ and $b=0.5$, and smoothness parameters
    $\alpha = 6 / 17$ (solid lines)
    and $2$ (dashed lines).
    We use two values of the characteristic lengthscale
    ${\cal R}_\ast \HH_\ast = 10^{-2}$ (red lines)
    and $10^{-1}$ (blue lines) as in \Fig{shapeGW_vortcomp},
    and a sourcing duration $\delta \tau_{\rm fin}/{\tau}_\ast = 100$.}
    \label{shapeGW_comp_phi}
\end{figure}

As discussed in \Sec{sec:smoothed_dbpl},
the power spectrum $E_\phi (k)$ for the scalar fields
in a vacuum first-order phase transition found in the 
simulations of refs.~\cite{Cutting:2018tjt,Cutting:2020nla} follows a broken
power law with $E_\phi (k) \sim k^{-2.5}$ at $k \gg k_\ast$, before falling
exponentially at scales around the wall thickness.
Since $E_\L^\phi (k) = k^2 E_\phi (k) \sim k^{-0.5}$, then we find that
the GW spectrum should decay as $\OmGW \sim k^{-2(1 + b)} = k^{-3}$,
according to the CiT model, which exactly
coincides with the bulk flow model.
We show the resulting
GW spectrum applying the CiT model to this case,
for different values of $\alpha$ and ${\cal R}_\ast {\cal H}_\ast = 1/(k_\ast \tini)$
in \Fig{shapeGW_comp_phi}, where the normalized ${\cal S}_{\rm comp}^{\phi}(K)$ is defined as in \Sec{gw_template_cit} [cf.~\Eq{OmGW_shape}] and refers to a purely compressional spectrum, as expected for the scalar field gradients.
Applying a similar analysis to the results from feebly interacting particles \cite{Jinno:2022fom}, we would find that their UV spectrum is compatible with
the CiT model if the anisotropic stress spectrum of the feebly interacting
particle has a small-scale slope $b_\Pi^{\rm comp} = 3$.
This corresponds to a flat
scalar field gradients spectrum $E_\phi^\L (k) \sim k^0$ at large $k$.
Hence, the scalar field spectrum decays as $E_\phi (k) \sim k^{-2}$.
Note that $E_\phi^\L (k)$ remains flat at large $k$ until the damping
scale is reached, i.e., for $k \ll k_{\rm D}$.
At larger $k$, the spectrum would decay exponentially, ensuring that the
scalar field gradient energy converges.

The power spectrum of the scalar field in these studies of GW production
should be computed for a direct comparison with the predictions of the CiT
model before being able to conclude that the CiT model applies to these
(or other) sources.
This remains to be checked in future work.

\section{Beyond the Constant-in-Time model: Coherent decay}
\label{beyond_cit}

\label{coherent_model}

In this section, we go beyond the CiT
model and include the decay of the source.
To take into account the decay,
we restrict our treatment to sources for which
the UETC of the anisotropic stresses is
coherent in time \cite{Caprini:2009fx},
\begin{equation}
\label{coherent_epi}
    E_\Pi (\tau_1, \tau_2, k) = \sqrt{E_\Pi(\tau_1, k)
    \, E_\Pi(\tau_2, k)}\,.
\end{equation}
This assumption allows us to introduce the time decay of the source
while keeping the
UETC as a separable function of $\tau_1$ and $\tau_2$,
simplifying the computation of the 
resulting GW spectrum.
This model of the UETC holds for sources of GWs that have
a long decorrelation timescale,
as discussed in \Sec{applications} for sound waves or decaying
MHD turbulence.
Indeed, the authors in ref.~\cite{Auclair:2022jod} found that the velocity field
followed Kraichnan decorrelation
and the resulting GW spectrum was accurately described by the CiT model,
such that only at relativistic speeds and large wave numbers
the decorrelation has an impact on the resulting GW spectrum.
In these cases, the decorrelation timescale is determined
by the correlation
velocity ($\cs$ or $u_{\rm dcorr}$) and the characteristic wave
number $k_\ast$ \cite{1965PhFl....8.1385K,1993PhFlA...5.2835K,Kosowsky:2001xp,Gogoberidze:2007an,Caprini:2009yp,Niksa:2018ofa,Auclair:2022jod}.
We show explicitly in a separate publication the comparison of the GW
background obtained only accounting for decay with the one also including Kraichnan
decorrelation on top of the decay of the source \cite{kraichnan}.

\subsection{Model}

In the case where the UETC of the stress spectrum is coherent
in time, as in \Eq{coherent_epi}, the GW energy density spectrum, given
in \Eq{eq:OmGWt}, is
\begin{align}
    \OmGW  (\tau, k) = 
    3 \, E_\Pi^\ast (k) \, F_{\rm GW}^0 \,
    \Delta^2_{\rm coh} (\tau, k)\,,
\end{align}
where
\begin{align}
    \Delta_{\rm coh} (\tau, & k) \nonumber \\ = &\, \int_{\tini}^{\bar \tau}
    \frac{\dd \tau'}{\tau'}
    \sqrt{\frac{E_\Pi (\tau', k)}
    {E_\Pi^\ast (k)}}
    \cos k(\tau - \tau')\,,
\label{delta_coh_formula}
\end{align}
is the kernel found after incorporating
the time dependence of the autocorrelator
$E_\Pi(\tau, k)$, and where $E_\Pi^\ast (k) = E_\Pi(\tini, k)$ is the initial spectrum of the anisotropic
stresses.

The assumption of coherent correlation in time allows us to compute the GW
spectrum from the time evolution of the
stress autocorrelation spectrum, 
without the need to know the full form of the UETC at
both times $\tau_1$ and $\tau_2$.
For non-linear decaying fields, e.g., decaying MHD turbulence,
the decay of the peak wave number and the energy density
of the sourcing field can be modeled with power laws
in time 
\cite{Brandenburg:2017neh},
\begin{subequations}
\label{eq:decayall}
\begin{align}\label{eq:decay}
    \kf (\tau) = &\  \kf \, \biggl(1 +
    \frac{\delta \tau}{\delta \tau_{\rm dc}} \biggr)^{-s},
    \\ 
    E_\ast (\tau) = &\ E_\ast \, \biggl(1 + \frac{\delta \tau}{\delta \tau_{\rm dc}} \biggr)^{-p},
    \label{eq:decay2}
\end{align}
\end{subequations}
where $\kf$ and $E_*$
refer to the peak wave number and amplitude of the source spectrum $E(k)$, 
$\delta \tau = \tau - \tini$, and $\delta \tau_{\rm dc} = \tau_{\rm dc}-\tau_\ast$ is the typical decay timescale of the source.
As an example, for MHD turbulence, taking the decay timescale as the eddy turnover time $\delta \te$ (see also discussion in \Sec{constant_model}) we have that, in the limit of infinite eddy turnover time,
the turbulent fields do not decay, and we recover
the CiT model.

The decay of the stress autocorrelation $E_\Pi(\tau, k)$, can be expressed as
\begin{align}\label{eq:EPi_decay}
    \frac{E_\Pi (\tau, k)}
    {E_\Pi (\tini, k) } = \zeta_\Pi^{\rm rat} (\tau, k) \,
    \biggl(1 + \frac{\delta \tau}{\delta \tau_{\rm dc}}
    \biggr)^{2m} \, ,
\end{align}
where $m = {3\over2}s - p$ is the power law decay of the
stress spectrum, always taking negative values for typical decaying sources \cite{Dahl:2024eup, Brandenburg:2024tyi}, with the factors $-2p$ and $3s$ respectively coming from the scaling of $E_\Pi$ with $E_\ast^2$ and $K^3$, where $K\equiv k/k_*$ [see \Eqs{tot_spec_vel}{EPi_u}]. Here $\zeta_\Pi^{\rm rat}$ is the ratio
of the normalized stress autocorrelation $\zeta_\Pi(K)$
at time $\tau$ to its value at $\tini$,
\begin{equation}
    \zeta_\Pi^{\rm rat} (\tau, k) \equiv \frac{\zeta_\Pi \bigl[k/\kf(\tau)
    \bigr]}
    {\zeta_\Pi (k/\kf)} \,.
\label{p_pi_rat}
\end{equation}

This yields the following expression for the GW
spectrum,
\begin{align}
    \OmGW &\, (\tau, k) = 3  \, E_\Pi^\ast (k) \, F_{\rm GW}^0\,
    \Delta_{ps}^2 (\tau, k; \delta \tau_{\rm dc})
    \nonumber \\ = &\, 3 \,
    K^3
    \frac{{\cal C}}{{\cal A}^2} 
    \frac{\tilde{\varphi}_{\rm rms}^4}{\bar{\rho}^2}
    \, \zeta_\Pi^\ast (K) \, F_{\rm GW}^0
    \Delta^2_{ps} (\tau, k; \delta \tau_{\rm dc})\,,
    \label{SGWB_coh}
\end{align}
where the function $\Delta_{ps} (\tau, k; \delta \tau_{\rm dc})$
is defined as
\begin{align}
    &\, \Delta_{ps}  (\tau, k; \delta \tau_{\rm dc}) = \nonumber \\ &\, \hspace{-2mm} 
    \int_{\tini}^{\tau}
    \frac{\dd \tau_1}{\tau_1} \biggl(1 + \frac{\delta \tau_1}
    {\delta \tau_{\rm dc}} \biggr)^{m}
    \cos k (\tau - \tau_1) \, \sqrt{\zeta_\Pi^{\rm rat} (\tau_1,k)}.
    \label{G_pq}
\end{align}
Notice that, since we have included the decay of the source, it is not necessary in this case to define a final sourcing time $\tau_{\rm fin}$, as we did for the CiT model (see \Sec{cit_model_sec}). 

However, if we included a final sourcing time $\tau_{\rm fin}$, then we would define a $\Delta_{ps}$ kernel where the upper limit of integration in \Eq{G_pq} would be replaced by $\bar{\tau} = \min(\tau, \tau_{\rm fin})$ and we would recover the CiT model result either in the case
$p = s = 0$ or in the limit of large decay timescale $\delta \tau_{\rm dc}/\tau_\ast \to \infty$, i.e., 
$\Delta(\tau, k) =  \Delta_{00}(\tau, k; \delta \tau_{\rm dc}) = \displaystyle\lim_{\delta \tau_{\rm dc}/\tau_\ast \rightarrow \infty} \Delta_{ps}(\tau, k; \delta \tau_{\rm dc})$ [see \Eq{G_constant}].
With such definition of $\Delta_{ps}$, if we considered the present-day observable $\tau = \tau_0$ and averaged over fast oscillations,
we would find
\begin{equation}
    \Delta_{ps, {\rm av}}^2 = \half \bigl[{\cal I}_{\sin}^2 (
    {\bar \tau}, k
    ) + {\cal I}_{\cos}^2 (
    {\bar \tau}, k
    )\bigr]\,,
\end{equation}
where
\begin{subequations}
\label{int_cos_sin_coh}
\begin{align}
    {\cal I}_{\cos} ({\bar \tau}, k) =&\, \int_{\tini}^{\bar \tau}
    \frac{\dd \tau_1}{\tau_1}\biggl(1 + \frac{\delta \tau_1}{\delta \tau_{\rm dc}}\biggr)^{m}
    \cos k \tau_1 \sqrt{\zeta_\Pi^{\rm rat} (\tau_1, k)}\,, \\
    {\cal I}_{\sin} ({\bar \tau}, k) =&\, \int_{\tini}^{\bar \tau}
    \frac{\dd \tau_1}{\tau_1}\biggl(1 + \frac{\delta \tau_1}{\delta \tau_{\rm dc}}\biggr)^{m}
    \sin k \tau_1 \sqrt{\zeta_\Pi^{\rm rat} (\tau_1, k)}\,.
\end{align}
\end{subequations}

In the following, except for \App{G_m_coherent}, we will consider $\Delta_{ps}$ defined as in \Eq{G_pq}. 
In general, we find that for 
small enough
values of $s$ the
effect in $\Delta_{ps}$ from the $\zeta_\Pi^{\rm rat}$ term
can be ignored [see \App{fast_q}] and the resulting $\Delta_{ps}(\tau, k; \delta \tau_{\rm dc})$ only depends on $m$ and $\delta \tau_{\rm dc}$. This happens when the timescale of decay of $\zeta_\Pi^{\rm rat}$ is larger than the saturation timescale of the GW spectrum, which happens when [cf.~\Eq{range_of_good_s_values}]
\begin{align}
     s < 1 +\frac{ \ln \bigl( k_\ast \, \tau_\ast \bigr)}{\ln \bigl( \tau/\tau_\ast \bigr)} \ \  \ {\rm for} \  \tau> \tau_\ast \,.
\label{range_of_good_s_values2}
\end{align}
Hence, from now on we will safely restrict to values $s \leq 1$, which allow us to ignore
this term and make the following approximation
\begin{align}
    \Delta_{ps}(\tau, k; \delta \tau_{\rm dc}) \simeq & \int_{\tau_\ast}^{\tau} \frac{\dd \tau_1}{\tau_1} \Bigl(1+\frac{\delta \tau_1}{\delta \tau_{\rm dc}} \Bigr)^{m} \cos k \bigl( \tau - \tau_1 \bigr) \nonumber \\
    & \equiv \Delta_{m}(\tau, k ; \delta \tau_{\rm dc})\,.
\label{coh_first_approx}
\end{align}

\begin{figure}
    \centering
    \includegraphics[width=\columnwidth]{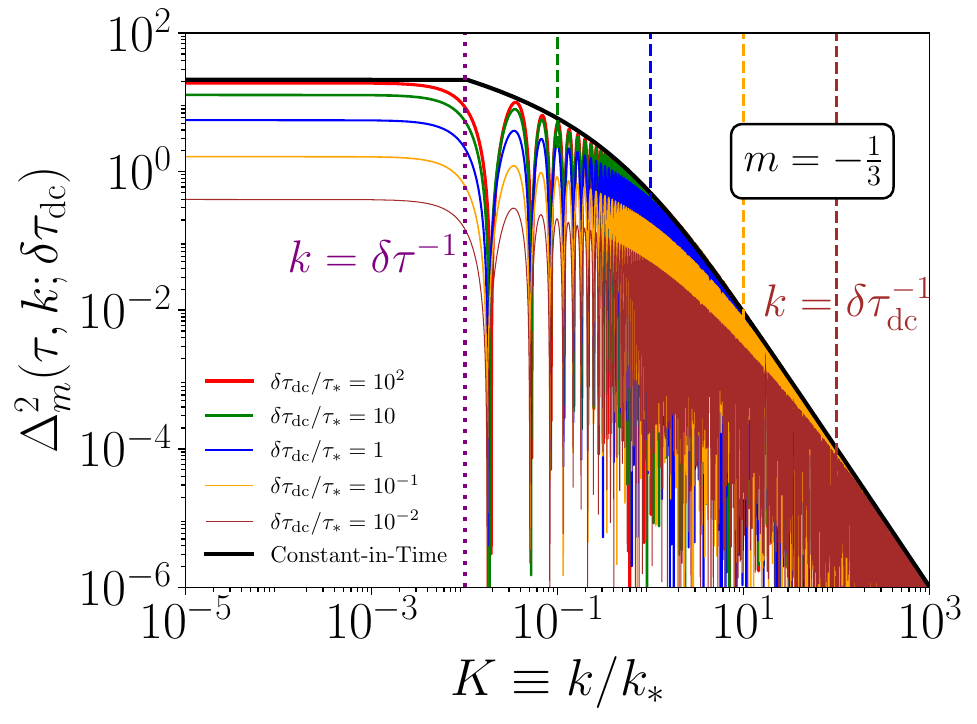}
    \caption{
    Spectra $\Delta^2_m(\tau, k; \delta \tau_{\rm dc})$ [cf.~\Eq{coh_first_approx}] for different choices of $\delta \tau_{\rm dc}/\tau_*$ for a negative value of $m=-\frac{1}{3}$. We see that as the decay timescale increases we recover, as expected, the result for $\Delta^2(\tau<\tau_{\rm fin}, k)$ in the CiT model [cf.~\Eq{G_constant}]. For shorter $\delta \tau_{\rm dc}/\tau_\ast$ the spectra $\Delta^2_m (\tau, k; \delta \tau_{\rm dc})$ deviate from the CiT result for values of $k < 1/\delta \tau_{\rm dc}$ (vertical dashed lines), while the transition from the $k \to 0$ to the $k \to \infty$ regime starts at $k \simeq 1/\delta \tau$ (vertical dotted line) for all values of $\delta \tau_{\rm dc}$.
    }
    \label{Delta_pq}
\end{figure}

We compare the above defined function for different choices of $\delta \tau_{\rm dc}/\tau_\ast$ in \Fig{Delta_pq} where we see that, as $\delta \tau_{\rm dc}/\tau_\ast$ increases, $\Delta_m(\tau, k; \delta \tau_{\rm dc})$ approaches the function $\Delta^2(\tau < \tau_{\rm fin}, k)$ in the CiT model [cf.~\Eq{G_constant}].

\begin{figure}
    \centering
    \includegraphics[width=\columnwidth]{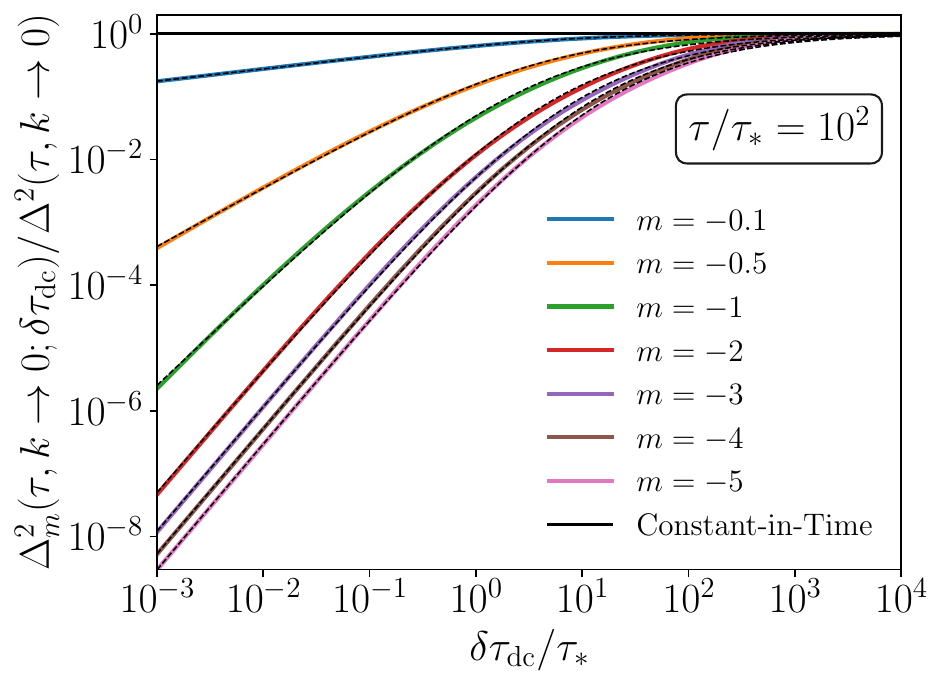}\\
    \includegraphics[width=\columnwidth]{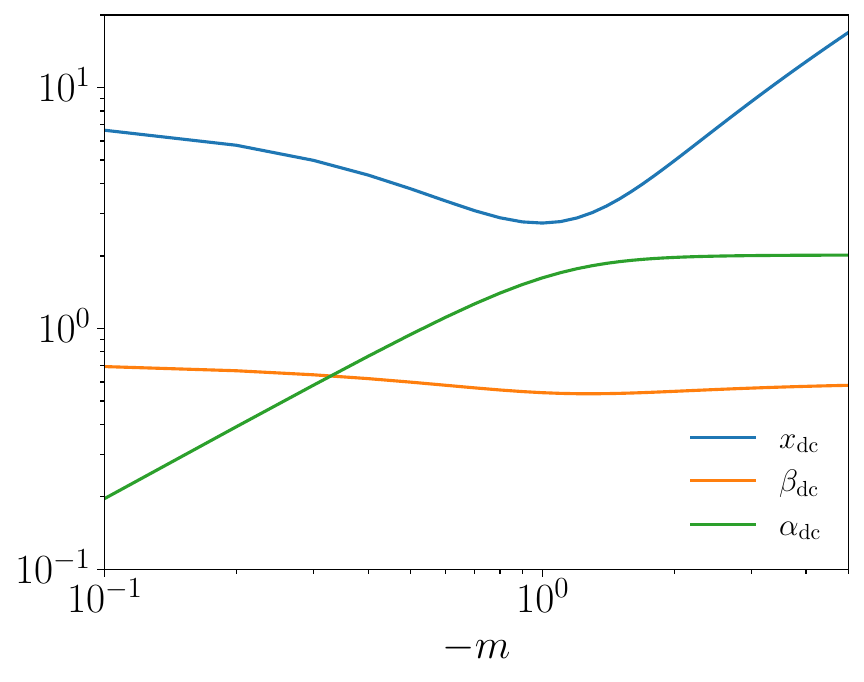}
    \caption{{\em Upper panel:} Amplitude in the $k \to 0$ limit of $\Delta^2_m(\tau, k; \delta \tau_{\rm dc})/\Delta^2(\tau, k)$ (solid lines) for $\tau/\tau_\ast =10^2$, different values of $\delta \tau_{\rm dc}$ and different choices of $m$. For all values of $-m \in [0.1, 5]$ we find the fit $r^2(\delta \tau_{\rm dc}/\tau_\ast; x_{\rm dc}, \alpha_{\rm dc}, \beta_{\rm dc})$ of \Eq{fit_delta2m_deltataudc_ktozero}, which is overplotted in dashed black lines. {\em Lower panel:} Values of the fitting parameters for $-m \in [0.1, 5]$.
    }
    \label{Delta2m_ktozero_amplitude}
\end{figure}

Moreover, in the upper panel of \Fig{Delta2m_ktozero_amplitude} we show the large-scale limit of $\Delta^2_m$ for different choices of $m$ and different values of $\delta \tau_{\rm dc}/\tau_\ast$, and we provide a fit (dashed black lines) of $\Delta^2_m/\Delta^2$, tested for all values of $\delta \tau_{\rm dc}/\tau_\ast \in [10^{-5}, 10^{5}]$ and $-m \in [0.1, 5]$, in the form of the following broken power law 
\begin{align}
\label{fit_delta2m_deltataudc_ktozero}
    r^2 (\delta \tau_{\rm dc}/\tau_\ast; x_{\rm dc}, \alpha_{\rm dc}, \beta_{\rm dc}) & = \Biggl[1 + \Bigl( \frac{x_{\rm dc}}{\delta \tau_{\rm dc}/\tau_\ast} \Bigr)^{\alpha_{\rm dc}} \Biggr]^{-\frac{\beta_{\rm dc}}{\alpha_{\rm dc}}}\,,
\end{align}
with the fitting parameters $x_{\rm dc}, \alpha_{\rm dc}, \beta_{\rm dc}$ given in the lower panel of \Fig{Delta2m_ktozero_amplitude}.

We can obtain an analytical expression for the resulting
kernel $\Delta^2_m$ in the particular case in which $\delta \tau_{\rm dc} = \tau_*$ which, in radiation-domination, is equivalent to $\delta \tau_{\rm dc} = \HH_*^{-1}$. In this case \Eq{coh_first_approx} reduces to
\begin{align}
    \Delta_{m}(\tau, k; \delta \tau_{\rm dc}=\tau_\ast) & =  \tau_\ast^{-m} \int_{\tau_\ast}^{{ \tau}} \, \tau_1^{m-1} \, \cos k(\tau - \tau_1) \dd \tau_1 \nonumber \\
     & \equiv {\Delta}_{m}(\tau, k)\,.
\label{delta_tdc_equal_tast}
\end{align}
The function $\Delta_m(\tau, k)$ is studied in 
\App{G_m_coherent}, \footnote{In \App{G_m_coherent} the function $\Delta_m$ is studied in the more general case in which the upper integration bound in \Eq{delta_tdc_equal_tast} is set to $\bar{\tau}$. However, all the results of this section can be obtained by the simple replacement $\bar{\tau} \to \tau$.} where
we show that it
can be computed in terms of the upper incomplete gamma function $\Gamma(x,y)\equiv \int_y^{\infty} t^{x-1} e^{-t} \dd t$ as [see \Eq{Gpq_coherent}]
\begin{align}
    {\Delta}_{m} &(\tau, k) = - \half \tau_\ast^{-m} 
    (ik)^{-m} \nonumber \\
    & \times \Biggl\{
    e^{ik\tau} 
   \Bigl[ \Gamma \bigl(m, ik\tau \bigr) - \Gamma \bigl(m, ik\tau_\ast \bigr) \Bigr]
    \nonumber \\
    & +  e^{-ik\tau} e^{i\pi m} \Bigl[
    \Gamma \bigl(m, -ik\tau \bigr) -\Gamma \bigl(m, -ik \tau_\ast \bigr) \Bigr] \Biggr\} \,.
    \label{Gpq_coherent_2}
\end{align}

For the cases in which $m$ is an integer, we can further
simplify the analytical expression. We give in \App{G_m_coherent} a method to find the analytical
expression for any $m$ and, in particular, the
expressions for $\Delta_{-1}$ and $\Delta_{-2}$
are explicitly given in \Eqs{Gm1}{Gm2},
\begin{align}
    {\Delta}_{-1} (\tau, k) = & \, - k \tau_\ast \cos k \tau \Bigl[\Si (k{\tau}) - 
    \Si (k \tini) \Bigr] \nonumber \\
    &\, + k \tau_\ast \sin k {\tau} \Bigl[\Ci (k{\tau}) - \Ci (k \tini)
    \Bigr] \nonumber \\
    &\, -  \frac{\tau_\ast}{{\tau}}  +  \cos k (\tau - \tini), \\
    {\Delta}_{-2} (\tau, k) = & \, - \frac{k^2\tau_\ast^2}{2}
    \cos k \tau \Bigl[\Ci (k{\tau}) - \Ci (k \tini) \Bigr] \nonumber \\
    &\, - \frac{k^2\tau_\ast^2}{2} \sin k \tau \Bigl[\Si (k{\tau}) - \Si (k \tini)
    \Bigr] \nonumber \\
    + \frac{1}{2} \bigl[ & \cos k (\tau - \tau_\ast) + k \tau_\ast \sin k (\tau - \tau_\ast)\bigr] 
    - \frac{1}{2}  \frac{\tau_\ast^2}{{\tau}^2}  \,. 
\end{align}
For comparison, the functions ${\Delta}_{1}$ and
${\Delta}_{2}$ are given in \Eqs{G1}{G2},
\begin{align}
    \Delta_1 (\tau, k) &= \frac{1}{k \tau_\ast} \sin k \bigl(\tau - \tau_\ast \bigr) \,,
    \label{G1} \\
    \Delta_2 ( \tau, k) & = \frac{1}{k^2 \tau_\ast^2} \Bigl[1- \cos k \bigl( \tau - \tau_\ast \bigr)\Bigr] \nonumber \\
     & + \frac{1}{k \tau_\ast} \sin k \bigl(\tau - \tau_\ast \bigr)\,.
     \label{G2}
\end{align}
Note that for positive values of $m$,
the autocorrelation stress spectrum increases with time,
which is not physically compatible with a decaying source (e.g., in the case of decaying MHD
turbulence).
However, these cases might be relevant for the phase of
turbulence production as long as
the UETC follows the coherent approximation.

In \App{G_m_coherent} we study the time evolution of the function $\Delta_m ( \tau, k)$ in the asymptotic limits $k \to 0$ and $k \to \infty$.

\begin{figure}
    \centering
    \includegraphics[width=\columnwidth]{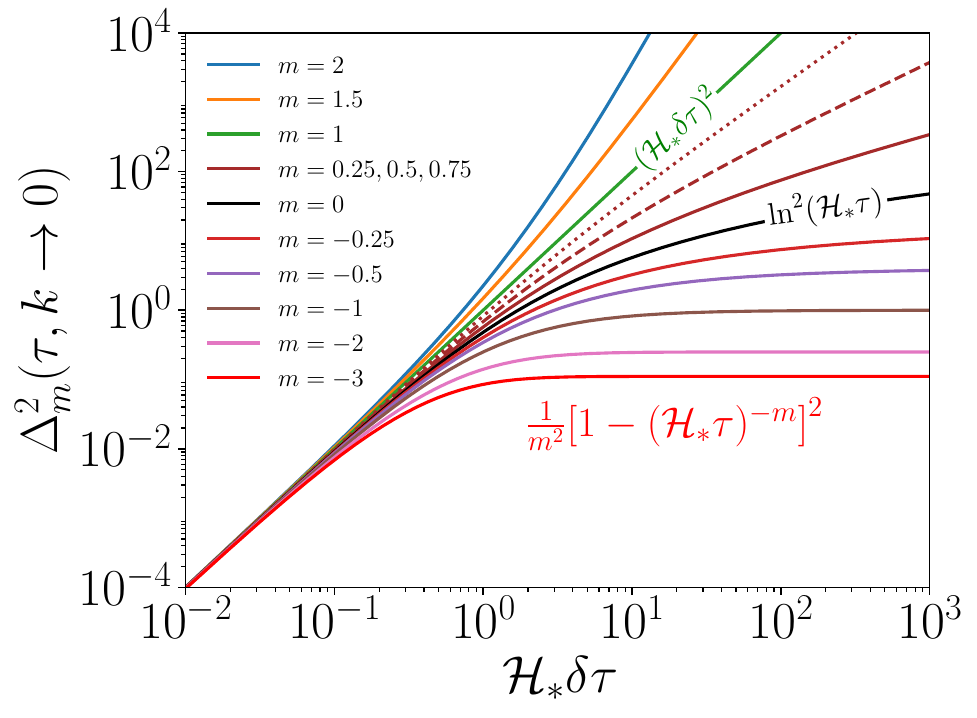}
    \caption{
    Time evolution of the integrals $\Delta_m^2 (\tau, k)$ in the limit $k \rightarrow 0$
    for different values of $m$.
    The time evolution for $m = -1$, $-2$, and $-3$
    can be analytically
    shown to follow the function in \Eq{lim_k0}.
    We find numerically that for non-integer values of $m = -n < 0$, this expression also holds
    (e.g., for $m = -0.25$, $-0.5$).
    The function ${\Delta}_1^2$ evolves as 
    $\bigl({\cal H}_* \delta \tau\bigr)^2$, while for values
    of $m \in (0, 1)$ (e.g., $0.25$, $0.5$, and $0.75$)
    $\Delta_m$ diverges
    for large $\delta \tau$ transitioning towards $\ln^2 (\HH_* \tau)$,
    which corresponds to the evolution of $\Delta^2$ in the CiT model [see \Eq{growing_delta2_cit}].
    For larger values of $m > 1$, the functions ${\Delta}_m^2$ diverge
    faster than $\delta \tau^2$.}
    \label{Gsq_k0}
\end{figure}

Let us first focus on the limit $k \to 0$. For negative integers $-n$ the function $\Delta_{-n}^2$ takes the asymptotic form [see \Eq{k_to_zero_gm}]
\begin{align}
    \lim_{k\rightarrow 0} {\Delta}_{-n}^2 (\tau, k)
    = \frac{1}{n^2} \biggl[1-\bigl( {\tau}/\tau_\ast \bigr) ^{-n} \biggr]^2 \,.
    \label{lim_k0}
\end{align}
The time evolution of $\Delta_m(\tau, k)$ for different values
of $m$ in the limit $k \rightarrow 0$ is shown in \Fig{Gsq_k0}.
For the values $m = -n < 0$, we find numerically that the
relation \Eq{lim_k0} holds even for non-integer $n$.
In the limit of large times $\tau \gg \tau_\ast$, $\Delta_{-n}^2(\tau, k \to 0)$ approaches the value [see \Eq{GG2mn}]
\begin{align}
    \lim_{\substack{k\rightarrow 0}}
    {\Delta}_{-n}^2 (\tau, k) = \frac{1}{n^2} \,.
\end{align}
This limit is shown in \Fig{Gsq_k0} at large $\delta \tau/\tau_\ast$.

The values of the limit $k \to 0$ for $\Delta_{m}(\tau, k)$ with $m=1,2$ are instead [see \Eqs{G1_k0}{G2_k0}]
\begin{subequations}
\label{delta_oneandtwo_ktozero}
\begin{align}
\lim_{k \rightarrow 0} \Delta_1^2 (\tau, k) &= (\delta \tau/\tau_\ast)^2\,,
\label{G1_1} \\
\lim_{k \rightarrow 0} \Delta_2^2 (\tau, k) &= \frac{1}{4} (\delta \tau/\tau_\ast)^4 + (\delta \tau/\tau_\ast)^3 \nonumber \\
& + (\delta \tau/\tau_\ast)^2 \,.
\label{G2_1}
\end{align}
\end{subequations}
The function ${\Delta}_1^2 (\tau, k \to 0)$ evolves quadratically
with $\delta \tau$ for all times, hence, it diverges faster
than $\ln^2 (\HH_* \tau)$, as expected.
On the other hand, the function
${\Delta}_2^2 (\tau, k\to 0)$ evolves as $\delta \tau^2$ at small $\delta \tau$
but as time increases, the term $\delta \tau^4$ dominates, making the function diverge even faster at large times.
We find numerically that for values of $m \in (0, 1)$ the 
dynamical evolution of ${\Delta}_m^2 (\tau, k \to 0)$ smoothly
transitions from the CiT value
$\ln^2 \bigl(\HH_* \tau\bigr)$ to $(\HH_* \delta \tau)^2$.
In particular, the values $m = 0.25$, $0.5$, and $0.75$
are shown in \Fig{Gsq_k0}.
Similarly, for values $m \in (1, 2)$ we find functions between
\Eqs{G1_1}{G2_1}, which diverge faster than $\delta \tau^2$
at late times.

\begin{figure}
    \centering
    \includegraphics[width=\columnwidth]{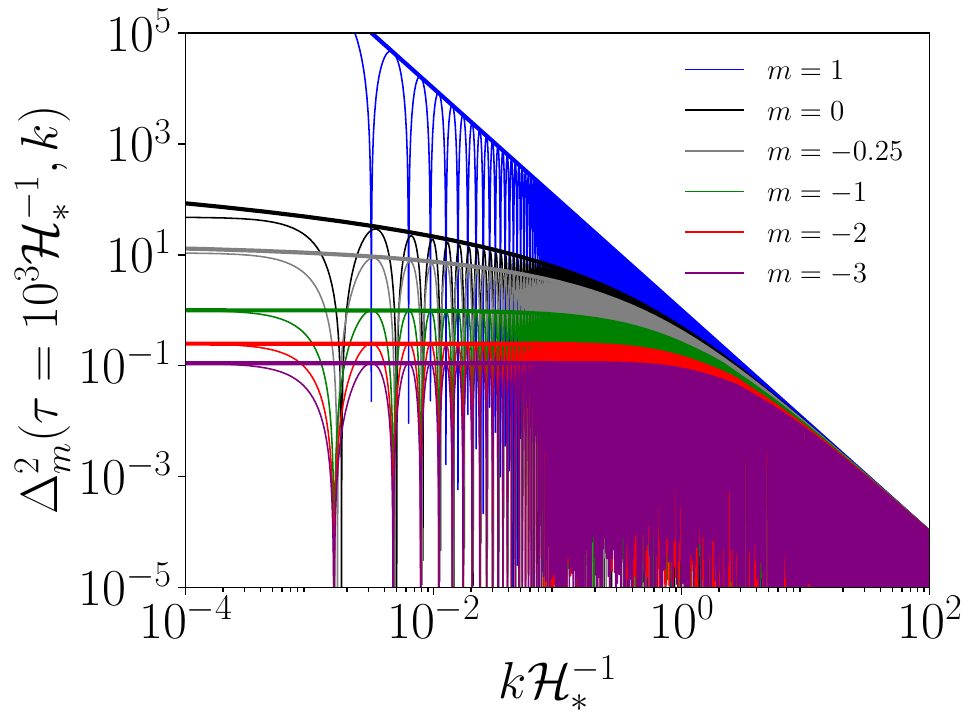}
    \caption{
    Spectra ${\Delta}_m^2 (\tau, k)$ for different values of
    $m$ at large times (in the figure $\HH_* \tau = 10^3$) compared to their
    envelope, given in \Eq{envelopes} for the cases $m \neq 0$ and in \Eq{saturated_delta2_cit} for the CiT case $m=0$.}
    \label{Gsq_envelope}
\end{figure}

In general, at wave numbers $k$, we find that each mode of the
spectrum $\Delta_m^2 (\tau, k)$ evolves following the dynamical
behavior shown in \Fig{Gsq_k0} until the expected saturated
value is reached at $\delta \tau \sim 1/k$ \cite{RoperPol:2022iel}.
Hence, the expected saturated spectra ${\Delta}_m^2 (\tau, k)$ can
be obtained substituting $\delta \tau = 1/k$ in \Eq{lim_k0} for $\Delta_{-n}$ and in \Eqq{delta_oneandtwo_ktozero} for $\Delta_m$ with $m=1,2$, thus giving 
\begin{align}
    \Delta_{-n}^2 (\tau, k) = &\, \frac{1}{n^2} \Biggl[
    1 - \frac{1}{\bigl(1 + \HH_*/k\bigr)^n} \Biggr]^2 \quad
    \forall n > 0, \nonumber \\
    {\Delta}_1^2 (\tau, k) = &\, \bigl(\HH_*/k)^2,
     \nonumber \\
    {\Delta}_2^2 (\tau, k) = &\, \fourth \bigl(\HH_*/k)^4
    + \HH_*^3 /k^3 + \HH_*^2/k^2,
    \label{envelopes}
\end{align}
to be compared to the CiT result $\Delta^2 (\tau, k) = \ln^2 \bigl(1 + \HH_*/k\bigr)$,
which was presented in ref.~\cite{RoperPol:2022iel}.
The spectra $\Delta_m^2 (\tau, k)$ at large times, such that all modes $k> 1/\delta \tau$ have saturated, are shown in \Fig{Gsq_envelope}, compared
to the envelope computed using \Eq{envelopes}.
From \Eq{envelopes}, we find that in the limit $k /\HH_* \gg 1$,
the envelope tends to $k^{-2}$ for all $n$, which is also seen in
\Fig{Gsq_envelope}.
We show in \App{G_m_coherent} that the behavior of $\Delta_m^2$ at large $k \to \infty$ is, in fact, [cf.~\Eq{k_to_inf_delta_m}]
\begin{align}
    \Delta_m(\tau, k) \underset{k \to \infty}{\simeq } & \,  \frac{1}{k \tau_\ast}  \sin k \bigl(\tau - \tau_\ast \bigr)\,,
\end{align}
for any real value of $m$, 
which coincides with the result in the CiT model [cf.~\Eq{delta_cit_inf_limits}]. Hence we see that the envelope of $\Delta^2_m$ in the limit $k \to \infty$ is $(\HH_\ast/k)^2$ for any value of $m$.

\begin{figure}
    \centering
    \includegraphics[width=\columnwidth]{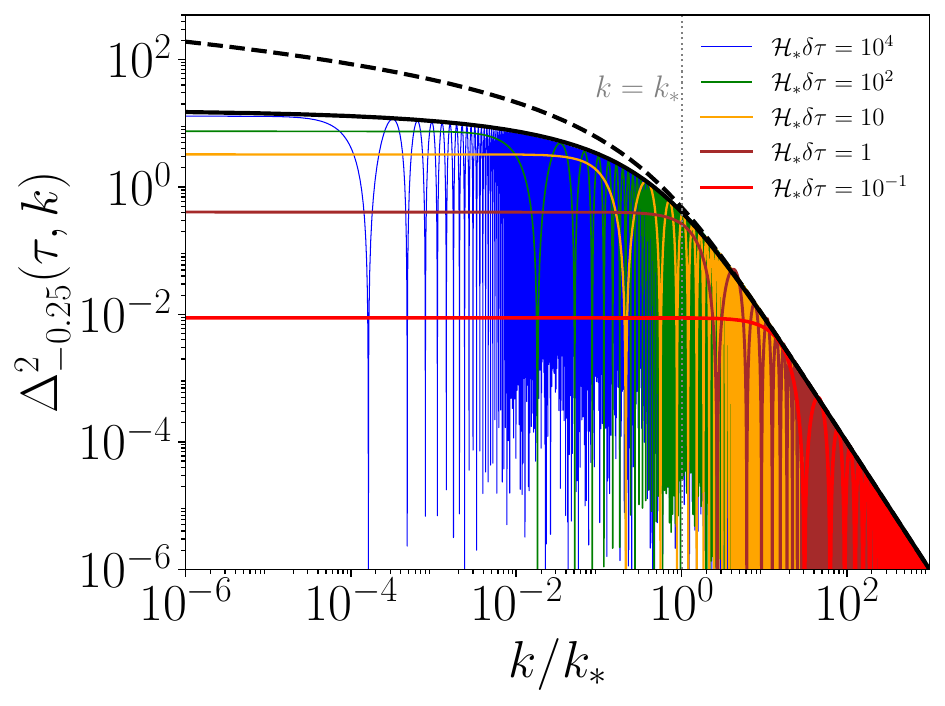}
    \caption{
    Spectra ${\Delta}_{m}^2(\tau, k)$, for $m=-1/4$, at different times
    $\HH_* \delta \tau = \{10^{-1}, 1, 10, 10^2, 10^4\}$ (colored solid
    lines).
    For comparison, we show the corresponding values of the envelope
    obtained using the CiT model
    \Eq{OmGW_envelope} (dashed black line) and the envelope
    predicted using \Eq{envelopes} (solid black line).}
    \label{Gsq_kt_p1_q05}
\end{figure}

In \Fig{Gsq_kt_p1_q05}, we show the spectra ${\Delta}_{m}^2 (\tau, k)$ 
at different times for $m=-1/4$, which can correspond, e.g.,
to the turbulent decay expected for nonhelical magnetic fields, with $p = 1$ and $s = \half$ \cite{Brandenburg:2014mwa}. 
The spectra are compared with those obtained in the CiT
model, given in \Eq{OmGW_envelope}.

\subsection{Gravitational wave template}
\label{gw_template_coh}

In this section we have generalized the CiT model of \Sec{cit_model_sec} to the case of coherent UETCs, described as in \Eq{coherent_epi}. We have furthermore assumed that we can describe the decay of the source autocorrelation function $E_\Pi(\tau, k)$ through the decay of its amplitude and peak wave number, as in \Eqq{eq:decayall}. This has led to the following expression for the Gravitational Wave spectrum [cf.~\Eq{SGWB_coh}]
\begin{align}
    \Omega_{\rm GW} (\tau, k) = 3 K^3 \frac{\cal C}{{\cal A}^2} \frac{\tilde{\varphi}_{\rm rms}^4}{\bar{\rho}^2} \zeta^*_\Pi(K) F^{0}_{\rm GW} \Delta^2_{ps} ( \tau, k; \delta \tau_{\rm dc})\,.
\end{align}
Here $\Delta^2_{ps}(\tau, k;\delta \tau_{\rm dc})$, given in \Eq{G_pq}, is a function which depends on both the decay exponents of the amplitude and peak wave number of $E_\Pi(\tau, k)$, respectively $p$ and $s$, and on the decay timescale $\delta \tau_{\rm dc}$. In the case of moderate values of $s<1$, as we show in \App{fast_q} (see \Eq{range_of_good_s_values} and \Fig{p_pi_ratios}) the dependence on $\zeta_\Pi^{\rm rat}$ in \Eq{G_pq} can be neglected, which leads to the approximations $\Delta^2_{ps} (\tau, k; \delta \tau_{\rm dc}) \simeq \Delta^2_{m}(\tau, k; \delta \tau_{\rm dc})$ [cf.~\Eq{coh_first_approx}], where $m = \frac{3}{2}s - p$. The function $\Delta^2_m(\tau, k; \delta \tau_{\rm dc})$ is shown in \Fig{Delta_pq} for different values of $\delta \tau_{\rm dc}$. In the case $\delta \tau_{\rm dc} = \tau_\ast$, moreover, the function $\Delta^2_m$, given in \Eq{delta_tdc_equal_tast}, can be computed analytically, and we derive in \App{G_m_coherent} its analytical expression, given in \Eq{Gpq_coherent_2}. 

As we have shown in this section (see \Fig{Gsq_envelope}), the function $\Delta^2_m(\tau, k)$ has the following envelope in the case of negative values of $m$ [cf.~\Eq{envelopes}]
\begin{align}
    \Delta^2_m (k) = \frac{1}{m^2} \Bigl[ 1 - (1+{\cal H}_*/k)^{m}\Bigr]^2 \ \forall m < 0\,.
\label{delta_coh_envelope_exact}
\end{align}
For this function we can then distinguish two regimes, respectively corresponding to large and small scales
\begin{subequations}
\begin{align}
\lim_{\substack{k \ll {\cal H}_\ast}}
    {\Delta}_{m}^2 (\tau, k)  & \simeq \frac{1}{m^2} \,,
    \label{coh_regime1}
    \\
    \lim_{\substack{k \gg {\cal H}_\ast}}
    {\Delta}_{m}^2 (\tau, k)  & \simeq \Bigl(\frac{{\cal H}_\ast}{k} \Bigr)^2\,.
    \label{coh_regime2}
\end{align}
\end{subequations}
Around the scale $k \simeq {\cal H}_\ast$ the envelope of $\Delta^2_m$ smoothly transitions from \Eq{coh_regime1} to \Eq{coh_regime2}, following the exact form in \Eq{delta_coh_envelope_exact}.

Hence, there are three different regimes for the Gravitational Wave spectrum.
\begin{enumerate}[\em i)]
    \item For $K < {\cal R}_\ast {\cal H}_\ast$ the Gravitational Wave spectrum $\Omega_{\rm GW}$ goes as $K^3$ with a $K \to 0$ amplitude proportional to $1/m^2$ [cf.~\Eq{coh_regime1}];
    \item For $K > {\cal R}_\ast {\cal H}_\ast$ the spectrum transitions to the regime where it is linear in $K$, due to the $1/k^2$ behavior of $\Delta^2_m$ [cf.~\Eq{coh_regime2}];
    \item For large $K > K_{\rm GW}$, with $K_{\rm GW}$ being the GW spectral peak, given in \Eq{KGW_fit} as for the CiT model case, the GW spectrum goes as $K^{1-b_{\Pi}}$.
\end{enumerate}

Hence the only changes with respect to the CiT model are at large scales, for $K < {\cal R}_\ast {\cal H}_\ast$, where the suppression of $\Omega_{\rm GW}$ is due to the decay of the source. At smaller scales the GW spectrum saturates with a timescale which is shorter than the decay timescale, hence we recover the same result as in the CiT model. 

Similarly to \Sec{gw_template_cit} we can define a normalized spectral shape ${\cal S}_{\rm coh}(K)$ as
\begin{align}
    {\cal S}_{\rm coh}(K) = {\cal N}_{\rm coh} \, K^3 \, \zeta_\Pi^*(K) \Delta^2_m(\tau, k) \,,
\end{align}
with 
\begin{align}
    {\cal N}^{-1}_{\rm coh} = \int_{-\infty}^{+\infty} K^3 \, \zeta_\Pi^*(K) \Delta^2_m(\tau, k) \dd \ln K\,,
\end{align}
and where we have restricted to the cases in which $\Delta^2_{ps}$ [cf.~\Eq{G_pq}] reduces to $\Delta^2_{m}(\tau, k)$ [cf.~\Eq{delta_tdc_equal_tast}].
Then we can define an efficiency factor 
\begin{align}
    \tilde{\Omega}_{\rm GW}^{\rm coh} = \frac{\cal C}{2{\cal A}^2} \frac{{\cal N}^{-1}_{\rm coh}}{({\cal R}_\ast {\cal H}_\ast)^2}\,,
\end{align}
such that the GW spectrum can be written as
\begin{align}
    \Omega_{\rm GW} = 3 \, F^0_{\rm GW} \, \tilde{\Omega}_{\rm GW}^{\rm coh} \frac{\tilde{\varphi}_{\rm rms}^4} {\bar{\rho}^2} \bigl({\cal R}_* {\cal H}_* \bigr)^2 \, {\cal S}_{\rm coh}(K) \,.
\end{align}

\begin{figure}
    \centering
    \includegraphics[width=0.5\textwidth]{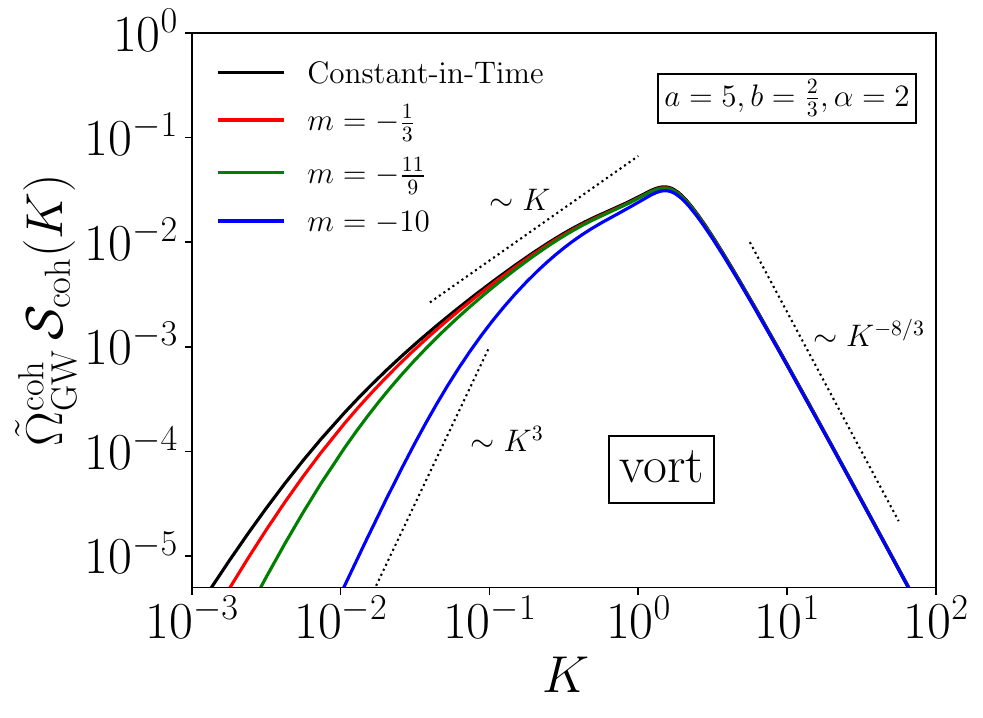}\\
    \includegraphics[width=0.5\textwidth]{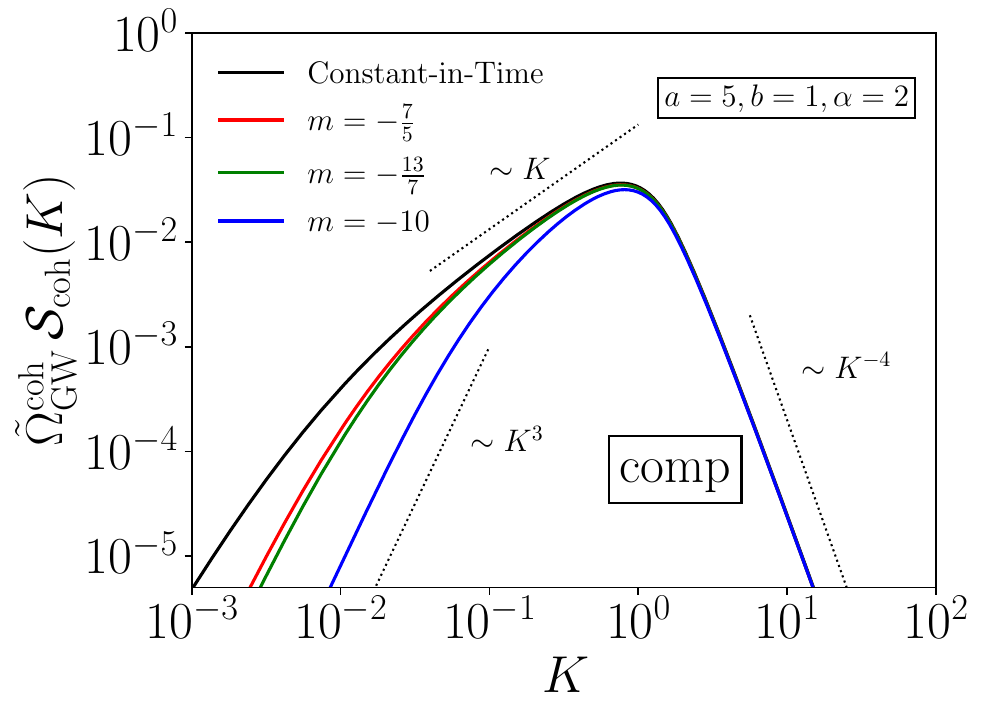}\\
    \caption{
    Normalized spectral shape of the GW spectrum multiplied by the efficiency $\tilde{\Omega}_{\rm GW}$, in the coherent case with $\delta \tau_{\rm dc} = \tau_\ast$ and for different values of $m$, both for a purely
    vortical (upper panel) and purely compressional (lower panel) source with $\alpha=2, a = 5$ and respectively $b=2/3$ and $b=1$ for the vortical and compressional cases.
    We consider ${\cal R}_\ast \HH_\ast = 10^{-2}$ and, for comparison, we show in black lines the result in the CiT model (see also \Fig{shapeGW_vortcomp}) with sourcing duration $\delta \tau_{\rm fin}/\tau_\ast = 100$.}
    \label{shapeGW_vortcomp_coh}
\end{figure}

We show in \Fig{shapeGW_vortcomp_coh} the normalized ${\cal S}_{\rm coh}(K)$ multiplied by $\tilde{\Omega}_{\rm GW}^{\rm coh}$ for different values of $m$ in the case $\delta \tau_{\rm dc} = \tau_\ast$ in which the function $\Delta^2_m$ can be computed analytically. Here we use for $\Delta^2_m$ its envelope which is given in \Eq{delta_coh_envelope_exact}. 

\begin{table}[]
    \centering
    \begin{tabular}{|c|c|c|c|c|}
    \hline 
    \rule{0pt}{3ex}
        \diagbox{$\ \ \ \ \ \alpha$}{$\vspace{-2mm} -m$} & $\frac{1}{3} \, ({\rm vort})$ & $\frac{11}{9} \, ({\rm vort})$ & $\frac{7}{5} \, ({\rm comp})$ & $\frac{13}{7} \, ({\rm comp})$ \\
        \hline
         0.05 & 2.287 & 1.826 & 0.397 & 0.364 \\
         0.1 & 0.466 & 0.432 & 0.167 & 0.161 \\
         0.2 & 0.190 & 0.183 & 0.107 & 0.105 \\
         6/17 & 0.125 & 0.121 & 0.090 & 0.089 \\
         0.75 & 0.088 & 0.086 & 0.080 & 0.079 \\
         1 & 0.080 & 0.078 & 0.078 &  0.077 \\
         1.5 & 0.071 & 0.070 & 0.074 & 0.073 \\
         2 & 0.067 & 0.065 & 0.072 & 0.071 \\
         3 & 0.062 & 0.061 & 0.070 & 0.069 \\
         4 & 0.059 & 0.058 & 0.068 & 0.067 \\
         5 & 0.058 & 0.057 & 0.067 & 0.066 \\ 
         10 & 0.054 & 0.053 & 0.064 & 0.064 \\
         \hline
    \end{tabular}
    \caption{
    Values of the GW efficiency in the coherent-decay model $\tilde{\Omega}_{\rm GW}^{\rm coh}$ for a purely vortical non-helical
    and a purely compressional source with slopes $a = 5$ and $b = 2/3$ for vortical (${\rm vort}$), while $a=5$ and $b=1$ for compressional (${\rm comp}$),
    and a range of values of the smoothness parameter $\alpha$. We use physically motivated values of $m = -1/3, -11/9$ for vortical \cite{Brandenburg:2024tyi} and $m=-7/5, -13/7$ for compressional \cite{Dahl:2024eup, Brandenburg:2024tyi}, as in \Fig{shapeGW_vortcomp_coh}. Here $\delta \tau_{\rm dc} = \tau_\ast$, ${\cal R}_\ast {\cal H}_\ast = 10^{-2}$. Note that the values of $\tilde{\Omega}_{\rm GW}^{\rm coh}$ are always smaller than those in the CiT model (see \Tab{KGW_values}), which they approach as $|m|$ decreases.}
    \label{tilde_omgw_coh}
\end{table}

Moreover, in \Tab{tilde_omgw_coh} we provide the GW efficiencies $\tilde{\Omega}_{\rm GW}^{\rm coh}$ for different values of $\alpha$ and $m$.
We consider 
physically motivated values of $m$.
In particular, $m = -1/3$ and $m = -11/9$ are expected
at late times
for magnetically-dominated turbulence when the magnetic
field is helical \cite{Brandenburg:2017neh} or non-helical
\cite{Brandenburg:2024tyi}, respectively, based on the
conservation of helicity or the Hosking invariant.
For kinetically-dominated turbulence, instead,
we expect $m=-7/5$ or  $m =-13/7$ for the IR slope
of the velocity spectrum $a = 3$ or $a = 5$,
respectively \cite{Brandenburg:2017neh}.
When the velocity field is purely irrotational (i.e.,
acoustic turbulence), then $a = 5$ and, hence, one expects
$m = -7/5$ \cite{Dahl:2021wyk,Dahl:2024eup}.

In the case of $\delta \tau_{\rm dc} \neq \tau_\ast$ the normalized ${\cal S}_{\rm coh}(K)$ would be modified at $k<1/\delta \tau_{\rm dc}$ with respect to the CiT model and the large-scale region of the GW spectrum would be suppressed by the factor $r^2(\delta \tau_{\rm dc}; x_{\rm dc}, \alpha_{\rm dc}, \beta_{\rm dc})$ which we have introduced in \Eq{fit_delta2m_deltataudc_ktozero} and shown in \Fig{Delta2m_ktozero_amplitude}. 

\begin{figure}
    \centering
    \includegraphics[width=0.5\textwidth]{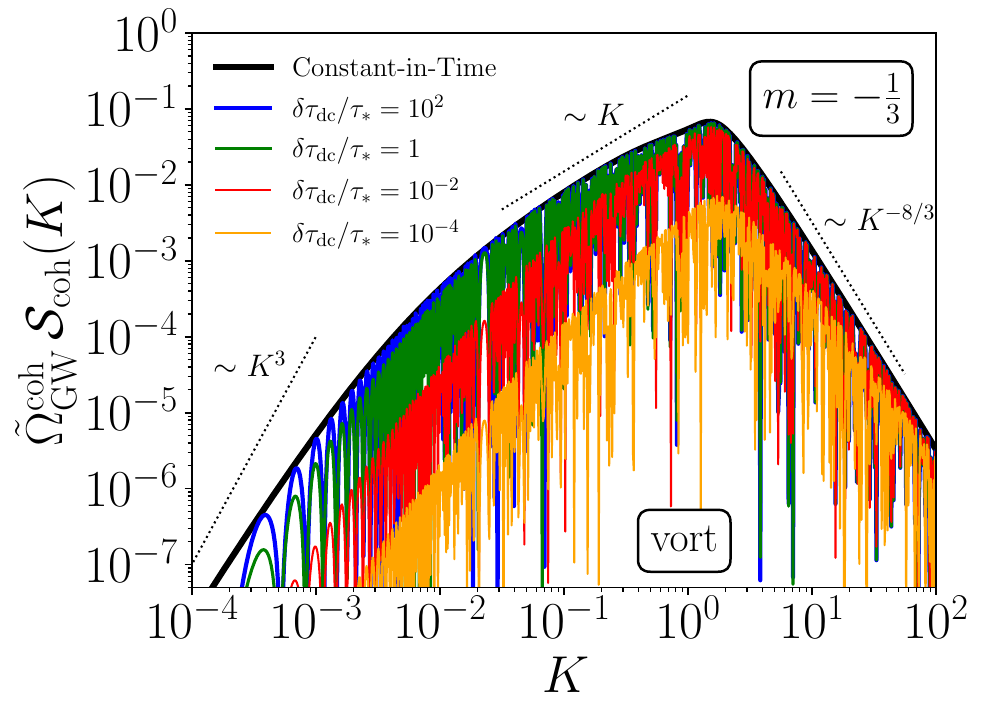}\\
    \includegraphics[width=0.5\textwidth]{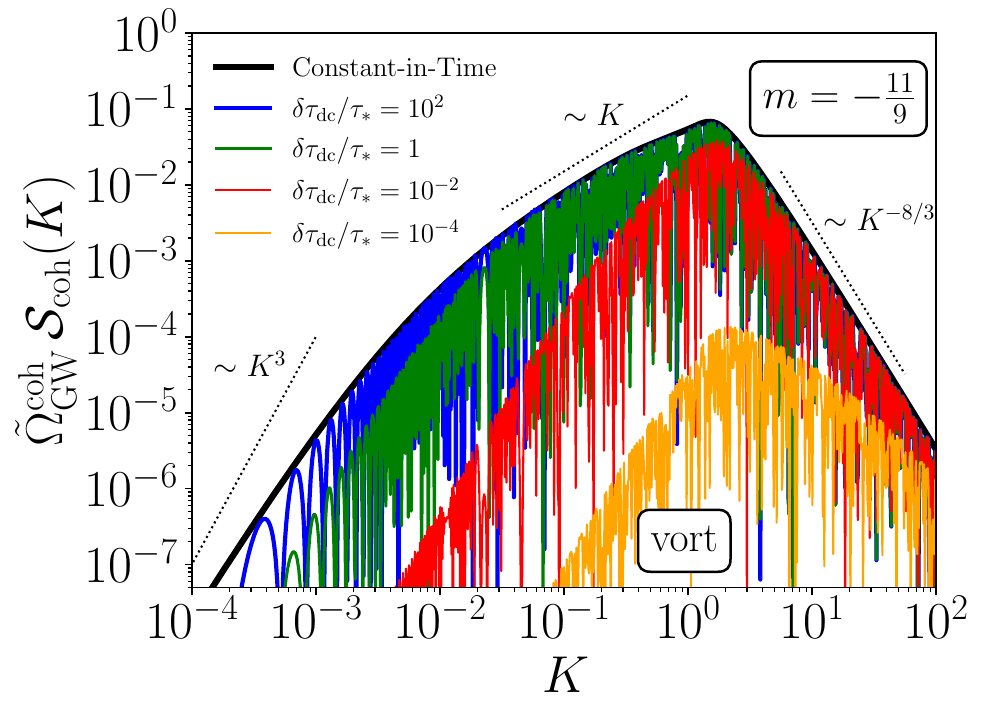}\\
    \caption{
    Normalized spectral shape of the GW spectrum multiplied by the efficiency $\tilde{\Omega}_{\rm GW}^{\rm coh}$ in the coherent case, for various $\delta \tau_{\rm dc}/\tau_\ast$ and respectively $m=-1/3$ (upper panel) and $m=-11/9$ (lower panel), for a purely
    vortical source with $\alpha=2, a = 5, b=2/3$.
    We consider ${\cal R}_\ast \HH_\ast = 10^{-2}$ and sourcing duration $\delta \tau_{\rm fin}/\tau_\ast = 100$. For comparison, we show in black lines the result in the CiT model (see also \Fig{shapeGW_vortcomp}).}
    \label{shapeGW_vort_coh_deltatau_dc}
\end{figure}

\begin{figure}
    \centering
    \includegraphics[width=0.5\textwidth]{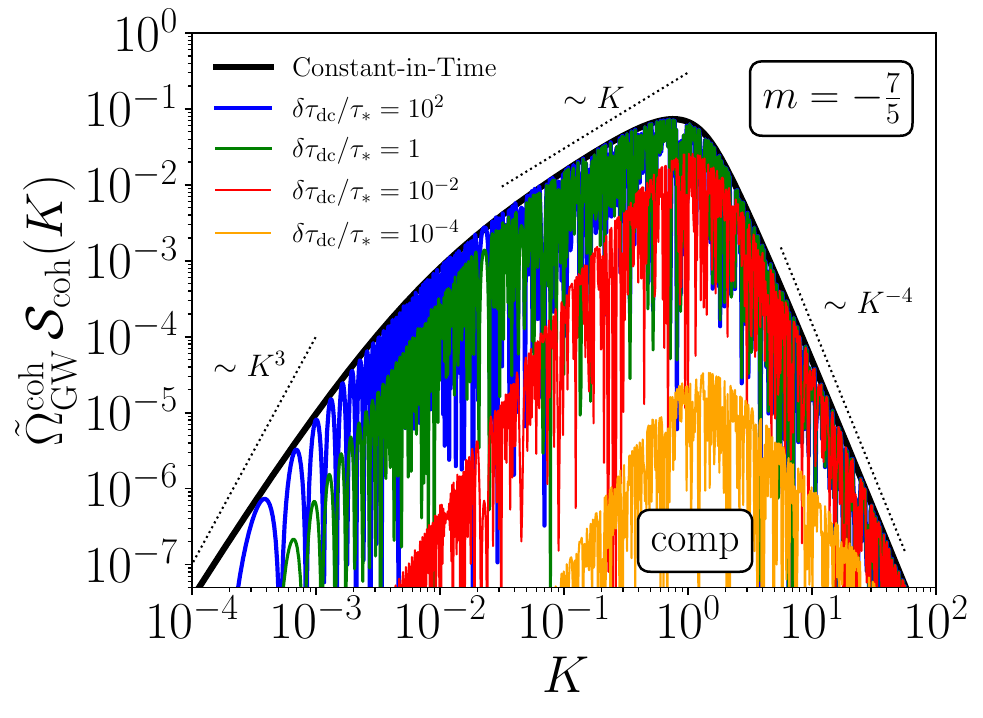}\\
    \includegraphics[width=0.5\textwidth]{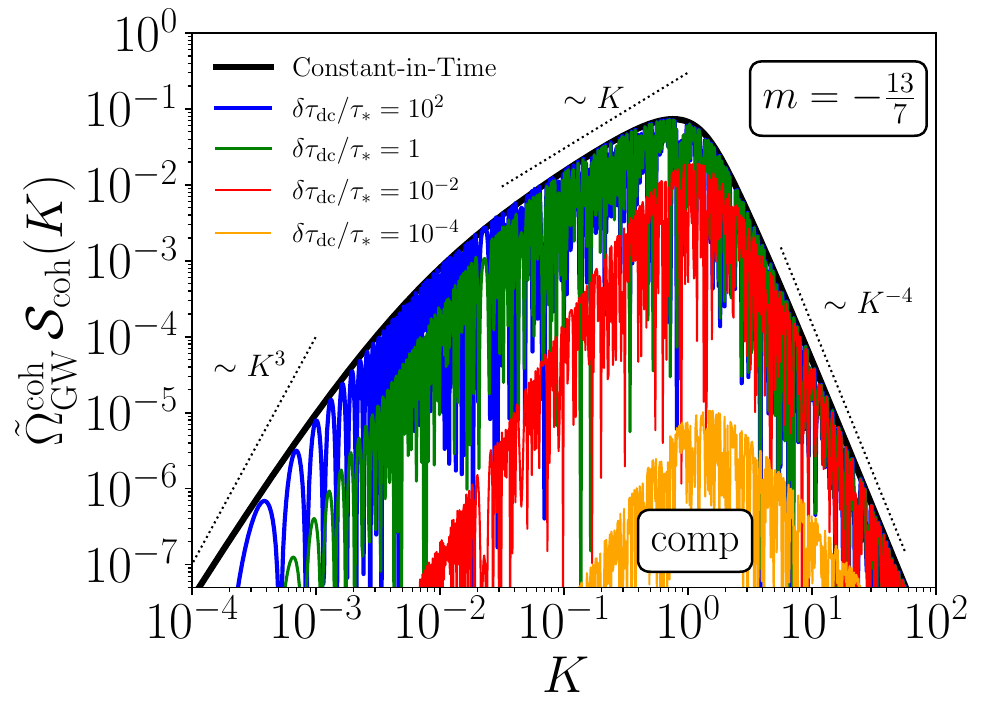}\\
    \caption{
    Normalized spectral shape of the GW spectrum multiplied by the efficiency $\tilde{\Omega}_{\rm GW}^{\rm coh}$ in the coherent case, for various $\delta \tau_{\rm dc}/\tau_\ast$ and respectively $m=-7/5$ (upper panel) and $m=-13/7$ (lower panel), for a purely
    compressional source with $\alpha=2, a = 5, b=1$.
    We consider ${\cal R}_\ast \HH_\ast = 10^{-2}$ and sourcing duration $\delta \tau_{\rm fin}/\tau_\ast = 100$. For comparison, we show in black lines the result in the CiT model (see also \Fig{shapeGW_vortcomp}).}
    \label{shapeGW_comp_coh_deltatau_dc}
\end{figure}

\begin{figure}
    \centering
    \includegraphics[width=0.5\textwidth]{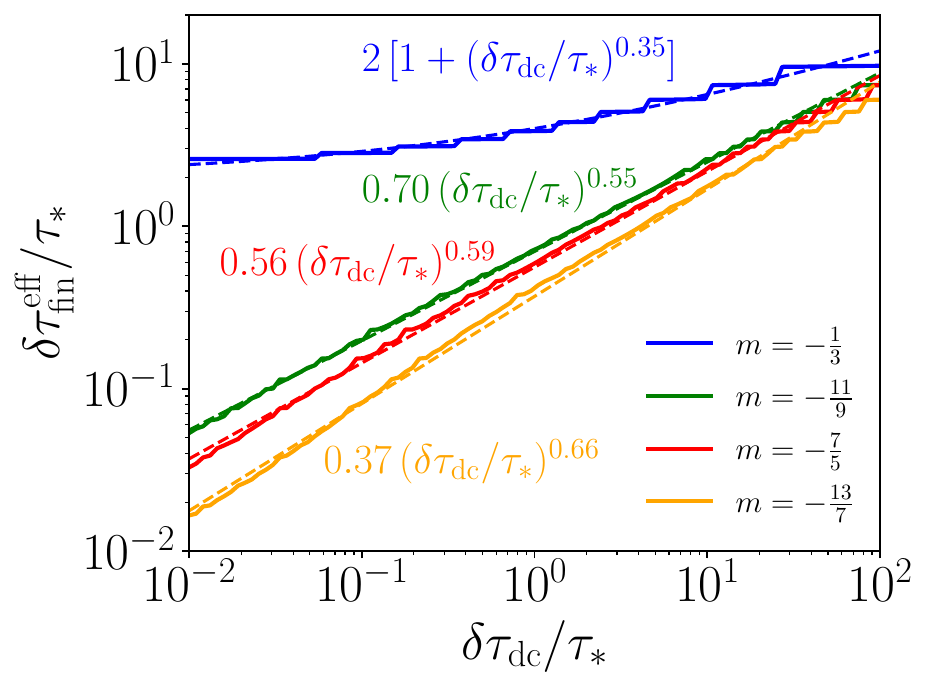}\\
    \caption{Effective final time of the sourcing $\delta \tau_{\rm fin}^{\rm eff}$ for the coherent-decay model (solid lines), as a function of the decay timescale, and for different values of $m$, corresponding to those of \Figs{shapeGW_vort_coh_deltatau_dc}{shapeGW_comp_coh_deltatau_dc}. Dashed lines represent the fitting functions found for each case.
    }
    \label{eff_fin_time_coh}
\end{figure}

We show in \Figs{shapeGW_vort_coh_deltatau_dc}{shapeGW_comp_coh_deltatau_dc} the normalized GW spectrum ${\cal S}_{\rm coh}(K)$ multiplied by the GW efficiency $\tilde{\Omega}_{\rm GW}^{\rm coh}$ for different values of $\delta \tau_{\rm dc}/\tau_\ast$ both for purely vortical and purely compressional fields, and the same values of $m$ as in \Tab{tilde_omgw_coh}. Here we see that the effect of having a shorter decay timescale is more relevant for larger values of the decay parameter $-m$.
Moreover, for large $\delta \tau_{\rm dc}/\tau_\ast$ and small $|m|$, we see that the large-scale $\sim K^3$ regime of the GW spectrum is only reached below $K \lesssim 10^{-4}$, corresponding to the prediction for the Constant-in-Time model, given that ${\cal R}_\ast {\cal H}_\ast = 10^{-2}$ and $\delta \tau_{\rm fin} = 10^2 \tau_\ast$, which implies $(k_\ast \delta \tau_{\rm fin})^{-1} = 10^{-4}$. However, as $|m|$ increases and as $\delta \tau_{\rm dc}/\tau_\ast$ decreases, the $\sim K^3$ regime is reached earlier, and we define as the effective final time of the sourcing $\delta \tau_{\rm fin}^{\rm eff}$ the inverse wave number at which this regime is reached. We show in \Fig{eff_fin_time_coh} the values of $\delta \tau_{\rm fin}^{\rm eff}/\tau_\ast$ for different choices of $m$, corresponding to those of \Figs{shapeGW_vort_coh_deltatau_dc}{shapeGW_comp_coh_deltatau_dc}, together with some simple power law fits for each value of $m$. 

\section{Conclusions}
\label{conclusions}

In this work, we have presented a general framework to describe the
production of stochastic gravitational wave (GW) backgrounds from slowly-evolving
sources in the early Universe, focusing in particular on the
role of the temporal evolution and decorrelation of the source. The central
quantity entering the computation of the GW spectrum is the unequal-time
correlator (UETC) of the anisotropic stresses, $E_{\Pi}(\tau_1,\tau_2,k)$.
Since the UETC is generally not known analytically, especially for nonlinear
sources such as turbulent velocity and magnetic fields, a suitable modelling of
its time dependence is essential to obtain reliable predictions for the
resulting GW background.

\paragraph*{\bf Constant-in-Time (CiT) model.}
We have first studied the CiT model, which assumes that the source evolves on
a timescale much longer than the characteristic timescale of GW production.
Under this assumption, the UETC can be approximated as constant during an
effective sourcing period $\delta\tau_{\rm fin} = \tau_{\rm fin} - \tau_\ast$,
\begin{equation}
    E_{\Pi}(\tau_1,\tau_2,k) \simeq E_{\Pi}^{*}(k) \equiv E_\Pi(\tau_\ast, \tau_\ast, k)\,,
\end{equation}
and the GW spectrum can be written as
\begin{equation}
    \Omega_{\rm GW}(\tau,k) = 3 E_{\Pi}^{*}(k) F_{\rm GW}^{0} \Delta^2(\tau,k)\,,
\end{equation}
where $\Delta(\tau,k)$ is the time kernel obtained by integrating the GW Green's
functions over the sourcing period [see \Eq{G_constant}], and $F^0_{\rm GW}$ accounts
for the redshift of the GW energy density from the moment of production to the present time.
The main advantage of this approximation is that it separates the spectral
properties of the source, encoded in $E_{\Pi}^{*}(k)$, from the time dependence
of GW production, encoded in $\Delta(\tau,k)$.

A characteristic feature of the CiT model is the resulting GW spectral shape,
which exhibits three different regimes. At sufficiently small wave numbers, the
spectrum follows the causal infrared (IR) behaviour $\Omega_{\rm GW}(k) \propto k^3$.
At intermediate wave numbers, between the inverse effective duration of the source
and the peak scale of $E_\Pi^\ast(k)$, the spectrum develops an approximately linear
dependence, $\Omega_{\rm GW}(k) \propto k$, which is, in the case of an expanding Universe,
more precisely given by $k^3 \ln^2 \bigl( 1+{\cal H}_\ast/k \bigr)$, with ${\cal H}_\ast$
being the conformal Hubble rate at the time of GW production.
Finally, at wave numbers above the peak of the UETC, the spectrum decreases with a
power-law slope determined by the small-scale behaviour of the anisotropic stress
spectrum. In particular, for a source whose anisotropic stress spectrum scales as
$E_{\Pi}(k)\propto k^{3-b_{\Pi}}$, the ultraviolet (UV) GW spectrum behaves as
$\Omega_{\rm GW}(k)\propto k^{1-b_{\Pi}}$.
As an example, for Kolmogorov turbulence, which is characterized by $b_{\Pi}=11/3$, this gives the
well-known scaling $\Omega_{\rm GW}(k)\propto k^{-8/3}$.
The UV behaviour is therefore not universal, but depends on the nature of the
source and on the spectral properties of its anisotropic stresses, while the
$k^3$ and linear regimes arise, respectively, from causality and from the Green's
function for GW production.

\paragraph*{\bf Applications of the CiT model.}
We have applied this description to velocity fields with generic compressibility and
fractional helicity, magnetic fields with generic fractional helicity, and scalar fields.
These cases encompass vortical and acoustic turbulence, sound waves, and
scalar-field gradients. Although the underlying source spectra can be
substantially different, the CiT framework provides a common description of their
GW production in the regime in which both their decay and decorrelation timescales are longer than that of GW production. To make this description easy to use, we have modelled the
source spectra with a smoothed broken power law, with slopes $a$ and $b$ at small
and large wave numbers and smoothness parameter $\alpha$, and provided the
corresponding analytical moments, as well as fits of the normalized anisotropic stresses
and of the characteristic wave number $k_{\rm GW} = K_{\rm GW}/{\cal R}_\ast$ of the GW peak, with ${\cal R}_\ast$ being the correlation length of the source.
This allows us to provide the following GW template,
\begin{equation}
    \Omega_{\rm GW} = 3 F_{\rm GW}^0 \,
    \frac{\tilde \varphi_{\rm rms}^4}{\bar \rho^2} \, ({\cal R}_\ast {\cal H}_\ast)^2 \, \tilde \Omega_{\rm GW} \, {\cal S} (K)\,,
\end{equation}
where the amplitude is determined by the GW efficiency $\tilde \Omega_{\rm GW}$, the energy density fraction
of the source $\varphi_i$, through the factor $\tilde \varphi_{\rm rms}^4/\bar\rho^2$ [see discussion below \Eq{EPi_norms}], and the characteristic length scale of the source
in units of the Hubble radius, ${\cal R}_\ast {\cal H}_\ast$, while the spectral shape ${\cal S}(K)$
is normalized such that $\int \dd \ln K \, {\cal S}(K) = 1$.
For helical sources, we have also computed, within the CiT model, the circular polarization of the GW background.

\paragraph*{\bf Decay of the source: coherent-decay model.}
An important limitation of the CiT approximation is that it does not describe
explicitly the decay of the source. In order to overcome this limitation, we have
introduced a coherent model for the UETC, in which this is factorized as
\begin{equation}
    E_{\Pi}(\tau_1,\tau_2,k) = \sqrt{E_{\Pi}(\tau_1,k) E_{\Pi}(\tau_2,k)}\,.
\end{equation}
This assumption describes sources whose anisotropic stresses remain correlated over
the relevant time interval of GW production. It allows us to account for the time
evolution of the source while retaining a separable expression for the UETC, so that
the GW spectrum can be computed analytically in certain regimes.
To model this evolution, we have considered power-law decays of both the
characteristic wave number and the amplitude of the source,
$k_\ast(\tau) \propto (1+\delta\tau/\delta\tau_{\rm dc})^{-s}$ and
$E_\ast(\tau) \propto (1+\delta\tau/\delta\tau_{\rm dc})^{-p}$,
where $\delta\tau_{\rm dc}$ is the characteristic decay timescale.
For moderate values of $s \lesssim 1$, we have shown that the dependence on the
change in the UV spectral shape of $E_\Pi$ can be approximately neglected, and
the time kernel $\Delta^2_{\rm coh}$ depends only on the combination $m=\frac{3}{2}s-p$, hence we call it $\Delta^2_m$ [see \Eqs{delta_coh_formula}{G_pq} and, also, \Eqs{coh_first_approx}{delta_tdc_equal_tast}].
This kernel can be computed analytically for the case
$\delta \tau_{\rm dc} = {\cal H}_\ast^{-1}$, and presents large-scale deviations from this
analytical expression for different values of $\delta \tau_{\rm dc}$ and $m$,
for which we have provided a fit.

An important consistency check of this model is that, in the limit of an infinitely
long decay timescale, the source becomes effectively constant in time and the CiT result
is recovered. For a finite decay timescale, instead, the source evolution modifies the
GW spectrum at large scales. In particular, when the decay timescale is longer than the
relevant GW production timescale, the impact of the decay is mainly restricted to
sufficiently small wave numbers, $k \lesssim \delta\tau_{\rm dc}^{-1}$, where the
amplitude of the $k^3$ regime is suppressed with respect to the CiT model. Moreover, in the coherent case, and for sufficiently short decay timescales, the intermediate linear regime disappears.
At larger wave numbers, GW production occurs on a timescale short enough that the
source can still be considered approximately constant, and the CiT model remains a good
description. The coherent-decay model therefore provides a consistent way to describe
the transition between the regime in which the source can be treated as constant and the
regime in which its decay becomes relevant. In particular, it avoids the need to
abruptly switch off the source after an effective finite duration
$\delta \tau_{\rm fin} = \tau_{\rm fin} - \tau_\ast$, as is done in the CiT prescription.

The application to decaying magnetohydrodynamic (MHD) turbulence illustrates the
relevance of this extension. In this case, the coherent-decay model reproduces the
expected spectral regions, including the causal $k^3$ tail, the intermediate linear
branch and the Kolmogorov ultraviolet scaling $k^{-8/3}$. However, the comparison
between the two descriptions shows that the CiT model can overestimate GW production
in the large-scale and intermediate regions. This
difference is associated with the time evolution of the UETC: while the CiT
approximation neglects its decay during the sourcing period, the coherent-decay model
explicitly accounts for the decrease of the source amplitude and the evolution of its
characteristic scale. The resulting GW spectrum is consequently sensitive to the decay
rate and to the timescale over which the source evolves.

\paragraph*{\bf Temporal decorrelation of the source.}
We have also discussed the role of temporal decorrelation. In stationary turbulence,
the UETC is expected to decorrelate in time because of the nonlinear dynamics. A commonly used description of this effect is the Kraichnan decorrelation model,
which we have applied to hydrodynamical vortical turbulence.
The CiT approximation is justified when the GW production timescale is much shorter than the source
decorrelation timescale. In this regime, the source remains correlated during the
relevant GW production interval and the CiT prediction is recovered. Conversely,
when decorrelation becomes relevant, it can modify the resulting GW spectrum.
We have found in particular that when the decorrelation velocity $u_{\rm dcorr}$ [see \Eq{UETC_kraichnan_formula} and discussion above \Eq{kraichnan_dcorr_fit}] is
\begin{align}
     u_{\rm dcorr} \, k_\ast \tau_\ast \Upsilon \gtrsim 1\,,
\end{align}
where $\Upsilon = \delta \tau_{\rm fin}{\cal H}_\ast/(1+\delta \tau_{\rm fin}{\cal H}_\ast)$ and $k_\ast = {\cal R}_\ast^{-1}$, then the effect of Kraichnan decorrelation has to be taken into account in the resulting
GW spectrum: a peak appears at a wave number proportional to $k_\ast u_{\rm dcorr}$ and
leads to a larger GW amplitude than in the CiT prediction. A similar condition holds for
stationary sound waves, for which we find that a peak structure which makes the GW
spectrum grow compared to the CiT prediction appears when the sound speed is
\begin{align}
    \cs \, k_\ast \tau_\ast \Upsilon \gtrsim 1\,.
\end{align}
Hence, one can think of
$\delta \tau_{\rm dcorr}^{\rm Kraichnan} \equiv 1/(k_\ast \, u_{\rm dcorr})$ and
$\delta \tau_{\rm dcorr}^{\rm sw} \equiv 1/(k_\ast \, \cs)$
as the typical timescales over which decorrelation becomes important, respectively for
turbulence and sound waves. Indeed, the CiT model applies whenever all relevant sourcing
times are much shorter than these typical decorrelation timescales, i.e., as long as
\begin{align}
    \delta \tau_{\rm fin} \ll  \delta \tau_{\rm dcorr}\,.
\end{align}

\paragraph*{\bf Outlook.}
The framework developed in this work provides a unified approach to the computation of
GW spectra from a broad range of slowly-evolving sources in the radiation-dominated era, including velocity, magnetic and scalar fields.
It also provides a systematic way to identify the regimes in which the CiT approximation
is reliable and those in which the time evolution of the source must be taken into
account. The resulting models have been implemented in the open-source
\href{https://github.com/AlbertoRoper/cosmoGW}{\sc CosmoGW} \cite{cosmogw_manual,CosmoGW_GH} code, which can be used for
numerical studies of different aspects of GW production from early Universe sources.

Several aspects deserve further investigation. In particular, a more detailed treatment
of the growth phase of the source is needed, since the present description focuses on GW
production during the source's decay phase. Moreover, especially in the case of turbulence, in which the decorrelation of the source becomes particularly relevant in the ultrarelativistic
nonlinear regime, a more complete description of realistic sources requires accounting simultaneously for both their decay and their
temporal decorrelation.
These improvements will be important for obtaining fully reliable predictions for the
stochastic GW background produced by realistic sources in the early Universe.

\makeatletter
\let\oldaddcontentsline\addcontentsline
\renewcommand{\addcontentsline}[3]{}
\makeatother

\begin{acknowledgments}
ASM and ARP are supported
through the Swiss National Science Foundation
SNSF Ambizione grant \href{https://data.snf.ch/grants/grant/208807}{208807}.
The authors acknowledge the hospitality of CERN
where part of this work was conducted.
The authors are grateful for the support
provided by Nordita during the program
``{\em Numerical Simulations of Early Universe Sources of Gravitational Waves},'' by the Bernoulli Center during
the program ``{\em Generation, evolution, and observations of cosmological magnetic fields},''
by the Centro de Ciencias de Benasque
Pedro Pascual during ``{\em The Dawn of Gravitational Wave Cosmology}'' conference,
and by CERN during the workshop
``{\em Advancing gravitational wave predictions from cosmological first-order phase transitions},''
where preliminary results of this work were presented.
ASM and ARP acknowledge the support of CERN during the 
2025 Pencil Code school
and user meeting,
where part of this work was conducted.
ARP acknowledges the hospitality of the Yukawa Institute for Theoretical Physics during the Gravity and Cosmology 2024
workshop.
\end{acknowledgments}

\bibliographystyle{unsrt}
\bibliography{library}

\let\addcontentsline\oldaddcontentsline

\appendix

\section{Moments of the smoothed broken power law model}
\label{appendix_sbpl}

The spectra of the velocity, magnetic, and scalar fields are
modeled using a
smoothed broken power law,
introduced in \Eq{powerlaw_model}.
In this section, we present analytical expressions
to compute the different moments
of the spectral shape
that are
relevant for our theoretical model description in \Sec{stress}.

The normalized spectral shape
$\zeta(K) \equiv E(K)/E_\ast$, being $K \equiv k/k_\ast$ the normalized wave number,
such that $\zeta (1) = 1$,
is
\begin{align}
    \zeta (K) =
    \frac{\bigl(b + a\bigr)^{1\over\alpha} K^a}{\left[b+a \, K^{\alpha(a + b)}
    \right]^{1\over\alpha}}\,,
    \label{EM_model}
\end{align}
with slopes $a$ and $-b$ at 
wave numbers $K \ll 1$
and $K\gg 1$ respectively.
Both $a$ and $b$ are taken positive to have a converging
value of the integral over $\ln k$,
\begin{equation}
    \frac{\varphi_{\rm rms}^2}{E_\ast}
   = {\cal A}_{ab} (\alpha) =
   \int_{-\infty}^\infty \zeta (K)
    \dd \ln K\,, \label{EE_zeta}
\end{equation}
as defined in \Eq{integ_OmM},
where $\varphi_{\rm rms}^2$ is the mean squared value of
the field $\varphi_i$ that is characterized by the spectral
function $E(K)$.
Therefore, we can define a spectrum with normalized
integral,
\begin{align}
    \zeta_{\rm int} (K) \equiv  \frac{\zeta(K)}{{\cal A}_{ab}
    (\alpha)} & \nonumber \\ & \hspace{-5mm}
    \text{with \ } 
    \int_{-\infty}^\infty
    \zeta_{\rm int} (K)
    \dd \ln K = 1\,.
\end{align}
Note that $E(K)$, $E_\ast$, and $\varphi_{\rm rms}^2$ have the same units, while $\zeta(K)$ and $\zeta_{\rm int} (K)$
are dimensionless.

The parameter $\alpha$ in \Eq{EM_model}
determines the smoothness of the transition between
the two regimes around $K \simeq 1$.
Alternatively, one can introduce the smoothness
parameter such that
\begin{equation}
    \zeta (K) = \frac{(b + a)^{a + b \over \alpha_2} K^a}{\bigl(b +  a \, K^{\alpha_2}\bigr)^{\frac{a + b}{\alpha_2}}}\,,
    \label{power_law_model_alp2}
\end{equation}
as done, e.g., in refs.~\cite{Auclair:2022jod,RoperPol:2023bqa,Caprini:2024hue}.
We note that this spectral
function is equivalent to the one in \Eq{EM_model} if we use 
$\alpha = \alpha_2/(a + b)$, so the results presented
in this section also apply to this definition of the generic broken
power law.
This motivates the use of $\alpha = 6/17$ considered
in refs.~\cite{RoperPol:2023bqa,Caprini:2024hue} that reproduces the von K\'arm\'an
spectrum with $a = 5$, $b = 2/3$, and $\alpha_2 = 2$,
\begin{equation}
    \zeta (K) = {\cal N} \frac{K^5}{(1 + 7.5\,
    K^2 )^{\frac{17}{6}}}\,, \label{von_Karman}
\end{equation}
with ${\cal N} = (1 + a/b)^{\frac{a + b}{\alpha_2}} =
(1 + 7.5)^{\frac{17}{6}}$.

\begin{figure*}
    \centering
    \includegraphics[width=\columnwidth]{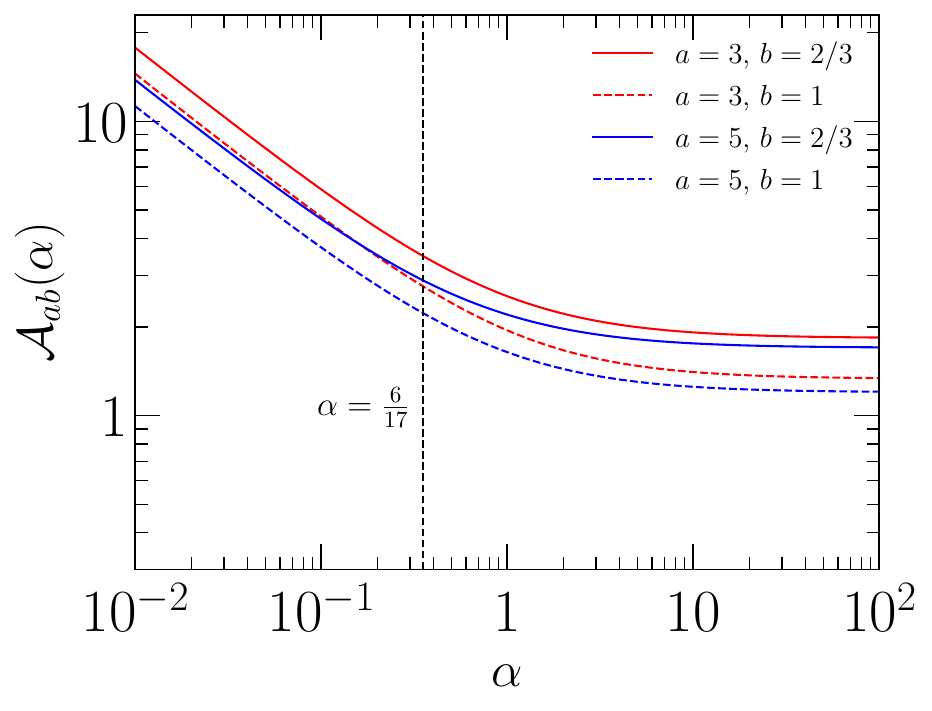}
    \includegraphics[width=\columnwidth]{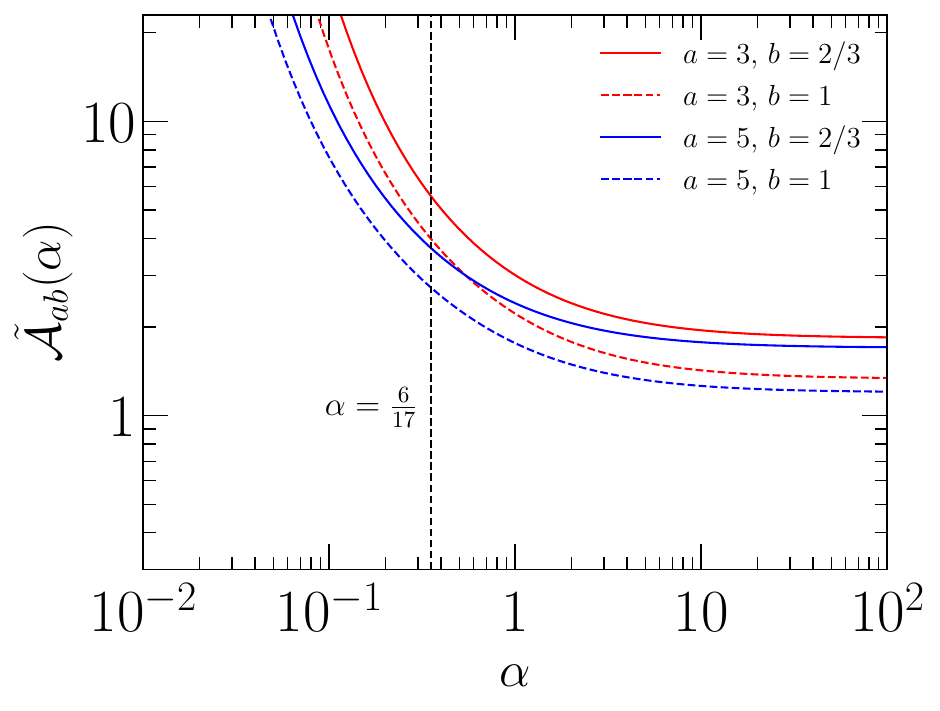}
    \caption{{\em Left panel:} Integrated amplitude
    ${\cal A}_{ab} (\alpha) \equiv
    \varphi_{\rm rms}^2/E_\ast$
    [see \Eq{EE_zeta}]
    for different values of $\alpha$ and for
    the slopes $a = 3$ (red) and
    $5$ (blue), and $b = 2/3$ (solid) and
    $1$ (dashed).
    The models
    used in refs.~\cite{RoperPol:2022iel} and refs.~\cite{Auclair:2022jod,RoperPol:2023bqa,Caprini:2024hue}
    correspond to $\alpha = 2$ and $\alpha_2 = 2$ (i.e., $\alpha = 6/17$), respectively.
    {\em Right panel}: Geometric factor $\tilde {\cal A}_{a b} (\alpha)$ found for the normalization of the spectrum
    $E_{-1} (K) \equiv \dd \varphi_{\rm rms}^2 / \dd K$
    used in previous work [see \Eq{tilde_A}]
    for different values of $\alpha$ and the
    same slopes as in the left panel.
    }
    \label{AA_ab_alp}
\end{figure*}

We now define the general integral
that describes the moments of the spectral
shape function $\zeta(K)$,
\begin{align}
    I_{ab; n} (\alpha) =&\,
    \int_0^\infty \zeta (K) \, K^n \dd K \nonumber \\
    = &\, \int_{-\infty}^\infty \zeta (K) \, K^{n+1} \dd \ln K\,,
    \label{Iabn_def}
\end{align}
which has the following analytical solution:
\begin{align}
    I_{ab; n} &\, (\alpha) = {\cal I}_{ab;n} (\alpha)
    \frac{\Gamma\bigl[{{a + n + 1} \over {\alpha(a+b)}}
    \bigr] \Gamma\bigl[{{b - n - 1} \over {\alpha(a+b)}}\bigr]}{\Gamma\bigl[{1 \over \alpha}
    \bigr]}\,. \label{Iabn}
\end{align}
The first term is a normalization factor
\begin{equation}
    {\cal I}_{ab;n} (\alpha) = \frac{1}{\alpha(a+b)}
    \biggl(\frac{a + b}{b} \biggr)^{1\over\alpha}
    \biggl(\frac{a}{b}\biggr)^{-{a + n + 1\over \alpha (a+b)}}\,,
\end{equation}
and the second term in \Eq{Iabn} is the complete beta function 
$B (a_\beta, b_\beta)$
with $a_\beta = (a + n + 1)/[\alpha (a+b)]$ and $b_\beta = (b - n - 1)/[\alpha (a+b)]$.

The integral $I_{ab; n}$ can be used to compute the different moments
of the spectral functions.
Let us start by considering the integrated spectrum of \Eq{EE_zeta},
\begin{align}
    {\cal A}_{ab} (\alpha) \equiv&\, I_{ab;-1} (\alpha)
    =   \frac{1}{\alpha(a+b)}
    \biggl(\frac{a + b}{b}\biggr)^{1\over\alpha} 
    \nonumber \\  \times &\, \biggl(\frac{a}{b}\biggr)^{-{a \over \alpha (a+b)}}  \frac{\Gamma\bigl[{a \over \alpha (a+b)}\bigr] \Gamma\bigl[{b\over \alpha (a+b)}\bigr]}{\Gamma\bigl[{1\over\alpha}\bigr]}\,.
    \label{Om_I_Ei}
\end{align}
The factor ${\cal A}_{ab} (\alpha)$ that relates the
integrated spectrum and the peak spectral
value is shown in \Fig{AA_ab_alp} for a few
choices of $a$ and $b$ as a function of $\alpha$.

\begin{figure}
    \centering
    \includegraphics[width=\columnwidth]{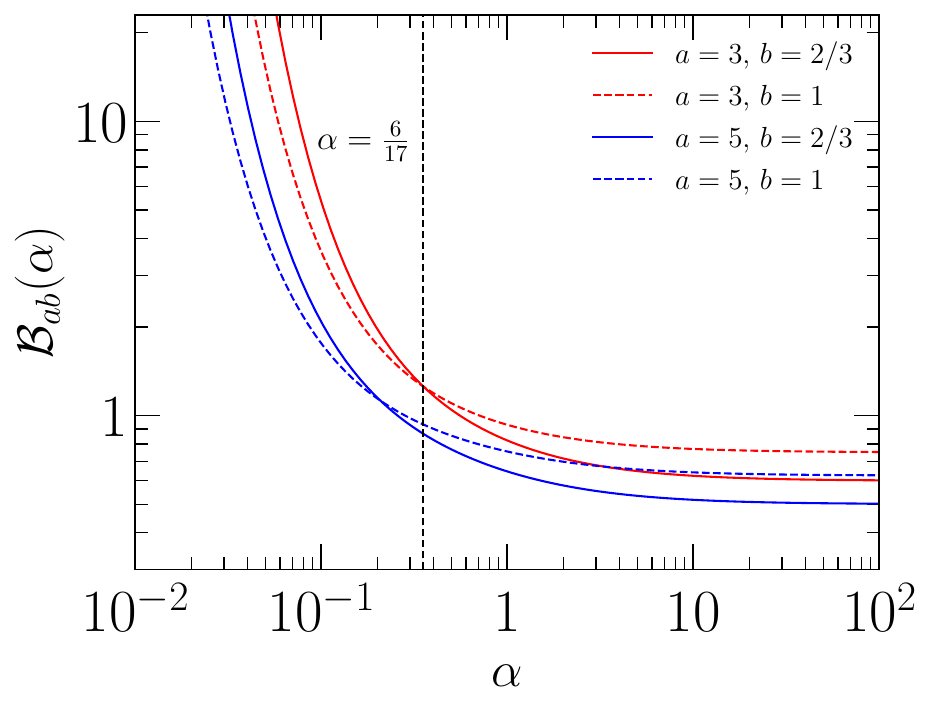}
    \caption{Normalized integral length scale $k_\ast \xi = {\cal B}_{ab} (\alpha)$
    for different values of $\alpha$ and the
    same slopes as in \Fig{AA_ab_alp}.}
    \label{BB_ab_alp}
\end{figure}

The integral length scale can be similarly computed as
\begin{align}
    k_\ast \xi&\, \equiv  {\cal B}_{ab} (\alpha)   \equiv
    \frac{1}{{\cal A}_{ab}(\alpha)}
    \displaystyle\int_{0}^\infty K^{-1}
    \zeta (K) \dd \ln K  \nonumber \\ 
    = &\, \frac{I_{ab; -2}}{{\cal A}_{ab} (\alpha)}
    = \biggl({a\over b}
    \biggr)^{
    \frac{1}{\alpha_2}}
    \frac{\Gamma\bigl[{{a - 1} \over {\alpha (a+b)}}
    \bigr]}
    {\Gamma\bigl[{{a} \over {\alpha (a+b)}}
    \bigr]} \frac{\Gamma\bigl[{{b + 1} \over {\alpha (a+b)}}
    \bigr]}{\Gamma\bigl[{{b} \over {\alpha (a+b)}}
    \bigr]}\,,
    \label{BB_ab}
\end{align}
which is shown in \Fig{BB_ab_alp}.

In general, we can define different characteristic length scales 
from the $n$-th moment of the spectrum
\begin{align} \label{generalized_Bab}
    k_\ast \xi^{(n)}  &\,\equiv   
    {\cal B}_{ab; n} (\alpha) \equiv
    \biggl[\frac{1}{{\cal A}_{ab}(\alpha)}
    \int_{0}^\infty \frac{\zeta 
    (K)}{K^{n}} \dd \ln K\biggr]^{{\frac{1}{n}}}
    \nonumber \\= &\, \biggl[\frac{I_{ab; -n - 1} (\alpha)}
    {{\cal A}_{ab} (\alpha)} \biggr]^{1 \over n}\nonumber \\ =
    &\, \!
    \biggl({a\over b}\biggr)^{1 \over \alpha (a+b)}
    \Biggl(\frac{
    \Gamma\bigl[ {a - n\over \alpha (a+b)} \bigr]}
    {\Gamma\bigl[ {a\over \alpha (a+b)} \bigr]}
    \frac{
    \Gamma\bigl[ {b + n\over \alpha (a+b)} \bigr]}
    {\Gamma\bigl[ {b \over \alpha (a+b)} \bigr]}
    \Biggr)^{{\frac{1}{n}}},
\end{align}
such that ${\cal B}_{ab} = {\cal B}_{ab; 1}$ recovers \Eq{BB_ab}.
For example,
${\cal B}_{ab; -1}$ and ${\cal B}_{ab; 1}$
relate the kinetic and magnetic helicity, respectively,
with the mean
squared value of a vector field [see \Eqq{helicities_length_scales}],
and
${\cal B}_{ab; 2}$ relates the scalar field
gradients energy density with the mean squared value of the scalar
field in \Eq{squared_xi_phi}.

At larger values of $\alpha$, the transition between
the power laws of the model in \Eq{EM_model} becomes sharper
(see \Fig{EM_alphas}).
In the limit $\alpha \rightarrow \infty$ (or $\alpha_2 \to \infty$), 
the model
becomes a piece-wise broken power law,
\begin{equation} \label{piecewise}
    \zeta (K) = \left\{\begin{array}{ll}
         K^a    & \quad  \text{for } K \leq 1\,,  \\
         K^{-b} & \quad \text{for } K > 1\,.
    \end{array}  \right.
\end{equation}
The integrated spectrum becomes
\begin{align}
    \lim_{\alpha \to \infty}
    {\cal A}_{ab} (\alpha) =
    \frac{a + b}{ab}\,,
\end{align}
which correspond to the constant values
that are found at large $\alpha$ in \Fig{AA_ab_alp}.

We can generalize this result to $I_{ab; n}$
taking the limit
$\alpha \rightarrow \infty$ in \Eq{Iabn}, by noting
$\Gamma (ax) \to 1/(ax)$ when $x \to 0$,
\begin{equation}
    \lim_{\alpha \to \infty} 
    I_{ab; n} (\alpha) = \frac{a + b}{(a + n + 1)(b - n - 1)}\,.
\end{equation}
Applying this result, the generic correlation length scale $\xi^{(n)}$
for a piece-wise broken power-law is
\begin{equation}
    \lim_{\alpha \to \infty} 
   {\cal B}_{ab;n} = \biggl[\frac{ab}{(a - n)(b + n)}\biggr]^{1 \over n}\,.
\end{equation}
For $n = 1$, $k_\ast \xi = {\cal B}_{ab; 1}
\to ab/[(a - 1)(b + 1)]$ when $\alpha \to \infty$,
which corresponds to the asymptotic values seen at large $\alpha$
in \Fig{BB_ab_alp}.

In some occasions, it is useful to rescale a function by a power of $K$.
The resulting $K^m \zeta (K)$ can be re-expressed as a new normalized
function, $\zeta_m (\tilde K)$,
\begin{equation}
    K^m \zeta (K)= \zeta_m^\ast  \,
    \zeta_m (\tilde K)\,,
    \label{rescaled_zeta_0}
\end{equation}
defined such that $\zeta_m (1) = 1$
as in \Eq{EM_model},
with the new slopes $m + a$ and $m - b$,
\begin{align} 
    \zeta_m (\tilde K) = \frac{
    (a + b)^{1\over \alpha} \tilde 
    K^{a + m}}{\bigl[b - m + (a + m) \,
    \tilde K^{\alpha_2}]^{1 \over \alpha}}\,, \label{rescaled_zeta}
\end{align}
where $\tilde K = K/K^{\ast}_{m}$ is the new normalized wave number, with
\begin{equation}
    K^\ast_m \equiv \frac{k_m^\ast}{k_\ast} = \biggl(\frac{b}{a}
    \frac{a + m}{b - m}\biggr)^{1\over\alpha (a+b)} \,, \label{Kast_m}
\end{equation}
indicating the location of the spectral peak after multiplying
the original spectrum by $K^m$.
The smoothness of the transition remains the same as that
of the
original $\zeta(K)$, characterized by $\alpha$.
The amplitude
of the function $K^m \zeta (K)$ at $K = K_m^\ast$ (i.e., $\tilde K = 1$) is
\begin{equation}
    \zeta_m^\ast = 
    \biggl(\frac{a + m}{a}\biggr)^{a + m \over \alpha (a+b)}
    \biggl(\frac{b - m}{b}
    \biggr)^{b - m \over \alpha (a+b)}\,. \label{zeta_m}
\end{equation}
Therefore, the moments of the rescaled function $\zeta_m (\tilde K)$ [see \Eq{Iabn_def}], expressed
with respect to the moments of the original $\zeta(K)$, are
\begin{align}
    I_{a+m,b-m; n} (\alpha) = &\,  \int_0^\infty \zeta_m(\tilde K) 
    \, \tilde K^{n} \dd \tilde K
    \nonumber \\ =  &\,
    \frac{I_{ab; n + m} (\alpha)}{[K_m^\ast]^{n + 1} \, 
    \zeta_m^{\ast}} \,. \label{rescaled_moments}
\end{align}
The difference between the two functions reduces as $\alpha$
increases, vanishing in the $\alpha \to \infty$ limit,
where $K_m^\ast$ and $\zeta_m^\ast$ approach 1.

\EEq{AA_alpha} yields ${\cal A}_{5, 2/3} (2)
\simeq 1.97$ and
${\cal A}_{5, 2/3} (6/17) \simeq 2.88$ for the values of $\alpha = 2$ and $6/17$ respectively
considered in ref.~\cite{RoperPol:2022iel} and
refs.~\cite{Auclair:2022jod,RoperPol:2023bqa,EPTA:2023xxk,Caprini:2024hue}.
However, these works
considered the following convention
to define the velocity or magnetic spectra,
\begin{equation}
    E_{-1} (K) \equiv
    \frac{\dd \varphi_{\rm rms}^2}{\dd K} =
    \frac{E(K)}{K} = E_\ast \frac{\zeta (K)}{K}\,.
    \label{rescaling_Zeta}
\end{equation}

Since both normalizations are related by a factor of $K$,
i.e., $E_{-1} (K) = E(K)/K$, we can apply the results
described above for $m = - 1$ to find [see \Eqss{rescaled_zeta_0}{zeta_m}]
\begin{align}
    \frac{E_{-1} (K)}{E_\ast} = \frac{\zeta (K)}{K} 
    = & \, \zeta_{-1}^\ast  \, 
    \, \zeta_{-1} (\tilde K)  \nonumber \\ 
    =  \biggl( \frac{a - 1}{a} \biggr)^{\frac{a - 1}{\alpha (a+b)}} 
    &\,
    \biggl( \frac{b + 1}{b} \biggr)^{\frac{b + 1}{\alpha (a+b)}}
    \zeta_{-1} (\tilde K)\,,
\end{align}
where
\begin{equation}
    \zeta_{-1} (\tilde K) = \frac{(a + b)^{\frac{1}{\alpha}} \tilde K^{a - 1}}
    {\bigl[(b + 1) + (a - 1) \, \tilde K^{\alpha (a+b)} \bigr]^{1 \over \alpha}}\,,
\end{equation}
and $\tilde K = K/K_{-1}^\ast$ with
\begin{equation} \label{Kstar_m1}
    K_{-1}^\ast = \biggl(\frac{b}{a} \frac{a - 1}{b + 1}
    \biggr)^{1 \over \alpha (a+b)}\,.
\end{equation}
The rescaling of $\zeta(K)$ with $1/K$ leads
to different moments of
the new normalized
spectral function $\zeta_{-1} (\tilde K)$
[see \Eq{rescaled_moments}],
\begin{align}
    I_{a - 1, b + 1; n} (\alpha) = &\, \int_0^\infty \zeta_{-1}(\tilde K)
    \,  \tilde K^n \dd \tilde K \nonumber \\
    = &\, \frac{I_{ab;n-1} (\alpha)}
    {[K_{-1}^\ast]^{n+1} \zeta_{-1}^\ast}\,.
\end{align}
In particular, the geometric factor $\tilde {\cal A}_{ab}$ found using the
convention $E_{-1} (K) = \dd \varphi_{\rm rms}^2 / \dd K$ to define the normalized spectrum
is
\begin{align} \label{tilde_A}
    \tilde {\cal A}_{ab} (\alpha) &  
    \equiv \int_0^{\infty} \zeta_{-1}(\tilde{K}) \dd \tilde{K}
    \nonumber \\
    & = I_{a-1, b+1;0} (\alpha)
    = \frac{{\cal A}_{ab} (\alpha)}{\zeta_{-1}^\ast \, K^\ast_{-1}}\,.
\end{align}
The resulting geometric factor is shown in \Fig{AA_ab_alp}
(right panel)
and allows us to recover the value of $\tilde {\cal A}_{5,2/3} (2) \simeq 2.064$
reported in ref.~\cite{RoperPol:2022iel} [see their Eq.~(9)],
which slightly differs with respect to ${\cal A}_{5,2/3}(2) \simeq 1.97$
given the different convention.
Of course, both conventions eventually yield the same results
for physical observables like the GW spectrum,
as we explicitly show when we present
it in \Eq{OmGW_envelope}.

\section{Spectrum of the anisotropic stresses}
\label{autocorr_EPi}

In \Sec{EPi_sec}, we have described the spectrum of the anisotropic stresses,
computed from the relevant terms in the stress-energy tensor, with these being
expressed as $T_{ij} (\xx) = \varphi_i (\xx) \, \varphi_j (\xx)$
where $\varphi_i = \{ a X_i,
i B_i, \partial_i \phi \}$ for velocity, magnetic, and scalar fields,
respectively [see \Eqs{Tij}{biproduct_Tij}].
Contracting \Eqq{twopoint_EPis} with $\Lambda_{ijlm}$ and ${\cal A}_{ijlm}$, we find
\begin{subequations}
\begin{align}
    \frac{1}{\bar \rho^2}
    \bra{T_{ij} (\kk) \, T_{lm}^\ast (\kk')} &\, \Lambda_{ijlm}  (\hat \kk)
    \nonumber \\ =  (2 \pi)^6 &\, \delta^3 (\kk - \kk') \frac{E_\Pi (k)}{4 \pi k^3}\,, \\ 
    \frac{1}{\bar \rho^2}\bra{T_{ij} (\kk) \, T_{lm}^\ast (\kk')}  &\, {\cal A}_{ijlm} (\hat \kk)
     \nonumber \\ =  (2 \pi)^6 &\, \delta^3 (\kk - \kk') \, i
     \, \frac{H_\Pi(k)}{4 \pi k^3}\,,
\end{align}
\end{subequations}
where we omit the arguments $\tau_1$ and $\tau_2$ for brevity
as the results of this
section are not affected by the times at which they are evaluated.
We have used the orthogonality between $\Lambda_{ijlm}$ and ${\cal A}_{ijlm}$,
such that $\Lambda_{ijlm} {\cal A}_{ijlm} = 0$,
and the norm
of each tensor,
${\cal A}_{ijlm} {\cal A}_{ijlm} = \Lambda_{ijlm} \Lambda_{ijlm} = 2$.
We have also used the following properties to relate the two-point
correlation function of $\Pi_{ij} = a^{-2} \Lambda_{ijlm} T_{lm}/\bar \rho$
with that of $T_{ij}$,
\begin{subequations}
\begin{align}
    \Lambda_{ijlm} \Lambda_{ijab} \Lambda_{lmcd} = &\, \Lambda_{abcd}\,,
    \\ 
    {\cal A}_{ijlm} \Lambda_{ijab} \Lambda_{lmcd} = &\, {\cal A}_{abcd}\,.
\end{align}
\end{subequations}

The two-point correlation function of the stresses $T_{ij}$ corresponding to the
source $\varphi_i$ is given
in \Eq{Tij_Fourier},
\begin{align}
    \bra{T_{ij} (\kk) \, T_{lm}^\ast (\kk')}  &\, = 
    \int\frac{\dd^3 \pp}{(2 \pi)^3}  \int \frac{\dd^3 \qq}{(2 \pi)^3}
     \nonumber \\ \times &\,
     \bra{\varphi_i (\tilde \pp) \, \varphi_j (\pp) \, \varphi_l^{\ast} (\tilde \qq)
    \, \varphi_m^{\ast} (\qq)}\,, \label{Fourier_Tij2}
\end{align}
where $\tilde \pp = \kk - \pp$ and $\tilde \qq = \kk' - \qq$.
Under the assumption that the sourcing fields $\varphi_i$
are statistically Gaussian,
Isserlis' theorem is applied to decompose the 
four-point correlation function in \Eq{Fourier_Tij2}
into a superposition of
products of two-point correlation functions.
Introducing the two-point correlation
of the source, ${\cal F}_{ij}^\varphi (\kk)$,
the resulting spectra
of the anisotropic stresses for a source $\varphi = \{ X, B, \phi \}$
are
\begin{subequations}
\label{Epi_Hpi_xy}
\begin{align}
    a^4 E_\Pi^\varphi (k) = \frac{k^3}{2 \pi \bar \rho^2}
    \int  \frac{\dd^3 \pp}{p^3 \tilde p^3} &\,  \Lambda_{ijlm} (\hat \kk)
    \nonumber \\ &\, \times 
    {\cal F}_{il}^\varphi (\pp) \, {\cal F}_{jm}^\varphi (\tilde \pp)  \,
    \,,
    \label{EPi_xy} \\ 
    a^4 H_\Pi^\varphi (k) =  \frac{k^3}{2 \pi \bar \rho^2}
    \int \frac{\dd^3 \pp}{p^3 \tilde p^3} &\, {\cal A}_{ijlm} (\hat \kk)
    \nonumber \\ &\,  \! \times 
    {\cal F}_{il}^\varphi (\pp) \, {\cal F}_{jm}^\varphi (\tilde \pp)\,
    \,,
    \label{HPi_xy} 
\end{align}
\end{subequations}
where we have omitted the homogeneous contribution proportional to $\delta^3(\kk)$,
and used the symmetries of the $\Lambda_{ijlm}$ and ${\cal A}_{ijlm}$
tensors under $j \leftrightarrow m$ and $i \leftrightarrow l$
transformations to combine the two remaining terms from Isserlis'
decomposition.

The tensor structure of the sourcing field $\varphi$
can in general be decomposed into a longitudinal,
a normal, and a helical component for any statistically homogeneous
and isotropic field that violates mirror symmetry [see \Eqq{two_point_Ev}],
\begin{align}
    {\cal F}_{ij}^\varphi (k) = \half P_{ij} (\hat \kk) E_\N^\varphi (k) &\,  +
    \hat k_i \hat k_j E_\L^\varphi (k) \nonumber \\ &\,
    + \tfrac{i}{2} \varepsilon_{ijl} \hat k_l H_\varphi (k)\,.
    \label{Fij_gen}
\end{align}
For velocity fields, we consider all the components while magnetic fields
are divergence-free and, hence, their longitudinal component
always vanishes, $E_\L^B (k) = 0$.
The gradients of the scalar field can only present a longitudinal component,
$E_\L^\phi (k) = k^2 E_\phi (k)$, so both $E_\N^\phi (k) = H_\phi (k) = 0$.

Introducing the general decomposition of \Eq{Fij_gen}
into the computation of the symmetric anisotropic stresses spectrum $E_\Pi (k)$
of a single source
[see \Eq{EPi_xy}], we find that all cross-terms between N or L components
and H vanish, so only NN (vortical), LL (compressional), NL (mixed), and
HH (helical) products survive, yielding the following decomposition
of $E_\Pi (k)$,
\begin{align}
    E_\Pi(k) = 
    E_\Pi^{\rm vort} (k)  &\, + E_\Pi^{\rm comp} (k)
    \nonumber \\ &\, + E_\Pi^{\rm mix} (k) + E_\Pi^{\rm hel} (k)\,. \label{decomp_Epi}
\end{align}
Hence,
the vortical (vort) and compressional (comp)
components are found when either $E_\N (k)$ or $E_\L (k)$ are
non-zero respectively.
The mixed (mix) component is found when both of them 
are non-zero, and
the helical (hel) one when $H (k) \neq 0$, which also requires
$E_\N (k) \neq 0$ due to the realizability condition $|H(k)| \leq E_\N (k)$.

Similarly, introducing the general decomposition of \Eq{Fij_gen}
into the computation of the antisymmetric $H_\Pi (k)$ [see \Eq{HPi_xy}],
only cross-terms of H with N or L do not vanish, yielding
the following split into vortical and compressional components,
\begin{equation}
    H_\Pi (k) = H_\Pi^{\rm vort} (k) + H_\Pi^{\rm comp} (k)\,.
\end{equation}
They are non-zero only when $H (k) \neq 0$, which also implies $E_\N(k) \neq 0$.
When $E_\L (k) \neq 0$, then we also find
the compressional component.

Due to statistical
isotropy, we convert the integrals over
$\dd^3 \pp$ in \Eqq{Epi_Hpi_xy} into integrals over $2 \pi \, p^2 \dd p
\dd z$, where $z = \hat \kk \cdot \pp$.
Then, since $\tilde p = \sqrt{p^2 + k^2 - 2p kz}$, we can convert the integral
over $z$ as an integral over $\tilde p$ using 
$\tilde p\dd \tilde p = - k p \dd z$
and noting that the integration limits $z \in [-1, 1]$ correspond to
$\tilde p \in [k + p, |k - p|]$.
This allows us to preserve the symmetry over $p$, $\tilde p$ in the integration.
After projecting the two-point correlators with $\Lambda_{ijlm}$
and ${\cal A}_{ijlm}$, respectively, we find the following
components of $E_\Pi (k)$ and $H_\Pi (k)$,
\begin{subequations}
\begin{widetext}
\begin{align}
    a^4 E_\Pi^{\rm vort}  (k)  = &\, \frac{k^2}{8 \bar \rho^2}
    \int_0^\infty \frac{\dd p}{p^2} E_\N (p)
    \int_{|k - p|}^{k + p} \frac{\dd \tilde p}{\tilde p^2} E_\N (\tilde p)
    (1 + z^2) (1 + \tilde z^2)
    \,,
    \label{EPi_vort} \\
    a^4 E_\Pi^{\rm comp}  (k)   = &\,  \frac{k^2}{2 \bar \rho^2}
    \int_0^\infty \frac{\dd p}{p^2} E_\L (p)
    \int_{|k - p|}^{k + p} \frac{\dd \tilde p}{\tilde p^2} E_\L (\tilde p)
    (1 - z^2) (1 - \tilde z^2)\,,
    \label{EPi_comp} \\ 
    a^4 E_\Pi^{\rm mix} (k) = &\,  \frac{k^2}{2 \bar \rho^2}
    \int_0^\infty \frac{\dd p}{p^2} E_\L (p)
    \int_{|k - p|}^{k + p} \frac{\dd \tilde p}{\tilde p^2} E_\N (\tilde p)
    \bigl(1 - z^2 \tilde z^2 \bigr)\,,
    \label{EPi_mix} \\
    a^4 E_\Pi^{\rm hel} (k) = &\,  \frac{k^2}{2 \bar \rho^2}
    \int_0^\infty \frac{\dd p}{p^2} H (p)
    \int_{|k - p|}^{k + p} \frac{\dd \tilde p}{\tilde p^2} H (\tilde p)
    \, z \tilde z \,, \label{EPi_hel} \\ 
    a^4 H_\Pi^{\rm vort}  (k)  = &\,  \frac{k^2}{4 \bar \rho^2}
    \int_0^\infty \frac{\dd p}{p^2} E_\N (p)
    \int_{|k - p|}^{k + p} \frac{\dd \tilde p}{\tilde p^2} H (\tilde p)
    \, z\, (1 + \tilde z^2)
    \,,
    \label{HPi_vort} \\
    a^4 H_\Pi^{\rm comp}  (k)   = &\, \frac{k^2}{2 \bar \rho^2}
    \int_0^\infty \frac{\dd p}{p^2} E_\L (p)
    \int_{|k - p|}^{k + p} \frac{\dd \tilde p}{\tilde p^2} H (\tilde p)
    \, z\, (1 - \tilde z^2)\,,
    \label{HPi_comp} 
\end{align}
\end{widetext}
\end{subequations}
where $z = \hat \kk \cdot \hat \pp$
and $\tilde z = \hat \kk \cdot \hat{\tilde \pp}$.
Explicitly in terms of $k$, $p$, and $\tilde p$, these cosines are
\begin{align}
    z = \frac{k^2 + p^2 - \tilde p^2}{2 k p} \,, \quad 
    \tilde z = \frac{k^2 + \tilde p^2 - p^2}{2 k \tilde p} \,.
\end{align}
We note that the terms corresponding to a product of different
spectral components, like $E_\Pi^{\rm mix}$, and $H_\Pi^j$ with $j = $ 
\{vort, comp\}, are not fully symmetric as each function of the product
is either evaluated at $p$ or at $\tilde p$.
However, since the integral is symmetric in $p \leftrightarrow \tilde p$,
we have combined the two products into a single integral.

We consider the normalization of the spectral components
as in \Eqq{norm_generic},
\begin{gather}
    E_i^\varphi (K) = \frac{\varphi_{\rm rms}^2}{{\cal A}_{ab} (\alpha)} \zeta_i^\varphi (K) \,,
\end{gather}
with $i =$ L, N, H, and $K = k/k_\ast$, and where we have defined the components $\zeta_i^{\varphi}(K)$ as
\begin{subequations}
\label{zeta_varphi_defs}
\begin{align}
    \zeta_\L^u(K) = &\,  \, q(K) \, \zeta_u (K)\,, 
    \label{def_comp_zeta}\\ 
    \zeta_\N^u(K) = &\, \, [1 - q (K)]
    \zeta_u (K)
    \,, \\ 
    \zeta_{\rm H}^u (K) = &\,
    \varepsilon_u (K) [1 - q (K)] \, 
     \zeta_u (K) \,.
\end{align}
\end{subequations}
Then, the normalized stress spectra $\zeta_\Pi (K)$ are
defined such that $|\zeta_\Pi (0)| = 1$ \cite{RoperPol:2022iel,RoperPol:2023dzg}
[see also \Eq{EPi_norms}],
\begin{align}
    \label{EPi_app}
    E_\Pi^{\varphi, j} (K) = &\, 
    \frac{\tilde \varphi_{\rm rms}^4}{\bar \rho^2} \frac{1}
    {{\cal A}_{ab}^2} \, K^3 \, {\cal C}_j^\varphi \, \zeta_\Pi^{\varphi, j} (K) \,,
\end{align}
where $j = $ \{vort, comp, mix, hel\}
and $\tilde \varphi_{\rm rms} = \varphi_{\rm rms}/a$.
The geometrical factor ${\cal C}_j$
is chosen
to ensure that $|\zeta_\Pi^j (0)| = 1$ 
\cite{RoperPol:2022iel,RoperPol:2023dzg}.
Similarly, the normalized spectrum
of the chiral stresses, $\zeta_\Pi^{\H, j}
(K)$ [see \Eq{normalized_HPi_i}],
are defined such that
\begin{align} \label{normalization_H_comps}
    H_\Pi^{\varphi, j} (K) = 
    \frac{\tilde \varphi_{\rm rms}^4}{\bar \rho^2}
    \frac{1}{{\cal A}_{ab}^2} K^4 \, {\cal C}_j^{\varphi, \H} \zeta_\Pi^{\varphi, \H, j} (K)\,,
\end{align}
with $j =$ \{vort, comp\}.
Note that the components of the antisymmetric (chiral)
stress spectrum are found to be proportional
to $K^4$ in the $K \to 0$ limit, compatible with the expected
causality condition \cite{Durrer:2003ja,Caprini:2009fx}.

All the symmetric and antisymmetric normalized 
components of the anisotropic
stress spectrum are
\begin{subequations}
\label{normalized_zetaPis}
\begin{widetext}
\begin{align}
    [{\cal C}\, \zeta_\Pi (K)]^{\rm vort}  = &\, {1 \over 8K}
    \int \frac{\dd P}{P^2} \zeta_\N (P)
    \int \frac{\dd \tilde P}
    {\tilde P^2}  \zeta_\N (\tilde P)
    (1 + z^2) (1 + \tilde z^2)\,,
    \label{zetaPi_vort} \\
    [{\cal C}\, \zeta_\Pi (K)]^{\rm comp}   = &\,
    \frac{1}{2K}
    \int \frac{\dd P}{P^2} \zeta_\L (P)
    \int \frac{\dd \tilde P}
    {\tilde P^2} \zeta_\L (\tilde P) (1 - z^2) (1 - \tilde z^2)\,,
    \label{zetaPi_comp} \\ 
    [{\cal C}\, \zeta_\Pi (K)]^{\rm mix} = &\, \frac{1}{2K}
    \int \frac{\dd P}{P^2} \zeta_\L (P)
    \int \frac{\dd \tilde P}
    {\tilde P^2} \zeta_\N (\tilde P)
    (1 - z^2 \tilde z^2)
    \,, \label{zetaPi_mix} \\
    [{\cal C}\, \zeta_\Pi (K) ]^{\rm hel} = &\, {1 \over 2K}
    \int \frac{\dd P}{P^2} \zeta_\H (P)
    \int \frac{\dd \tilde P}
    {\tilde P^2} \zeta_\H (\tilde P) \,
    z \tilde z
    \,, \label{zetaPi_hel} \\
    [{\cal C} \, \zeta_\Pi (K)]_{\H}^{\rm vort} = &\, 
    \frac{1}{8K^2}
    \int \frac{\dd P}{P^2} \zeta_\N (P)
    \int \frac{\dd \tilde P}
    {\tilde P^2} \zeta_\H (\tilde P) 
    \bigl[z (1 + \tilde z^2) + \tilde z (1 + z^2) \bigr]
    \,, \label{zetaPi_Hvort} \\
     [{\cal C} \, \zeta_\Pi (K)]_{\H}^{\rm comp}  = &\, \frac{1}{4K^2}
    \int \frac{\dd P}{P^2} \zeta_\L (P)
    \int \frac{\dd \tilde P}
    {\tilde P^2} \zeta_\H (\tilde P)
    \bigl[z (1 - \tilde z^2) + \tilde z (1 - z^2) \bigr]\,, \label{zetaPi_Hcomp}
\end{align}
\end{widetext}
\end{subequations}
where we have also defined $P = p/k_\ast$ and $\tilde P = \tilde p/k_\ast$.
The normalized components $\zeta_\Pi^j (K)$ of the symmetric spectrum
are shown in \Fig{p_pi_alpha}, while the sum of the components $\zeta_\Pi^{\H, j} (K)$
of the antisymmetric spectrum is shown in \Fig{zetaPihel_vel}.

\begin{figure*}
    \centering
    \includegraphics[width=\columnwidth]{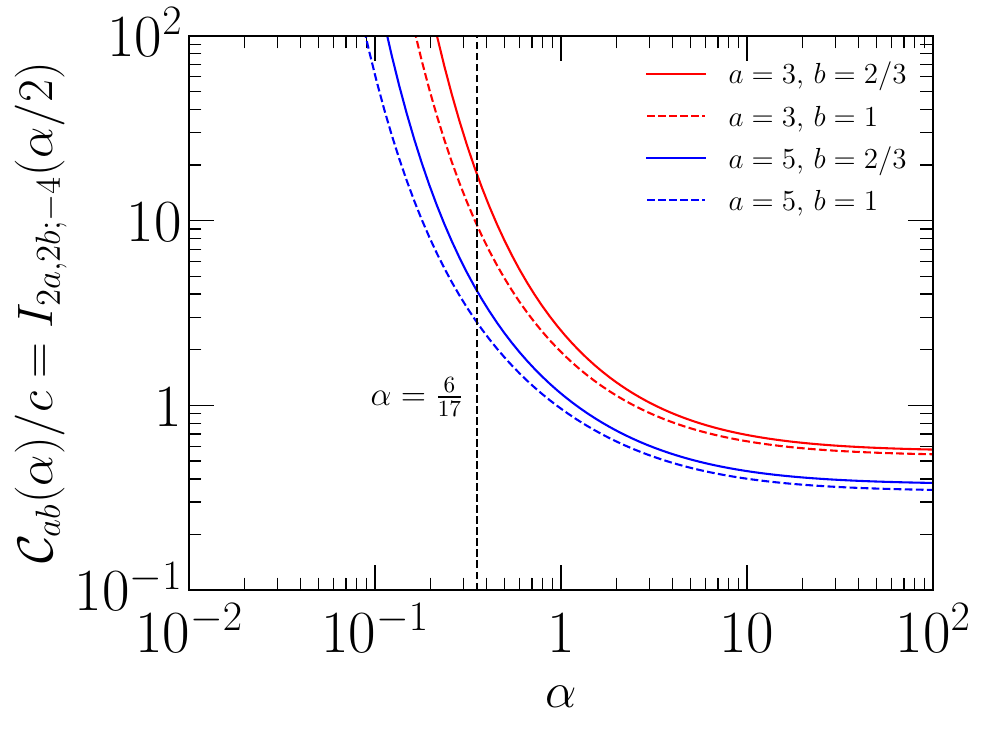}
    \includegraphics[width=\columnwidth]{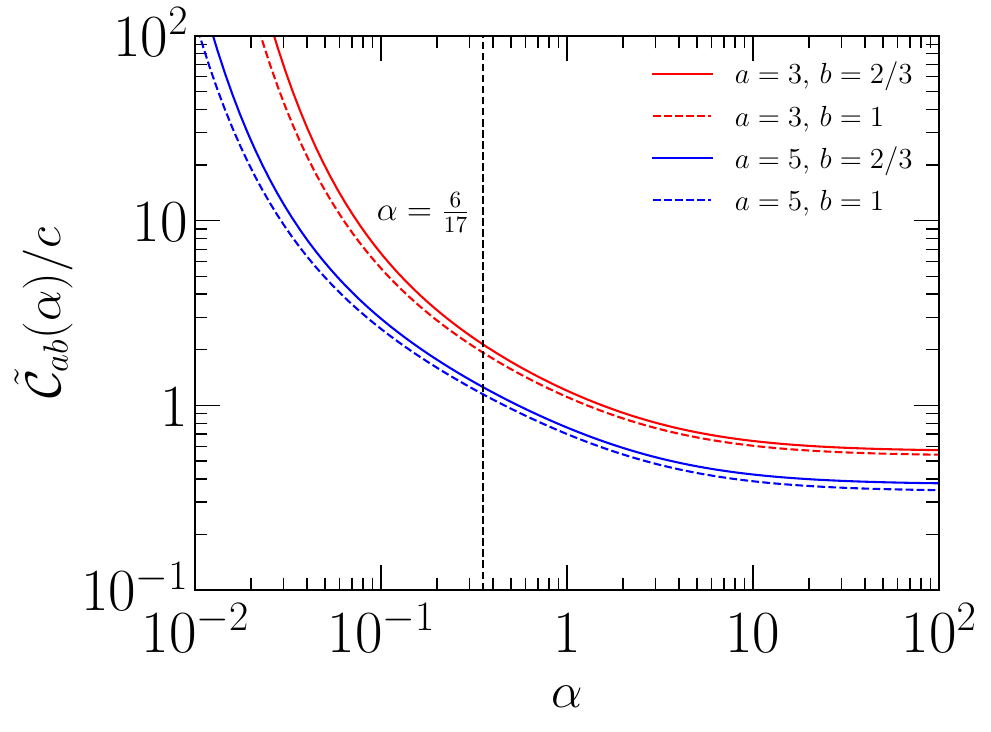}
    \caption{Geometrical factors ${\cal C}_{ab} (\alpha)/c = I_{2a, 2b; -4} (\alpha/2)$ (left panel) and $\tilde {\cal C}_{ab} (\alpha)/c \equiv I_{2(a  - 1), 2(b + 1); -2} (\alpha/2)$ (right panel)
    that determine the amplitude of the anisotropic
    stresses in the $K \to 0$ limit, respectively when using $\zeta (K)$ and $\zeta_{-1} (\tilde K)$,
    for different values of $\alpha$ and the
    same slopes as in \Fig{AA_ab_alp}.
    Notice that the ratios ${\cal C}^j_{ab}/c_j$ and $\tilde{\cal C}^j_{ab}/c_j$ do not depend on $j$, hence we can omit the index here.}
    \label{CC_ab_alp}
\end{figure*}

The geometrical factors ${\cal C}_j$ correspond to the
absolute values of the
limits of the symmetric components ${\cal C}_j \zeta_\Pi^j (K)$ when $K \to 0$.
They can be directly found taking $K \to 0$,
$\tilde P \to P$, and $\tilde z \to - z$
in \Eqs{zetaPi_vort}
{zetaPi_hel}, and transforming the integrals over $\tilde P$
into integrals over $z$,
\begin{subequations}
\label{C_pis}
\begin{align}
    {\cal C}_{\rm vort} = &\, \frac{7}{15} \int_0^\infty
    \frac{\zeta_\N^2 (P)}{P^4} \dd P\,, \label{Cvort} \\ 
    {\cal C}_{\rm comp} = &\, \frac{8}{15} 
    \int_0^\infty
    \frac{\zeta_\L^2 (P)}{P^4} \dd P\,, \label{Ccomp} \\ 
    {\cal C}_{\rm mix} = &\, \frac{4}{5} 
    \int_0^\infty \frac{\zeta_\L(P) \, \zeta_\N (P)}{P^4}
    \dd P\,, \label{Cmix} \\ 
    {\cal C}_{\rm hel} = &\, \frac{1}{3} \int_0^\infty
    \frac{\zeta_\H^2 (P)}{P^4} \dd P\,. \label{Chel}
\end{align}
\end{subequations}
All of them have a different coefficient $c_j$ [see also \Eqq{c_components}]
multiplying the integral of $\zeta_i^2 (P)/P^4$ over $P$.
Note that, unlike the other contributions,
$\zeta_\Pi^{\rm hel} (K) \to - 1$ in the $K \to 0$ limit.

Using the smoothed broken-power law of \App{appendix_sbpl}, we first note that
any power of the normalized function $\zeta^n_i (K)$ can be expressed using
\Eq{EM_model} with the mapping
$a \to an$, $b \to bn$, $\alpha \to \alpha/n$.
Then, the integrals in \Eqq{C_pis} become
\begin{align}
    \frac{{\cal C}^j_{ab} (\alpha)}{c_j}
    = &\, I_{2a, 2b; -4} \, (\alpha/2)=
    \biggl( \frac{a + b}{b}  \biggr)^{2\over\alpha}
    \biggl(\frac{a}{b}\biggr)^{-{2a - 3 \over 
    \alpha(a+b)
    }} \nonumber \\ &\, \hspace{15mm} \times
    \frac{\Gamma\bigl[{{2 a  - 3} \over {\alpha(a+b)
    }}
    \bigr] \Gamma\bigl[{{2 b + 3} \over {\alpha(a+b)
    }}\bigr]}{\alpha(a+b)
    \, \Gamma\bigl[{2 \over \alpha}
    \bigr]}\,,
    \label{Cu2}
\end{align}
with $c_j = \tfrac{7}{15} (1 - q)^2$, $\tfrac{8}{15} \, q^2$, $\tfrac{4}{5} \, q (1 - q)$, and $\tfrac{1}{3} \, \varepsilon^2 (1 - q)^2$,
for the vortical,
compressional, mixed, and helical components, respectively.
We have assumed that the degree of compressibility $q$ and the fractional helicity
$\varepsilon$ are
constants, such that all components have the same spectral shape.
The value of $I_{2a, 2b; -4} (\alpha/2)$ is shown in \Fig{CC_ab_alp}
as a function of $\alpha$
for specific choices of the slopes $a$ and $b$.
In particular, when the broken power law becomes a piece-wise
function [see \Eq{piecewise}], we find
\begin{equation}
    \lim_{\alpha \to \infty} \frac{{\cal C}_{ab}^j (\alpha)}{c_j} = 
    \frac{2 \, (a + b)}{(2 a - 3) (2 b + 3)}\,.
\end{equation}

In previous work \cite{RoperPol:2022iel,RoperPol:2023dzg,RoperPol:2023bqa,Caprini:2024hue}, the normalized spectral function of the source was considered
with a different normalization, such that their $\zeta (K)$ corresponds
to our $\zeta_{-1} (\tilde K)$ [see \Eq{rescaled_zeta}], defined
such that
\begin{align}
    \frac{\varphi_{\rm rms}^2}{E_\ast} = &\, \int_{-\infty}^\infty \zeta (K) \dd \ln K \nonumber \\ 
    =  &\, K_{-1}^\ast \zeta^*_{-1}
    \int_0^\infty \zeta_{-1} (\tilde K) \dd \tilde K\,. \label{EE_East}
\end{align}
Based on these normalizations, we can relate the moments of a power of the
normalized function $\zeta^l (K)$ with those of the rescaled function $\zeta_m^l (\tilde K)$,
\begin{align}
    I_{la, lb; lm + n} &\, (\alpha/l) =  \int_0^\infty \zeta^l (K) K^{ml + n} \dd K
    \nonumber \\ = &\, [K_m^\ast]^{n + 1} \zeta_m^{\ast l} \,
    I_{l(a +m), l(b -m); n} (\alpha/l)\,.
\end{align}
Therefore, setting $l = 2$ and $m = -1$, the integrals in \Eq{Cu2} can be expressed
as
\begin{align}
    {\cal C}_{ab}^j(\alpha)
    & =  c_j \frac{[\zeta_{-1}^*]^2}{K_{-1}^*} I_{2(a-1), 2(b+1); -2}(\alpha/2) \nonumber \\ &
    \equiv \frac{[\zeta_{-1}^*]^2}{K_{-1}^*} \tilde{\cal C}_{ab}^j(\alpha) \,,
    \label{normal_Cuu}
\end{align}
where $\tilde {\cal C}_{ab}(\alpha)$ is the geometrical factor
that one finds when using $\zeta_{-1} (\tilde K)$
as done in previous work \cite{RoperPol:2022iel,RoperPol:2023bqa,RoperPol:2023dzg,Caprini:2024hue}.
$\tilde {\cal C}_{ab}(\alpha)$
shown in \Fig{CC_ab_alp} (right panel), for comparison with ${\cal C}_{ab}(\alpha)$
(left panel).

The geometric factor ${\cal C}$ was already introduced in ref.~\cite{RoperPol:2022iel} and it becomes
${\cal C}^{\rm vort}_{5,2/3} (2) \simeq 0.34$ for a fully vortical non-helical
von K\'arm\'an spectrum with $a = 5$, $b = 2/3$, and $\alpha = 2$,
and ${\cal C}_{5,2/3}^{\rm vort} (6/17) \simeq 1.96$ when $\alpha = 6/17$ instead
\cite{Auclair:2022jod,RoperPol:2023bqa,Caprini:2024hue}.
For purely compressional motion, for example, sound waves studied in 
ref.~\cite{RoperPol:2023dzg}, with
$a = 5$, $b = 1$, and $\alpha = 2$, we find ${\cal C}_{5,1}^{\rm comp} (2) \simeq 0.34$.
Since previous work
used $\zeta_{-1} (\tilde K)$
instead of our $\zeta (K)$, defined in  \Eq{rescaled_zeta},
as the normalized spectral shape [see discussion around
\Eq{rescaling_Zeta}],
the modified prefactor
is given in \Eq{normal_Cuu},
\begin{align}
    \frac{\tilde {\cal C}_{ab}^j (\alpha)}{c_j} =  \biggl(\frac{a + b}{b + 1}
    \biggr)^{2 \over \alpha}  &\, 
    \biggl(\frac{a - 1}{b + 1} \biggr)^{-{2 a - 3 \over \alpha(a+b)}}
    \nonumber \\  &\, \hspace{-5mm} \times
    \frac{\Gamma\bigl[{{2 a  - 3} \over {\alpha(a+b)}}
    \bigr] \Gamma\bigl[{{2 b + 3} \over {\alpha(a+b)
    }}\bigr]}{\alpha(a+b)
    \, \Gamma\bigl[{2 \over \alpha}
    \bigr]}\,. \label{Cuutilde_2}
\end{align}
Using this expression, we recover for vortical turbulence
$4 \, \tilde {\cal C}_{5,2/3}^{\rm vort} (2) \simeq 1.1$
reported in ref.~\cite{RoperPol:2022iel} (see their Eq.~(22)
and note that our velocity and magnetic spectra have a factor of $1/2$ when compared
to theirs, introducing a factor of $1/4$ in $\tilde {\cal C}$), and $4\, \tilde {\cal C}_{5,2/3}^{\rm vort} (6/17) \simeq 2.35$.
For comparison with previous work,
this value of $\tilde {\cal C}_{5,2/3}^{\rm vort}$ when $\alpha = 6/17$ recovers the
values $2 \, \tilde {\cal C}/\tilde {\cal A}^2 \simeq 0.085$ (see footnote 1 in ref.~\cite{Caprini:2024hue}) and $\tilde {\cal C}/(2 \pi^2 \tilde {\cal A}^2) \simeq 2 \times 10^{-3}$ (see footnote 3 in ref.~\cite{RoperPol:2023dzg}).

The geometrical factors ${\cal C}^\H_j$
of the chiral components, with $j =$ \{vort, comp\},
correspond to the $K \to 0$ limits of \Eqs{zetaPi_Hvort}{zetaPi_Hcomp}.
However, expanding $\tilde P \to P$
and $\tilde z \to - z$, i.e.,
up to zero-th order in $K$, is not enough, as the resulting integrand
exactly vanishes.
This is expected from the causality condition
for the spectrum of the anisotropic chiral stresses \cite{Durrer:2003ja,Caprini:2009fx},
and
motivated our choice to normalize $H_\Pi (K)$ in \Eq{normalization_H_comps}
with $K^4$, introducing a $1/K^2$ factor in \Eqs{zetaPi_Hvort}{zetaPi_Hcomp}.

To find the leading-order values when $K \to 0$,
we then need to expand
$\tilde z$ up to first order in $K$,
$\tilde z \to - z + \frac{K}{P} (1 - z^2)$
and keep the second term in the expansion.
We also expand $\tilde P \to P - K z$
and $\zeta(\tilde P) \to \zeta(P) - K z \, \zeta' (P)$
up to first order
in $K$.
Substituting the expansion of $\tilde z$ up
to first order in $K$, transforming
integrals \Eqs{zetaPi_Hvort}
{zetaPi_Hcomp} into integrals over $z$,
and integrating by parts,
we find
\begin{subequations}
\label{zero_K_ampl_H}
\begin{align} \label{C_vort_H}
    {\cal C}^{\rm vort}_\H =
    \frac{2}{15}  &\, \biggl|  \int_0^\infty 
    \frac{\dd P}{P^5} \zeta_\H (P) \, \zeta_\N (P)
     \nonumber \\ &\, \hspace{9mm} \times \bigl[1 + P \,
     \zeta_\N (P) \, \varepsilon' (P)\bigr]
    \biggr|\,, \\ 
    {\cal C}^{\rm comp}_\H = 
     \frac{2}{5}  &\, \biggl| \int_0^\infty 
     \frac{\dd P}{P^5}  \zeta_\H (P)  \, \zeta_\L (P)
     \nonumber \\ &\, \hspace{9mm} \times \bigl[1 
     + P \, \zeta_\L (P) \, \varepsilon_\L' (P)
     \bigr] \biggr|\,,
\end{align}
\end{subequations}
where we have defined $\varepsilon_\L (K) \equiv H(K)
/ E_\L (K) = \varepsilon(K) \bigl[1 - q(K)\bigr] / q(K)$.
Note that for constant fractional helicity and degree
of compressibility, $\varepsilon'(P) = \varepsilon'_\L (P) = 0$
and the term in brackets for both terms reduces to one.

\begin{figure}
    \centering
    \includegraphics[width=\columnwidth]{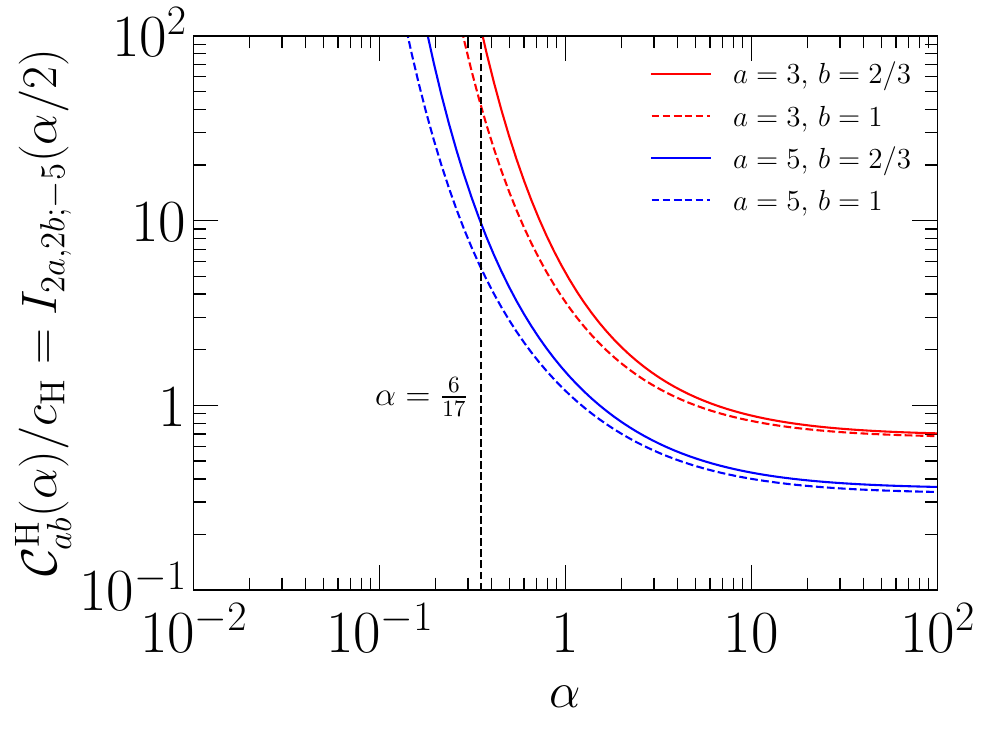}
    \caption{Geometrical factors ${\cal C}_{ab}^\H
    (\alpha)/c = I_{2a, 2b; -5} (\alpha/2)$
    that determine the amplitude of the helical anisotropic
    stresses in the $K \to 0$ limit
    for different values of $\alpha$ and the
    same slopes as in \Fig{AA_ab_alp}.
    Notice that the ratio ${\cal C}^{j, \H}_{ab}/c_j^\H$
    does not depend on $j$, hence we can omit the index here.}
    \label{CC_H_ab_alp}
\end{figure}

As before, assuming that $\zeta_i (P)$ is given by the smoothed
broken power law of \App{appendix_sbpl}, and taking constant $q$ and $\varepsilon$, we can express the
geometrical factors ${\cal C}_j^\H$ using \Eq{Iabn},
\begin{align} \label{CH_0_app}
    \frac{{\cal C}^{j, \H}_{ab} (\alpha)}
    {c_j^\H} = &\, I_{2a, 2b; -5} (\alpha/2)
    = 
    \biggl(\frac{a +b}{b} \biggr)^{2 \over \alpha}
    \biggl(\frac{a}{b}\biggr)^{-\frac{2a - 4}{\alpha(a + b)}}
    \nonumber \\ &\, \hspace{14mm} \times 
    \frac{\Gamma\bigl[\frac{2a - 4}{\alpha (a + b)}
    \bigr] \Gamma\bigl[\frac{2b + 4}{\alpha (a + b)}
    \bigr]}{\alpha (a + b) \, \Gamma\bigl[\frac{2}{\alpha}\bigr]}\,,
\end{align}
with $c_j^\H = \tfrac{2}{15}(1 - q)^2 |\varepsilon|$, $\tfrac{2}{5}
q (1 - q) \, |\varepsilon|$ for vortical and compressional components,
respectively.
The value of $I_{2a, 2b; -5} (\alpha/2)$ is 
shown in \Fig{CC_H_ab_alp}.
In the $\alpha \to \infty$ limit,
\begin{equation}
    \lim_{\alpha \to \infty} \frac{{\cal C}_{ab}^{j, \H} (\alpha)}{c_j^\H} = \frac{2\, (a + b)}
    {(2a - 4) (2 b + 4)}\,.
\end{equation}

For comparison, let us now
consider the spectrum of the
unprojected stresses for each source,
$E_T^\varphi (k)$, obtained from the trace
of ${\cal F}_{ij, lm}^{T} (\kk)$,
\begin{equation}
    \bra{T_{ij} (\kk) \,
    T_{ij}^\ast (\kk')} = (2 \pi)^6 \delta^3 (\kk -
    \kk') \frac{E_T (k)}{4 \pi k^3}\,.
\end{equation}
Introducing Isserlis' decomposition
into \Eq{Fourier_Tij2}, we find
(see ref.~\cite{Brandenburg:2018ptt}
for the case of helical magnetic fields)
\begin{align} \label{ET_Fourier}
    E_T^\varphi (k) = \frac{k^3}{4 \pi}
    \Biggl[ \int \frac{\dd^3 \pp}{p^3 \tilde p^3}
    \bigl[&\, {\cal F}_{ii}^\varphi (\pp) \, {\cal F}_{jj}^\varphi
    (\tilde \pp) \nonumber \\  &\, +
    {\cal F}_{ij}^\varphi (\pp) \,
    {\cal F}_{ji}^\varphi (\tilde \pp) \bigr] \nonumber \\
    + \, \frac{1}{3} \delta^3(\kk) \biggl( 
    &\, \int  
    \frac{\dd^3 \pp}{p^3}   {\cal F}_{ii}^\varphi (\pp)
    \biggr)^2 \Biggr]\,.
\end{align}
The decomposition of $E_T^\varphi (k)$
can be split into the same
components as $E_\Pi^\varphi (k)$
[see 
\Eq{decomp_Epi}],
\begin{align}
\label{ET_components}
    E_T^i (k) &= k^2 \int_0^\infty 
    \frac{\dd p}{p^2} \int_{|k - p|}^{k + p}
    \frac{\dd \tilde p}{\tilde p^2} \nonumber \\
    &\, \times
    E_j (p) \, E_l (\tilde p)
    \, {\cal K}_i (\mu)\,, \nonumber \\
    & \hspace{-1cm}
    \, + \, \frac{4 \pi}{3} \, c_i \, k^3 \delta^3 (\kk)
    \int_0^\infty \frac{\dd p}{p} \, E_j (p) \int_0^\infty \frac{\dd \tilde p}{\tilde p} \, E_l (\tilde p)\,,
\end{align}
where $j = l = \N$ for $i =$ vort, $j = l = \L$
for $i =$ comp, $j = \N$ and $l = \L$
for $i =$ mix, and $j = l = \H$ for $i =$ hel.
The last term of \Eq{ET_components} is the contribution from the homogeneous ($\kk = 0$) mode.
The coefficients are $c_i = 1, 1, 2, 0$ for $i =$ vort, comp, mix, hel, respectively; the factor 2 in the
mixed component accounts for the two orderings of N and L, which are combined in a single integral, and the helical component does not contribute due to antisymmetry. 
The kernels ${\cal K}_i(\mu)$ are
\begin{subequations}
\label{kernels_ET}
\begin{align}
    {\cal K}_{\rm vort} (\mu) = &\, \frac{1}{8}
    (5 + \mu^2) \,,\\
    {\cal K}_{\rm comp} (\mu) = &\, \frac{1}{2}
    (1 + \mu^2)\,,\\
    {\cal K}_{\rm mix} (\mu) = &\, \frac{1}{2}
    (3 - \mu^2)\,, \\
    {\cal K}_{\rm hel} (\mu) = &\, \frac{1}{4}\mu\,,
\end{align}
\end{subequations}
where
$\mu = \hat \pp \cdot \hat{\tilde \pp}$,
explicitly in terms of $p$, $\tilde p$,
and $k$, is
\begin{equation}
    \mu = \frac{k^2 - p^2 - \tilde p^2}{2 p \tilde p}\,.
\end{equation}

Again introducing the same normalization as in
\Eq{EPi_app}, we obtain
\begin{equation}
    E_T^{\varphi, j} (K) =
    \frac{\tilde \varphi_{\rm rms}^4}{\bar \rho^2} \frac{1}
    {{\cal A}_{ab}^2} \, K^3 \, {\cal C}_T^{ \varphi, j} \,
    \zeta_T^{\varphi, j} (K) \,.
\end{equation}
The geometrical factors are computed taking
the $k \to 0$ limit of the expressions in \Eqq{ET_components},
such that $\tilde p \to p$ and $\mu \to - 1$,
\begin{subequations}
\label{CT_components}
    \begin{align}
        {\cal C}_T^{\rm vort} = &\,
        \frac{3}{2} \int_0^\infty 
        \frac{ \zeta_\N^2 (P)}{P^4} \dd P\,, \\ 
        {\cal C}_T^{\rm comp} = &\,
        2 \int_0^\infty 
        \frac{\zeta_\L^2 (P)}{P^4} \dd P \,, \\ 
        {\cal C}_T^{\rm mix} = &\,
        2 \int_0^\infty 
        \frac{\zeta_\L (P) \, \zeta_\N (P)}{P^4} \dd P \,, \\ 
        {\cal C}_T^{\rm hel} = &\,
        \frac{1}{2} \int_0^\infty 
        \frac{\zeta_\H^2 (P) }{P^4} \dd P \,.
    \end{align}
\end{subequations}

The average squared value of the stress-energy tensor
of a source $\varphi$
can be directly computed from the two-point
correlator function in physical
space evaluated at $\rr = 0$,
\begin{align}
    a^4 \, \bar \rho_T^\varphi \equiv  &\,
    \bra{T_{ij}^\varphi (\xx)\, T_{ij}^\varphi (\xx)}
    =\bra{\varphi^4 (\xx)} \nonumber \\ 
     &\, \hspace{15mm} = \int_{-\infty}^\infty
    E_T^\varphi (K) \dd \ln K \,.
\label{rho_T_forms}
\end{align}
Decomposing the left-hand side using Isserlis'
theorem,
and noting that, due to isotropy, $\bra{\varphi_i \varphi_j}
= \frac{1}{3} \delta_{ij} \varphi_{\rm rms}^2$,
we find the mean squared value of $T_{ij}^\varphi$,
\begin{equation} \label{mean_rhoT}
    \bar \rho_T^\varphi \equiv
    a^{-4} \bra{T_{ij}^\varphi (\xx)\, T_{ij}^\varphi (\xx)}
    = \frac{5}{3} \tilde \varphi_{\rm rms}^4\,.
\end{equation}
This result is also found integrating 
$E_T(k)$, as the sum over its components in \Eq{ET_components}, over $\ln k$, as in \Eq{rho_T_forms}, and using the change of variables $k \dd k \to p \, \tilde{p} \dd \mu$,
\begin{align}
    a^4 \, \bar \rho_T^\varphi &=
    \int E_T^\varphi (k)
    \dd \ln k 
    = \int_0^\infty \frac{\dd p}{p}
    \int_0^\infty \frac{\dd \tilde p}{\tilde p}
    \nonumber \\ &\times
    \int_{-1}^1 \dd  \mu 
    \sum_i  E_j (p)\, E_l (\tilde p)
    \, {\cal K}_i (\mu) \nonumber \\
    &  \, + \, \frac{1}{3} \biggl( \int_0^\infty \frac{\dd p}{p} \bigl[ E_\N (p) + E_\L (p) \bigr] \biggr)^2\,,
\end{align}
where $j$ and $l$ are the corresponding
components of the source spectrum contributing
to the stresses component $i$, as indicated below
\Eq{ET_components}.
Using the kernels in \Eqq{kernels_ET}, we find that
the contribution to $\rho_T^\varphi$
from the helical component vanishes, and the remaining
ones become a separable product of $p$ and $\tilde p$,
\begin{align} \label{mean_inhom_rhoT}
    \bar \rho_T^\varphi = &\, \frac{5}{3}  a^{-4}
    \biggl(\int_{-\infty}^\infty \bigl[E_\N (P) +
    E_\L (P) 
    \bigr] \dd \ln P\biggr)^2 \nonumber \\ = &\, 
    \frac{5}{3} \tilde \varphi_{\rm rms}^4 \,,
\end{align}
as found 
in \Eq{mean_rhoT}.

\section{Dynamical evolution of the Constant-in-Time kernel}
\label{dynamical_G}

In this section, we study the dynamical evolution of
the kernel $\Delta^2 (\tau, k)$ that characterizes the
GW spectrum in the Constant-in-Time (CiT)
model described in
\Sec{cit_model_sec} [see \Eq{G_constant}].

We consider the two regimes described in \Sec{cit_model} separately.
Let us start with
times $\tau \leq \tfin$, when the kernel takes the following form
\begin{align}
    \Delta (\tau, k) = &\, \cos (k \tau) \Delta \Ci(k\tau)
    \nonumber \\ &\, + \sin (k \tau) \Delta \Si(k\tau)\,,
    \label{kernel_aux}
\end{align}
where $\Delta \Ci (k \tau) = \Ci (k \tau) - \Ci (k \tau_\ast)$ and $\Delta \Si (k \tau) = \Si (k \tau) - \Si (k \tau_\ast)$ [see \Eqq{Delta_Ci_Si}].

Let us consider the asymptotic limits $k \tau \ll 1$
and $k\tau \gg 1$ during the sourcing times $\tau \leq \tfin$.
In the case when $k \tau =
k \tini + k \delta \tau \ll 1$,
which implies $k \tini \ll 1$
and $k \delta \tau \ll 1$,
the cosine and sine integral functions become
\begin{subequations}
\label{lim_Cis}
\begin{align}
    \lim_{k\tau \ll 1}
    \Ci (k \tau) = &\, \ln (k\tau) + \gamma\,,
    \label{lim_Ci}
    \\
     \lim_{k\tau\ll 1}
    \Si (k \tau) = &\, k\tau\,,
    \label{lim_Si}
\end{align}
\end{subequations}
where $\gamma$ is the Euler-Mascheroni constant.
Hence, the kernel becomes
\begin{equation}
\lim_{k\tau\ll 1}
    \Delta^2 (\tau, k) = \ln^2 (1 + \delta \tau/\tini)\,.
\label{log_cit_small_args}
\end{equation}
Note that this result is valid only for a radiation-dominated Universe \cite{RoperPol:2022iel}.
This limit of the kernel represents the regime {\em i)} listed in \Sec{cit_model},
when the GW modes are growing in time
(see right panel of \Fig{G2_kt}).
For short durations, $\delta \tau/\tini \ll 1$, we
recover the quadratic growth valid when the expansion of
the Universe is neglected,
\begin{equation}
\lim_{k\tau\ll 1}
    \Delta^2_{\rm flat} (\tau, k) = (\delta \tau/\tini)^2\,.
\end{equation}

In the opposite
limit, $k \tau \gg 1$,
\begin{subequations}
\begin{align}
\label{delta_ci_large_delt}
    \lim_{k  \tau \gg 1}
    \Delta \Ci(k\tau) = &\,
    \frac{\sin k\tau}{k  \tau} -\Ci (k \tini)\,, \\
\label{delta_si_large_delt}
    \lim_{k \tau \gg 1}
    \Delta \Si(k \tau)  
    = &\,
    \frac{\pi}{2} - \frac{\cos k \tau}{k  \tau} - \Si (k\tini) \,,
\end{align}
\end{subequations}
where we have used $\Si(x) \simeq \frac{\pi}{2} - \frac{\cos x}{x}$ and $\Ci(x) \simeq \frac{\sin x}{x}$, valid for large $x$.
We can now further consider two limits,
corresponding to $k\tini \gg 1$ and $k \tini \ll 1$, which describe modes inside and outside
the Hubble scale, respectively.
When $k \tini \gg 1$, we find
\begin{subequations}
\begin{align}
    \lim_{\tau, \tini \gg 1/k} \Delta \Ci(k \tau) =&\, \frac{\sin k\tau}{k \tau} - \frac{\sin k\tini}{k \tini}\,, \\
    \lim_{\tau, \tini \gg 1/k} \Delta \Si(k \tau) =&\, -\frac{\cos k \tau}
    {k \tau} + \frac{\cos k\tini}{k\tini} \,.
\end{align}
\end{subequations}
Hence, the kernel in this limit becomes,
\begin{equation}
    \lim_{\tau, \tini \gg 1/k} \Delta^2 (\tau, k) = \frac{\sin^2 (k\delta \tau)}{(k \tini)^2}\,,
\label{delta_cit_inf_limits}
\end{equation}
which is an oscillatory function with an envelope amplitude $1/k^2$,
describing the regime {\em ii)} of \Sec{cit_model}.
Since it describes sub-horizon
modes $k \tini \gg 1$, this is the
limit
of the kernel that applies in
flat space-time [see left panel of \Fig{G2_kt}
and \Eq{Delta_flat_1}].

However, when we
consider super-horizon modes
at the time of GW production, $k \tau_\ast \ll 1$,
using \Eq{lim_Cis},
we recover the $k \tau \gg 1$ limit found in
ref.~\cite{RoperPol:2022iel},
\begin{equation}
    \lim_{k \delta \tau \gg 1} \Delta^2(\tau, k) = \sin^2 (k \delta\tau)
    \ln^2 \bigl[1 + (k\tini)^{-1}\bigr]\,.
\end{equation}
We note that this expression also applies in the $k \tau_\ast \gg 1$
limit since $\ln [1 + (k \tau_\ast)^{-1}]$ reduces to $1/(k\tau_\ast)$
in \Eq{delta_cit_inf_limits}.
Actually, we find numerically that this is a good approximation 
at all $k \tau_\ast$.

Let us now consider the case when $\tau \geq \tfin$.
In this case, the kernel is
\begin{align}
    \Delta (\tau, k) = &\, \cos (k \tau) \Delta \Ci(k\tfin) \nonumber \\ &\, +
    \sin (k \tau)
    \Delta \Si (k \tfin)\,,
\end{align}
and we see that a new timescale, $\tfin$, appears.
Hence, let us again consider the limit $k \tini \leq
k \tfin \leq k \tau \ll 1$.
In this limit, the kernel becomes
\begin{equation}
    \lim_{k\tau \ll 1} \Delta^2 (\tau, k) = \ln^2 (1 + 
    \delta \tfin/\tini)\,,
\end{equation}
where $\delta \tfin = \tfin - \tini$.

In the limit $k \tfin \geq k \tini \gg 1$, the function
becomes
\begin{align}
    \lim_{k\tini \gg 1} 
    \Delta^2  (\tau, k) &\, \nonumber \\ =   \frac{1}{(k \tini)^2}
    \biggl[ \cos &\, k \tau \biggl(\frac{\tini}{\tfin} \sin k\tfin -
    \sin k\tini \biggr) \nonumber \\ - \sin   k & \tau \biggl(\frac{\tini}{\tfin} \cos k \tfin - \cos k\tini \biggr) \biggr]^2.
\end{align}
This shows that the $k^{-2}$ branch is multiplied by
the oscillatory term in brackets.
The amplitude of these oscillations depends on the
specific values of $\tau$, $\tini$, and $\tfin$.
However, in general, the amplitude is bounded by the
following function
\begin{equation}
    \lim_{k\tini \gg 1} \Delta^2 (\tau, k)
    \leq \frac{{\cal F}^2 (\tfin/\tini)}{(k\tini)^2}
    \,,
\label{upper_bound_cit}
\end{equation}
where we have defined
\begin{equation}
    {\cal F} (x) = 1 + x^{-1} \geq 1\,.
    \label{F_factor}
\end{equation}
This function is always larger than one and it shows that
for $\tau > \tfin$, the function $\Delta^2 (\tau, k)$ presents
oscillations with an amplitude larger than
that at earlier times, when $\tau \leq \tfin$, in the
limit $k \tau \geq k \tau_\ast \gg 1$.
As argued in \Sec{cit_model}, this is a consequence of the discontinuity
in the source at $\tfin$ and it is not expected to be found if the
source decays smoothly in time (see refs.~\cite{RoperPol:2022iel,Auclair:2022jod} and \Sec{coherent_model}).
As $\tfin$ increases, this
effect is less significant and in the limit $\tfin \rightarrow \infty$, we recover the case $\tau \leq \tfin$ [see \Eq{delta_cit_inf_limits}].

\section{Integration of the Green's function in the coherent-decay
case}
\label{G_m_coherent}

We present in this section some calculations that are used
when we compute the power spectrum of the
GW background produced by velocity or magnetic
fields, under the assumption of coherent unequal
time correlation of the stresses (see \Sec{beyond_cit}).

We consider here that we can omit the factor $\zeta_\Pi^{\rm rat}(\tau, k)$ appearing in \Eq{G_pq}, which means that we restrict to cases with moderate $s \lesssim 1$ (see \App{fast_q}, in particular \Eq{range_of_good_s_values} and \Fig{p_pi_ratios}).
Under this assumption, and with the additional hypothesis that $\delta \tau_{\rm dc} = \tau_\ast$, the integral in \Eq{G_pq} can be
written as
\begin{align}
    \Delta_{ps} & ( \tau, k; \tau_\ast, \tau_{\rm fin}, \delta \tau_{\rm dc} = \tau_\ast) \nonumber \\
    & \simeq  \tau_\ast^{-m} \int_{\tau_\ast}^{{\bar \tau}} \, \tau_1^{m-1} \, \cos k(\tau - \tau_1) \dd \tau_1 \nonumber \\
     & \equiv {\Delta}_{m}(\tau, k; \tau_\ast, \tau_{\rm fin})\,,
\label{new_delta_pq_def}
\end{align}
where $\bar{\tau} = \min (\tau, \tau_{\rm fin})$, we use the subindex $m=\frac{3}{2}s-p$ instead of the
subindices $ps$ [see discussion below \Eq{eq:EPi_decay}], and we have defined the function
\begin{align}
    {\Delta}_{m} (\tau, k; \tau_\ast, \tau_{\rm fin}) \equiv & \, \tau_\ast^{-m} \int_{\tau_*}^{\bar \tau}    \tau_1^{m - 1} \cos k \bigl(\tau - \tau_1 \bigr) \dd \tau_1 \,.
\label{def_delta_m}
\end{align}
Taking into account that $\cos x = \half
\bigl(e^{ix} + e^{-ix}\bigr)$, $\Delta_m$ can be expressed in the following form,
\begin{align}
    \frac{1}{2} \tau_\ast^{-m} \int_{\tau_*}^{\bar \tau} \tau_1^{m - 1} &\,
    \Bigl[e^{ik(\tau - \tau_1)}
    + e^{-ik (\tau - \tau_1)} \Bigr] \dd \tau_1 = \nonumber \\
    &\, \half \tau_\ast^{-m} e^{ik\tau} \int_{\tau_*}^{\bar \tau}
    \tau_1^{m - 1} e^{-ik\tau_1} \dd \tau_1 \nonumber \\ +&\, \half \tau_\ast^{-m} e^{-ik\tau} 
    \int_{\tau_*}^{\bar \tau} \tau_1^{m - 1} e^{ik\tau_1} \dd \tau_1 \,.
    \label{aux}
\end{align}
To study these integrals, we use
the upper incomplete gamma function $\Gamma(m, x)$,
\begin{equation}
    \Gamma(m, x) = \int_x^\infty t^{m - 1} e^{-t} \dd t \,.
\end{equation}
Using $x = \pm ikz$ and the change of variable $t = \pm ikp$,
we find
\begin{align}
    (i k)^{-m} \, \Gamma(m, ikz) = &\, 
    \int_{z}^{-i \, \infty} p^{m - 1} e^{-i k p} \dd p\,,
    \label{Gamma_ikz} \\
    (- i k)^{-m} \, \Gamma(m, - ikz) = &\, 
    \int_{z}^{+i \, \infty} p^{m - 1} e^{i k p} \dd p\,.
    \label{Gamma_mikz}
\end{align}
By substituting these integrals in \Eq{aux}, we find an
expression for $\Delta_m$ in terms of the
incomplete Gamma function
\begin{align}
    &{\Delta}_{m} (\tau, k; \tau_\ast, \tau_{\rm fin}) = - \half \tau_\ast^{-m}
    (ik)^{-m} \nonumber \\
    & \times \Biggl\{
    e^{ik\tau} 
   \Bigl[ \Gamma \bigl(m, ik\bar{\tau} \bigr) - \Gamma \bigl(m, ik\tau_\ast \bigr) \Bigr]
    \nonumber \\
    & +  e^{-ik\tau} e^{i\pi m} \Bigl[
    \Gamma \bigl(m, -ik\bar{\tau} \bigr) -\Gamma \bigl(m, -ik \tau_\ast \bigr) \Bigr] \Biggr\} \,,
\label{Gpq_coherent}
\end{align}
where $(\pm i k)^{-m} =k^{-m} e^{\mp i \pi m/2}$.

Let us now show that for $m = 0$ we recover \Eq{G_constant}.
In first place, we use the general expression \Eq{Gpq_coherent}
to get
\begin{align}
    \Delta_0 (\tau, k; \tau_\ast, \tau_{\rm fin}) = - \half \cos  &\, k\tau \Bigl[
    \Gamma\bigl(0, ikt \bigr) +
    \Gamma \bigl(0, -ik t \bigr) \Bigr]
    \Bigr|_{\tini}^{{\bar \tau}} \nonumber \\
    - \half \, i \sin k\tau \Bigl[&\, \Gamma\bigl(0, ikt \bigr) 
    -  \Gamma \bigl(0, -ik t \bigr) \Bigr] \Bigr|_{\tini}^{{\bar \tau}} \,,
    \label{aux3}
\end{align}
where the notation $f(t)|_{\tini}^\tau = f (\tau) - f(\tini)$ is used for compactness.
Using \Eqs{Gamma_ikz}{Gamma_mikz}, we find
\begin{align}
    \Gamma(0, \pm ik\tau) = &\, \int_\tau^{\mp i \infty} \frac{e^{\mp ikp}}
    {p}\dd p \nonumber \\
    = &\, \int_\tau^{\mp i \infty} \frac{\cos kp}{p} \dd p \mp i 
    \int_\tau^{\mp i\infty} \frac{\sin kp}{p} \dd p \,.
    \label{aux2}
\end{align}
Hence, it follows that
\begin{align}
    \Bigl[\Gamma \bigl(0, ik t \bigr) + & \, \Gamma \bigl(0, -ik t \bigr) \Bigr]
    \Bigr|_{\tini}^{\bar{\tau}}
    = -2 \Bigl[\Ci (k \bar{\tau}) - \Ci (k \tini) \Bigr], \label{aux5}  \\
    \Bigl[\Gamma \bigl(0, ik t \bigr) - &\, \Gamma \bigl(0, -ik t \bigr)
    \Bigr]\Bigr|_{\tini}^{\bar{\tau}}
    = 2 i \Bigl[\Si (k\bar{\tau}) - \Si (k\tini) \Bigr].
    \label{aux4}
\end{align}
Finally, substituting \Eqs{aux5}{aux4} into \Eq{aux3}, we find
\Eq{G_constant}, q.e.d.

We can now consider a few special cases of $m = {3\over2}s-p $.
In particular, if we consider negative integers $m = -n$, the incomplete
$\Gamma$ function at $\pm ik\tau$ satisfies the following recurrence relation
\begin{align}
    \Gamma  \bigl(-n,  ik\tau  \bigr)  = 
     \frac{1}{n!} &\, \biggl[  (-1)^n \Gamma \bigl(0, ik \tau 
    \bigr) \nonumber \\
    &\,  +
    e^{- ik \tau} \sum_{q = 0}^{n - 1} (-1)^q
    A_{n - q} (\tau)
    \biggr], \\
    \Gamma  \bigl(-n, -ik\tau  \bigr)  =
    \frac{1}{n!} &\, \biggl[
    (-1)^n \Gamma \bigl(0, - ik \tau 
    \bigr) \nonumber \\ + &\, (-1)^n  e^{ik \tau} \sum_{q = 0}^{n - 1}
    A_{n - q} (\tau) \biggr],
    \label{recursive_negG}
\end{align}
\begin{equation}
    \text{with} \quad A_a = \frac{\bigl(a - 1\bigr)!}{(ik\tau)^{a}}.
\end{equation}
In this case, we can substitute the values of the incomplete
$\Gamma$ into \Eq{Gpq_coherent},
\begin{align}
     {\Delta}_{-n} &\, (\tau, k; \tau_\ast, \tau_{\rm fin}) = \nonumber \\ &\,
     - \frac{1}{2} \tau_\ast^{n} \frac{i^n k^n}{n!}
     \Bigl[  {\Delta}_{-n, A}  + {\Delta}_{-n, B} \Bigr],
     \label{aux6}
\end{align}
where, for simplicity, we have split the integral in the terms
proportional to $\Gamma(0, \pm ik\tau)$ (A), 
and the rest (B), which can be computed separately.
Hence,
\begin{align}
    {\Delta}_{-n, A} (\tau, k; \tau_\ast, \tau_{\rm fin}) = &\, (-1)^n e^{ik\tau}  \, \Bigl[ \Gamma
    \bigl(0, ikt \bigr) \Bigr] \bigr|_{\tini}^{\bar{\tau}} \nonumber \\
    &\, + e^{-ik\tau} \, \Bigl[ \Gamma \bigl(0, -ik t \bigr) \Bigr] \bigr|_{\tini}^{\bar{\tau}}
    \nonumber \\
    =  \cos k \tau \Bigl[(-1)^n \, &\, \Gamma \bigl(0, ik t \bigr)
    + \Gamma \bigl(0, -ik t\bigr) \Bigr] \Bigr|_{\tini}^{\bar{\tau}} \nonumber
    \\ + i \, \sin k \tau \Bigl[(-1)^n \, &\, \Gamma \bigl(0, ik t \bigr)
    - \Gamma \bigl(0, -ik t \bigr) \Bigr] \Bigr|_{\tini}^{\bar{\tau}}.
\end{align}
Note that ${\Delta}_{-n, A}$ will
be a function of the sine and cosine integrals [see \Eqs{aux5}{aux4}],
depending on the specific values of $n$.
For different values of $n$, we have
\begin{align}
    {\Delta}_{-1, A} = &\, - 2 i \cos k \tau \Bigl[\Si (k\bar{\tau}) - \Si (k\tini)
    \Bigr] \nonumber \\ &\, + 2 i \sin k \tau \Bigl[\Ci(k\bar{\tau}) - \Ci (k\tini)
    \Bigr], \\
    {\Delta}_{-2, A} = &\, - 2 \cos k \tau \Bigl[\Ci(k\bar{\tau}) - \Ci(k\tini)
    \Bigr] \nonumber \\
    &\, -  2 \sin k \tau \Bigl[ \Si (k\bar{\tau}) - \Si (k\tini) \Bigr].
\end{align}
For other values of $n$, we note that ${\Delta}_{-n, A}$ will be equal
to ${\Delta}_{-1, A}$ if $n$ is odd and equal to ${\Delta}_{-2, A}$
if $n$ is even.
The other term that appears in \Eq{aux6} is
\begin{align}
    {\Delta}_{-n, B}  &\,  (\tau, k; \tau_\ast, \tau_{\rm fin}) =\Bigl[e^{ik(t - \tau)}
    \sum_{q = 0}^{n - 1}  A_{n-q} (t) \Bigr]\Bigr|_{\tini}^{\bar{\tau}} \nonumber \\
    &\, \hspace{8mm} + \Bigl[ e^{-ik(t - \tau)} \sum_{q = 0}^{n - 1} (-1)^q
    A_{n-q} (t) \Bigr] \Bigr|_{\tini}^{\bar{\tau}} \nonumber \\
    =  \Bigl[2 &\, \cos k (t - \tau) \sum_{q = 0}^{\bigl\lfloor \half (n - 1)
    \bigr\rfloor } A_{n - 2q} (t) \Bigr] \Bigr|_{\tini}^{\bar{\tau}} \nonumber \\
     + \Bigl[  2  i  &\, \sin k (t - \tau) \sum_{q = 0}^{\bigl\lfloor \half (n - 2)
    \bigr\rfloor}  A_{n - 2q - 1} (t) \Bigr] \Bigr|_{\tini}^{\bar{\tau}},
    \label{G_n_B}
\end{align}
where $\lfloor x \rfloor$ is the floor function.
Hence, we will have terms proportional to inverse power laws of $k\tau$ up to order $-n$.
We can now compute the terms ${\Delta}_{-n, B}$ for different
values of $n$,
\begin{align}
    {\Delta}_{-1, B} = &\, - \frac{2 i}{k} \biggl[\frac{\cos k (\tau - \bar{\tau})}{\bar{\tau}} -
    \frac{\cos k (\tau - \tini)}{\tini} \biggr], \\
    {\Delta}_{-2, B} = &\, \frac{2}{k^2} \biggl[
    \frac{\cos k (\tau - \tini)}{\tini^2} - \frac{\cos k (\tau - \bar{\tau})}{\bar{\tau}^2}  \nonumber \\
    &\, \hspace{6mm} + \frac{k}{\tini}  \sin k (\tau - \tini) - \frac{k}{\bar{\tau}}  \sin k (\tau - \bar{\tau}) \biggr]\,.
\end{align}
Finally, combining ${\Delta}_{-n, A}$ and ${\Delta}_{-n, B}$ allows us
to compute the values of the integral ${\Delta}_{-n}$ in \Eq{aux6},
\begin{align}
    {\Delta}_{-1} (\tau, k; \tau_\ast, \tau_{\rm fin}) = & \, - k \tau_\ast \cos k \tau \Bigl[\Si (k\bar{\tau}) - 
    \Si (k \tini) \Bigr] \nonumber \\
    &\, + k \tau_\ast \sin k {\tau} \Bigl[\Ci (k\bar{\tau}) - \Ci (k \tini)
    \Bigr] \nonumber \\
    &\, -  \frac{\tau_\ast}{\bar{\tau}} \cos k (\tau - \bar{\tau}) +  \cos k (\tau - \tini), \label{Gm1} \\
    {\Delta}_{-2} (\tau, k; \tau_\ast, \tau_{\rm fin}) = & \, - \frac{k^2\tau_\ast^2}{2}
    \cos k \tau \Bigl[\Ci (k\bar{\tau}) - \Ci (k \tini) \Bigr] \nonumber \\
    &\, - \frac{k^2\tau_\ast^2}{2} \sin k \tau \Bigl[\Si (k\bar{\tau}) - \Si (k \tini)
    \Bigr] \nonumber \\
    + \frac{1}{2} \bigl[ & \cos k (\tau - \tau_\ast) + k \tau_\ast \sin k (\tau - \tau_\ast)\bigr] \nonumber \\
    - \frac{1}{2} \Bigl[ & \frac{\tau_\ast^2}{\bar{\tau}^2} \cos k (\tau - \bar{\tau}) + \frac{\tau_\ast}{\bar{\tau}} k \tau_\ast \sin k (\tau - \bar{\tau})\Bigr]\,. 
    \label{Gm2} 
\end{align}
Let us now consider the cases with $m = n$, being $n$ positive
integers.
The case $m = 1$ is straightforward
\begin{align}
    {\Delta}_1 (\tau, k; \tau_\ast, \tau_{\rm fin}) = &\, \tau_\ast^{-1} \int_{\tini}^{\bar{\tau}} \cos k \bigl(\tau - \tau_1\bigr)
    \dd \tau_1 \nonumber \\ = &\, \frac{1}{k\tau_\ast} \bigl[ \sin k \bigl(\tau -\tini\bigr) - \sin k \bigl(\tau - \bar{\tau} \bigr) \bigr] \,, \label{G1_app}
\end{align}
which can also be computed from \Eq{Gpq_coherent} using
\begin{align}
    \Gamma(1, \pm ik\tau ) = &\, \pm ik \int_\tau^{\mp i \, \infty} e^{\mp 
    ik p}\dd p = e^{\mp ik \tau}.
\end{align}
Larger values of $n$ can be computed using the recursive formula
for the upper incomplete gamma,
\begin{align}
    \Gamma(n, \pm ik \tau) = &\, (n - 1) \,
    \Gamma(n - 1, \pm ik \tau) \nonumber \\
    + &\, \bigl(\pm ik \tau \bigr)^{n - 1} e^{\mp ik \tau}.
\end{align}
In particular, for $n = 2$, we have
\begin{align}
    \Gamma\bigl(2, \pm ik \tau\bigr) = &\, e^{\mp i k\tau} \Bigl(1  \pm ik \tau \Bigr),
\end{align}
and hence, from \Eq{Gpq_coherent},
\begin{align}
    {\Delta}_2 & (\tau, k; \tau_\ast, \tau_{\rm fin}) =  \half  (k\tau_\ast)^{-2} \nonumber \\
    &\times \Bigl[e^{-ik(t - \tau)}
    \bigl(1 + i k t \bigr)
    \, + e^{ ik(t - \tau)} \bigl(1 - ik t \bigr) \Bigr]
    \Bigl|_{\tini}^{\bar{\tau}} \nonumber \\
    & = \frac{1}{k^2\tau_\ast^2} \Bigl[  \cos k \bigl(\tau - \bar{\tau} \bigr)  - \cos k \bigl(\tau -  \tini \bigr) 
    \Bigr]  \nonumber \\
    & + \frac{1}{k \tau_\ast} \bigl[ \sin k \bigl(\tau - \tini \bigr) - \frac{\bar{\tau}}{{\tau}_\ast}\sin k \bigl(\tau - \bar{\tau} \bigr)  \bigr].
    \label{G2_app}
\end{align}

\subsubsection*{Asymptotic values of ${\Delta}_m(\tau, k; \tau_\ast, \tau_{\rm fin})$ when $k \rightarrow 0$}

Let us now study the values of ${\Delta}_m$ in the asymptotic limits
$k \rightarrow 0$ and $k \rightarrow \infty$.

We start by considering the case with negative integers
$m = -n$, given in \Eq{aux6} as
\begin{align}
    {\Delta}_{-n} = - \frac{1}{2} \tau_\ast^{n}
    \frac{i^n k^n}
    {n!} \bigl[&\,{\Delta}_{-n, A} + {\Delta}_{-n, B} \bigr]\,,
    \label{aux8}
\end{align}
which shows that the asymptotic behavior of ${\Delta}_{-n}$
will be determined by that of the
functions inside the square brackets multiplied by $k^n$.

In the limit $k \rightarrow 0$, for even $n = 2l$, the
leading order term of the contribution from ${\Delta}_{-2l, A}$ is
\begin{align}
    - \half
    \frac{i^{2l} k^{2l} \tau_\ast^{2l}}{(2l)!} &\,
    {\Delta}_{-2l, A}  =
    \nonumber \\ \frac{(-1)^l 
    k^{2l}\tau_\ast^{2l}}{(2l)!}
    \Bigl\{ &\cos k \tau \Bigl[\Ci (k \bar{\tau}) - \Ci (k \tini)
    \Bigr]  \nonumber \\
    &\, \hspace{-2mm} + \sin k \tau \Bigl[\Si (k \bar{\tau}) - \Si (k\tini)
    \Bigr] \biggr\}
    \nonumber \\ 
     \underset{k \to 0}{=} &  \frac{(-1)^{l}  (k \tau_\ast)^{2l}}{(2l)!} \Bigl[ \ln \bigl(\bar{\tau}/\tau_\ast \bigr)
    + {\cal O}[(k\tau_\ast)^2] \Bigr] \,,
    \label{limit_A_k0}
\end{align}
where the functions $\Si(k x)$ and $\Ci(kx)$ in the limit $kx \ll 1$
are given in \Eqs{lim_Ci}{lim_Si}.
On the other hand, in the limit $k \rightarrow 0$, the leading
order term of ${\Delta}_{-2l, B}$ is given by [see \Eq{G_n_B}],
\begin{align}
    - \half &\, \frac{i^{2l} k^{2l}\tau_\ast^{2l}}{(2l)!}
    {\Delta}_{-2l, B}   \nonumber \\
    & \underset{k \to 0}{=} \frac{(-1)^{l + 1} k^{2l} \tau_\ast^{2l}}{(2l)!}
    \Bigl\{
    A_{2l} (\bar{\tau}) - A_{2l} (\tini) \nonumber \\
    & \hspace{3.5cm} + {\cal O} [(k \tau_\ast)^{- 2l + 1}] \Bigr\} \nonumber
     \\ &\, \hspace{2mm} =   \frac{1}{2 l} 
      \biggl[1
     -\bigl(\bar{\tau}/\tau_\ast \bigr)^{-2l} \biggr] + {\cal O} (k\tau_\ast).
     \label{limit_k0}
\end{align}
This shows that for even $n$, the value of ${\Delta}_{-n}$, which
is the combination of the terms in \Eqs{limit_A_k0}{limit_k0}, is
dominated by the latter term in the limit
$k \rightarrow 0$,
as long as $l > 0$, and its squared value is
\begin{align}
    \lim_{k \rightarrow 0} &\, {\Delta}_{-2l}^2 ( \tau, k; \tau_\ast, \tau_{\rm fin}) =
    \nonumber \\ &\,
     \frac{1}{(2 l)^2} 
      \biggl[1
     -\bigl(\bar{\tau}/\tau_\ast \bigr)^{-2l} \biggr]^2\,.
\end{align}
At large ${\bar \tau}/\tau_\ast \gg 1$, given that $l>0$, we find the following asymptotic value
\begin{equation}
    \lim_{\substack{k \rightarrow 0 \\ \bar{\tau}/{\tau_\ast} \to \infty}}
    {\Delta}_{-2l}^2 (\tau, k; \tau_\ast, \tau_{\rm fin}) = \frac{1}{(2l)^2} \,.
\end{equation}

For odd $n = 2l - 1$ with $l > 0$,
the leading order term in $\Delta_{-2l+1, A}$ is, in the limit $k \rightarrow 0$,
\begin{align}
    - &\, \half  \frac{i^{2l - 1} k^{2l - 1} \tau_\ast^{2l-1}}
    {(2l - 1)!}
    \, {\Delta}_{-2l + 1, A} =  \nonumber \\
    &\, 
    \frac{(-1)^{l} (k\tau_\ast)^{2l - 1}}{(2l - 1)!}
    \biggl\{ \cos k \tau \Bigl[\Si(k\bar{\tau}) - \Si (k \tini)\Bigr] \nonumber \\
    &\, \hspace{26mm} 
    - \sin k \tau \Bigl[\Ci (k\bar{\tau}) - \Ci (k\tini)\Bigr] \biggr\} 
    \nonumber \\
    &\,  \underset{k \to 0}{=} \frac{(-1)^{l} (k\tau_\ast)^{2l}}
    {(2l - 1)!}
    \Bigl[\frac{\bar\tau}{\tau_\ast} -1 - \frac{\tau}{\tau_\ast} \ln \bigl(\bar{\tau}/\tau_\ast \bigr)
    +{\cal O}(k\tau_\ast) \Bigr].
\end{align}
The leading-order term in ${\Delta}_{-n, B}$ for $n=2l-1$ with $l > 0$
is, as before, proportional to $A_n$
\begin{align}
    &\, - \half \frac{i^{2l - 1} (k \tau_\ast)^{2l - 1}}
    {(2l - 1)!}
    \, {\Delta}_{-2l + 1, B} \underset{k \to 0}{=}  \nonumber \\
    &\, \hspace{-3mm}
    \frac{1}{2l - 1}  \biggl[1
    -\bigl(\bar{\tau}/\tau_\ast \bigr)^{-2l + 1}\biggr] + {\cal O} (k\tau_\ast),
    \label{lim_G2n}
\end{align}
which, as previously, dominates over the ${\Delta}_{-n, A}$ contribution, which is of order $k^{2l}$ for all 
$l > 0$.

Hence, for any $m=-n$, we find that the value of ${\Delta}_{m}$, in the limit $k \rightarrow 0$, is
\begin{equation}
\label{k_to_zero_gm}
    \lim_{k\rightarrow 0} {\Delta}_{-n}^2 (\tau, k; \tau_\ast, \tau_{\rm fin})
    = \frac{1}{n^2} \biggl[1-\bigl( \bar{\tau}/\tau_\ast \bigr) ^{-n} \biggr]^2 \,,
\end{equation}
which, at large $\bar{\tau}/\tau_\ast \gg 1$, yields the following asymptotic value
\begin{equation}
    \lim_{\substack{k\rightarrow 0 \\ \bar{\tau}/\tau_\ast \to \infty }}
    {\Delta}_{-n}^2 (\tau, k; \tau_\ast, \tau_{\rm fin}) = \frac{1}{n^2} \,.
    \label{GG2mn}
\end{equation}

For positive integers $m = n$ we find that, in the limit $k \rightarrow 0$, the functions $\Delta_n$ for $n=1,2$ take the values [see \Eqs{G1_app}{G2_app}]
\begin{align}
    \lim_{k \rightarrow 0} \Delta_1^2 (\tau, k; \tau_\ast, \tau_{\rm fin}) = &\, (\delta \bar\tau/\tau_\ast)^2\,,
    \label{G1_k0_gen} \\
    \lim_{k \rightarrow 0} \Delta_2^2 (\tau, k; \tau_\ast, \tau_{\rm fin}) = &\,
    \Bigl(\frac{\delta \bar{\tau}}{\tau_\ast}\Bigr)^2
    \Bigl[ 1 + \frac{\delta \bar{\tau}}{\tau_\ast}
    + \frac{1}{4} \Bigl(\frac{\delta \bar{\tau}}{\tau_\ast}\Bigr)^2 \Bigr] \,,
    \label{G2_k0_gen} 
\end{align}
where $\delta \bar{\tau} \equiv \bar{\tau} - \tini$. For $\tau < \tau_{\rm fin}$ we have $\bar{\tau} = \tau$ and these expressions reduce to
\begin{align}
    \lim_{k \rightarrow 0} \Delta_1^2 (\tau, k; \tau_\ast, \tau_{\rm fin}) \underset{\tau < \tau_{\rm fin}}{=}  &\, (\delta \tau/\tau_\ast)^2\,,
    \label{G1_k0} \\
    \lim_{k \rightarrow 0} \Delta_2^2 (\tau, k; \tau_\ast, \tau_{\rm fin}) \underset{\tau < \tau_{\rm fin}}{=} &\,
    \fourth \bigl(\delta \tau/\tau_\ast\bigr)^4 + (\delta \tau/\tau_\ast)^3 \nonumber \\
    + &\,  (\delta \tau/\tau_\ast)^2 \,. 
    \label{G2_k0}
\end{align}

\subsubsection*{Asymptotic values of ${\Delta}_m (\tau, k; \tau_\ast, \tau_{\rm fin})$ when
$k \rightarrow \infty$}

We now study the $k \rightarrow \infty$ limit.
To find the asymptotic form for $k \to \infty$ of $\Delta_{m}$ we can start directly from its definition in \Eq{def_delta_m} and integrate by parts
\begin{align}
    {\Delta}_{m} &(\tau, k; \tau_\ast, \tau_{\rm fin}) = \tau_\ast^{-m} \int_{\tau_*}^{\bar \tau}    \tau_1^{m - 1} \cos k \bigl(\tau - \tau_1 \bigr) \dd \tau_1 \nonumber \\
    = & -\frac{\tau_\ast^{-m}}{k} \bigl[ \tau_1^{m-1} \sin k \bigl(\tau - \tau_1 \bigr) \bigr]^{\bar{\tau}}_{\tau_\ast} \nonumber \\
    + & \frac{(m-1)\tau_\ast^{-m}}{k} \int_{\tau_\ast}^{\bar \tau} \tau_1^{m-2} \sin k \bigl(\tau - \tau_1 \bigr) \dd \tau_1 \nonumber \\
    \underset{k \to \infty}{\simeq } & \,  \frac{1}{k \tau_\ast} \Bigl[ \sin k \bigl(\tau - \tau_\ast \bigr) - \Bigl(\frac{\bar{\tau}}{\tau_\ast}\Bigr)^{m-1} \sin k \bigl(\tau - \bar{\tau} \bigr) \Bigr]\,,
\label{k_to_inf_delta_m}
\end{align}
where the boundary term from the integration by parts goes as $k^{-1}$, while the remaining integral can be shown to be, through an additional integration by parts, of ${\cal O} (k ^{-2})$.

We then see that, for $m=0$, \Eq{k_to_inf_delta_m} recovers the result for the function $\Delta$ in the CiT model [see \Sec{dynamical_G} and in particular \Eq{delta_cit_inf_limits}]. For $m \neq 0$, instead, the envelope of the function for $\tau> \tau_{\rm fin}$ is modified and bounded by
\begin{align}
    \bigl[\Delta_m^2\bigr]_{\rm env}  \underset{k \to \infty}{ \leq } \frac{1}{k^2 \tau_\ast^2} \Bigl[ 1 + \Bigl(\frac{\tau_{\rm fin}}{\tau_\ast}\Bigr)^{m-1} \Bigr]^2\,,
\end{align}
which reduces to \Eqs{upper_bound_cit}{F_factor} for $m=0$. However, the additional contribution to the envelope for $\tau > \tau_{\rm fin}$ is an artifact of the abrupt switching off of the sourcing at the time  $\tau_{\rm fin}$. Hence,  if we neglect this factor, we see that the large wave numbers are not affected by the decay of the source and we recover, in the $k \to \infty$ limit, the same behavior as in the CiT model.

\section{Fast evolution of the spectral peak in the coherent-decay model}
\label{fast_q}

\begin{figure*}[t!]
    \centering
    \includegraphics[width=\columnwidth]{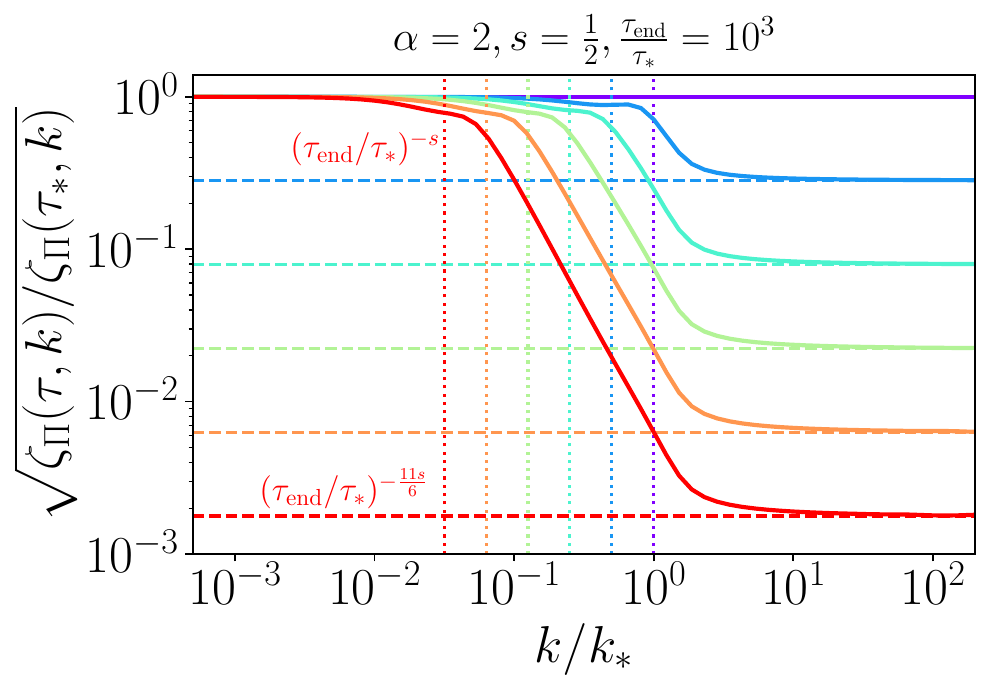}
    \includegraphics[width=\columnwidth]{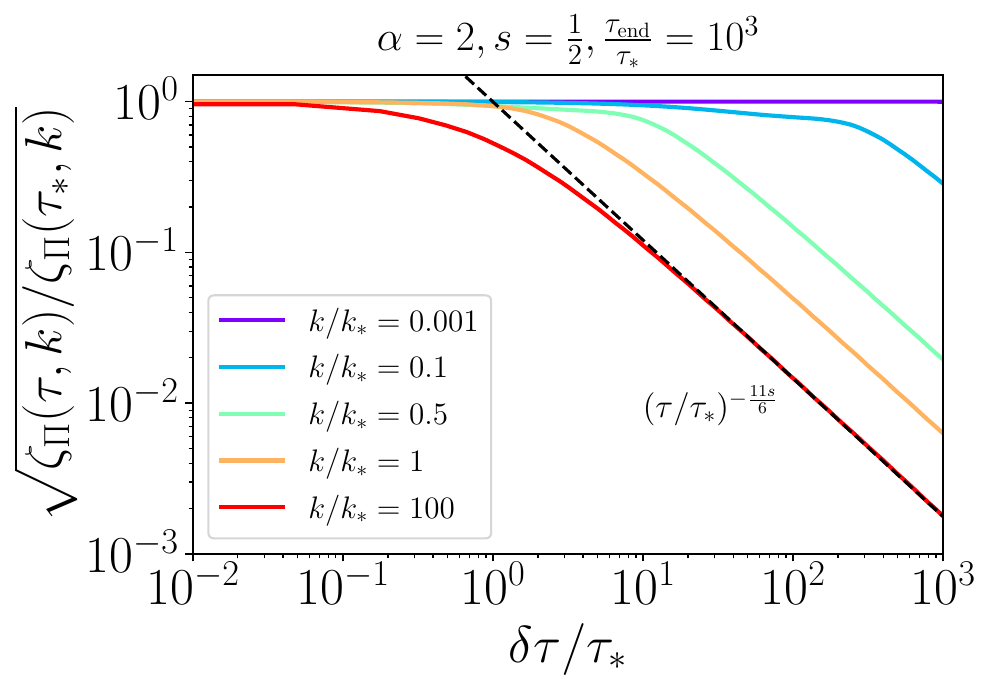}\\
    \includegraphics[width=\columnwidth]{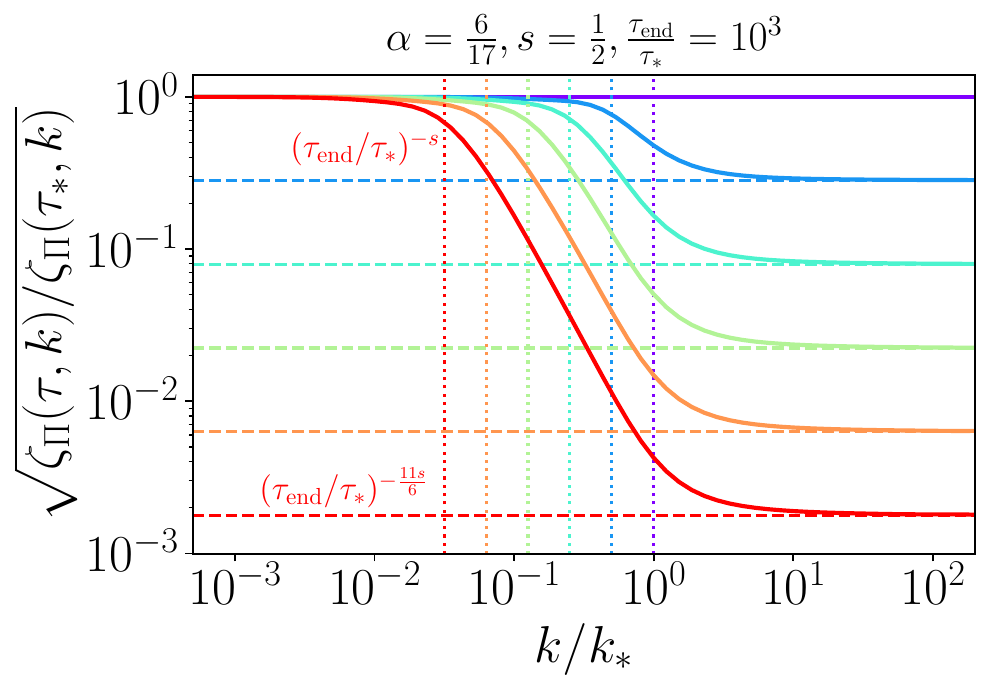}
    \includegraphics[width=\columnwidth]{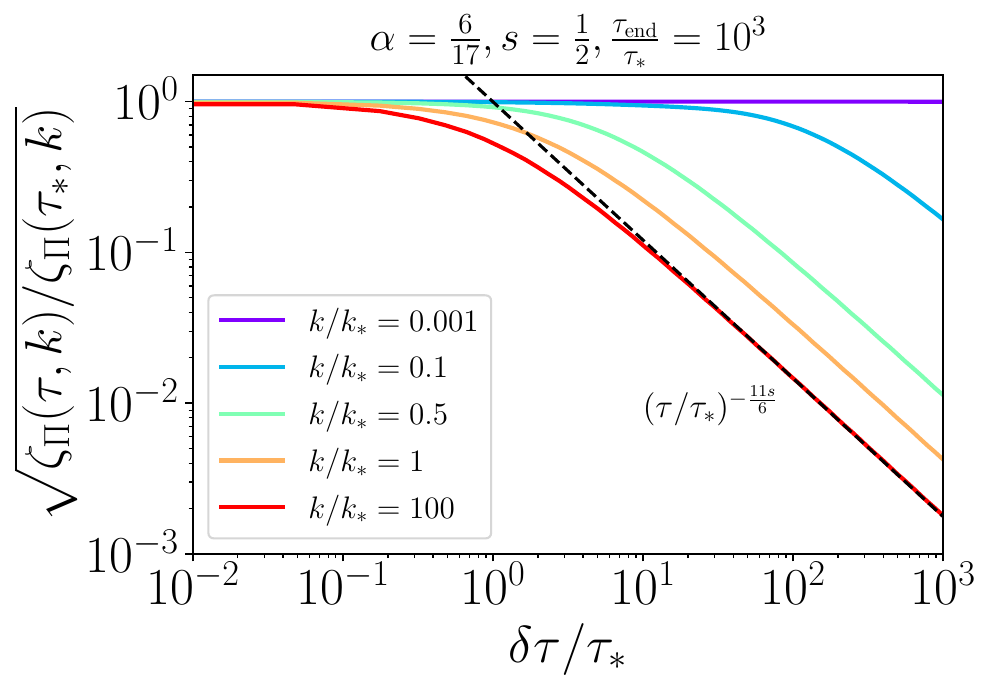}
    \caption{
    Time evolution of the normalized stress
    spectrum $\sqrt{\zeta_\Pi^{\rm rat} ( \tau, k)}=\sqrt{\zeta_\Pi(\tau, k)/\zeta_\Pi(\tau_\ast, k)}$ [see \Eqs{eq:EPi_decay}{p_pi_rat}] for $\alpha = 2$ (upper 
    panels) and $\alpha = 6/17$ (lower panels), when the wave
    number of the peak evolves as $\kf (\tau) = \kf (\tau/\tau_\ast)^{-s}$ (cf.~\Eq{eq:decay} with $\delta \tau_{\rm dc} = \tau_\ast$),
    with $s = \half$, and for a purely vortical non-helical field with $a=5$ and $b=2/3$. 
    Left panels show the stress spectra, normalized to their values at initial time $\tini$, at logarithmically spaced times between $\tau_\ast$ (purple) and $\tau_{\rm end} = 10^3 \tau_\ast$ (red).
    For each different time, the horizontal dashed line indicates the amplitude of $\sqrt{\zeta_\Pi^{\rm rat}}$ at large wave numbers, going as $(\tau/\tau_\ast)^{-\frac{11s}{6}}$, while the vertical dotted line shows the position of the spectral
    peak, going as $(\tau/\tau_\ast)^{-s}$. The right panels show the time evolution at specific
    wave numbers. We see that at large times all wave numbers eventually evolve as $(\tau/\tau_\ast)^{-\frac{11s}{6}}$.}
    \label{p_pi_ratios}
\end{figure*}

Let us now study the time evolution
of $\zeta_{\Pi}^{\rm rat}(\tau, k)$, i.e., the function in \Eq{p_pi_rat} with $k_\ast (\tau)$ evolving as in \Eq{eq:decay}. In the following, we will restrict ourselves, for simplicity, to the case $\delta \tau_{\rm dc} = \tau_\ast$, as in \App{G_m_coherent}.
\FFig{p_pi_ratios} shows the time evolution of this function
for $\alpha = 2$ and $\alpha = 6/17$, while the broken power law parameters $a,b$ of $\zeta(K)$ are always chosen to be $5$ and $2/3$ and we restrict to the case of a fully vortical non-helical vector field.
We observe that, at time $\tau$, the wave numbers smaller than
$\kf (\tau/\tau_\ast)^{-s}$ are still
not significantly affected by the time evolution of the peak and $\zeta_\Pi^{\rm rat}$ has an approximately constant value.
In the large wave numbers region (i.e., $k \gg \kf$), the asymptotic branch of $\zeta_\Pi^{\rm rat}$, proportional to $(k \tau^s \tau_\ast^{-s}/\kf)^{-{11\over3}}$, decays indeed as
$(\tau/\tau_\ast)^{-{11s\over3}}$ (c.f.~the black dashed line in the right panels of
\Fig{p_pi_ratios}).
The time evolution of $\zeta_\Pi^{\rm rat}$ at the intermediate wave numbers
$k/\kf \in (\tau^{-s} \tau_\ast^s, 10)$ smoothly transitions from constant at
$\tau \lesssim \tau_\ast(k/\kf)^{-{1\over s}}$ to $(\tau/\tau_\ast)^{-{11s\over3}}$ as $k/k_\ast$ increases. 

We find that, in the case with $ p = 1$ and $s = \half$, 
shown in \Fig{Gsq_kt_p1_q05}, the inclusion of the time evolution of  $\zeta_\Pi^{\rm rat}$ does not lead to a significant difference in
the resulting $\Delta^2_{ps}$ [cf.~\Eq{G_pq}]. Note that, in fact, the evolution of $\zeta_\Pi^{\rm rat}$ is mostly relevant around the peak
wave number $k_\ast(\tau)$, and that the time evolutions of $\kf(\tau)$
and ${E_\ast}(\tau)$ [cf.~\Eqq{eq:decayall}]
are still being taken into account in the amplitude of $E_\Pi$ via the $\tau_1^{m}$ factor in $\Delta_{ps}$. Moreover, comparing the wave number range $k > k_* \bigl( \tau/\tau_\ast)^{-s}$ affected by the decay, to the range in which the GW is already saturated $k > 1/\tau$, we find that the evolution of $\sqrt{\zeta_{\Pi}^{\rm rat}}$ can be neglected when
\begin{align}
    & k_* \Bigl( \frac{\tau}{\tau_\ast} \Bigr)^{-s} > \frac{1}{\tau}  \nonumber \\
     & \rightarrow s < 1 +\frac{ \ln \bigl( k_\ast \, \tau_\ast \bigr)}{\ln \bigl( \tau/\tau_\ast \bigr)} \ \  \ {\rm for} \  \tau> \tau_\ast \,.
\label{range_of_good_s_values}
\end{align}
Hence, for all subhorizon $k_\ast > {\cal H}_\ast$, the evolution of $\zeta_\Pi^{\rm rat}$ can be neglected for all values $s < 1$, i.e., when the evolution of the peak scale of $\zeta_\Pi^{\rm rat}$ is slower than the GW saturation.

\end{document}